\documentclass[aps,rmp,reprint,amsmath,amssymb,longbibliography]{revtex4-1}

\usepackage{hyperref}
\usepackage{graphicx}
\usepackage{xcolor}
\usepackage{bbm}

\usepackage{calrsfs}
\DeclareMathAlphabet{\pazocal}{OMS}{zplm}{m}{n}	
\usepackage{dutchcal}

\newcommand{\imagefileprefix}{pezze_fig}	
\newcommand{\imagefileextension}{}

\definecolor{mycitecolor}{rgb}{0,0.65,0}	
\definecolor{mylinkcolor}{rgb}{1,0,0}		
\definecolor{myurlcolor}{rgb}{0,0,1}		

\hypersetup{
  colorlinks,
  linkcolor={mylinkcolor},
  citecolor={mycitecolor},
  urlcolor={myurlcolor}
}

\newcommand{\vect}[1]{\boldsymbol{#1}}
\newcommand{\op}[1]{\hat{#1}}
\newcommand{\vectop}[1]{\vect{\op{#1}}}

\newcommand{\be}{\begin{equation}}
\newcommand{\ee}{\end{equation}}
\newcommand{\bra}[1]{{\ensuremath{\langle#1\rvert}}}
\newcommand{\ket}[1]{{\ensuremath{\lvert#1\rangle}}}
\newcommand{\mean}[1]{{\ensuremath{\langle#1\rangle}}}
\newcommand{\abs}[1]{{\ensuremath{\lvert#1\rvert}}}
\newcommand{\norm}[1]{{\ensuremath{\lVert#1\rVert}}}
\newcommand{\scp}[2]{{\langle#1|#2\rangle}}
\newcommand{\me}[3]{{\langle#1\lvert#2\rvert#3\rangle}}

\newcommand{\ps}{\theta}  
\newcommand{\rhops}{\op{\rho}_{\ps}}     
\newcommand{\J}{\op{J}}

\newcommand{\Sp}{\op{S}}
\newcommand{\lab}{l}
\newcommand{\di}{m}
\newcommand{\res}{\mu}
\newcommand{\est}{\Theta}
\newcommand{\estmom}{\Theta\si{mom}}
\newcommand{\Fish}{F}
\newcommand{\m}{\nu}
\newcommand{\POVM}{\op{E}(\res)}
\newcommand{\hell}{d\si{H}}
\newcommand{\bures}{d\si{B}}
\newcommand{\f}{F}
\newcommand{\mf}{m_F}
\newcommand{\mass}{M}
\newcommand{\xiN}{\xi\si{N}}
\newcommand{\xiS}{\xi\si{S}}
\newcommand{\xiR}{\xi\si{R}}
\newcommand{\xiD}{\xi\si{D}}
\newcommand{\C}{\mathcal{C}}

\newcommand{\dd}{\text{d}}
\newcommand{\ii}{\text{i}}
\newcommand{\dagg}{^{\dagger}}
\newcommand{\1}{\ensuremath{\mathbbm{1}}}
\newcommand{\si}[1]{_{\text{#1}}}
\newcommand{\se}[1]{^{\text{#1}}}
\newcommand{\ie}{\emph{i.e.}}
\newcommand{\eg}{\emph{e.g.}}

\newcommand{\as}{\op{\mathcal{a}}}
\newcommand{\sa}{\op{\mathcal{s}}}

\DeclareMathOperator{\Tr}{Tr}
\DeclareMathOperator{\argmax}{argmax}
\DeclareMathOperator{\Cov}{Cov}
\DeclareMathOperator{\Var}{Var}
\DeclareMathOperator{\artanh}{artanh}

\begin{document}

\title{Quantum metrology with nonclassical states of atomic ensembles} 
\date{\today}

\author{Luca Pezz\`e}
\affiliation{QSTAR, INO-CNR and LENS, Largo Enrico Fermi 2, 50125 Firenze, Italy}

\author{Augusto Smerzi}
\affiliation{QSTAR, INO-CNR and LENS, Largo Enrico Fermi 2, 50125 Firenze, Italy}

\author{Markus K.\ Oberthaler}
\affiliation{Kirchhoff-Institut f\"ur Physik, Universit\"at Heidelberg, Im Neuenheimer Feld 227, 69120 Heidelberg, Germany}

\author{Roman Schmied}
\author{Philipp Treutlein}
\affiliation{Department of Physics, University of Basel, Klingelbergstrasse 82, 4056 Basel, Switzerland}


\begin{abstract}
Quantum technologies exploit entanglement to revolutionize computing, measurements, and communications. 
This has stimulated the research in different areas of physics to engineer and manipulate fragile many-particle entangled states. 
Progress has been particularly rapid for atoms. Thanks to the large and tunable nonlinearities and the well developed 
techniques for trapping, controlling and counting, many groundbreaking experiments have demonstrated the generation 
of entangled states of trapped ions, cold and ultracold gases of neutral atoms. 
Moreover, atoms can couple strongly to external forces and light fields, which makes them ideal for ultra-precise sensing and time keeping. 
All these factors call for generating non-classical atomic states designed for phase estimation in atomic clocks and atom interferometers, 
exploiting many-body entanglement to increase the sensitivity of precision measurements. 
The goal of this article is to review and illustrate the theory and the experiments with 
atomic ensembles that have demonstrated many-particle entanglement and quantum-enhanced metrology. 
\end{abstract}

\maketitle
\tableofcontents{}

\section{Introduction}
\label{Sec.Introduction}

\begin{figure}[b!]
\begin{center}
\includegraphics[width=\columnwidth]{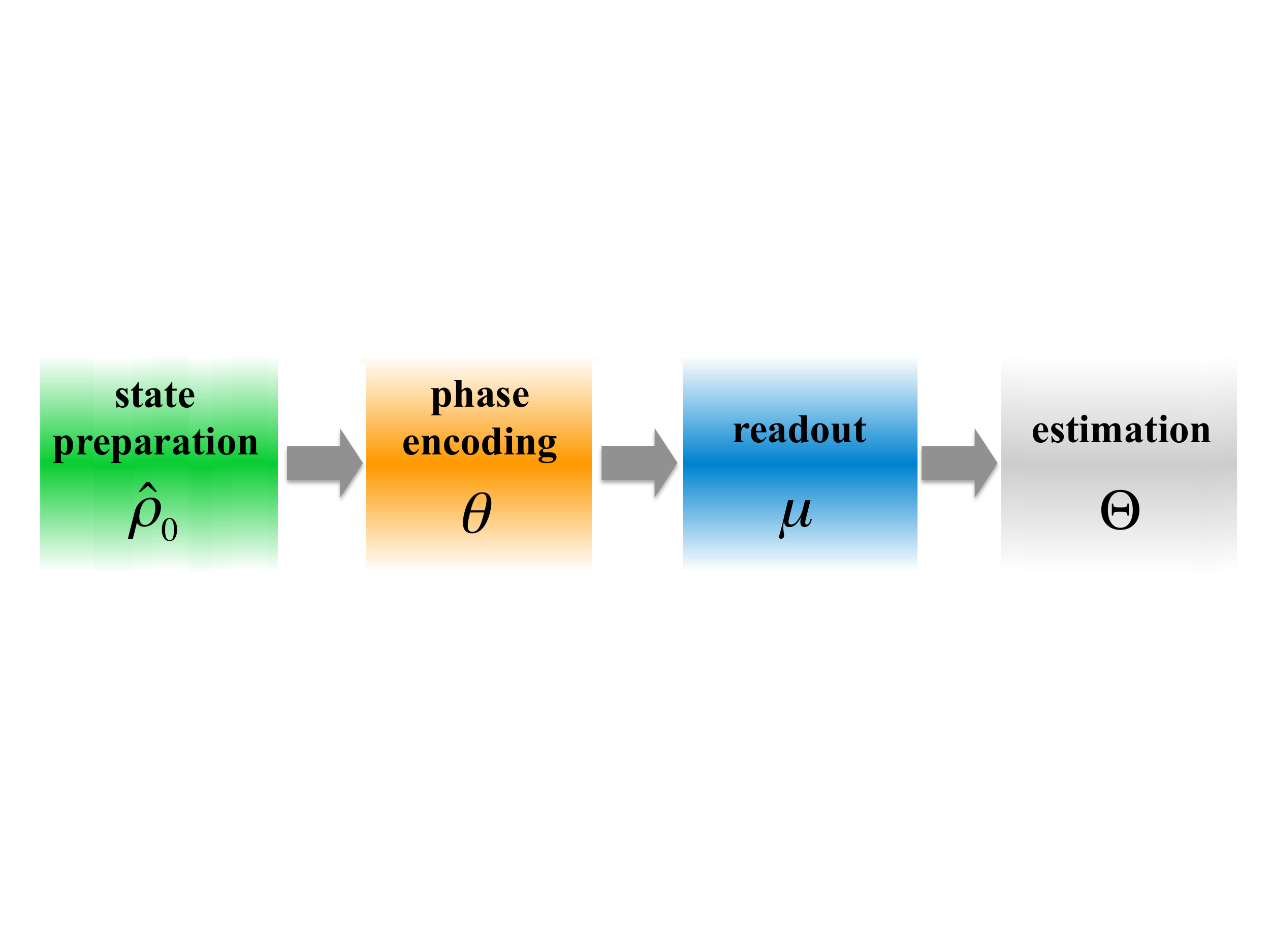}
\end{center}
\caption{{\bf Building blocks of phase estimation.} 
Phase estimation follows four steps: 
$i$) preparation of the probe state $\op{\rho}_0$;
$ii$) encoding of a phase shift $\ps$ that depends on the physical quantity of interest;
$iii$) readout, $\res$ indicating a generic measurement result; 
$iv$) deriving a phase estimate $\est(\res)$ from the measurement results.
The uncertainty $\Delta \ps$ of the estimation depends crucially on all of these operations.}
\label{Fig1}
\end{figure} 

\begin{figure*}[t!]
\begin{center}
\includegraphics[width=0.9\textwidth]{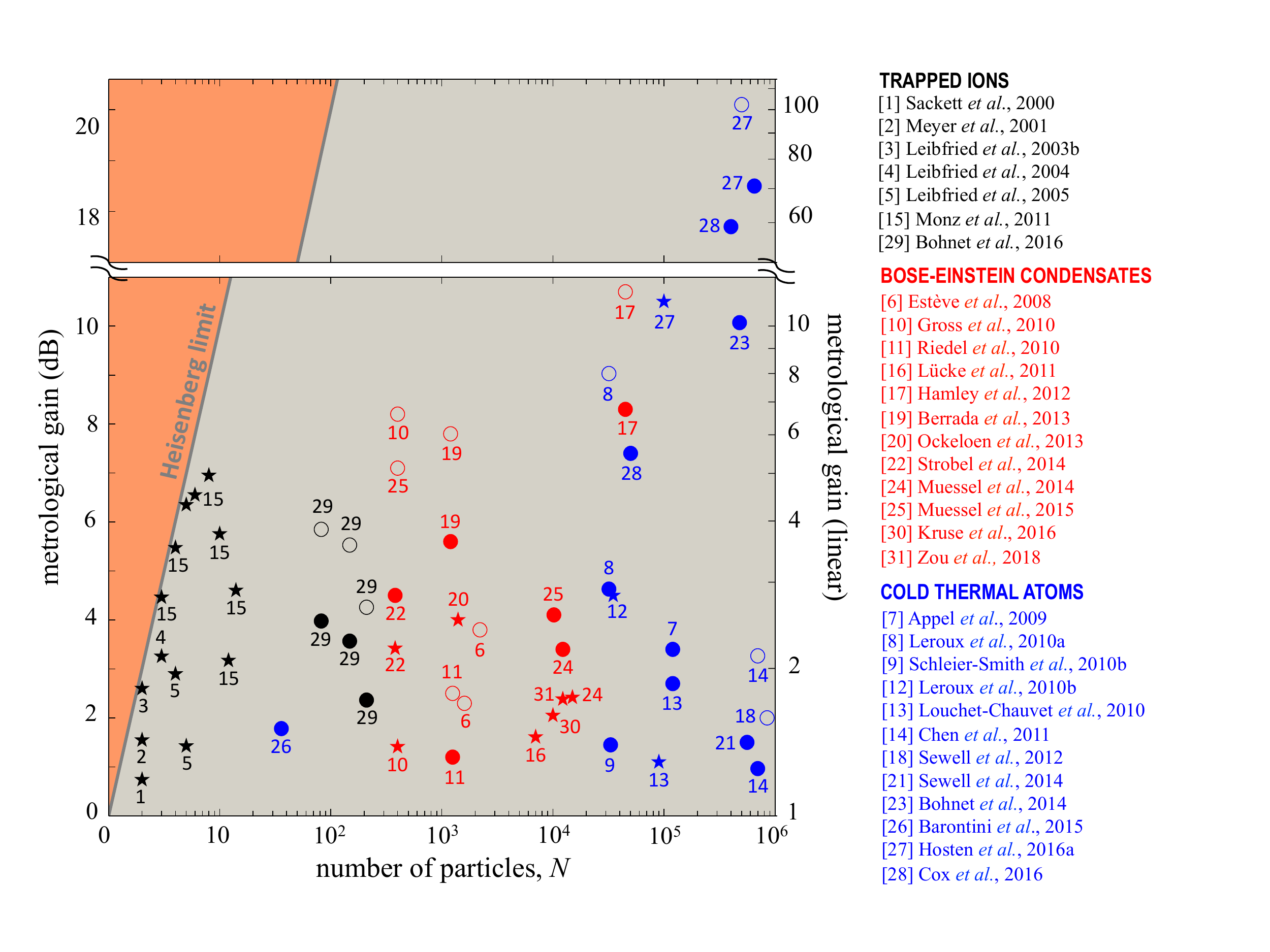}
\end{center}
\caption{{\bf Summary of experimental achievements.}
Gain of phase sensitivity over the standard quantum limit $\Delta \ps\si{SQL}=1/\sqrt{N}$ achieved 
experimentally with trapped ions (black symbols), Bose-Einstein condensates (red) and cold thermal ensembles (blue). 
The gain is shown on logarithmic [left, dB, $10\log_{10} (\Delta \ps\si{SQL} / \Delta \ps)^2$] and linear 
[right, $(\Delta \ps\si{SQL} / \Delta \ps)^2$] scale. The solid thick line is the Heisenberg limit $\Delta \ps\si{HL}=1/N$. 
Stars refer to directly measured phase sensitivity gains, performing a full phase estimation experiment. Circles are expected gains based on a characterization of the quantum state, \eg, calculated as $\Delta \ps = \xiR/\sqrt{N}$, where $\xiR$ is the spin-squeezing parameter, or 
as $\Delta \ps=1/\sqrt{\Fish\si{Q}}$, where $\Fish\si{Q}$ is the quantum Fisher information, see Sec.~\ref{Sec.Fundamentals}. Filled (open) circles indicate results obtained without (with) subtraction of technical and/or imaging noise. Every symbol is accompanied by a number (in chronological order) corresponding to the reference reported in the side table. Here $N$ is the total number of particles or, in presence of fluctuations, the mean total.}
\label{Fig_summary}
\end{figure*} 

The precise measurement of physical quantities such as the strength of a field, a force, or time, plays a crucial role in the advancement of physics.
Precision measurements are very often obtained by mapping the physical quantity to a phase shift that can be determined using interferometric techniques.
Phase estimation is thus a unifying framework for precision measurements. 
It follows the general scheme outlined in Fig.~\ref{Fig1}: a probe state $\op{\rho}_0$ of $N$ particles is prepared, acquires a phase shift $\ps$, and is finally detected. 
From the measurement outcome $\res$ an estimate $\est(\res)$ of the phase shift is obtained. 
This conceptually simple scheme is common to all interferometric sensors: from gravitational wave detectors to atomic clocks, gyroscopes, and gravimeters, just to name a few.
The goal is to estimate $\ps$ with the smallest possible uncertainty $\Delta \ps$ given finite resources such as time and number of particles. The noise that determines $\Delta \ps$ can be of a technical (classical) or fundamental (quantum) nature \cite{HelstromBOOK1976, BraunsteinPRL1994, HolevoBOOK1982}. Current two-mode atomic sensors are limited by the so-called standard quantum limit, $\Delta\ps\si{SQL} = 1/\sqrt{N}$, inherent in probes using a finite number of uncorrelated \cite{GiovannettiPRL2006}
or classically-correlated  \cite{PezzePRL2009} particles. 
Yet, the standard quantum limit is not fundamental \cite{CavesPRD1981, YurkePRA1986, BondurantPRD1984}. 
Quantum-enhanced metrology studies how to exploit quantum resources, such as squeezing and entanglement, 
to overcome this classical bound \cite{GiovannettiNATPHOT2011, GiovannettiSCIENCE2004, PezzeBOOK2014, TothJPA2014}. 
Research on quantum metrology with atomic ensembles also sheds new light on fundamental 
questions about many-particle entanglement \cite{AmicoRMP2008, GuhnePHYSREP2009, HorodeckiRMP2009} and related concepts, 
such as Einstein-Podolsky-Rosen correlations \cite{ReidRMP2009} and Bell nonlocality \cite{BrunnerRMP2014}.

Since the systems of interest for quantum metrology often contain thousands or even millions of particles, 
it is generally not possible to address, detect, and manipulate all particles individually. 
Moreover, the finite number of measurements limits the possibility to fully reconstruct the generated quantum states. 
These limitations call for conceptually new approaches to the characterization of entanglement that rely on a finite number of coarse-grained measurements. 
In fact, many schemes for quantum metrology require only collective manipulations and measurements on the entire atomic ensemble. 
Still, the results of such measurements allow one to draw many interesting conclusions about the underlying quantum correlations between the particles.

\subsection{Entanglement and interferometric sensitivity enhancement: exemplary cases}

We illustrate here how the standard quantum limit arises in interferometry and how it can be overcome using entanglement. Consider a Ramsey interferometer sequence \cite{RamseyBOOK1990} between two quantum states $\ket{a}$ and $\ket{b}$, as used e.g.\ in an atomic clock, see Sec.~\ref{Sec.Fundamentals}.
Let us discuss first the case of a single atom initially prepared in the probe state $\ket{a}$. 
The atom is transformed to $(\ket{a} + \ket{b})/\sqrt{2}$ by a resonant $\pi/2$-pulse corresponding to the first beam splitter of the interferometer. 
During the subsequent interrogation time, $\ket{a}$ and $\ket{b}$ acquire a relative phase $\theta$, such that the state evolves to 
$(e^{-\ii \theta/2} \ket{a} + e^{\ii \theta/2} \ket{b})/\sqrt{2}$. The phase $\theta$ encodes the quantity to be measured, such as frequency in the case of an atomic clock or the strength of an external field in the case of an atom interferometer. To convert it into an observable population difference, a second resonant $\pi/2$-pulse (the second beam splitter) is applied so that the final state is
$\cos \tfrac{\theta}{2} \ket{a} + \sin \tfrac{\theta}{2} \ket{b}$.
The phase $\theta$ can now be estimated, for instance, by measuring the population difference $\hat{M} = \ket{a} \bra{a} - \ket{b} \bra{b}$ between the two states. In this simple example the interferometer signal is the expectation value $\mean{\hat{M}} = \cos \theta$ while the noise at the output is quantified by the variance $\Delta^2 \hat{M} = \sin^2 \theta$. 

If we now repeat the same interferometric procedure with $N$ uncorrelated atoms, the signal $\mean{\hat{M}}$,
where $\hat{M} = \sum_{i=1}^N \ket{a}_i \bra{a} - \ket{b}_i \bra{b}$ is now the population difference of $N$ atoms,
will be simply given by $N$ times that of a single atom.
Because the atoms are uncorrelated, the variance $\Delta^2 \hat{M}$ will also be multiplied by a factor $N$ and correspondingly the standard deviation $\Delta \hat{M}$ will increase by $\sqrt{N}$. 
Overall, this results in a phase uncertainty of $\Delta \theta = \Delta \hat{M}/|d \mean{\hat{M}}/d\theta| = 1/\sqrt{N}$. This is precisely the standard quantum limit $\Delta\ps\si{SQL}$, which arises from the binomial statistics of the $N$ uncorrelated particles.

Overcoming this sensitivity limit requires entanglement between the particles. 
One possibility, suggested by the above formula, is to engineer quantum correlations that lead to sub-binomial statistics $\Delta^2 M < N$ 
at the point of maximum slope of the signal, while keeping that slope ({\it i.e.}\ the interferometer contrast) of the order of $N$. In this way, a phase uncertainty of $\Delta \theta < 1/\sqrt{N}$ can be achieved.
States that satisfy these conditions are called spin-squeezed\footnote{Indeed we will later see that $d \mean{\hat{M}}/d\theta$ and $\Delta^2 \hat{M}$ can be written in terms of mean and variance of collective spin operators, see Sec.~\ref{Sec.Fundamentals.spin}} \cite{WinelandPRA1992, WinelandPRA1994} 
and are an important class of useful states in quantum metrology.
Spin-squeezed states can be created by making the atoms interact with each 
others for a relatively short time~\cite{KitagawaPRA1993} generating entanglement between them~\cite{SorensenPRL2001, SorensenNATURE2001}.
For instance, in the case of two atoms, such interactions (known as two-axis counter-twisting) lead to the state 
$\ket{\psi(\alpha)}=\cos(\alpha)\times\big(\tfrac{\ket{a}+\ket{b}}{\sqrt{2}}\big)^{\otimes 2}+\sin(\alpha)\times\big(\tfrac{\ket{a}-\ket{b}}{\sqrt{2}}\big)^{\otimes 2}$ that 
is entangled and spin-squeezed, reaching $\Delta \theta =\sqrt{\frac{1-\sin(2\alpha)}{\cos^2(2\alpha)}} < \Delta \theta_{\rm SQL}$ for  $0<\alpha<\pi / 2$. 

Spin-squeezed states are only a small subset of the full class of entangled states that are useful for quantum-enhanced metrology.
A prominent example is the Greenberger-Horne-Zeilinger (GHZ) state $\ket{\text{GHZ}} = \tfrac{\ket{a}^{\otimes N}+\ket{b}^{\otimes N}}{\sqrt{2}}$
[also indicated as NOON state when considering bosonic particles], 
which is not spin-squeezed but can nevertheless provide phase sensitivities beyond 
the standard quantum limit~\cite{BollingerPRA1996, LeeJMO2002}. 

\subsection{Entanglement useful for quantum-enhanced metrology}
\label{sec:useful}
In the context of phase estimation, the idea that quantum correlations are necessary to overcome the 
standard quantum limit emerged already in pioneering works \cite{KitagawaPRA1993, YurkePRA1986, WinelandPRA1992}.
In recent years, it has been clarified that only a special class of quantum correlations can be exploited to
estimate an interferometric phase with sensitivity overcoming $\Delta\ps\si{SQL}$. 
This class of entangled states is fully identified by the quantum Fisher information, $\Fish\si{Q}$.
The quantum Fisher information is inversely proportional to the maximum phase sensitivity 
achievable for a given probe state and interferometric transformation---the so-called quantum Cram\'er-Rao bound,
$\Delta \ps\si{QCR}=1/ \sqrt{\Fish\si{Q}}$ \cite{HelstromPLA1967, BraunsteinPRL1994}.
It thus represents the figure of merit for the sensitivity of a generic parameter estimation problem 
involving quantum states and will be largely discussed in this review. 
The condition $\Fish\si{Q} > N$ \cite{PezzePRL2009} is sufficient for entanglement and \emph{necessary and sufficient for the 
entanglement useful for quantum metrology}: it identifies the class of states characterized by $\Delta \ps\si{QCR} < \Delta \ps\si{SQL}$, \ie, those that can be used to  
overcome the standard quantum limit in any two-mode interferometer where the phase shift is generated by a local Hamiltonian.
Spin-squeezed, GHZ and NOON states fulfill the condition $\Fish\si{Q} > N$.
Phase uncertainties down to $\Delta \ps = 1/\sqrt{k N}$ can be obtained with metrologically useful $k$-particle entangled states \cite{HyllusPRA2012,TothPRA2012}. 
In the absence of noise, the ultimate limit is $\Delta\ps\si{HL} = 1/N$, the so-called Heisenberg limit \cite{GiovannettiPRL2006, YurkePRA1986, HollandPRL1993},
which can be reached with metrologically useful genuine $N$-particle entangled states ($k=N$). 

\subsection{Generation of metrologically useful entanglement in atomic ensembles}

A variety of techniques have been used to generate entangled states useful for quantum metrology with atomic ensembles. 
The crucial ingredient is interaction between the particles, for instance
atom-atom collisions in Bose-Einstein condensates, atom-light interactions in cold thermal ensembles 
(including experiments performed with warm vapors in glass cells), 
or combined electrostatic and ion-light interaction in ion chains.    
Figure~\ref{Fig_summary} summarizes the experimental achievements (gain of phase sensitivity relative to the standard quantum limit) 
as a function of the number of particles. 
Stars in Fig.~\ref{Fig_summary} show the measured phase-sensitivity gain obtained 
after a full interferometer sequence using entangled states as input to the atom interferometer.
Filled circles report witnesses of metrologically useful entanglement (\ie, spin squeezing and Fisher information) 
measured on experimentally generated states, representing potential improvement in sensitivity.
Open circles are inferred squeezing, being obtained after subtraction of detection noise. 
The Heisenberg limit has been reached with up to $\sim$10 trapped ions. 
Attaining this ultimate bound with a much larger number of particles 
is beyond current technology as it requires the creation and protection of large amounts of entanglement. 
Nevertheless, metrological gains up to $\sim$100 have been reported with large atomic ensembles \cite{HostenNATURE2016}.
A glance at Fig.~\ref{Fig_summary} reveals how quantum metrology with atomic ensembles is a very active area of research in physics. 
Moreover, the reported results prove that the field is now mature enough to take the step from proofs-of-principle to technological applications. 

\subsection{Outline}

This Review presents modern developments of phase-estimation techniques in atomic systems aided by quantum-mechanical entanglement,
as well as fundamental studies of the associated entangled states.
In Sec.~\ref{Sec.Fundamentals}, we give a theoretical overview of quantum-enhanced metrology.
We first discuss the concepts of spin squeezing and Fisher information considering spin-1/2 particles. 
We then illustrate different atomic systems where quantum-enhanced phase estimation---or, at least, 
the creation of useful entanglement for quantum metrology---has been demonstrated. 
Sections~\ref{Sec.Atom-Atom} and~\ref{Sec.Spin-Mix} review the generation of entangled states in Bose-Einstein condensates.
Section~\ref{Sec.Atom-Light} describes the generation of entangled states of many atoms through the common coupling to an external light field. 
Section~\ref{Sec.Ions} describes metrology with ensembles of trapped ions. 
Finally, Sec.~\ref{Sec.Working-Entanglement} gives an overview 
of the experimentally realized entanglement-enhanced interferometers and the realistic perspective
to increase the sensitivity of state-of-the-art atomic clocks and magnetometers. 
This section also discusses the impact of noise in the different interferometric protocols. 

\section{Fundamentals}
\label{Sec.Fundamentals}

\begin{figure}[t!]
\begin{center}
\includegraphics[width=\columnwidth]{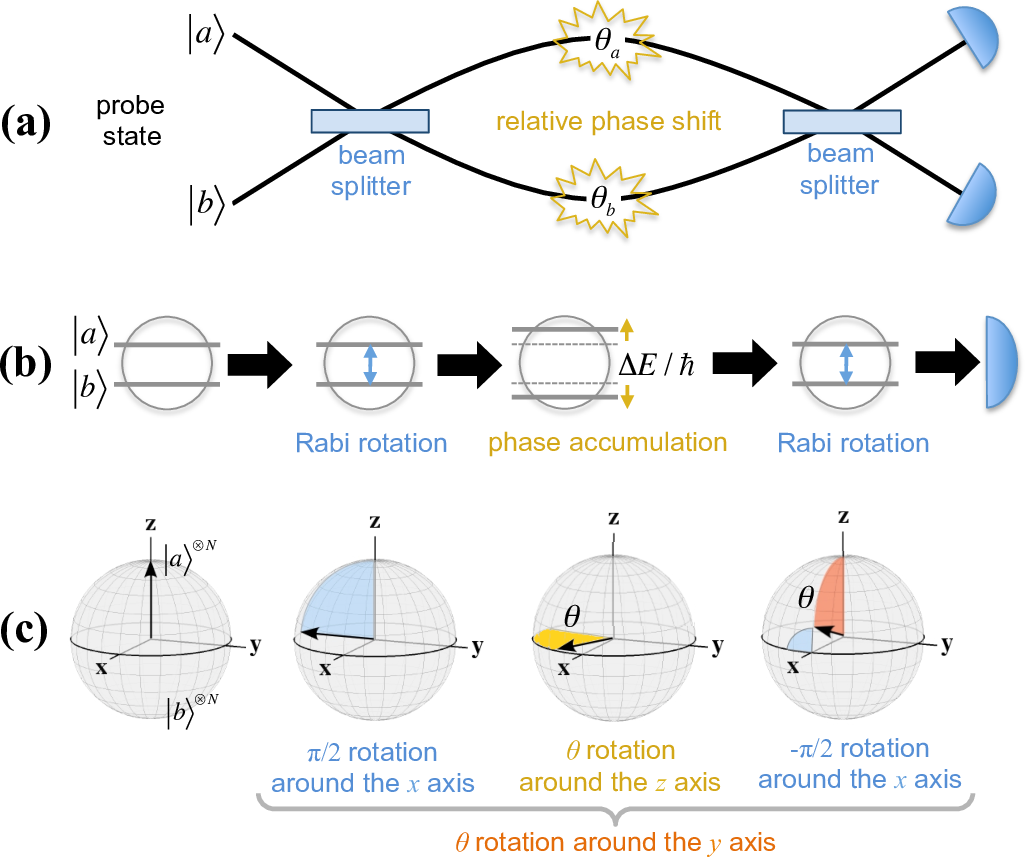}
\end{center}
\caption{{\bf Two-mode interferometers.}
In a Mach-Zehnder interferometer (a) 
two spatial modes $\ket{a}$ and $\ket{b}$ are combined on a balanced beam splitter, followed by a relative phase shift $\ps=\ps_a-\ps_b$ between the two arms, 
and finally recombined on a second balanced beam splitter. 
In a Ramsey interferometer (b) a resonant Rabi rotation creates a balanced superposition between two internal states $\ket{a}$ and $\ket{b}$, 
followed by a relative phase shift given by the energy difference between these states multiplied by the interrogation time, 
$\ps = (\Delta E/\hbar) \times T\si{R}$. Finally a second resonant Rabi rotation recombines the two modes.
(c) Equivalent representation of Mach-Zehnder and Ramsey interferometer operations as rotations of the collective spin on the generalized Bloch sphere.
The initial state here, $\ket{a}^{\otimes N}$, is pointing toward the north pole.
The full sequence is equivalent to the rotation of an angle $\ps$ around the $y$ axis.}
\label{Fig:2A}
\end{figure}

In this Review we consider systems and operations involving $N$ particles and assume that all of their degrees of freedom are restricted to only two modes
(single-particle states) that we identify as $\ket{a}$ and $\ket{b}$.
These can be two hyperfine states of an atom, as in a Ramsey interferometer \cite{RamseyBOOK1990}, 
two energy levels of a trapping potential,
or two spatially-separated arms, as in a Mach-Zehnder interferometer \cite{Zehnder1891, Mach1892}, see Fig.~\ref{Fig:2A}. 
The interferometer operations are collective, acting on all particles in an identical way.  
The idealized formal description of these interferometer models is mathematically equivalent \cite{WinelandPRA1994, LeeJMO2002}:
as discussed in Sec.~\ref{Sec.Fundamentals.spin}, it corresponds to the rotation of a collective spin.  

\subsection{Collective spin systems}
\label{Sec.Fundamentals.spin}

\paragraph{Single spin.} 
By identifying mode $\ket{a}$ with spin-up and mode $\ket{b}$ with spin-down, a (two-mode) atom 
can be described as an effective spin-1/2 particle: a qubit \cite{NielsenBOOK, PeresBOOK}. 
Any pure state of a single qubit can be written as $\ket{\vartheta, \varphi} =  \cos \tfrac{\vartheta}{2} \ket{a} + e^{\ii \varphi} \sin \tfrac{\vartheta}{2} \ket{b}$, 
with $0 \leq \vartheta \leq \pi$ and $0 \leq \varphi < 2\pi$ the polar and azimuthal angle, respectively, in the Bloch sphere.
Pure states satisfy $\mean{\vectop{\sigma}} = \vect{s}$, 
where $\vectop{\sigma} = \{ \op{\sigma}_x, \op{\sigma}_y, \op{\sigma}_z \}$ is the Pauli vector
and $\vect{s}  = \{  \sin \vartheta \cos \varphi,  \sin \vartheta \sin \varphi, \cos \vartheta \}$
is the mean spin direction. 
Mixed qubit states can be expressed as $\op{\rho}=(\1+r\vect{s}\cdot\vectop{\sigma})/2$, and 
have an additional degree of freedom given by the length of the spin vector $0 \leq r \leq 1$, 
such that the effective state vector $r\vect{s}$ lies inside the Bloch sphere.

\paragraph{Many spins.}
To describe an ensemble of $N$ distinguishable qubits, 
we can introduce the collective spin vector $\vectop{J} = \{ \J_x, \J_y, \J_z\}$, where 
\begin{align} \label{Eq.collspin}
	\J_x &= \frac{1}{2}\sum_{\lab=1}^N \op{\sigma}_{x}^{(\lab)}, &
	\J_y &= \frac{1}{2}\sum_{\lab=1}^N \op{\sigma}_{y}^{(\lab)}, &
	\J_z &= \frac{1}{2}\sum_{\lab=1}^N \op{\sigma}_{z}^{(\lab)},
\end{align}
and $\vectop{\sigma}^{(\lab)}$ is the Pauli vector of the $\lab$th particle.
In particular, $\op{J_z}$ is half the difference in the populations of the two modes.
The operators~\eqref{Eq.collspin} satisfy the angular-momentum commutation relations
\begin{align} \label{Eq.J_commutation}
	\big[\J_x, \J_y\big] &= \ii \J_z, & \big[\J_z, \J_x \big] &= \ii \J_y, & \big[\J_y,\J_z\big] &= \ii \J_x,
\end{align}
and have a linear degenerate spectrum spanning the $2^N$-dimensional Hilbert space. The well-known set of states $\ket{J,M}$ forms a basis, where $(\J_x^2+\J_y^2+\J_z^2)\ket{J,M}=J(J+1)\ket{J,M}$ and $\J_z\ket{J,M}=M\ket{J,M}$, and $J\in\{N/2, N/2-1, \ldots\}$ as well as $M\in\{-J,-J+1,\ldots,+J\}$ \cite{ZareBOOK}.

\paragraph{Many spins in a symmetrized state.} 
The Hilbert space spanned by many-qubit states symmetric under particle exchange is that of total spin $J=N/2$, which is the maximum allowed spin
length for $N$ particles.
It has dimension $N+1$, linearly increasing with the number of qubits.
Symmetric qubit states are naturally obtained for $N$ indistinguishable bosons and 
are described by the elegant formalism developed by Schwinger in the 1950s \cite{BiederharnBOOK}.
Angular momentum operators are expressed in terms of bosonic creation, $\op{a}^\dag$ and $\op{b}^\dag$, and 
annihilation, $\op{a}$ and $\op{b}$, operators for the two modes $\ket{a}$ and $\ket{b}$: 
\begin{align} \label{Eq.Joperators_symmetric}
	\J_x &= \frac{\op{a}^\dag \op{b} + \op{b}^\dag \op{a}}{2}, &
	\J_y &= \frac{\op{a}^\dag \op{b} - \op{b}^\dag \op{a}}{2 \ii}, &
	\J_z &= \frac{\op{a}^\dag \op{a} - \op{b}^\dag \op{b}}{2}.
\end{align}
They satisfy the commutation relations~\eqref{Eq.J_commutation} and commute with the total number of particles
$\op{N}=\op{a}^{\dag}\op{a}+\op{b}^{\dag}\op{b}$.
The common eigenstates of $\J_z$ and  $\vect{\J}^2 = (\op{N}/2)(\op{N}/2+1)$ are called Dicke states \cite{DickePHYSREV1954} or two-mode Fock states,
\begin{eqnarray}  \label{Eq.DickeState}
	\ket{\di_z} &=& \ket{N/2+\di}_a \ket{N/2-\di}_b \nonumber \\
	 &=& \frac{(\op{a}^\dag)^{N/2+\di}}{\sqrt{(N/2+\di)!}}\frac{(\op{b}^\dag)^{N/2-\di}}{\sqrt{(N/2-\di)!}}\ket{\text{vac}},
\end{eqnarray} 
where $\ket{\text{vac}}$ is the vacuum. 
They correspond to the symmetrized combinations of $N/2+\di$ particles in mode $\ket{a}$
and $N/2-\di$ particles in mode $\ket{b}$, where $\di = -N/2, -N/2+1, ..., N/2$. 
The eigenstates $\ket{\di_{\vect{n}}}$ along an arbitrary spin direction $\J_{\vect{n}}=\vect{n}\cdot\vectop{J}$ can be obtained by a proper rotation of $\ket{\di_z}$:
$\ket{\di_x} = e^{-\ii \frac{\pi}{2}\op{J}_y} \ket{\di_z}$ and 
$\ket{\di_y} = e^{\ii \frac{\pi}{2} \op{J}_x } \ket{\di_z}$, for instance.
Finally, it is useful to introduce raising and lowering operators, 
$\J_{\pm}=\J_x \pm \ii \J_y$ ($\J_+=\op{a}^{\dag}\op{b}$ and $\J_-=\op{b}^{\dag}\op{a}$), transforming the Dicke states as 
$\J_{\pm} \ket{ \di_z } = \sqrt{(N/2)(N/2+1)-\di(\di\pm1)} \, \ket{ (\di \pm 1)_z }$.

\paragraph{Collective rotations.}
Any unitary transformation of a single qubit is a rotation $e^{-\ii \frac{\ps}{2} \op{\sigma}_{\vect{n}}}$ on the Bloch sphere, where 
$\vect{n}$ and $\ps$ are the rotation axis and rotation angle, respectively. 
With $N$ qubits, each locally rotated about the same axis $\vect{n}$ and angle $\ps$, 
the transformation is $\otimes_{\lab=1}^N e^{-\ii \frac{\ps}{2} \op{\sigma}_{\vect{n}}^{(\lab)} } = e^{-\ii \ps \J_{\vect{n}}}$, 
where $\J_{\vect{n}}$ is the generation of the collective rotation.
This is the idealized model of most of the interferometric transformations discussed in this Review.
In the collective-spin language, a balanced beam splitter is described by $e^{-\ii \frac{\pi}{2} \J_x}$, and a relative phase shift by $e^{-\ii \ps \J_z}$.
Combining the three transformations, $e^{\ii \frac{\pi}{2} \J_x} e^{-\ii \ps \J_z} e^{-\ii \frac{\pi}{2} \J_x} = e^{-\ii \ps \J_y}$,
the whole interferometer sequence (Mach-Zehnder or Ramsey), is equivalent to a collective rotation around the $y$-axis 
on the generalized Bloch sphere of maximum radius $N/2$ \cite{YurkePRA1986}, see Fig.~\ref{Fig:2A}(c).

\subsection{Phase estimation}
\label{Sec.Fundamentals.phase-est}

Broadly speaking, an interferometer is any apparatus that 
transforms a probe state $\op{\rho}_0$ depending on the value of an unknown phase shift $\ps$, see Fig.~\ref{Fig1}. 
The parameter $\ps$ cannot be measured directly and its
estimation proceeds from the results of measurements performed on identical copies of the output state $\rhops$. 
There are good and bad choices for a measurement observable. 
Good ones (that we will quantify and discuss in more details below) are those characterized by a statistical distribution 
of measurement results that is maximally sensitive to changes of $\ps$. 
We indicate as $P(\res | \ps)$ the probability of a result\footnote{\label{POVMfootnote} In a simple scenario, $\res$ is the eigenvalue of an observable. 
In a more general situation, the measurement is described by 
a positive-operator-values measure (POVM). A POVM is
a set of Hermitian operators $\{\POVM\}$ parametrized by $\res$ \cite{NielsenBOOK} that
are positive, $\POVM \geq 0$, to guarantee non-negative probabilities $P(\res | \ps)=\Tr[\rhops \POVM]\ge0$, and
satisfy $\sum_\res \POVM = \1$, to ensure normalization $\sum_{\res}P(\res|\ps)=1$.}
$\res$ given that the parameter has the value $\ps$.
The probability of observing the sequence $\vect{\res} = \{ \res_1, ..., \res_{\m} \}$ of $\m$ independent measurements is 
$P(\vect{\res} | \ps) = \prod_{i=1}^{\m} P(\res_i | \ps)$.
An estimator $\est(\vect{\res})$ is a generic function associating each set of measurement outcomes $\vect{\res}$ with an estimate of $\ps$. 
Interference fringes of a Ramsey interferometer are a familiar example of such an estimation (they belong to a more general estimation technique known as the method of moments, 
discussed in \ref{SubSec.Mom}).
Since the estimator is a function of random outcomes, it is itself a random variable. 
It is thus characterized by a $\ps$-dependent statistical mean value
$\bar{\est}= \sum_{\vect{\res}} P(\vect{\res} | \ps) \, \est(\vect{\res})$ and variance 
\be \label{Eq.variance}
(\Delta \ps)^2 = \sum_{\vect{\res}} P(\vect{\res} | \ps) \, \big[ \est(\vect{\res}) - \bar{\est} \big]^2,
\ee
the sum extending over all possible sequences of measurement results.
Different estimators can yield very different results when applied to the same measured data.  
In the following, we will be interested in locally-unbiased estimators, \ie, those for which  
$\bar{\est}=\ps$ and $ \partial \bar{ \est } /  \partial \ps = 1$, so that the statistical average yields the true parameter value. 

\subsubsection{Cram\'er-Rao bound and Fisher information}

How precise can a statistical estimation be?
Are there any fundamental limits?
A first answer came in the 1940s with the works of \textcite{CramerBOOK1946}, \textcite{Rao1945}, and \textcite{Frechet1943},
who independently found a lower bound to the variance~\eqref{Eq.variance} of any arbitrary estimator.
The Cram\'er-Rao bound is one of the most important results in parameter-estimation theory.
For an unbiased estimator and $\m$ independent measurements, the Cram\'er-Rao bound reads
\begin{equation} \label{Eq:CRLB}
	\Delta \ps \ge \Delta \ps\si{CR} = \frac{1}{\sqrt{\m \Fish(\ps)}}, 
\end{equation}
where
\be \label{Eq.FI}
\Fish(\ps) = 
\sum_{\res} \frac{1}{P(\res | \ps)} \bigg( \frac{\partial P(\res | \ps)}{\partial \ps} \bigg)^2
\ee
is the Fisher information \cite{Fisher1922, Fisher1925}, 
the sum extending over all possible values of $\res$.
The factor $1/\sqrt{\m}$ in Eq.~\eqref{Eq:CRLB} is the statistical improvement when performing independent 
measurements on identical copies of the probe state. 
The Cram\'er-Rao bound assumes mild differentiability properties of the likelihood function $P(\vect{\res} | \ps)$ and thus holds under
very general conditions,\footnote{The Cram\'er-Rao theorem follows from 
$\sum_{\vect{\mu}} \partial_\theta P(\vect{\mu}| \theta) =0$, 
that implies $\partial_\theta  \bar{\est} = \sum_{\vect{\mu}} [\est(\vect{\mu}) - \bar{\est}] \partial_\theta P(\vect{\mu}| \theta)$, 
and the Cauchy-Schwarz inequality
$(\partial_\theta  \bar{\est})^2  \leq (\Delta \theta)^2 \times \sum_{\vect{\mu}} P(\vect{\mu}| \theta) [\partial_\theta \log P(\vect{\mu}| \theta)]^2$.
The equality is obtained if and only if $\partial_\theta \log P(\vect{\mu}| \theta) = \lambda [\est(\vect{\mu}) - \bar{\est}]$ with 
$\lambda$ independent on $\vect{\mu}$.
Equation~\eqref{Eq:CRLB} is recovered for unbiased estimators, \ie, $\partial_\theta  \bar{\est} = 1$, using the 
additivity of the Fisher information, $\sum_{\vect{\mu}} P(\vect{\mu}| \theta) [\partial_\theta \log P(\vect{\mu}| \theta)]^2 = \nu F(\theta)$.} 
see for instance \textcite{KayBOOK1993}. 
No general unbiased estimator is known for small $\m$.
In the central limit, $\m \gg 1$, at least one efficient and unbiased estimator exists in general: 
the maximum of the likelihood, see Sec.~\ref{SubSec.MaxLik}.

\subsubsection{Lower bound to the Fisher information}
\label{SubSec.LowerBoundFI}

A lower bound to the Fisher information can be obtained from the rate of change with $\ps$ of specific moments of 
the probability distribution \cite{PezzePRL2009}:
\be \label{Eq.MomentIneq}
F(\ps) \geq 
\frac{1}{(\Delta \mu)^2}
\bigg( \frac{\dd \bar{\mu}}{\dd \ps} \bigg)^2,
\ee
where $\bar{\mu} = \sum_\res P(\res | \ps) \res$, and
$(\Delta \mu)^2 = \sum_\res P(\res | \ps) (\res -\bar{\mu})^2$.
The Fisher information is larger because it depends on the full probability distribution rather than some moments.

Lower bounds to the Fisher information can be also obtained from reduced probability distributions.
These are useful, for instance, when estimating the phase shift encoded in a many-body distribution 
from the reduced one-body density, 
\eg, from the intensity of a spatial interference pattern \cite{ChwedenczukNJP2012}. One find
\be
F(\ps) \geq \frac{N F_1(\ps)}{1 + (N-1)C/F_1(\ps)},
\ee
where $F_1(\ps) = \int \dd x \tfrac{1}{P_1(x | \ps)} ( \tfrac{\dd P_1(x | \ps)}{\dd \ps})^2$
is the Fisher information corresponding to the one-body density $P_1(x | \ps)$, and the coefficient
 $C =  \int \dd x_1 \dd x_2 \tfrac{P_2(x_1,x_2 | \ps)}{P_1(x_1 | \ps) P_1(x_2 | \ps)} 
 \tfrac{\dd P_1(x_1 | \ps)}{\dd \ps} \tfrac{\dd P_1(x_2 | \ps)}{\dd \ps}$ 
 further depends on the two-body density $P_2(x_1,x_2 | \ps)$.
 Notice that $C=0$ in absence of correlations, namely $P_2(x_1,x_2 | \ps) = P_1(x_1 | \ps) P_1(x_2 | \ps)$.

\subsubsection{Upper bound to the Fisher information: the quantum Fisher information}

An upper bound to the Fisher information is obtained by maximizing Eq.~\eqref{Eq.FI} over all possible generalized measurements 
in quantum mechanics \cite{BraunsteinPRL1994}, $\Fish\si{Q}[\rhops] =  \max_{ \{ \op{E} \} } \Fish(\ps)$, called the quantum Fisher information
(see footnote \ref{POVMfootnote} for the notion of generalized measurements and their connection to conditional probabilities).
We have $F(\ps) \leq \Fish\si{Q}[\rhops]$, and the corresponding bound on the phase sensitivity for unbiased estimators and $\m$ independent measurements is 
\be \label{Eq.QCRLB}
\Delta \ps\si{CR} \ge \Delta \ps\si{QCR} = \frac{1}{\sqrt{\m \Fish\si{Q}[\rhops]}},
\ee
called the quantum Cram\'er-Rao bound \cite{HelstromPLA1967}.
The quantum Fisher information and the quantum Cram\'er-Rao bound are fully determined by the interferometer output state $\rhops$.
Hence they allow to calculate the optimal phase sensitivity of any given 
probe state and interferometer transformation \cite{HelstromBOOK1976, HolevoBOOK1982}, 
for recent reviews see \textcite{ParisIJQI2009,GiovannettiNATPHOT2011,PezzeBOOK2014}.
In general, the quantum Fisher information can be expressed as the variance $\Fish\si{Q}[ \rhops] = (\Delta  \op{L})^2$ 
of a $\ps$-dependent Hermitian operator $\op{L}$ called the
symmetric logarithmic derivative and defined as the solution of 
$\partial_\theta \rhops = (\rhops \op{L} + \op{L} \rhops)/2$ \cite{HelstromPLA1967}.
A general expression of the quantum Fisher information can be found in terms of the spectral decomposition of the output state 
$\op{\rho}_\ps=\sum_{\kappa} q_{\kappa} \ket{\kappa}\bra{\kappa}$
where both the eigenvalues $q_{\kappa} \geq 0$ and the associated eigenvectors $\ket{\kappa}$ depend on $\theta$ \cite{BraunsteinPRL1994}:
\be
\Fish\si{Q}[\rhops] = \sum_{\substack{\kappa, \kappa' \\ q_{\kappa}+q_{\kappa'}>0}} \frac{2}{q_{\kappa}+q_{\kappa'}} 
\abs{\me{\kappa'}{\partial_{\ps} \rhops}{\kappa}}^2,
\ee
showing that $\Fish\si{Q}[\rhops]$ depends solely on $\rhops$ and its first derivative $\partial_{\ps} \rhops$.
We can decompose this equation as 
\be \label{Eq.QFI_twoterms}
\Fish\si{Q}[\rhops] = \sum_{\kappa} \frac{(\partial_\theta q_{\kappa})^2}{q_{\kappa}}
+ 2 \sum_{\substack{\kappa, \kappa' \\ q_{\kappa}+q_{\kappa'}>0}} \frac{(q_{\kappa}-q_{\kappa'})^2}{q_{\kappa}+q_{\kappa'}} 
\abs{\scp{\kappa'}{\partial_{\theta} \kappa}}^2.
\ee
The first term quantifies the information about $\ps$ encoded in $q_{\kappa}$ and
corresponds to the Fisher information obtained when projecting over the eigenstates of $\op{\rho}_\ps$.
The second term accounts for change of eigenstates with $\ps$ 
(we indicate $\ket{\partial_{\theta} \kappa} \equiv \partial_\ps \ket{\kappa}$).
For pure states, $\rhops = \ket{\psi_\ps} \bra{\psi_\ps}$, the first term in Eq.~\eqref{Eq.QFI_twoterms} vanishes, while 
the second term simplifies dramatically to $\Fish\si{Q}[ \ket{\psi_\ps} ] = 4(\scp{\partial_\ps \psi_\ps}{\partial_\ps \psi_\ps}  - 
\abs{\scp{\partial_\ps \psi_\ps}{\psi_\ps}}^2)$.

For unitary transformations generated by some Hermitian operator $\op{H}$, we have $\partial_\ps \rhops = \ii [\rhops, \op{H}]$, and
Eq.~\eqref{Eq.QFI_twoterms} becomes\footnote{We use the notation $\Fish\si{Q}[\rhops]$ to indicate the quantum Fisher 
information for a generic transformation of the probe state, 
and $\Fish\si{Q}[\op{\rho}_0, \op{H}]$ for unitary transformations.} 
\cite{BraunsteinPRL1994, BraunsteinANNPHYS1996}
\be \label{Eq.QFI2}
\Fish\si{Q}[\op{\rho}_0, \op{H}] = 2 \sum_{
\substack{ \kappa,\kappa' \\ q_{\kappa}+q_{\kappa'} > 0 }
} 
\frac{(q_{\kappa} - q_{\kappa'})^2}{q_{\kappa} + q_{\kappa'}} \abs{\me{\kappa'}{\op{H}}{\kappa}}^2.
\ee
For pure states $\ket{\psi_0}$, Eq.~\eqref{Eq.QFI2} reduces to
$\Fish\si{Q}[\ket{\psi_0}, \op{H}] = 4 ( \Delta \op{H} )^2$.
For mixed states, $\Fish\si{Q}[\op{\rho}_0, \op{H}] \leq 4 ( \Delta \op{H} )^2$.
It is worth recalling here that $4 (\Delta \op{H})^2 \leq (h\si{max} - h\si{min})^2$, where $h\si{max}$ and $h\si{min}$ are the 
maximum and minimum eigenvalues of $\op{H}$ with eigenvectors $\ket{h\si{max}} $ and $\ket{h\si{min}} $, respectively. 
This bound is saturated by the states $(\ket{h\si{max}} + e^{\ii \phi} \ket{h\si{min}})/\sqrt{2}$, with arbitrary real $\phi$, which are 
optimal input states for noiseless quantum metrology.
In presence of noise, the search for optimal quantum states is less straightforward, as discussed in Sec.~\ref{Sec.Working-Entanglement.decoherence}.

\paragraph{Convexity and additivity.} 

The quantum Fisher information is convex in the state:
\be \label{Eq.ConvexityQFI}
\Fish\si{Q}\big[ p \op{\rho}^{(1)}_\ps + (1-p) \op{\rho}^{(2)}_\ps \big] \leq p \Fish\si{Q}\big[ \op{\rho}^{(1)}_\ps \big] + (1-p) \Fish\si{Q}\big[ \op{\rho}^{(2)}_\ps \big],
\ee
with $0\leq p \leq 1$.
This expresses the fact that mixing quantum states cannot increase the achievable estimation sensitivity. 
The inequality~\eqref{Eq.ConvexityQFI} can be proved using the fact that the Fisher information 
is convex in the state \cite{CohenIEEE1968, PezzeBOOK2014}.

The quantum Fisher information of independent subsystems is additive:
\be \label{Eq.AdditivityQFI}
\Fish\si{Q}\big[  \op{\rho}^{(1)}_\ps \otimes \op{\rho}^{(2)}_\ps \big] =\Fish\si{Q}\big[ \op{\rho}^{(1)}_\ps \big] + \Fish\si{Q}\big[ \op{\rho}^{(2)}_\ps \big]
\ee 
In particular, for an $m$-fold tensor product of the system, $\op{\rho}^{\otimes m}_\ps$, we obtain an $m$-fold
increase of the quantum Fisher information: $\Fish\si{Q}[  \op{\rho}^{\otimes m}_\ps ]  = m  \Fish\si{Q}[  \op{\rho}_\ps ]$. 
A demonstration of Eq.~\eqref{Eq.AdditivityQFI} can be found in \textcite{PezzeBOOK2014}.

\paragraph{Optimal measurements.} 
The equality $F(\ps)=F\si{Q}[\rhops]$ can always be achieved by optimizing over all possible measurements. 
A possible optimal choice of measurement for both pure and mixed states is given by the 
set of projectors onto the eigenstates of $\op{L}$ \cite{BraunsteinPRL1994}. 
This set of observables is necessary and sufficient for the saturation of the quantum Fisher information whenever $\rhops$ is invertible, and only sufficient otherwise. 
In particular, for pure states and unitary transformations, the quantum Cram\'er-Rao bound can be saturated, in the limit $\ps\to 0$,
by a dichotomic measurement given by the projection onto the probe state itself, $\ket{\psi_0} \bra{\psi_0}$, 
and onto the orthogonal subspace, $\1-\ket{\psi_0} \bra{\psi_0}$ \cite{PezzeBOOK2014}. 
It should be noted that the symmetric logarithmic derivative, and thus also the optimal measurement, generally depends on $\ps$, even for unitary transformations.
Nevertheless, without any prior knowledge of $\ps$, the quantum Cram\'er-Rao bound can be saturated in the asymptotic limit 
of large $\m$ using adaptive schemes \cite{HayashiBOOK, FujiwaraJPA2006}.

\paragraph{Optimal rotation direction.}
Given a probe state, and considering a unitary transformation generated by $\op{H}=\J_{\vect{n}}$, 
it is possible to optimize the rotation direction $\vect{n}$ in order to maximize the quantum Fisher information \cite{HyllusPRA2010}. 
This optimum is given by the maximum eigenvalue of the ${3\times 3}$ matrix 
\be
[\vect{\Gamma}\si{Q}]_{ij} = 2 \sum_{
\substack{ \kappa,\kappa' \\ q_{\kappa}+q_{\kappa'} > 0 }}
 \frac{(q_{\kappa} - q_{\kappa'})^2}{ q_{\kappa} + q_{\kappa'} }\bra{\kappa'} \op{J}_i \ket{\kappa}  \bra{\kappa} \op{J}_j \ket{\kappa'}, 
\ee 
with $i,j=x,y,z$, 
and the optimal direction by the corresponding eigenvector.
For pure states, $[\vect{\Gamma}\si{Q}]_{ij} =  2( \mean{\op{J}_i\op{J}_j} + \mean{\op{J}_j\op{J}_i})  - 4\mean{\op{J}_i} \mean{\op{J}_j}$.

\begin{figure*}
\begin{center}
\includegraphics[width=\textwidth]{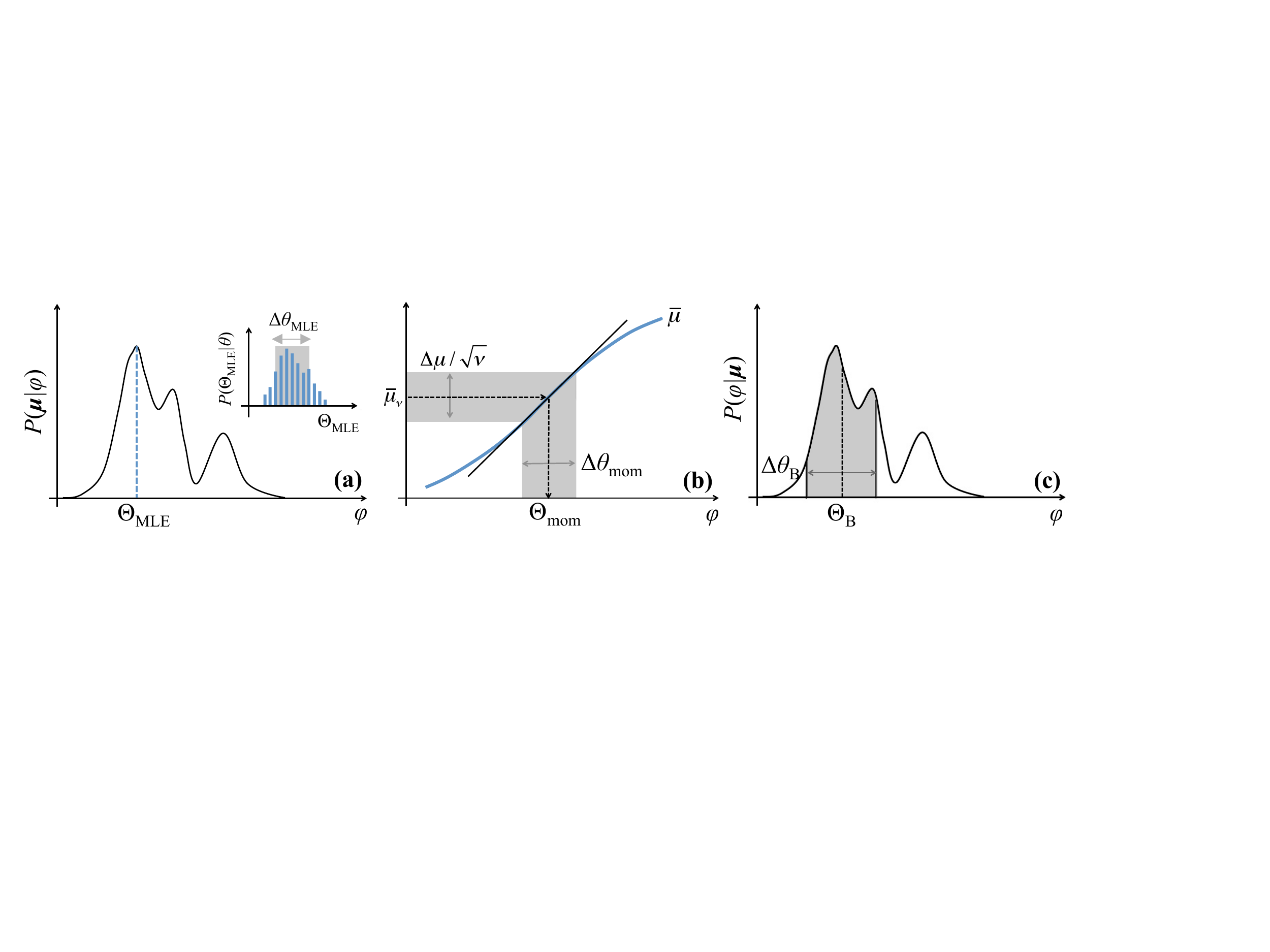}
\end{center}
\caption{{\bf Schematic representation of different phase estimation protocols.} 
(a) \emph{Maximum likelihood estimation.} 
$\est\si{MLE}$ is the absolute maximum (dashed line) of the likelihood function $P(\vect{\mu}| \varphi)$ (solid line) 
corresponding to the observed sequence of results $\vect{\mu}$.
As schematically shown in the inset, the phase sensitivity $\Delta\ps\si{MLE}$ is identified as the root-mean-square fluctuation 
(shaded region) of the statistical distribution of $\est\si{MLE}$, obtained by repeating the measurements several times (at a fixed value of $\ps$).
(b) \emph{Method of moments.} 
The blue line is $\bar{\mu}$ as a function of the parameter $\varphi$.
Via this functional monotonic behavior, it is possible to associate  the phase estimate $\estmom$ with the measured mean value $\bar{\mu}_{\m}$.
The phase uncertainty $\Delta \ps\si{mom} $ follows from the statistical uncertainty $\Delta \mu/\sqrt{\m}$ of $\bar{\mu}_{\m}$ (grey region).
For $\m\gg 1$, $\Delta \ps\si{mom} $ has a simple expression, Eq.~\eqref{MomCRLB}, obtained by error 
propagation, locally approximating $\bar{\mu}$ by the tangent to the curve (thin black line). 
(c) \emph{Bayesian estimation.} Posterior probability distribution $P(\varphi | \vect{\mu})$ (solid line). 
The phase estimate $\Theta\si{B}$ can be chosen as the weighted averaged phase. 
The integral $\int_{\Delta \ps\si{B}} \dd\varphi P(\varphi | \vect{\mu})$ gives the probability that the 
true phase value falls into the confidence interval $\Delta \ps\si{B}$ (grey region) around $\Theta\si{B}$.} 
\label{Fig:estimation}
\end{figure*} 

\subsubsection{Phase sensitivity and statistical distance} 
\label{SubSecStatDist}

Parameter estimation is naturally related to the problem of distinguishing neighboring 
quantum states along a path in the parameter space \cite{WoottersPRD1981, BraunsteinPRL1994}.
Heuristically, the phase sensitivity of an interferometer can be understood as the smallest 
phase shift for which the output state $\rhops$ of the interferometer can be distinguished from the input $\op{\rho}_0$. 
We introduce a statistical distance between probability distributions, 
\be \label{Eq.Hell}
\hell^2(P_0, P_\ps) = 1- \mathcal{F}\si{cl}(P, P_\ps),
\ee
called the Hellinger distance, where $\mathcal{F}\si{cl}(P_0, P_\ps) \equiv \sum_\mu \sqrt{P(\mu| 0) P(\mu| \ps)}$ 
is the statistical fidelity, or overlap, between probability distributions, also known as Bhattacharyya coefficient \cite{BhattacharyyaBCMS1943}.
$\hell$ is non-negative, $0 \leq \hell \leq 1$, and its Taylor expansion reads
\be \label{Eq.Hell.expansion}
\hell^2(P_0, P_\ps) = \frac{\Fish(0)}{8} \ps^2 + \mathcal{O}(\ps^3).
\ee
This equation reveals that the Fisher information is the square of a statistical speed, 
$\upsilon\si{H} = \partial \hell/\partial \ps = \sqrt{\Fish(0)/8}$.
It measures the rate at which a probability distribution varies when tuning the phase parameter $\ps$.
Equation~\eqref{Eq.Hell.expansion} has been used to extract the Fisher information experimentally \cite{StrobelSCIENCE2014}, see Sec.~\ref{Sec.Atom-Atom.twistNturn}.
As Eq.~\eqref{Eq.Hell} depends on the specific measurement, 
it is possible to associate different statistical distances to the same quantum states.  
This justifies the introduction of a distance between quantum states by maximizing  
$\hell^2(P_0, P_\ps)$ over all possible generalized measurements (\ie, over all POVM sets, see footnote \ref{POVMfootnote}),
$\bures^2(\op{\rho}_0, \op{\rho}_\ps) = \max_{\{\op{E}\}} \hell^2(P_0, P_\ps)$ \cite{Fuchs1995}, 
called the Bures distance \cite{BuresTAMS1969}.
\textcite{HubnerPLA1992} showed that
\be \label{Eq.Bures}
\bures^2(\op{\rho}_0, \rhops) = 1 -\mathcal{F}\si{Q} (\op{\rho}_0, \rhops), 
\ee
where $\mathcal{F}\si{Q}(\op{\rho}_0, \op{\rho}_\ps) = \Tr[\sqrt{\sqrt{\op{\rho}_\ps} \op{\rho}_0 \sqrt{\op{\rho}_\ps}}]$ is the transition probability \cite{UhlmannRMP1976} or
the quantum fidelity between states \cite{JozsaJMO1995},
see \textcite{BengtssonBOOK2006, SpehnerJMP2014} for reviews. 
Uhlmann's theorem \cite{UhlmannRMP1976} states that 
$\mathcal{F}\si{Q}(\op{\rho}, \op{\sigma}) = \max_{\ket{\psi},\ket{\phi}} \abs{\scp{\psi}{\phi}}$, 
where the maximization runs over all purifications $\ket{\psi}$ of $\op{\rho}$ and $\ket{\phi}$ of $\op{\sigma}$ \cite{NielsenBOOK}.
In particular, $\mathcal{F}\si{Q}(\ket{\psi}, \ket{\phi})=\abs{\scp{\psi}{\phi}}$ for pure states.
A Taylor expansion of Eq.~\eqref{Eq.Bures} for small $\ps$ gives
\be \label{Eq.Bures.expansion}
\bures^2(\op{\rho}_0, \op{\rho}_\ps)  = \frac{F\si{Q}[\op{\rho}_0]}{8} \ps^2 + \mathcal{O}(\ps^3).
\ee
The quantum Fisher information is thus the square of a quantum statistical speed, 
$\upsilon\si{Q} = \partial \bures/\partial \ps = \sqrt{F\si{Q}[\op{\rho}_0]/8}$,
maximized over all possible generalized measurements.
The quantum Fisher information has  also been related to the dynamical susceptibility \cite{HaukeNATPHYS2016},
while lower bounds have been derived by \textcite{ApellanizPRA2015} and \textcite{FrerotPRB2016}.

\subsubsection{The maximum likelihood estimator}
\label{SubSec.MaxLik}

The maximum likelihood estimator is the phase value that maximizes the likelihood of the observed measurement sequence $\vect{\mu}$, see Fig.~\ref{Fig:estimation}(a):
$\est\si{MLE}(\vect{\mu}) = \argmax_{\varphi} P(\vect{\res} | \varphi)$.
The key role played by $\est\si{MLE}(\vect{\mu})$ in parameter estimation is due to its asymptotic properties for independent measurements. 
For sufficiently large $\m$, the distribution of the maximum likelihood estimator tends to a Gaussian 
centered at the true value $\ps$ and of variance equal to the inverse Fisher information \cite{LehmanBOOK}:
$P(\est\si{MLE}  | \ps) = \sqrt{\tfrac{\m F(\ps)}{2 \pi}} e^{- \frac{\m F(\ps)}{2} (\ps - \est\si{MLE})^2}$.
Therefore, the maximum likelihood estimator is asymptotically unbiased and its variance 
saturates the Cram\'er-Rao bound: $\Delta\ps\si{MLE} = 1 / \sqrt{\m F(\ps)}$.
In the central limit, any estimator is as good as---or worse than---the maximum likelihood estimate.

\subsubsection{Method of moments}
\label{SubSec.Mom}

The method of moments exploits the variation of collective properties of the probability distribution---such as the mean value 
$\bar{\mu}$ and variance $(\Delta \mu)^2$---with the phase shift $\ps$.
Let us  take the average $ \bar{\mu}_{\m} = \frac{1}{\m} \sum_{i=1}^{\m} \res_i$ of $\m$ measurements results $\mu_1, ..., \mu_{\m}$.
The estimator $\estmom$ is the value for which $\bar{\mu}$
is equal to $\bar{\mu}_{\m}$, see Fig.~\ref{Fig:estimation}(b).
Applying this method requires $\bar{\mu}$ to be a monotonous function of the parameter $\ps$, at least in a local region of parameter values determined from prior knowledge.
The sensitivity of this estimator can be calculated by error propagation,\footnote{A Taylor expansion of $\bar{\mu}_{\m}$ around the true value $\ps$ gives 
$\bar{\mu}_{\m} \approx \bar{\mu} + \frac{\dd \bar{\mu} }{\dd \ps}(\estmom-\ps)$. 
We obtain Eq.~\eqref{MomCRLB} by identifying $\bar{\mu}_{\m} - \bar{\mu} \approx \Delta \mu/\sqrt{\m}$ (valid for $\m \gg 1$) and $ \estmom-\ps \approx \Delta \ps\si{mom}$.} giving 
 \be \label{MomCRLB}
\Delta \ps\si{mom} = \frac{ \Delta \mu}{\sqrt{\m} \abs{\dd \bar{\mu} / \dd \ps}}, 
\ee
As expected on general grounds and proved by Eq.~\eqref{Eq.MomentIneq}, the method of moments 
is not optimal in general, $\Delta \ps\si{mom} \leq \Delta \ps\si{CR}$, with no guarantee of saturation even in the central limit.  
The equality $\Delta \ps\si{mom} = \Delta \ps\si{CR}$ is obtained when the probability distribution is Gaussian, 
$P(\res | \ps) = \tfrac{e^{-(\mu - \bar{\mu})^2/2(\Delta \mu)^2}}{\sqrt{2 \pi (\Delta \mu)^2}}$,
and $\tfrac{d (\Delta \mu)}{d \theta} \ll \tfrac{d \bar{\mu}}{d \theta}$, such that
the changes of the complete probability distribution are fully captured by the shift of its mean value \cite{PezzeBOOK2014}.
Nevertheless, due to its simplicity, Eq.~\eqref{MomCRLB} 
is largely used in the literature to calculate the phase sensitivity of an interferometer for various input states and 
measurement observables \cite{WinelandPRA1994, DowlingPRA1998, YurkePRA1986}.
For instance, in the case of unitary rotations generated by $\op{H}=\J_y$ (as in Ramsey and Mach-Zehnder interferometers) and
taking $\J_z$ as measurement observable, 
Eq.~\eqref{MomCRLB} in the limit $\ps \approx 0$ can be rewritten as
\be \label{Eq.ErrProp}
\Delta \ps\si{mom}  = \frac{ \Delta \J_z}{ \sqrt{\m} \abs{\mean{\J_x}}}.
\ee
This equation is useful to introduce the concept of metrological spin-squeezing, see Sec.~\ref{SubSec.SpinSq}.
We recall that Eqs.~\eqref{MomCRLB} and \eqref{Eq.ErrProp} are valid for a sufficiently large number of measurements.

Finally, there are many examples in the literature where a small $\Delta \ps\si{mom}$ is obtained for phase values where
$\Delta \mu, \dd \bar{\mu} / \dd \ps \to 0$, while the ratio $\Delta \mu / \abs{\dd \bar{\mu} / \dd \ps}$ remains finite \cite{YurkePRA1986, KimPRA1998}.
These ``sweet spots'' are very sensitive to technical noise: an infinitesimal amount of noise may prevent $\Delta \mu$ to vanish, 
while leaving unchanged  $\dd \bar{\mu} / \dd \ps$, such that $\Delta \ps\si{mom}$ diverges \cite{LuckeSCIENCE2011}.

\subsubsection{Bayesian estimation}

The cornerstone of Bayesian inference is Bayes' theorem.
Let us consider two random variables $\mathcal{x}$ and $\mathcal{y}$.
Their joint probability density can be expressed as $P(\mathcal{x}, \mathcal{y}) = P(\mathcal{x} | \mathcal{y}) P(\mathcal{y})$ in terms of the 
conditional probability $P(\mathcal{x} | \mathcal{y} )$ and the marginal probability distribution $P( \mathcal{y} ) = \int \dd \mathcal{x} P(\mathcal{x},\mathcal{y})$.
Bayes' theorem
\be \label{Eq.Bayes_theorem}
P(\mathcal{x} | \mathcal{y})
= \frac{ P( \mathcal{y} | \mathcal{x}) P(\mathcal{x}) }{P( \mathcal{y})}
\ee
follows from the symmetry of the joint probability $P(\mathcal{x}, \mathcal{y}) = P(\mathcal{y} | \mathcal{x}) P(\mathcal{x})$.

In the Bayesian subjective interpretation of probabilities, $\varphi$ and $\vect{\mu}$ are both considered as random variables with 
$P(\varphi|\vect{\res})$ as the posterior probability distribution given the measurement results $\vect{\mu}$.
$P(\varphi)$ is the prior probability distribution that quantifies our (subjective) ignorance of the true value of the interferometric phase, 
\ie, before any measurements were done. 
One often has no prior knowledge on the phase (maximum ignorance), which is expressed by a flat prior distribution $P(\varphi)=1/(2\pi)$.
Bayes' theorem $P (\varphi | \vect{\res} ) = P( \vect{\res} | \varphi ) P(\varphi) / P( \vect{\res})$
allows to update our knowledge about the interferometric phase $\ps$ by 
including measurement results, since $P(\vect{\res}|\varphi)$ can be calculated directly 
(see the introduction of Sec.~\ref{Sec.Fundamentals.phase-est}) and $P(\vect{\res})$ is determined by the 
normalization $\int_0^{2\pi} P(\varphi|\vect{\mu})\dd\varphi=1$. Bayesian probabilities express our (lack of) 
knowledge of the interferometric phase as a probability distribution $P(\varphi|\vect{\res})$.
This is radically different from the standard frequentist view where
the probability is defined as the infinite-sample limit of the outcome frequency of an observed event. 
Having the posterior distribution $P(\varphi|\vect{\mu})$, we can consider any phase $\varphi$ as the estimate. 
In practice, it is convenient to choose
the weighted averaged $\int_0^{2\pi}\varphi P(\varphi|\vect{\res})\dd\varphi$,
or the phase corresponding to the maximum of the probability $\argmax_{\varphi}  P(\varphi|\vect{\res})$,
since the corresponding mean square fluctuations saturate the Cram\'er-Rao bound (see below). 
We can further calculate the probability 
that the chosen estimate falls into a certain interval $[\ps_1,\ps_2]$ by integrating $\int_{\ps_1}^{\ps_2}P(\varphi|\vect{\mu})\dd\varphi$.
To take into account the periodicity of the probability, quantities like $\int_0^{2\pi} e^{\ii\varphi}P(\varphi|\vect{\res})\dd\varphi$
can be calculated.
Remarkably, Bayesian estimation is asymptotically consistent: 
as the number of measurements increases, the posterior probability distribution assigns more weight 
in the vicinity of the true value.
The Laplace-Bernstein-von Mises theorem \cite{LehmanBOOK, PezzeBOOK2014, GillBOOK2008} demonstrates that, under quite general conditions,
$P\big(\varphi | \ps \big) = \sqrt{\frac{\m \Fish(\ps)}{2 \pi}} \, e^{ - \frac{\m \Fish(\ps)}{2} (\ps - \varphi)^2}$,
to leading order in $\m$, for $\m \gg 1$.
In this limit, the posterior probability becomes normally distributed, centered at the true value of the parameter, and with 
a variance inversely proportional to the Fisher information. 
See \textcite{VanTreesBOOK} for a review of bounds in Bayesian phase estimation.

\subsection{Entanglement and phase sensitivity}
\label{Sec.Fundamentals.entanglement}

In this section we show how entanglement can offer a precision enhancement in quantum metrology. 
We start with the formal definition of multiparticle entanglement and then clarify, via the Fisher information introduced in the previous section, the 
notion of useful entanglement for quantum metrology. 

\subsubsection{Multiparticle entanglement}

Let us consider a system of $N$ particles (labeled as $\lab=1,2,...,N$), 
each particle realizing a qubit.
A pure quantum state is separable in the particles if it can be written as a product 
\be \label{sep}
\ket{\psi\si{sep}} = \ket{\psi^{(1)}} \otimes \ket{\psi^{(2)}} \otimes \dotsm \otimes \ket{\psi^{(N)}},
\ee
where $\ket{\psi^{(\lab)}}$ is the state of the $\lab$th qubit.
A mixed state is separable if it can be written as a mixture of product states \cite{WernerPRA1989}, 
\be \label{seps}
 \op{\rho}\si{sep} = \sum_q p_q \ket{\psi_{\text{sep},q}} \bra{\psi_{\text{sep},q}},
\ee
with $p_q \geq 0$ and $\sum_q p_q =1$.
States that are not separable are called entangled \cite{HorodeckiRMP2009, GuhnePHYSREP2009}.
In the case of $N=2$ particles, any quantum state is either separable or entangled. 
For $N>2$, we need further classifications \cite{DurPRA2000}. 
Multiparticle entanglement is quantified by
the number of particles in the largest non-separable subset.
In analogy with Eq.~\eqref{sep}, a pure state of $N$ particles is $k$-separable (also indicated as $k$-producible in the literature) if 
it can be written as 
\be \label{eq:ksep1} 
\ket{ \psi\si{$k$-sep} } = \ket{\psi_{N_1}} \otimes \ket{\psi_{N_2}} \otimes ... \otimes \ket{\psi_{N_M}}, 
\ee
where $\ket{\psi_{N_\lab}}$ is the state of $N_\lab \leq k$ particles and $\sum_{\lab=1}^M N_{\lab} = N$.
A mixed state is $k$-separable if it can be written as a mixture of $k$-separable pure states \cite{GuhneNJP2005}
\be \label{eq:ksep2}
\op{\rho}\si{$k$-sep} = \sum_q p_q \ket{\psi\si{$k$-sep,$q$}} \bra{\psi\si{$k$-sep,$q$}}.
\ee
A state that is $k$-separable but not $(k-1)$-separable is called $k$-particle entangled:
it contains at least one state of $k$ particles that does not factorize.
Using another terminology \cite{SorensenPRL2001}, it has an entanglement depth larger than $k-1$.
In maximally entangled states ($k=N$) each particle is entangled with all the others.
Finally, note that $k$-separable states form a convex set containing the set of $k'$-separable states with $k'<k$ \cite{GuhnePHYSREP2009}.

\subsubsection{Sensitivity bound for separable states: the standard quantum limit}
\label{Sec:separable}

The quantum Fisher information of any separable state of $N$ qubits is upper-bounded \cite{PezzePRL2009}:
\be \label{Fishsepsep}
F\si{Q}\big[\op{\rho}\si{sep}, \J_{\vect{n}} \big] \leq N. 
\ee
This inequality follows from the convexity and additivity of the quantum Fisher information and uses $4 (\Delta \op{\sigma}_{\vect{n}})^2 \leq 1$ \cite{PezzeBOOK2014}.
As a consequence of Eqs.~\eqref{Eq.QCRLB} and~\eqref{Fishsepsep}, the maximum phase sensitivity achievable with separable states is \cite{GiovannettiPRL2006}
\be \label{SNbound}
\Delta \ps\si{SQL} =  \frac{1}{\sqrt{ N \m} },
\ee 
generally indicated as the shot-noise or standard quantum limit.
This bound is independent of the specific measurement and estimator, and refers to unitary collective 
transformations that are local in the particles. 
In Eq.~\eqref{SNbound} $N$ and $\nu$ play the same role:
repeating the phase estimation $\nu$ times with one particle has the same sensitivity bound 
as repeating the phase estimation one time with $N=\nu$ particles in a separable state.

\begin{figure}[t!]
\begin{center}
\includegraphics[width=\columnwidth]{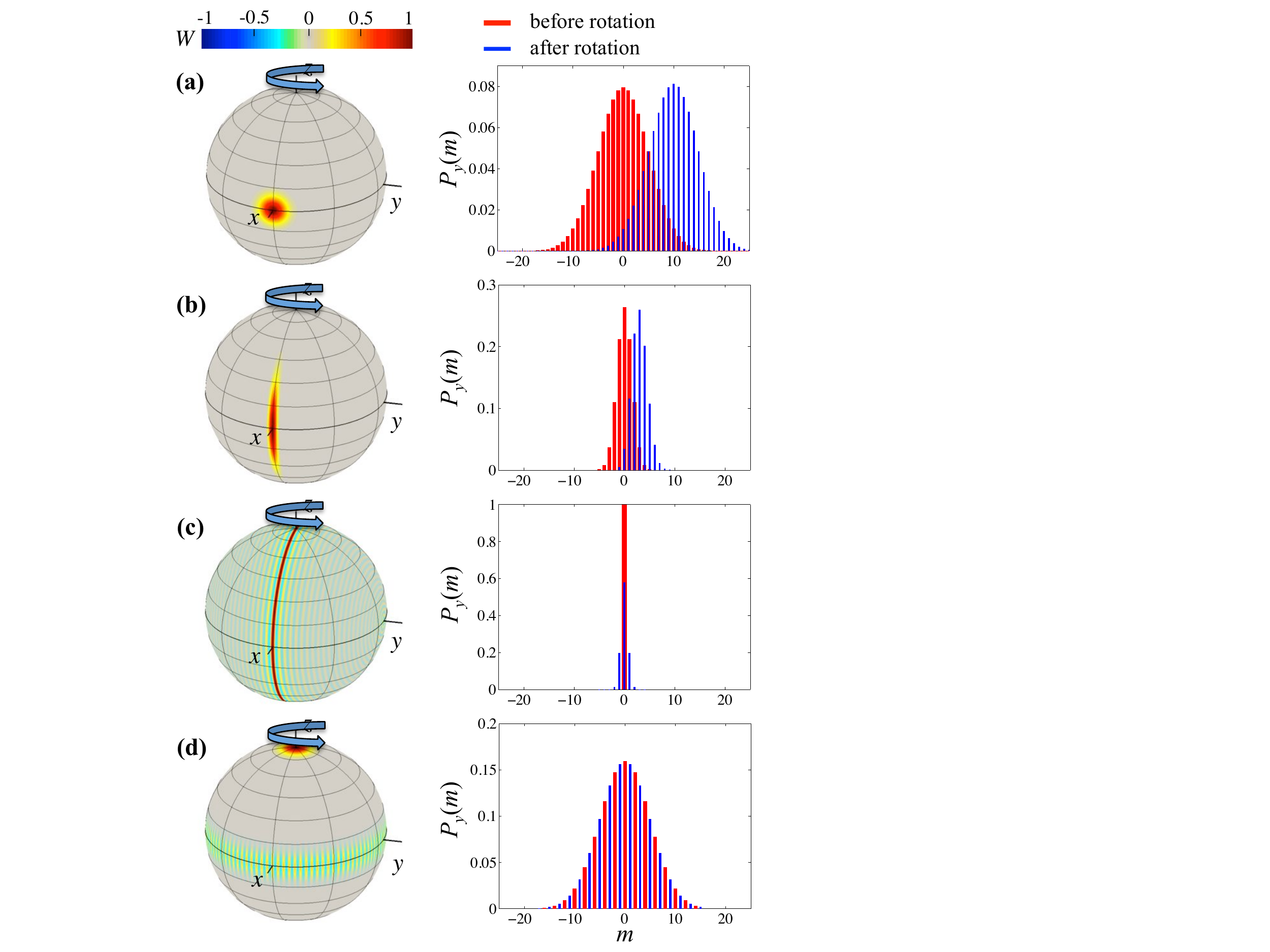}
\end{center}
\caption{{\bf Rotation of different quantum states.} 
Wigner distribution (normalized to its maximum value -- left column), see Sec.~\ref{Sec.Fundamentals.tomography}, and spin probability
$P_y(\di) = \abs{\scp{\di_y}{\psi}}^2$ along the $y$ direction (thick red histograms -- right column)
of different quantum states $\ket{\psi}$:
(a) a coherent spin state pointing along the positive $x$-axis, $\ket{\pi/2, 0,N}$, (Sec.~\ref{SubSecCSS});
(b) a spin-squeezed state with $\xiR^2=0.1$ (Sec.~\ref{SubSec.SpinSq});
(c) a twin-Fock state (Sec.~\ref{sec:DickeStates}) and 
(d) a NOON state (Sec.~\ref{sec:NOONstate}).
The thin blue histogram is $P_y(\di) = \abs{\me{\di_y}{e^{-\ii \ps \J_z}}{\psi}}^2$
obtained after a rotation of the state by an angle $\ps=2/\sqrt{N}$ in (a), 
$\ps = 2\xiR/\sqrt{N}$ in (b), $\ps = 2/N$ in (c), and 
$\ps = \pi/N$ in (d). 
Here $N=100$.}
\label{Fig:rotations}
\end{figure}

\subsubsection{Coherent spin states}
\label{SubSecCSS}

The notion of coherent spin states was introduced by \textcite{ArecchiPRA1972, RadcliffeJPA1971}
as a generalization of the field coherent states first discussed by \textcite{GlauberPR1963}, see \textcite{ZhangRMP1990} for a review. 
Coherent spin states are constructed as the product of $N$ qubits (spins-1/2) in pure states all pointing along the same 
mean-spin direction $\vect{s}=\{\sin \vartheta \cos \varphi ,\sin \vartheta \sin \varphi,\cos \vartheta \}$:
\be \label{Eq.CSS}
\ket{ \vartheta, \varphi,N } = \bigotimes_{\lab=1}^N \bigg[ \cos \frac{\vartheta}{2} \ket{a}_{\lab} + e^{\ii \varphi} \sin \frac{\vartheta}{2} \ket{b}_{\lab}  \bigg].
\ee
Equation~\eqref{Eq.CSS} is the eigenstate of $\J_{\vect{s}}$ with the maximum eigenvalue of $N/2$.
The coherent spin state is a product state and no quantum entanglement is present between the particles.
$\ket{\vartheta,\varphi,N}$ can also be written as a binomial sum of Dicke states with 
$\scp{\di_z}{\vartheta,\varphi,N} = \sqrt{\binom{N}{N/2+\di}}  (\cos\frac{\vartheta}{2})^{N/2-\di} (\sin \frac{\vartheta}{2})^{N/2+\di} e^{-\ii (\di+N/2) \varphi}$ \cite{ArecchiPRA1972}.
When measuring the spin component of $\ket{ \vartheta, \varphi, N }$ along any direction $\perp$ orthogonal to $\vect{s}$, 
each individual atom is projected with equal probability into
the up and down eigenstates along this axis, with eigenvalues $\pm1/2$, respectively:
we thus have $\mean{\J_{\perp}}=0$, and $( \Delta \J_{\perp}  )^2=N/4$ \cite{ItanoPRA1993,YurkePRA1986}.

Coherent spin states are optimal separable states for metrology. 
They saturate the equality sign in Eq.~\eqref{Fishsepsep} and thus reach the standard quantum limit. 
Let us consider the rotation of $\ket{\vartheta,\varphi,N}$ around a direction $\vect{n}$ perpendicular to the mean spin direction $\vect{s}$
(here $\vect{s}$, $\vect{n}$ and $\perp$ are mutually orthogonal).
This rotation displaces the coherent spin state on the surface of the Bloch sphere, see Fig.~\ref{Fig:rotations}(a). 
The initial and final states become distinguishable after rotating by an angle $\ps\si{min}$ 
heuristically giving the phase sensitivity of the state. 
This rotation angle can be obtained from a geometric reasoning \cite{YurkePRA1986}:
we have $\Delta \J_{\perp} \approx \mean{\J_{\vect{s}}} \sin \ps\si{min}$, 
giving $\ps\si{min} \approx 1/\sqrt{N}$ for $N \gg 1$. 
More rigorously, the squared Bures distance, Eq.~\eqref{Eq.Bures}, between $\ket{\vartheta,\varphi, N}$ and 
the rotated $e^{-\ii \ps \J_{\vect{n}}} \ket{\vartheta,\varphi, N}$ is 
\be \label{Eq.overlapCSS}
\bures^2\big( \ket{\vartheta, \varphi, N }, e^{-\ii \ps \J_{\vect{n}}} \ket{ \vartheta, \varphi, N }\big) = 1-\cos^N(\ps/2), 
\ee
that is $\bures^2 = N \ps^2/8 + \mathcal{O}(\ps^4)$ for small values of $\ps$.
According to Eq.~\eqref{Eq.QFI2} we obtain a quantum Fisher information 
$\Fish\si{Q}[\ket{\vartheta, \varphi, N}, \J_{\vect{n}}] = 4 (\Delta \J_{\vect{n}})^2 = N$. 
With the method of moments, Eq.~\eqref{Eq.ErrProp}, we find a phase sensitivity 
$\Delta \ps\si{mom} = \Delta \J_{\perp}/ (\sqrt{\m} \abs{\mean{\J_{\vect{s}}}}) =1/ \sqrt{\m N}$ \cite{ItanoPRA1993, YurkePRA1986}: while $\abs{\mean{\J_{\vect{s}}}}=N/2$ reaches
its maximum value, it is the quantum projection noise
of uncorrelated atoms, $(\Delta\J_{\perp})^2 = N/4$, that limits the
achievable sensitivity \cite{ItanoPRA1993,WinelandPRA1992,WinelandPRA1994}.
For any rotation around an axis orthogonal to the mean spin direction, 
coherent spin states thus satisfy $\Delta \ps\si{QCR} = \Delta \ps\si{SQL}$.

\begin{figure}
\begin{center}
\includegraphics[width=\columnwidth]{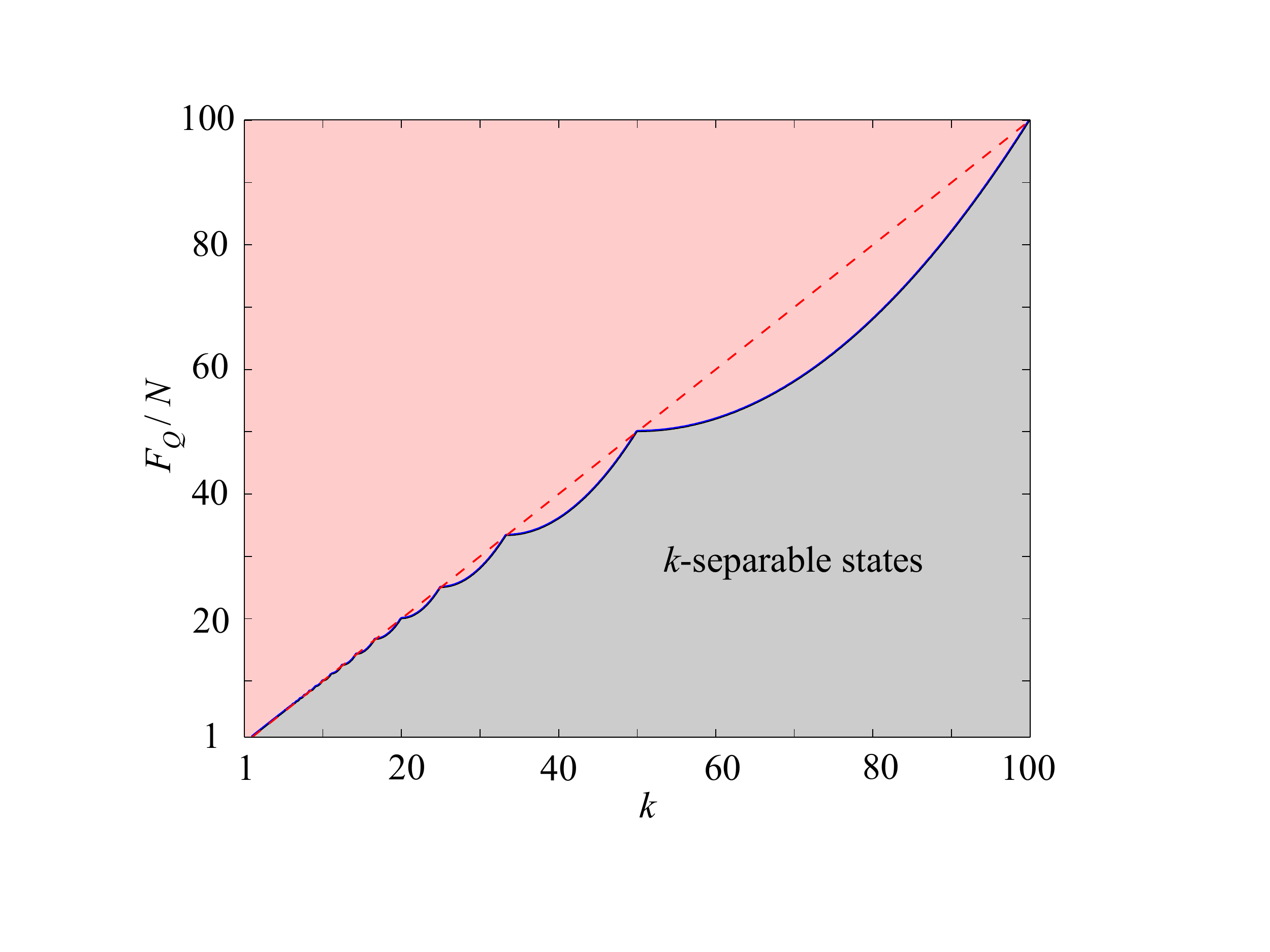}
\end{center}
\caption{{\bf Useful $k$-particle entanglement for quantum metrology.} 
$k$-separable states have a quantum Fisher information bounded by the  
solid blue line, Eq.~\eqref{Eq.FQIk}.
The dashed line is $F\si{Q}/N=k$.
Here, $N=100$. Adapted from \textcite{HyllusPRA2012}.} 
\label{Fig:multipart}
\end{figure}

\subsubsection{Useful entanglement for quantum metrology}
\label{Sec:usefulentangled}

The violation of Eq.~\eqref{Fishsepsep}, \ie, 
\be \label{Fishent}
\Fish\si{Q}\big[\op{\rho}, \J_{\vect{n}} \big] > N, 
\ee
is a sufficient condition for particle-entanglement in the state $\op{\rho}$. 
To be more precise, the inequality~\eqref{Fishent} is the condition of useful entanglement for quantum metrology: 
it is a necessary and sufficient condition for a quantum state to be useful in the estimation of a phase shift $\ps$---with an interferometer implementing the transformation $e^{-\ii \ps \J_{\vect n}}$---with a sensitivity overcoming the standard quantum limit \cite{PezzePRL2009}.
Not all entangled states are useful for quantum metrology. 
Yet, useless entangled states for quantum metrology might be useful for other quantum technologies.
It should also be noted that not all useful entangled states for quantum metrology are equally useful: large quantum Fisher information requires large entanglement depth.
For states of type \eqref{eq:ksep2}, we have \cite{HyllusPRA2012,TothPRA2012}
\be \label{Eq.FQIk}
\Fish\si{Q}\big[\op{\rho}\si{$k$-sep}, \J_{\vect n} \big]  \leq \mathfrak{s} k^2 + \mathfrak{r}^2,
\ee
where 
$\mathfrak{s}=\lfloor \frac{N}{k}\rfloor$ is the integer part of $\frac{N}{k}$, and $\mathfrak{r}=N-\mathfrak{s} k$
(note that $\mathfrak{s} k^2 + \mathfrak{r}^2 = N k$ when $N$ is divisible by $k$).
If the bound~\eqref{Eq.FQIk} is surpassed, then the probe state contains metrologically useful $(k+1)$-particle entanglement:
when used as input state of the interferometer defined by the transformation 
$e^{-\ii \ps\J_{\vect{n}}}$, this state enables a phase sensitivity better than any $k$-separable state.
The bound~\eqref{Eq.FQIk} increases monotonically with $k$ (see Fig.~\ref{Fig:multipart}), in particular $\Fish\si{Q}\big[\op{\rho}\si{$k$-sep}, \J_{\vect n} \big]  \leq Nk$.
 The maximum value of the quantum Fisher information is obtained for genuine $N$-particle entangled states, $k=N$, giving \cite{PezzePRL2009}
\be \label{Fishentent}
\Fish\si{Q}\big[\op{\rho}, \J_{\vect{n}} \big] \leq N^2.
\ee 
Equation~\eqref{Fishentent} defines
the ultimate Heisenberg limit\footnote{The name ``Heisenberg limit'' was first introduced in \textcite{HollandPRL1993} referring to the heuristic
number-phase Heisenberg uncertainty relation $\Delta \ps \Delta N \geq 1$. 
We refer to the Heisenberg \emph{scaling} of phase sensitivity when 
$\Delta\ps=\mathcal{O}(N^{-1})$.} 
\footnote{It is possible to maximize the phase sensitivity by optimizing the number of particles $N$ entering into the interferometer 
multiplied by the times $\m$ that the measurement is performed \cite{LanePRA2003, PezzePRA2013, BraunsteinPRL1992}. 
This provides a definition of Heisenberg limit $1/N_T$, where $N_T = N \times \m\si{opt}$, and 
$\m\si{opt}$ is the optimal number of measurements that maximize the phase sensitivity for a fixed number of particles $N$.
Since $\m\si{opt}$ may depend on $N$, there might be, in principle, states having a Fisher information larger than $N$ but a phase variance 
above the standard quantum limit $1/\sqrt{N_T}$.}
of phase sensitivity \cite{GiovannettiPRL2006}, 
\be \label{HLbound}
\Delta \ps\si{HL} =  \frac{1}{N \sqrt{\m}}.
\ee
The difference between Eq.~\eqref{SNbound} and Eq.~\eqref{HLbound} is a faster scaling 
of the phase sensitivity with the number of particles, which cannot be obtained by exploiting classical 
correlations among the qubits.
Still the standard quantum limit can be surpassed using separable states at the expense of other resources \cite{GiovannettiPRL2006} 
such as, for instance, exploiting a multiround protocol \cite{HigginsNATURE2007}.

We note that the quantum Fisher information is bounded by $\Fish\si{Q}[\op{\rho}, \J_{\vect{n}}] \leq 
4 (\Delta \J_{\vect{n}})^2 \leq (2J)^2 \leq N^2$, where 
$\sqrt{J(J+1)}$ is the spin length.
This shows that the most sensitive states lie in a subspace with maximum spin $J = N/2$, namely those
symmetric under particle exchange (see Sec.~\ref{Sec.Fundamentals.spin}).

Equations~\eqref{SNbound} and~\eqref{HLbound} can be generalized to transformations $e^{-\ii \ps \sum_{\lab=1}^N \op{h}^{(\lab)}}$, 
where $\op{h}^{(\lab)}$ is an arbitrary local Hamiltonian for the $\lab$th particles (that can be a generic qudit).
Taking $\op{h}^{(\lab)}= \op{h}$ for all $N$ particles, 
we have $\Delta \ps\si{SQL} =  1/\delta h \sqrt{ N \m}$, and 
$\Delta \ps\si{HL} =  1/\delta h N\sqrt{\m}$ \cite{GiovannettiPRL2006}, 
where $\delta h = \abs{h\si{max}-h\si{min}}$, and 
$h\si{max}$ and $h\si{min}$ are the maximum and minimum eigenvalues of $\op{h}$, respectively.

We finally note that not all $N$-particle entangled states reach the Heisenberg limit, as exemplified by the 
generalized $W$ state $\ket{W} = \ket{{\pm N/2 \mp 1}}_z$, which corresponds to a Dicke state with one excitation. 
While the $W$ state is $N$-particle entangled \cite{DurPRA2000}, its quantum Fisher information only amounts to $3N-2$.

\subsubsection{Metrological spin squeezing}
\label{SubSec.SpinSq}

Spin-squeezed states are a class of states having squeezed spin variance along a certain direction, 
at the cost of anti-squeezed variance along an orthogonal direction. 
Spin squeezing is one of the most successful approaches to witness large-scale quantum entanglement beating the standard quantum limit in interferometry.

Let us consider the unitary rotation of a state on the Bloch sphere around an axis $\vect{n}$ perpendicular to the mean spin direction $\vect{s}$, see Fig.~\ref{Fig:rotations}(b), 
and calculate the phase sensitivity according to 
the error propagation formula, Eq.~\eqref{Eq.ErrProp}.
We can write $\Delta \ps\si{mom} = \xiR / \sqrt{\nu N}$,
\footnote{In the literature, it is possible to find different notations for the metrological spin-squeezing parameter (\eg, $\xi$ and $\xi\si{S}$ are also commonly used).
Here we follow the notation first introduced by \textcite{WinelandPRA1994} and used in a previous review \cite{MaPHYSREP2011}.
In particular, in this review $\xiS$ refers to the \textcite{KitagawaPRA1993} spin-squeezing parameter, see Eq.~\ref{Eq.SSKU}.} 
where
 \be \label{Eq:xiWineland}
\xiR^2 = \frac{ N (\Delta \J_{\perp})^2 }{\mean{\J_{\vect{s}}}^2},
\ee
and $\perp$ is orthogonal to both $\vect{s}$ and $\vect{n}$.
$\xiR$ is the spin-squeezing parameter introduced by \textcite{WinelandPRA1992, WinelandPRA1994}. 
If $\xiR^2<1$ holds, the state is said to be (metrologically) spin squeezed along the $\perp$-axis \cite{WinelandPRA1992, WinelandPRA1994} and 
it can be used to overcome the standard quantum limit (\ie, reaching $\Delta \ps\si{mom} < 1/\sqrt{N}$).
This requires states having spin fluctuations orthogonal to the rotation axis 
smaller than the projection noise of uncorrelated atoms, \ie, $( \Delta \J_{\perp}  )^2 < N/4$, and sufficiently 
large spin length $\mean{\J_{\vect{s}}}$.

\textcite{HeNJP2012} have generalized this criterion to systems of fluctuating numbers of particles by introducing scaled spin operators $\vectop{\jmath}=\vectop{J}\op{N}^+$ in terms of the Moore-Penrose pseudoinverse $\op{N}^+$ of the particle number operator.

\paragraph{Optimal spin-squeezed states.} 
\label{Sec.OptimalSS}
Optimal spin-squeezed states are searched among the so-called minimum uncertainty states \cite{AragoneJPA1974, RashidJMP1978, WodkiewiczJOSAB1985}. 
These states saturate the Heisenberg uncertainty relation
\begin{equation}
	\label{Eq.Heisenberg}
	(\Delta \J_{\vect{n}})^2 (\Delta \J_{\perp})^2 \geq \abs{\mean{\J_{\vect{s}}}}^2/4,
\end{equation}
since $[\J_{\vect{n}},\J_{\perp}]=\ii\J_{\vect{s}}$. 
We thus have $\xiR^2 = N/[4 (\Delta \J_{\vect{n}})^2]$ and a lower bound to $\xiR^2$ is obtained by maximizing 
$(\Delta \J_{\vect{n}})^2$, giving \cite{HilleryPRA1993, AgarwalPRA1994} 
\be \label{Eq.boundxiR}
\xiR^2 \geq \frac{2}{N+2}.
\ee
This bound can be saturated by the state $\sqrt{1-\alpha^2} \ket{0_\perp} + \alpha (\ket{1_\perp}+\ket{-1_\perp})/\sqrt{2}$ in the limit $\alpha \mapsto 0$ \cite{BrifPRA1996}, 
where $\ket{m_\perp}$ are Dicke states defined in Sec.~\ref{Sec.Fundamentals.spin}.
Notice that spin-squeezed states can achieve a Heisenberg scaling of phase sensitivity, $\Delta\ps\si{mom}=\mathcal{O}(N^{-1})$, but not the Heisenberg limit~\eqref{HLbound} for $N > 2$.
Optimal spin-squeezed states for even values of $N$  and fixed values of $\mean{\J_{\vect{s}}}$ are given by the ground state of the Hamiltonian 
$\J_{\perp}^2 - \lambda \J_{\vect{s}}$, where $\lambda \geq 0$ is a Lagrange multiplier \cite{SorensenPRL2001}.

\begin{figure}
\begin{center}
\includegraphics[width=\columnwidth]{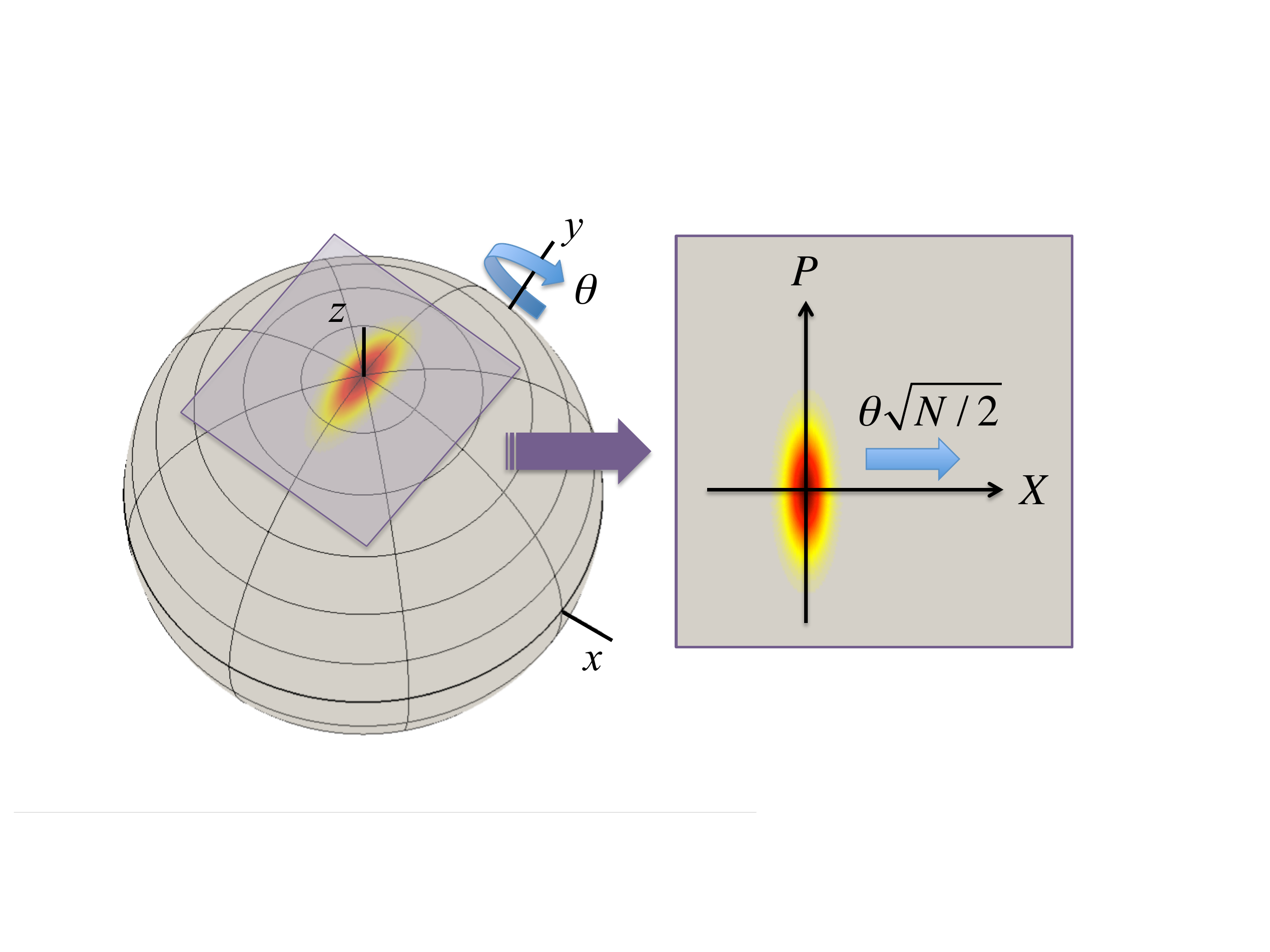}
\end{center}
\caption{{\bf Relation between spin squeezing and quadrature squeezing.} 
For $\mean{\J_z} \approx N/2 \gg 1$, the Bloch sphere can be 
approximated locally as a plane orthogonal to the mean spin direction (here, the $z$-axis). 
Spin squeezing along the $\J_x$ axis is equivalent to squeezing of the $\op{X}$ quadrature. 
A rotation around the $y$-axis is equivalent to a displacement in the $X$ direction.} 
\label{Fig:quadrature}
\end{figure}

\paragraph{Spin squeezing and bosonic quadrature squeezing.}
\label{Sec.QuadSq}
In the case of probe states having $N\gg 1$ and a strong population imbalance between the two modes, 
spin squeezing can be well approximated by single-mode quadrature-squeezing.  
Let $\ket{a}$ be the highly-populated mode (continuous-variable limit, $\mean{\op{a}^\dag \op{a}} \approx N \gg 1$) and 
perform the Holstein-Primakoff transformation \cite{WangPRA2003b, MadsenPRA2004, DuanPRA2002} 
$\J_+ / \sqrt{N} \mapsto \op{b}$, $\J_- / \sqrt{N} \mapsto \op{b}^\dag$, and $2\J_z/N \mapsto 1$, 
formally equivalent to the mean-field replacement $\op{a} \mapsto \sqrt{N}$. 
Within this approximation, the rescaled spin operators $\sqrt{2/N} \J_x$ and $\sqrt{2/N} \J_y$ 
map onto the position 
$\sqrt{2/N}\J_x  \mapsto (\op{b} + \op{b}^\dag)/\sqrt{2} = \op{X}$
and momentum 
$ \sqrt{2/N}\J_y \mapsto (\op{b} - \op{b}^\dag)/(\ii\sqrt{2}) = \op{P}$
quadrature operators
\cite{ScullyBOOK}, respectively. 
We thus find
\be \label{xiQ}
\xiR^2 =  2(\Delta \op{Q})^2,
\ee
where $\op{Q} = (\op{b} e^{- \ii \phi} + \op{b}^\dag e^{\ii \phi})/\sqrt{2} = \op{X} \cos \phi  + \op{P} \sin \phi $ and $0\leq \phi < 2 \pi$. 
Equation~\eqref{xiQ} shows the equivalence between the metrological spin-squeezing parameter and the quadrature variance, within the approximations.
In particular, the rotation $e^{-\ii \ps \J_y}$ maps onto a displacement of the state along the $X$ direction in the quadrature plane by an amount $\ps \sqrt{N/2}$, 
described by $e^{-\ii \ps \sqrt{N/2} \op{P}}$, see Fig.~\ref{Fig:quadrature}.
When squeezing the quadrature variance below the vacuum noise limit, \ie, $(\Delta \op{Q})^2<1/2$, it is possible to 
overcome the standard quantum limit of phase sensitivity, \ie, $\xiR^2<1$.
The sensitivity of interferometers using a probe state with all atoms in a single mode can be increased by feeding the other mode with a 
quadrature-squeezed state, as first proposed by \textcite{CavesPRD1981} for an optical interferometer.

\paragraph{Spin squeezing, entanglement and Fisher information.}
Spin squeezing $\xiR<1$ is a sufficient condition for useful particle entanglement in metrology \cite{SorensenNATURE2001}. 
Furthermore, \textcite{SorensenPRL2001} showed that the degree of spin squeezing is related to metrologically useful $k$-particle entanglement:
for a given spin length, smaller and smaller values of $\xiR$ can only be obtained by increasing the entanglement depth, see Fig.~\ref{Fig:SMplot}.
The quantum Fisher information detects entanglement in a larger number of 
states than those recognized by metrological spin squeezing \cite{PezzePRL2009}:
\be \label{FishVsSS}
\frac{N}{\Fish\si{Q}[\op{\rho}, \J_{\vect{n}}]} \leq \xiR^2.
\ee
This inequality, which follows from Eq.~\eqref{Eq.MomentIneq}, shows that  
if a state is spin squeezed, $\xiR^2<1$, 
it also satisfies the condition of metrologically useful entanglement, $\Fish\si{Q}[\op{\rho}, \J_{\vect{n}}] > N$.
The contrary is not true: there are states that are not spin squeezed and yet entangled and useful for quantum metrology. 
The Dicke and NOON states, discussed below, are important examples.

\subsubsection{Dicke states}
\label{sec:DickeStates}

Dicke states, Eq.~\eqref{Eq.DickeState}, have a precise relative number of particles between the two modes and a completely undefined phase.
They are not spin squeezed \cite{WangEPJD2002}. 
A direct calculation of the quantum Fisher information gives 
\be \label{Eq.QFITF}
\Fish\si{Q}\big[\ket{\di_z}, \J_{\perp}\big] = \frac{N^2}{2} - 2 \di^2 + N
\ee
for any rotation direction $\perp$ orthogonal to the $z$ axis.
Dicke states with $\di=\pm N/2$ are coherent spin states; those with $\di \neq \pm N/2$ are metrologically usefully entangled. 
From the perspective of quantum metrology, the most interesting Dicke state is the twin-Fock state \cite{HollandPRL1993, SandersPRL1995}, 
$\ket{N/2}_a \ket{N/2}_b  = \ket{0_z}$, 
corresponding to $N/2$ particles in each mode.
It can be visualized as a ring on the equator of the Bloch sphere, see Fig.~\ref{Fig:rotations}(c).
A rotation around any axis in the $x$-$y$ plane converts the well-defined 
number difference into a well defined relative phase between the two modes.
It should be noted that the twin-Fock state has zero mean spin length.
Therefore, the metrologically useful entanglement of the twin-Fock state cannot be exploited when measuring the 
relative number of particles.
A possible phase-sensitive signal is the variance of the relative population \cite{KimPRA1998}, see Sec.~\ref{SubSec.TF}. 
The phase sensitivity calculated via the method of moments
strongly depends on $\ps$: for $\ps \approx 0$ and $N \gg 1$, we have
$(\Delta \ps)^2 =  \frac{2}{\m(N^2 + 2N) } + \mathcal{O}(\ps^2)$, which is a factor two above the Heisenberg limit at $\ps=0$ and remains
below the standard quantum limit for $\ps \lesssim 1/\sqrt{N}$.
Similar results can be obtained with error propagation when estimating the phase shift from the measurement of 
the parity operator \cite{CamposPRA2003, GerryPRL2004}. Parity measures the difference in populations between 
even and odd eigenstates of $\J_z$ and can be difficult to implement for large spins.
A $\ps$-independent phase sensitivity can be reached when measuring the number of particles at the output ports of the interferometer and
using a maximum likelihood estimator or a Bayesian method \cite{KrischekPRL2011, HollandPRL1993, PezzePRA2006}. 

Squeezing the number of particles at both inputs of the interferometer is not necessary to overcome the 
standard quantum limit \cite{PezzePRL2013}.  
Let us consider a probe state $\op{\rho} = \op{\rho}_{a} \otimes \ket{N/2}_{b} \bra{N/2}$, where $\op{\rho}_{a}$ is an 
arbitrary state in mode $\ket{a}$ with mean particle number $N_a$
and $\ket{N/2}_{b}$ is a Fock state of $N/2$ particles in mode $\ket{b}$. 
We find
\be
\Fish\si{Q}\big[ \op{\rho}_{a} \otimes \ket{N/2}_b \bra{N/2}, \J_y \big] = N N_a + \frac{N}{2} + N_a.
\ee
Heisenberg scaling is achieved when $N_a = N/2$, without any assumptions on $\op{\rho}_{a}$.
In particular, existing interferometers that operate with uncorrelated atoms can be improved by simply replacing 
the vacuum state in one of the two input ports by a Fock state. 

\subsubsection{NOON states}
\label{sec:NOONstate}

The Heisenberg limit can be saturated by the state
\be \label{MaxEnt}
\ket{\text{NOON}} = \frac{ 
\ket{N}_a \ket{0}_b + e^{\ii \phi} \ket{0}_a \ket{N}_b}{\sqrt{2}},
\ee
given by a coherent superposition of all particles in mode $\ket{a}$ and all particles in mode $\ket{b}$, where $\phi$ is an arbitrary phase.
This state is called NOON state \cite{LeeJMO2002} when considering indistinguishable bosonic particles.
When considering distinguishable particles, as ions in a Paul trap for instance, see Sec.~\ref{Sec.Ions}, the state~\eqref{MaxEnt} is generally called a
``Schr\"odinger cat'' \cite{BollingerPRA1996, LeibfriedNATURE2006} or Greenberger-Horne-Zeilinger state \cite{MonzPRL2011},
originally introduced in \textcite{GreenbergerAJP1990} for three particles.
A look at the Wigner distribution of the NOON state, see Fig.~\ref{Fig:rotations}(d,left), reveals substructures of angular size $1/N$ 
given by spherical harmonic contributions $Y_k^q$ with the maximum allowed value $k=N$ \cite{SchmiedNJP2011}.
Rotating the NOON state around the $z$-axis, the initial and final states becomes distinguishable after a rotation angle $\ps\si{min} \approx 1/N$.
The squared Bures distance between the probe and the rotated state is
\be
\bures^2\big( \ket{\text{NOON}}, e^{-\ii \ps \J_z} \ket{\text{NOON}} \big) = 1 - \cos (N \ps / 2 ), 
\ee
which oscillates in phase $N$ time faster than the corresponding overlap for a coherent spin state, Eq.~\eqref{Eq.overlapCSS}, and 
is $\bures^2 = N^2 \ps^2/8 + \mathcal{O}(\ps^4)$ for small $\ps$ \cite{PezzeEPL2007}.
Note that the relative spin probability distribution of the NOON state, $P(\di_y) = \abs{\me{\di_y}{e^{-\ii \ps \J_z}}{\text{NOON}}}^2$, 
see Fig.~\ref{Fig:rotations}(d,right), shows a comb-like structure as a function of $m$.
These substructure change quickly with $\ps$:
for $\ps=2\pi n /N$ only even values of $m$ are populated, for $\ps=\pi(1+2n)/N $ only odd values of $m$ are populated, with $n=0,1,...,N-1$.
According to Eq.~\eqref{Eq.QFI2}, we obtain $\Fish\si{Q}[ \ket{\text{NOON}}, \J_z] = 4 (\Delta \J_z)^2 = N^2$, 
and thus $\Delta \ps\si{QCR} = \Delta \ps\si{HL}$ for the NOON state.
It is possible to reach this sensitivity via the method of moments by measuring the parity of the relative number of particles among 
the two modes \cite{BollingerPRA1996}, see also Sec.~\ref{GHZions}.

\subsubsection{Further notions of spin squeezing and their relation to entanglement}
\label{Sec.FurtherSS}

When it is not possible to address individual qubits, or in presence of low counting statistics, entanglement criteria based 
on the measurement of collective properties---as the condition $\xiR<1$ introduced in Sec.~\ref{SubSec.SpinSq}---are 
experimentally important.
Moreover, states of a large number of particles cannot be characterized via full state tomography:
the reconstruction of the full density matrix is hindered and finally prevented by the exponential increase in the required number of measurements. 
In the literature, different definitions of spin squeezing for collective angular momentum operators can be found \cite{MaPHYSREP2011, TothJPA2014}.
In the following we review the ones most relevant for the present context. 

\begin{figure}[t!]
\begin{center}
\includegraphics[width=\columnwidth]{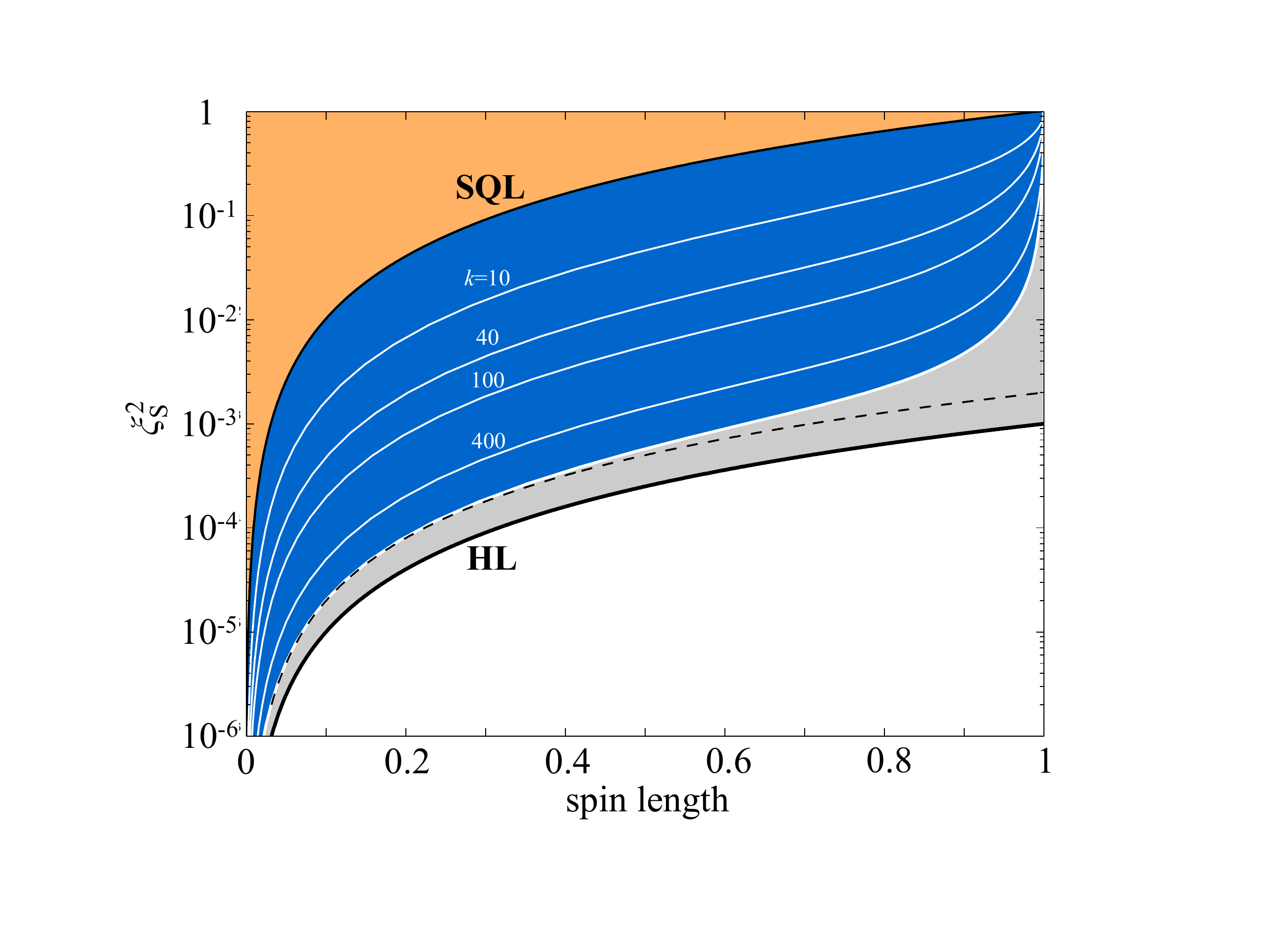}
\end{center}
\caption{{\bf $\xiS^2$ versus $\xiR^2$.}
The metrological spin-squeezing parameter $\xiR^2$ is given by the ratio between the
Kitagawa-Ueda spin-squeezing parameter $\xiS^2$ and the squared spin length (Ramsey contrast) $2 \abs{\mean{\J_{\vect{s}}}}/N$.
$\xiR^2<1$ is found in the blue/dark region: the upper limit is the standard quantum limit,
while the lower limit is the bound obtained for optimal spin-squeezed states, see Sec.~\ref{Sec.OptimalSS}.
The dashed line is Eq.~\eqref{Eq.boundxiR}, which is tight in the limit $\mean{\J_{\vect{s}}} \to 0$. 
The lower thick black line is the Heisenberg limit.
The orange region above the SQL highlights the regime of parameters showing $\xiS^2<1$ but no metrological spin squeezing (\ie, $\xiR^2\geq 1$). 
The white lines are lower bounds to $\xiR^2$ obtained for $k$-particle entangled states \cite{SorensenPRL2001}. 
The grey region is not accessible by metrological spin-squeezed states.
Here $N=1000$.}
\label{Fig:SMplot}
\end{figure}

\paragraph{Squeezing parameter of Kitagawa and Ueda.}
A spin-1/2 particle is characterized by isotropic spin 
fluctuations, equal to 1/4, along any direction orthogonal to the mean spin direction $\vect{s}$.
By adding $N$ uncorrelated spins all pointing along $\vect{s}$ (as in a coherent spin state), 
we have $(\Delta \J_{\perp})^2 =N/4$, where $\perp$ is an arbitrary direction orthogonal to $\vect{s}$.
Quantum correlations between spins may result in reduced fluctuations in one direction, $(\Delta \J_{\perp})^2 < N/4$, at the expense of enhanced 
fluctuations along the other direction orthogonal to $\vect{s}$.
This suggests the introduction of the spin-squeezing parameter \cite{KitagawaPRA1993}
\be \label{Eq.SSKU}
\xiS^2 = \frac{4 \min_\perp (\Delta \J_{\perp})^2}{N},
\ee
$\xiS^2 < 1$ being the spin-squeezing condition.
Equation~\eqref{Eq.SSKU} is related to metrological spin squeezing via the relation
$\min_\perp \xiR^2 = [N/(2\mean{\J_{\vect{s}}})]^2 \xiS^2$.
Since $\abs{\mean{\J_{\vect{s}}}}\leq N/2$, we obtain 
\be
\xiS^2 \leq \xiR^2.
\ee
In other words, metrological spin squeezing, $\xiR^2 < 1$, implies spin squeezing according to the definition of Kitagawa and Ueda. 
The converse is not true: there is no direct relation between $\xiS^2<1$ and the improvement of metrological sensitivity, 
as illustrated in Fig.~\ref{Fig:SMplot}. 
It is worth noting that the minimum in Eq.~\eqref{Eq.SSKU} is given by the smallest eigenvalue of the covariance matrix 
$\Cov(\J_{\vect{n}_i}, \J_{\vect{n}_j} ) = \mean{\{\J_{\vect{n}_i},\J_{\vect{n}_j}\}}/2 - 
\mean{\J_{\vect{n}_i}} \mean{\J_{\vect{n}_j}}$, where $\vect{n}_1,\vect{n}_2$ are two mutually orthogonal directions
in the plane perpendicular to $\vect{s}$ and $\{ \J_{\vect{n}_1}, \J_{\vect{n}_2} \} = \J_{\vect{n}_1}\J_{\vect{n}_2}+\J_{\vect{n}_2}\J_{\vect{n}_1}$. 
Taking, without loss of generality,  $\mean{\op{J}_{\vect{n}_{1}}} = \mean{\op{J}_{\vect{n}_{2}}} = 0$, we have \cite{WangPRA2003}
\be
\xiS^2 = \frac{ \mean{\J_{\vect{n}_1}^2 + \J_{\vect{n}_2}^2} - \sqrt{ \mean{\J_{\vect{n}_1}^2 - \J_{\vect{n}_2}^2}^2 + \mean{\{\J_{\vect{n}_1}, \J_{\vect{n}_2}\}}^2   }}{N/2}.
\ee

\paragraph{Entanglement witnessed by mean values and variances of spin operators.}
For separable states~\eqref{seps}, the inequalities 
\begin{subequations} \label{ineq}
\begin{align} 
        & N (\Delta \J_{\vect{n}_1})^2  \geq  \mean{\J_{\vect{n}_2}}^2+\mean{\J_{\vect{n}_3}}^2, \label{Ineq1}\\
	& (\Delta \J_{\vect{n}_1})^2 + (\Delta \J_{\vect{n}_2})^2 + (\Delta \J_{\vect{n}_3})^2 \geq \frac{N}{2}, \label{Ineq2}\\
	& (N-1) (\Delta \J_{\vect{n}_1})^2 \geq \mean{\J^2_{\vect{n}_2}}+\mean{\J^2_{\vect{n}_3}} - \frac{N}{2}, \label{Ineq3}\\
	& (N-1) \big[ (\Delta \J_{\vect{n}_1})^2 + (\Delta \J_{\vect{n}_2})^2\big] \geq \mean{\J^2_{\vect{n}_3}} + \frac{N(N-2)}{4} \label{Ineq4}
\end{align}
\end{subequations}
are all fulfilled, where $\vect{n}_1$, $\vect{n}_2$ and $\vect{n}_3$ are three mutually orthogonal directions. 
The violation of at least one of the above inequalities signals that the state is entangled. 
Equation~\eqref{Ineq1} is equivalent to $\xiR^2 \geq 1$, see Sec.~\ref{SubSec.SpinSq}, and was introduced by \textcite{SorensenNATURE2001}. 
The inequalities~\eqref{Ineq2}-\eqref{Ineq4} have been introduced by \textcite{TothPRL2007, TothPRA2009}.
A violation of the condition~\eqref{Ineq2} can be used to detect entanglement in singlet states \cite{TothNJP2010, BehboodPRL2014}.
The third condition, Eq.~\eqref{Ineq3}, can be rewritten as $\xiD^2 \geq 1$, where
\be \label{Eq.xiE}
\xiD^2 = \frac{ N (\Delta \J_{\vect{n}_1})^2 } {\mean{\vect{\J}^2} - N/2 - \mean{\J_{\vect{n}_1}}^2}.
\ee
In particular, the condition $\xiD^2<1$ can be used to detect entanglement close to Dicke states \cite{TothPRL2007}, see also \textcite{RaghavanOPTCOMM2001}.
The detection of multiparticle entanglement close to Dicke states for spin-1/2 particles has been studied by \textcite{DuanPRL2011, LueckePRL2014}
and for spin-$J$ particles with $J>1/2$ by \textcite{VitaglianoNJP2017}.
The inequalities~\eqref{Ineq2}-\eqref{Ineq4} and the further inequality 
$\mean{\vect{\J}^2} \leq (N/2)(N/2+1)$, which is valid for all quantum states
(not only for separable states), 
form a system of conditions that defines a polytope in the three dimensional 
space with coordinates $\mean{\J_{\vect{n}_1}^2}$, $\mean{\J_{\vect{n}_2}^2}$, and $\mean{\J_{\vect{n}_3}^2}$ \cite{TothPRL2007, TothPRA2009}.
The polytope encloses all separable states. 
It has been demonstrated that Eqs.~\eqref{ineq} form a complete set \cite{TothPRL2007,TothPRA2009}, meaning that
it is not possible to add new entanglement conditions based on mean values and variances of spin moments that detect more entangled states. 
The inequalities~\eqref{ineq} have been generalized to arbitrary spin systems \cite{VitaglianoPRL2011, VitaglianoPRA2014}
and to systems of fluctuating numbers of particles \cite{HyllusPRA2012b}. 
Furthermore, \textcite{KorbiczPRL2005, KorbiczPRA2006} have shown that, if the inequality 
\begin{multline}
\label{xiL}
\bigg( \mean{\J_{\vect n_1}^2} + \frac{N(N-2)}{4}\bigg)^2\\
< \bigg( \mean{\J_{\vect n_2}^2} + \mean{\J_{\vect n_3}^2} - \frac{N}{2}\bigg)^2 + (N-1)^2 \mean{\J_{\vect n_1}}^2
\end{multline}
holds, then the state possesses pairwise entanglement, 
\ie, entanglement in the two-qubit reduced density matrix $\op{\rho}_{i,j} = \Tr_{\{1,...,N\}\setminus\{i,j\}}[\op{\rho}]$
obtained by tracing the $N$ qubit state $\op{\rho}$ over all particles except the $i$th and $j$th.

\paragraph{Spin squeezing and entanglement of symmetric states.}
We emphasize that none of the entanglement witnesses above require any assumptions on the symmetry of the state.
For states that are symmetric under particle exchange, we have $\mean{\vect{\J}^2} = (N/2)(N/2+1)$.
In this case, Eq.~\eqref{xiL} can be rewritten as
\be \label{Eq.xiNineq}
\xiN^2 = \frac{4 (\Delta \J_{\vect n})^2}{N} < 1 - \frac{4 \mean{\J_{\vect n}}^2}{N^2},
\ee
where $\xiN^2$ is called the number-squeezing parameter.
The inequality~\eqref{Eq.xiNineq} is necessary and sufficient for pairwise entanglement \cite{KorbiczPRL2005, KorbiczPRA2006}.
It should be noted that if $\xiS^2<1$ holds, then the inequality~\eqref{Eq.xiNineq} is satisfied as well (taking $\vect{n}=\perp$ and $\mean{\J_{\vect n}}=0$). 
Hence, symmetric spin-squeezed states possess two-qubit entanglement.
The converse is not true: since $\vect{n}$ in Eq.~\eqref{Eq.xiNineq} is not necessarily orthogonal to the mean spin direction $\vect{s}$, 
number squeezing ($\xiN^2<1$) does not imply spin squeezing ($\xiS^2<1$).

The relationship between Kitagawa-Ueda spin squeezing and pairwise 
entanglement has also been studied by \textcite{Ulam-OrgikhPRA2001, WangPRA2003}.
For an arbitrary symmetric state of $N$ qubits, the spin-squeezing parameter~\eqref{Eq.SSKU}
can be written in terms of the two-spin correlation function \cite{Ulam-OrgikhPRA2001}
\be
\xiS^2 = 1+(N-1) \mean{\op{\sigma}_{\perp}^{(i)} \otimes \op{\sigma}_{\perp}^{(j)}}, \quad i\neq j.
\ee
This equation shows that spin squeezing $\xiS^2<1$ is equivalent to negative pairwise spin-spin correlations that, in turn, are sufficient for pairwise
entanglement.\footnote{The two-qubit reduced density matrix of a separable symmetric state of $N$ qubits has positive pairwise correlations: 
$\mean{\op{\sigma}_{\perp}^{(i)} \otimes \op{\sigma}_{\perp}^{(j)}}>0$ for any $i\neq j$ \cite{WangPRA2003}.}
Furthermore, for symmetric pure states of two qubits, there is a direct correspondence between $\xiS^2$ and the concurrence 
$\mathcal{C}$ \cite{Ulam-OrgikhPRA2001, WangPRA2003}: $\xiS^2 = 1 - \C$.
We recall that $\C > 0$ is a necessary and sufficient condition of---and quantifies---entanglement of a pair of qubits \cite{HillPRL1997, WoottersPRL1998}.
Symmetric pure states of two qubits are entangled if and only if they satisfy $\xiS^2<1$ \cite{Ulam-OrgikhPRA2001}.
For symmetric states of $N$ qubits that fulfill $\xiS^2 \leq 1$ and other conditions,\footnote{Equation~\eqref{xiSC} has been derived in \textcite{WangPRA2003} for symmetric states having $\mean{\J_{\vect{s}}}\neq 0$ and 
$\mean{\J_{\vect{n}_i}} = \mean{\J_{\vect{s}} \J_{\vect{n}_i}}=\mean{\J_{\vect{n}_i} \J_{\vect{s}} } = 0$, $i=1,2$, 
where $\vect{n}_1$ and $\vect{n}_2$ are vectors orthogonal to the mean spin direction $\vect{s}$.} the equality  
\be \label{xiSC}
\xiS^2 = 1 - (N-1) \C
\ee
holds \cite{WangPRA2003}, where $\C$ is calculated from the two-particles reduced density matrix.
Equantion~\eqref{xiSC} tells us that $\xiS^2 < 1$ implies $\C>0$ and thus pairwise entanglement. 
When $\xiS^2 > 1$, Eq.~\eqref{xiSC} breaks down and we cannot draw any 
conclusion about pairwise entanglement: for example,
Dicke states can be pairwise entangled even though they are not spin squeezed \cite{WangEPJD2002}.

\paragraph{Planar spin-squeezed states.}
\label{Sec.PQSintro}
While many useful spin-squeezed states have reduced quantum fluctuations along a single spin direction (with a corresponding increase in fluctuations along a perpendicular direction), the spin commutation relations make it possible to reduce the fluctuations along two orthogonal spin directions simultaneously while increasing those along a third direction. Specifically, an initially coherent state along the $\vect{n}_3$ direction can be squeezed in the perpendicular plane such that simultaneously $(\Delta \J_{\vect{n}_1})^2 < J/2$ and $(\Delta \J_{\vect{n}_2})^2 < J/2$ \cite{PuentesNJP2013}, which does not violate Heisenberg's uncertainty relation, Eq.~\eqref{Eq.Heisenberg}, if $\mean{\J_{\vect{n}_3}}$ is reduced at the same time.
In general, such \emph{planar spin-squeezed states} reduce the variance sum $(\Delta\J_{\parallel})^2=(\Delta \J_{\vect{n}_1})^2 + (\Delta \J_{\vect{n}_2})^2$ below the coherent-state value of $(\Delta \J_{\parallel}\se{coh})^2=J$, ultimately limited by \cite{HePRA2011}
\begin{equation}
	(\Delta \J_{\parallel})^2 \ge C_J \approx
	0.595275J^{2/3}+\mathcal{O}(J^{1/3}).
\end{equation}
Planar spin-squeezed states are useful for interferometric phase measurements where the phase fluctuations and the number fluctuations are squeezed simultaneously (see Sec.~\ref{Sec.QND}), while the spin length fluctuates significantly. Note that not all planar spin-squeezed states are entangled \cite{HeNJP2012,PuentesNJP2013,VitaglianoPRA2018}; see Eqs.~\eqref{Ineq2} and~\eqref{Ineq4} for relevant entanglement criteria.

\subsubsection{Einstein-Podolsky-Rosen entanglement and Bell correlations}
\label{Sec.EPRBELL}

\paragraph{Continuous variable and Einstein-Podolsky-Rosen entanglement}
\label{sec:EPR}
Let us consider two bosonic modes, $\ket{a_+}$ and $\ket{a_-}$, and introduce the corresponding annihilation $\op{a}_{\pm 1}$ and creation $\op{a}_{\pm 1}^\dag$ operators. 
Mode-separable quantum states are defined as $\op{\rho}\si{msep} = \sum_q p_q \op{\rho}_q^{(+1)} \otimes \op{\rho}_q^{(-1)}$, 
where $p_q>0$, $\sum_q p_q=1$, and $\op{\rho}_q^{(\pm1)}$ is the state of the $\ket{a_{\pm 1}}$ mode.
Mode entanglement, \ie, $\op{\rho} \neq \op{\rho}\si{msep}$, can be revealed by correlations between 
bosonic position $\op{X}_{\pm 1}(\phi) = (\op{a}_{\pm 1} e^{-\ii \phi} + \op{a}_{\pm 1}^\dag e^{\ii \phi})/\sqrt{2}$ 
and momentum $\op{P}_{\pm 1}(\phi) = \op{X}_{\pm 1}(\phi+\pi/2)$ quadratures \cite{ReidRMP2009}.
Mode-separable states fulfill \cite{DuanPRL2000cv, SimonPRL2000}
\be \label{Eq.SEPcriterion}
V_{X(\phi)}^{\pm} + V_{P(\phi)}^{\mp} \geq 2,
\ee 
where $V_{X(\phi)}^{\pm} = \Var[\op{X}_{+ 1}(\phi) \pm \op{X}_{- 1}(\phi)]$ and $V_{P(\phi)}^{\mp} = \Var[\op{P}_{+ 1}(\phi) \mp \op{P}_{- 1}(\phi)] $ are variances.
A violation of this condition detects entanglement between the $\ket{a_{\pm 1}}$ modes.
It is also a necessary and sufficient condition for mode entanglement in Gaussian states \cite{DuanPRL2000cv, SimonPRL2000}, see also
\cite{GiovannettiPRA2003, WalbornPRL2009, ShchukinPRL2005, GessnerPRA2016, GessnerPRA2017}
for further (and sharper) conditions.
Mode entanglement finds several applications in quantum technologies \cite{BraunsteinRMP2005}.

Correlations between quadrature variances are at the heart of the Einstein-Podolsky-Rosen (EPR) paradox \cite{EPR1935}.
When the quadratures $\op{X}_{+1}$ and $\op{P}_{+1}$ are measured in independent realizations of the same state, the correlations 
allow for a prediction of $\op{X}_{-1}$ and $\op{P}_{-1}$ with inferred variances violating the Heisenberg uncertainty relation 
$\Delta \op{X}_{-1}\se{inf} \Delta \op{P}_{-1}\se{inf} < 1/2$, known as EPR criterion \cite{ReidPRA1989, ReidRMP2009}. 
This extends the original EPR discussion that was limited to perfect quadrature correlations.
Non-steerable states, including separable states, fulfill \cite{ReidPRA1989}
\be \label{Eq.EPRcriterion}
V_{X(\phi)}^- \times V_{P(\phi)}^+ \geq 1/4.
\ee 
The violation of this condition witnesses a strong form of entanglement (``EPR entanglement'') necessary to fulfill the EPR criterion. 
With atoms, continuous-variable entanglement has been first proved with room-temperature vapor cells \cite{JulsgaardNATURE2001}. 
With spinor Bose-Einstein condensates, mode entanglement \cite{GrossNATURE2011} and 
EPR entanglement \cite{PeiseNATCOMM2015} have been demonstrated, see Sec.~\ref{Sec.Spin-Mix.squeezing}.

\paragraph{Bell correlations.}
\label{sec:BellCorrelations}

The strongest form of correlations between particles are those that violate a Bell inequality \cite{BellPHYSICS1964}. 
The existence of Bell correlations has profound implications for the foundations of physics and underpins quantum technologies such as quantum key distribution and certified randomness generation \cite{BrunnerRMP2014}. Bell correlations have been observed in systems of at most a few (usually two) 
particles \cite{FreedmanPRL1972,ZhaoPRL2003, EiblPRL2003,HofmannSCIENCE2012,LanyonPRL2014,HensenNATURE2015,AspectPRL1982,GiustinaPRL2015,ShalmPRL2015,RosenfeldPRL2017,MatsukevichPRL2008}, 
but their role in many-body systems is largely unexplored \cite{TuraSCIENCE2014}.

Entanglement is necessary but not sufficient for Bell correlations \cite{WernerPRA1989,BrunnerRMP2014}.
Therefore, entanglement criteria as those discussed above,
cannot be used to determine whether a system could violate a Bell inequality.
In \textcite{SchmiedSCIENCE2016} a criterion in the spirit of a spin-squeezing parameter is derived to 
determine whether Bell correlations are present in an $N$-particle quantum system. 
For any two axes $\vect{n}_1$ and $\vect{n}_2$, the inequality
\begin{equation}
	\label{eq:BellWitnessFull}
	\mathcal{W}=-\frac{\abs{\mean{\J_{\vect{n}_2}}}}{N/2} + (\vect{n}_1\cdot\vect{n}_2)^2
	\frac{\mean{\J_{\vect{n}_1}^2}}{N/4}
	+1-(\vect{n}_1\cdot\vect{n}_2)^2 \ge 0
\end{equation}
is satisfied for all states that are not Bell correlated. 
States that satisfy $\mathcal{W}<0$ can violate the many-particle Bell inequality of \textcite{TuraSCIENCE2014}, which is a 
statement about the strength of two-body correlations, but does not imply the violation of a two-particle Bell inequality for every pair of particles.
By optimizing Eq.~\eqref{eq:BellWitnessFull} over the angle between the two axes, a criterion follows that facilitates comparison with well-known spin-squeezing criteria: for any two axes $\vect{n}_1$ and $\vect{n}_2$ perpendicular to each other,
\begin{equation}
	\label{eq:BellWitness}
	\frac{\mean{\J_{\vect{n_1}}^2}}{N/4} \ge \frac12 \left[ 1-\sqrt{1-\left(\frac{\mean{\J_{\vect{n}_2}}}{N/2}\right)^2}\right]
\end{equation}
is satisfied for all states that are not Bell correlated.
A similar criterion that is violated more easily was derived by \textcite{WagnerPRL2017},
\begin{equation}
	\label{eq:BellWitness2}
	\frac{\mean{\J_{\vect{n_1}}^2}}{N/4} \ge 1-\Big(\frac{\mean{\J_{\vect{n}_2}}}{N/2}\Big)/\artanh\Big(\frac{\mean{\J_{\vect{n}_2}}}{N/2}\Big).
\end{equation}
Detecting Bell correlations by violating inequality~\eqref{eq:BellWitnessFull}, \eqref{eq:BellWitness}, or~\eqref{eq:BellWitness2} requires only collective manipulations and measurements on the entire $N$-particle system.
While this does not provide a loophole-free and device-independent Bell test, it is a powerful tool for characterizing correlations in many-body systems state-independently.
Bell correlations according to Eqs.~\eqref{eq:BellWitnessFull}, \eqref{eq:BellWitness}, and~\eqref{eq:BellWitness2} have been observed with Bose-Einstein condensates, see Sec.~\ref{SubSec.BellCorr}.

\subsection{Tomography of spin states}
\label{Sec.Fundamentals.tomography}

\subsubsection{Spin-noise tomography}
\label{sec:SpinNoiseTomo}

Spin-noise tomography is a widely-used technique to gain information about spin-squeezed states, whose main characteristics are captured by their second spin moments along the squeezing and anti-squeezing directions.
For this tomography, the state is rotated by an angle $\vartheta$ using resonant Rabi rotations, followed by projective spin measurements along the $z$-axis:
$\op{J}_z$ defined in Eqs.~\eqref{Eq.Joperators_symmetric} is measured by counting the numbers of particles in the two modes, see Sec.~\ref{sec:detection}.
The $k$th moment of these spin projections can be fit to a linear combination of $\cos(n\vartheta)$ and $\sin(n\vartheta)$ with $n\in\{k,k-2,k-4,\ldots\}$, 
which allows interpolating these projective measurements to arbitrary angles $\vartheta$.
Also, technical noise sources can be characterized and their influence subtracted from the resulting moments; 
any spin squeezing concluded from these inferred moments will be called \emph{inferred} spin squeezing.
Since this method estimates spin projection moments separately, they do not necessarily fulfill all consistency criteria imposed by the positive-semi-definiteness of the density operator \cite{SchmiedJMO2016}. 
In practical situations concerning spin-squeezed states, however, this restriction is often irrelevant.
The information captured from low-order spin moments through spin-noise tomography 
may be insufficient to characterize and visualize general spin states. 
Different techniques have been developed for full state tomography \cite{BlumeKohoutNJP2010,QuantumStateEstimation,SchmiedNJP2011}, 
estimating all spin moments up to order $N$.

\begin{figure}
\begin{center}
\includegraphics[width=\columnwidth]{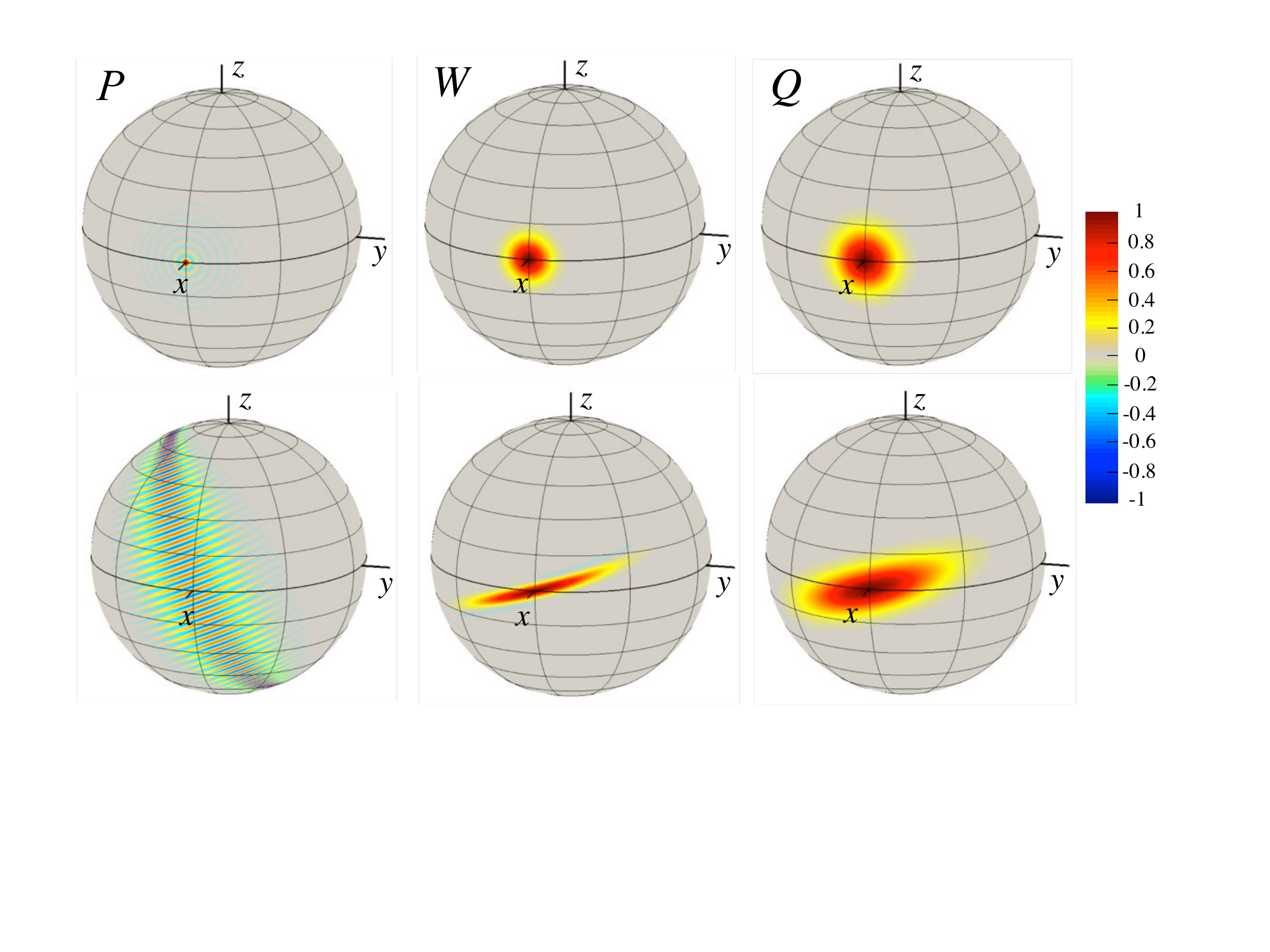}
\end{center}
\caption{{\bf Quasi-probability representations.}
Upper row: $P$, $W$, and $Q$ representations of a coherent state pointing in the $+x$ direction, see Table~\ref{Tab:dist}, with $\vect{s}=\{1,0,0\}$.
Lower row: $P$, $W$, and $Q$ representations of a spin-squeezed state reached by one-axis twisting with $\chi t=0.01\pi$ (see Sec.~\ref{Sec.Atom-Atom.twisting}).
The color scale for each panel is normalized to its maximum value. Here $N=100$.}
\label{FigPWQ}
\end{figure} 

\subsubsection{Visualizing spin states}
\label{sec:WignerFunction}

State representations on the Bloch sphere are very useful to 
gain an intuition about the properties of quantum states.
In the following we consider
symmetric states of $N$ spin-1/2 particles (see Sec.~\ref{Sec.Fundamentals.spin}).
There are various representations in the form of pseudo-probability densities \cite{SchleichBook}. These are
based on the decomposition of a general density operator 
$\op{\rho}=\sum_{k=0}^{2J} \sum_{q=-k}^k \rho_{k q}\op{T}_{k q}^{(J)}$
into the ortho-normalized spherical tensor (or multipole) operators
$\op{T}_{k q}^{(J)}=\sum_{\mu,\mu'=-J}^J (-1)^{J-\mu'} \scp{J,\mu;J,-\mu'}{k,q} \ket{J,\mu_z}\bra{J,\mu'_z}$
defined in terms of Clebsch-Gordan coefficients, with  $\rho_{k q}=\Tr [\op{\rho}(\op{T}_{k q}^{(J)})\dagg ]$ 
\cite{ArecchiPRA1972,AgarwalPRA1981,DowlingPRA1994,SchmiedNJP2011}.
This decomposition has properties similar to the Fourier transform in Euclidean space: it separates low-frequency components 
(small $k$) from high-frequency components (large $k$), which are affected very differently by experimental noise.
Further, it allows us to define the family of spherical functions \cite{AgarwalPRA1981}
\begin{equation} \label{eq:sphericalquasiprobability}
	f(\vartheta,\varphi) =
	\sqrt{\frac{N + 1}{4 \pi}}
	\sum_{k=0}^{N} f_k \sum_{q=-k}^k \rho_{k q} Y_{k q}(\vartheta,\varphi)
\end{equation}
in terms of spherical harmonics $Y_{k q}(\vartheta,\varphi)$. All of these functions are linear in the density operator.
The following representations (corresponding to different choices for the coefficients $f_k$) are often used, with 
examples given in Table~\ref{Tab:dist} and Fig.~\ref{FigPWQ}.

\begin{table}
\caption{ 
{\bf Distributions of a coherent spin state.} 
Different representations of a coherent spin state with mean spin direction $\vect{s}$; 
the last column is the approximate expression for $N \gg 1$.
Here $x= \{ \sin \vartheta \cos \varphi, \sin \vartheta \sin \varphi, \cos \vartheta \} \cdot \vect{s}$, 
$P_k (x)$ are Legendre polynomials, $J_1(x)$ is a Bessel function of the first kind, and $\zeta=\cos^{-1}(x)$.} 
\label{Tab:dist}
\begin{tabular}{lll}
\hline\hline
 & exact & $N\gg1$\\
\hline
$P(x)$ & ${=}\sum_{k=0}^N \frac{2k+1}{4\pi} P_k(x)$ & ${\approx}\frac{(N+1)^2}{2\pi}\frac{J_1[(N+1)\zeta]}{(N+1)\zeta}$\\
$W(x)$ & ${=}\sum_{k=0}^N \frac{2k+1}{4\pi} \sqrt{\frac{N!(N+1)!}{(N-k)!(N+k+1)!}} P_k(x)$ & ${\approx}\frac{N+\frac32}{2\pi}\exp[-\frac{N+\frac32}{2}\zeta^2]$\\
$Q(x)$ & ${=}\frac{N+1}{4\pi} \left( \frac{1+x}{2} \right)^{N}$ & ${\approx}\frac{N+2}{4\pi}\exp[-\frac{N+2}{4}\zeta^2]$\\
\hline\hline
\end{tabular}
\end{table}

\paragraph{Wigner distribution.}
The Wigner quasi-probability distribution $W(\vartheta,\varphi)$ \cite{WignerPR1932} 
corresponds to the case $f_k=1$ in Eq.~\eqref{eq:sphericalquasiprobability}.
It is found by replacing the spherical tensor operators in the decomposition of $\op{\rho}$ by spherical harmonics of the same order, which obey the same ortho-normalization. 
Because of this close similarity, the Wigner quasi-probability distribution is equivalent to the density operator.
$W(\vartheta,\varphi)$ is not a true probability density, 
as it can take negative values \cite{LeibfriedPRL1996, LvovskyPRL2001, McConnellNATURE2015}. 
For continuous variables, this is often understood as a sign of nonclassical behavior.
Note however that in the present finite-dimensional space, even the Wigner distribution of a coherent spin state 
has negative parts of amplitude ${\sim}2^{-N}$, exponentially decreasing with the number of particles.

\paragraph{Husimi-Kano representation.}
The Husimi-Kano Q representation $Q(\vartheta,\varphi)=\frac{N+1}{4\pi}\me{\vartheta,\varphi, N}{\op{\rho}}{\vartheta,\varphi, N}$ \cite{HusimiPMSJ1940,HaasSCIENCE2014,BarontiniSCIENCE2015} 
corresponds to the case $f_k = \sqrt{ \tfrac{N! (N+1)!}{(N-k)!(N+k+1)!} }$ in Eq.~\eqref{eq:sphericalquasiprobability}.
It is nonnegative and proportional to the probability of finding the system in the coherent spin state $\ket{\vartheta, \varphi, N}$.
Since $Q(\vartheta,\varphi)$ is the convolution of $W(\vartheta,\varphi)$ with the Wigner distribution of a coherent spin state \cite{AgarwalPRA1981}, in practice the former contains 
much less information than the latter. Furthermore, recovering the Wigner distribution (and hence the density matrix) from an experimentally determined Q representation is generally not feasible.

\paragraph{Glauber-Sudarshan representation.}
The Glauber-Sudarshan P representation $P(\vartheta,\varphi)$ \cite{GlauberPR1963, SudarshanPRL1963, KieselPRA2008}
is obtained for $f_k = \sqrt{ \tfrac{(N-k)!(N+k+1)!}{N ! (N+1)!} }$ in Eq.~\eqref{eq:sphericalquasiprobability}.
It is the deconvolution of $W(\vartheta,\varphi)$ with the Wigner distribution of a coherent spin state \cite{AgarwalPRA1981}, and serves to construct the density operator from coherent spin states:
$\op{\rho}=\int_0^{\pi} \dd \vartheta \sin \vartheta  \int_0^{2\pi}\dd\varphi P(\vartheta,\varphi)\,\ket{\vartheta,\varphi,N}\bra{\vartheta,\varphi,N}$.
While in infinite-dimensional Hilbert spaces the P representation is often of limited practical use because of its singular behavior \cite{ScullyBOOK}, 
this is not the case for the representation of a spin.
Indeed, in this case, partial-wave contributions are limited to 
angular momenta (spherical harmonics) $k\le N$ in Eq.~\eqref{eq:sphericalquasiprobability} 
in order to match the number of degrees of freedom of the density
operator; the amplitudes of the partial waves with $k>N$ are not determined by the density matrix, and may be set to zero. 
However, if higher-order partial waves ($k>N$) are added, then the P representation of any symmetric separable state can be chosen
nonnegative \cite{KorbiczPRL2005}. 
In this case, the entanglement condition~\eqref{Fishent} is
sufficient for non-classicality \cite{RivasPRL2010}. 
In general, the indeterminacy of the P representation does not allow the conclusion that a 
given P representation with negative regions implies either non-classicality or entanglement:
separable states may have a P representation with negative parts,
as can be seen in Table~\ref{Tab:dist}.

\subsection{Detection of atomic states}
\label{sec:detection}

Quantum metrology requires the detection of large ensembles of $N$ atoms with a
resolution in atom number that is significantly better than $\sqrt{N}$. 
In particular, reaching the Heisenberg limit requires counting the $N$ atoms with single-atom resolution
(we note that this requirement can be relaxed by nonlinear detection, see
\ref{sec:finitedetectionefficiency}).
Traditionally, techniques that provide single-atom detection have only been applied
to systems with at most a few atoms: for quantum metrology, single-atom detectivity
must be combined with a much larger dynamic range.

\subsubsection{Atom counting}
For atomic qubits, there are two principal destructive methods using (near-)resonant
light for counting the number of atoms in one level:
\paragraph{Resonant absorption imaging.} The absorption of a narrow-linewidth
laser beam is measured quantitatively and converted to an absolute atom number
\cite{ReinaudiOL2007}. The precision of this technique on mesoscopic ensembles is currenty
at the level of four atoms
\cite{OckeloenPRA2010,MuesselPRA2015,SchmiedSCIENCE2016,MuesselAPB2013} (standard
deviation on the detection of hundreds atoms).
However, it is state-selective and can be used to measure both $N_a$ and $N_b$ in a
single atomic ensemble, \ie, in a single run of the experiment.
\paragraph{Resonant fluorescence imaging.}
The intensity of atomic fluorescence is
converted to an absolute atom number. This method is used especially for ions
\cite{RoweNATURE2001} but also finds application for atomic ensembles. 
        In free space, single-atom resolution has been achieved in ensembles of up to about
$N=1200$ \cite{HumePRL2013}; however, it is challenging to measure $N_a$ and $N_b$
separately \cite{StroescuPRA2015}.
Very high sensitivity in fluorescence detection of many atoms has been shown by
spatially resolving each atom in an optical lattice \cite{BakrNATURE2009,
ShersonNATURE2010, NelsonNATPHYS2007}.
This technique can image and count individual atoms 
but does not determine the exact atom number as atom pairs are quickly lost due to
light-assisted collisions. \\

In order to count the atom numbers $N_a$ and $N_b$ in the two modes separately,
different strategies have been employed.
If the two modes \ket{a} and \ket{b} are
localized at different points in space, then spatially resolved imaging can yield
mode-selective atom counts \cite{StroescuPRA2015}. Spatial separation can also be
achieved by time-of-flight imaging if two initially overlapping modes are given
different momentum kicks, usually by a state-selective force as in a Stern-Gerlach
experiment \cite{LuckeSCIENCE2011}. This method is often used when the two modes
are hyperfine levels with equal total angular momentum $F$ but different
projections $M_F$.
If the modes are spectrally separated by more
than the atomic linewidth, they can be addressed individually with a laser and
counted separately by absorption or fluorescence. Particularly for states in
different hyperfine $F$ levels this option is used frequently
\cite{RiedelNATURE2010,GrossNATURE2010}. The detection of level \ket{b} can occur
at a different time than level \ket{a} by the same absorption or fluorescence
technique. The population of one level is counted first, followed by a population
exchange or transfer between the levels \ket{a} and \ket{b}, after which the same
level is counted again but now representing the original population of the other
level.

\subsubsection{Quantum non-demolition measurements of atom number}
Off-resonant light can be used to perform non-destructive measurements of the atom numbers \cite{HammererRMP2010, RitschRMP2013}. 
Quantum non-demolition measurement can also be used to entangle the atoms (see Section~\ref{Sec.Atom-Light.nd}).
Specific techniques are:
\paragraph{Faraday effect:} A probe-light beam's polarization rotates slowly
around the polarization direction of an atomic ensemble
\cite{WasilewskiPRL2010,SewellPRL2012}. By measuring the effected rotation angle,
the ensemble's polarization and thus its value of $\op{J}_z$ is determined.
\paragraph{Dispersion:} The refractive index of an atomic ensemble depends on the
atomic populations; an off-resonant probe-light beam thus picks up a
$\op{J}_z$-dependent phase shift \cite{KuzmichPRL2000,AppelPNAS2009} that can be
measured in an optical interferometer (usually of Mach-Zehnder type).
\paragraph{Cavity-enhanced detection:} Atoms that are coupled to a
high-finesse optical cavity make its transmission depend on the atoms' internal
state \cite{KimblePS1998} and allow $\op{J}_z$ of the atoms to be measured
\cite{HostenNATURE2016,McConnellNATURE2015,Schleier-SmithPRL2010,ZhangPRL2012}. For
small atom numbers, this method can resolve single excitations
\cite{HaasSCIENCE2014}.

\subsubsection{Inhomogeneous spin coupling and effective spin}
\label{sec:effectivespin}
The definition in Eq.~\eqref{Eq.collspin} assumes that each atom contributes to the collective spin with the same weight. This assumption is not always satisfied: in experiments exploiting atom-light interactions, the coupling is inhomogeneous if the atoms are trapped in a standing wave whose wavelength is incommensurate with that of the probe field \cite{LerouxPRL2010b,McConnellNATURE2015,TanjiSuzukiAAMOP2011}, if the atoms are trapped in a large volume that samples the spatial profile of the probe field \cite{AppelPNAS2009}, or if the atoms move in space \cite{HammererRMP2010}. In these situations, Eq.~\eqref{Eq.collspin} is modified so that each atom contributes to the collective spin with a weight given by its coupling $g_i$ to the cavity mode. The internal-state dynamics of the effectively addressed atoms can be described by an effective spin operator
\begin{equation}
	\vect{\J}\si{eff} = \sum_i \frac{g_i}{g\si{eff}} \frac{\vectop{\sigma}_i}{2}
\end{equation}
and an effective atom number 
\begin{equation}
	N\si{eff} = \sum_i \frac{g_i}{g\si{eff}}=\frac{(\sum_ig_i)^2}{\sum_ig_i^2}, 
\end{equation}
where $g\si{eff} = (\sum_i g_i^2)/(\sum_i g_i)$ is the effective coupling strength \cite{HuPRA2015}. $\vect{\J}\si{eff}$ satisfies the usual angular momentum commutation and uncertainty relations as long as the average total spin remains large ($\norm{\mean{\vect{\J}\si{eff}}} \approx N\si{eff}/2 \gg 1$) and the detection does not resolve single spins, \ie, in the limit where the Holstein-Primakoff approximation is valid. In this limit, the effective spin can be treated in the same way as a real spin of length $N\si{eff}/2$, and the metrological methods described above remain valid. Special care is required for conclusions about the correlations between real (not effective) atoms, such as the entanglement depth \cite{HuPRA2015,McConnellNATURE2015}.


\section{Entanglement via atomic collisions: the Bosonic Josephson Junction}
\label{Sec.Atom-Atom}

\begin{figure}[t!]
\begin{center}
\includegraphics[width=\columnwidth]{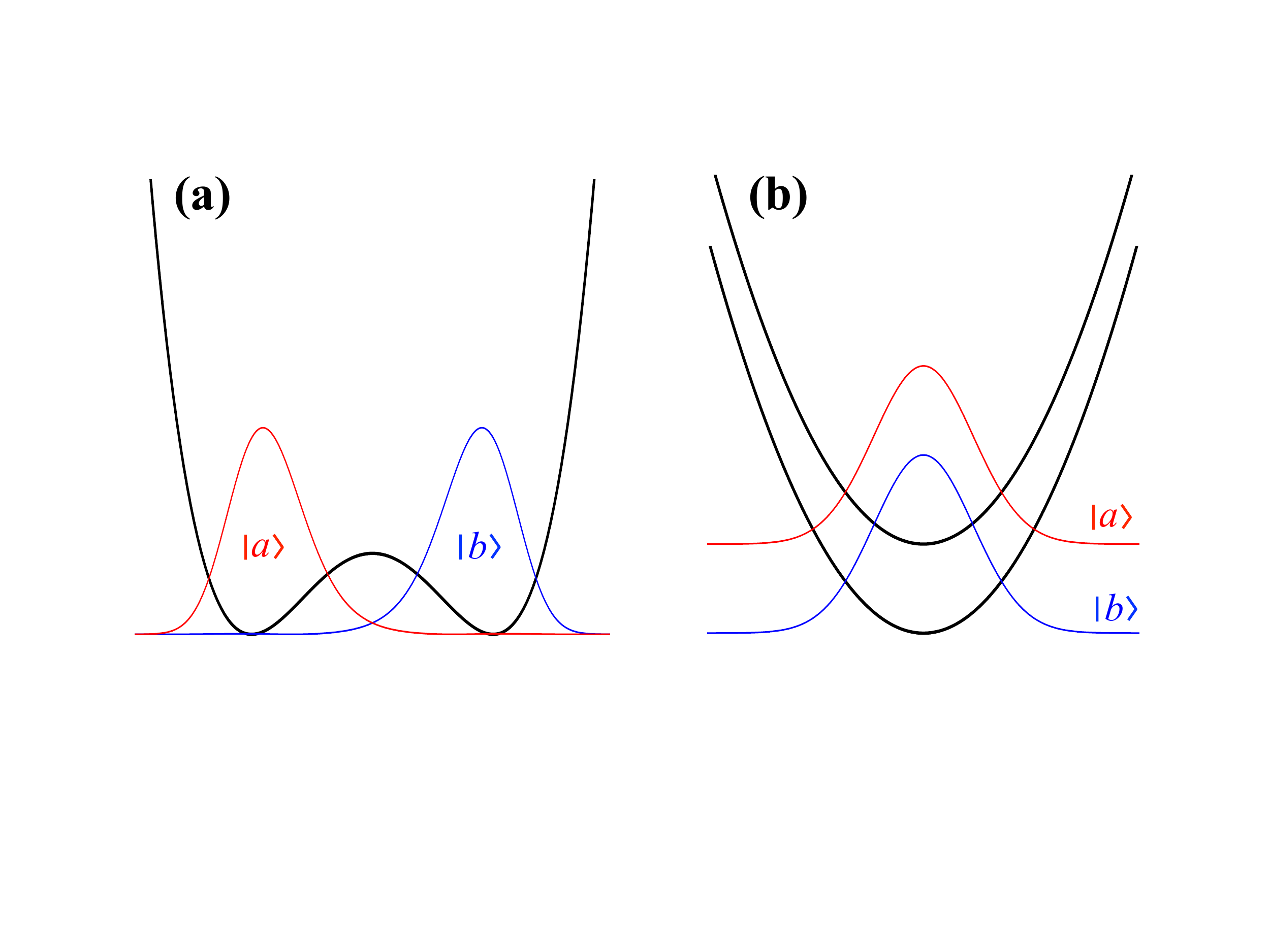}
\end{center}
\caption{\textbf{Bosonic Josephson junction with an atomic Bose-Einstein condensate.} 
(a) External bosonic Josephson junction realized by an ultracold gas trapped in a double-well potential (thick black line).
The thin colored lines are the amplitudes of the mean-field wavefunctions (see text) localized on the left and right wells. 
(b) Internal bosonic Josephson junction made by a trapped Bose-Einstein condensate in two different hyperfine states.
For each state, the thick black line is the harmonic trap and the thin colored line is the amplitude of the single-particle ground state.}  
\label{Fig:DW}
\end{figure}

Tunable elastic atom-atom collisions are naturally present in Bose-Einstein condensates and represent 
a well-established method to generate metrologically useful entanglement in these systems. 
Furthermore, Bose-Einstein condensates have a weak coupling to the environment and can be restricted to occupy two modes only. 
These can be two ``internal'' hyperfine atomic states or two ``external'' spatial states of a trapping potential, see Fig.~\ref{Fig:DW}. Two-mode Bose-Einstein condensates 
can be described by the bosonic Josephson junction model\footnote{The Hamiltonian~\eqref{Eq.BJJ} 
belongs to a class of models first introduced by Lipkin, Meshkov and Glick  
in nuclear physics \cite{LipkinNP1965a, LipkinNP1965b, LipkinNP1965c},
see \textcite{UlyanovPR1992} for a review. This corresponds to a fully connected Ising Hamiltonian of spin-1/2 particles 
where each spin interacts with all the other spins.}
\be \label{Eq.BJJ}
\op{H}\si{BJJ} = - \hbar \Omega \op{J}_x + \hbar \chi \op{J}_z^2 + \hbar \delta \J_z.
\ee
Here $\op{J}_x$ and $\J_z$ describe a linear coupling and an energy imbalance between the two modes, respectively.
The term $\op{J}_z^2$ accounts for the interaction of each atom with all the other particles in the system.
The parameters $\chi$ and $\Omega$ (in the following we assume $\Omega\geq 0$, without loss of generality) 
can be precisely and independently tuned and, in particular, switched on and off at will \cite{PethickBOOK}.
Furthermore, Bose-Einstein condensates offer the possibility to control the trapping geometry and to count atoms
using established techniques such as absorption or fluorescence imaging.
Useful entanglement for quantum metrology can be found in the ground state of Eq.~\eqref{Eq.BJJ},
for $\chi\neq 0$ (either positive or negative), see Sec.~\ref{Sec.Atom-Atom.splitting}. 
The ground state can be experimentally addressed by adiabatically tuning interaction and/or coupling parameters.
The nonlinear interaction can also be exploited for the dynamical 
generation of entanglement starting from particles prepared in a separable state, see Secs.~\ref{Sec.Atom-Atom.twisting} and~\ref{Sec.Atom-Atom.twistNturn}.
In particular, for $\Omega,\delta=0$, Eq.~\eqref{Eq.BJJ} is equivalent to the one-axis twisting Hamiltonian first introduced by \textcite{KitagawaPRA1993}.
A limitation is that the contact interaction, which is the ingredient to entangle the atoms, is also---via particle losses induced by non-elastic scattering---a main source of decoherence in these systems. 

The external bosonic Josephson junction can be realized with a dilute Bose-Einstein condensate confined in a double-well 
potential $V\si{dw}(\vect{r})$ \cite{JavanainenPRL1986, SmerziPRL1997, MilburnPRA1997}, see Fig.~\ref{Fig:DW}(a).
For a sufficiently high barrier and weak interaction, we can 
describe the system in a two-mode approximation.
The two modes are given by
the first spatially symmetric, $\ket{\psi\si{g}}$, and first antisymmetric, $\ket{\psi\si{e}}$,
solutions of the Gross-Pitaevskii equation in the double-well trap \cite{ZapataPRA1998, RaghavanPRA1999}.
Spatial modes localized in the left and right well are given by 
$\ket{a} = (\ket{\psi\si{g} } + \ket{\psi\si{e} })/\sqrt{2}$ and $\ket{b} = (\ket{\psi\si{g} } - \ket{\psi\si{e} })/\sqrt{2}$,
respectively, see Fig.~\ref{Fig:DW}(a).
The parameters in Eq.~\eqref{Eq.BJJ} are then identified as \cite{AnanikianPRA2006}
\begin{subequations}
	\begin{align}
		\hbar \Omega & = \mu\si{e} - \mu\si{g}, \\   
		\hbar \chi & = 2 g  \int \dd^3 \vect{r} \, \psi^2\si{g}(\vect{r}) \psi\si{e}^2(\vect{r}), 
	\end{align}
\end{subequations}
where 
\be
\mu\si{g} = \int \dd^3 \vect{r} \, \psi\si{g}(\vect{r}) 
\left[ -\tfrac{\hbar^2}{2\mass} \nabla^2 + V\si{dw}(\vect{r}) \right] \psi\si{g}(\vect{r})  + g N \psi\si{g}^4(\vect{r}) 
\ee
is the chemical potential (and analogous definition for $\mu\si{e}$),
$g=4 \pi \hbar^2 a\si{s}/\mass$,
$a\si{s}$ is the s-wave scattering length
(positive for repulsive interactions and negative for attractive interactions) and $\mass$ is the atomic mass.
In the derivation of Eq.~\eqref{Eq.BJJ} we have taken $\psi\si{g,e}(\vect{r})$ real and normalized to one, and assumed 
$\int \dd^3 \vect{r} \, \psi\si{g}^2 \psi\si{e}^2 \approx \int \dd^3 \vect{r} \, \psi\si{g}^4 \approx  \int \dd^3 \vect{r} \, \psi\si{e}^4$,
which is valid for a sufficiently high tunneling barrier.

Experimentally, the external weak link has been realized on atom 
chips \cite{HallPRL2007, SchummNATPHYS2005, JoPRL2007, LeBlancPRL2011, MaussangPRL2010} 
as well as in optical traps \cite{ShinPRL2004, AlbiezPRL2005, HadzibabicNATURE2006}.
The spatial separation allows for sensing of a variety of fields and forces \cite{CroninRMP2009, InguscioBOOK}. 
The experimental challenges are the required high degree of stability of the external potential, 
as well as the precise control of the tunneling coupling between the two wells \cite{GatiPRL2006, LevyNATURE2007, SpagnolliPRL2017}.

The internal bosonic Josephson junction is created with a trapped Bose-Einstein condensate 
in two different hyperfine states \cite{SteelPRA1998, CiracPRA1998}, see Fig.~\ref{Fig:DW}(b). 
The Josephson-like coupling is provided by an electromagnetic field 
that coherently transfers particles between the two states via Rabi rotations \cite{HallPRL1998a, HallPRL1998b, BohiNATPHYS2009}.
Assuming that the external motion of the atoms is not influenced by the internal dynamics, 
one can apply a single-mode approximation for each atomic species.
The many-body Hamiltonian becomes Eq.~\eqref{Eq.BJJ} with coefficients
\begin{subequations}
	\begin{align}
		\hbar \Omega &= \hbar\Omega\si{R} \int \dd^3 \vect{r} \, \psi_a^*(\vect{r}) \psi_b(\vect{r}),  \\
		\hbar \chi &= U_{aa} + U_{bb} -2 U_{ab}, \label{Eq.chiBJJ}  \\
		U_{ij} & = \frac{2\pi \hbar^2 a\si{s}^{(i,j)}}{M}\int \dd^3 \vect{r} \, \abs{\psi_i(\vect{r})}^2  \abs{\psi_j(\vect{r})}^2, \label{Eq.UBJJ}
	\end{align}
\end{subequations}
where $\Omega\si{R}$ is the Rabi frequency, 
$a\si{s}^{(a,a)}$,  $a\si{s}^{(b,b)}$ and $a\si{s}^{(a,b)}$ are the intra- and inter-species s-wave scattering lengths, and  
$\psi_{a,b}(\vect{r})$ are single-particle mode functions of the two internal states, 
which can be determined in a mean-field description from the Gross-Pitaevskii equation. 
A more accurate value for $\chi$ is obtained if one also takes into account the change of the mode functions with particle number, see \textcite{LiEPJB2008, SmerziPRA2003}.
Since the phase and amplitude of the coupling field $\Omega\si{R}$ can be switched on and off on nanosecond time scales, 
it is possible to implement arbitrary rotations on the Bloch sphere that
are helpful to read out and characterize the internal state.
Furthermore, during Rabi coupling pulses it is possible to reach the regime  $\Omega \gg N \chi$, where interaction effects can be neglected. 

\subsection{Metrologically useful entanglement in the ground state of the bosonic Josephson junction}
\label{Sec.Atom-Atom.splitting}

Several approaches to the bosonic Josephson junction model Hamiltonian~\eqref{Eq.BJJ} have been discussed in the literature. 
A semiclassical (mean-field) approximation provides useful insights \cite{RaghavanPRA1999, SmerziPRL1997}. 
It is obtained by replacing mode operators with complex numbers:
$\op{a} \mapsto \sqrt{N_a} e^{-\ii \varphi_a}$ and $\op{b} \mapsto \sqrt{N_b} e^{-\ii \varphi_b}$, 
where $N_{a,b}$ and $\varphi_{a,b}$ are the numbers of particles and phases of the condensate in the $\ket{a}$ and $\ket{b}$ modes, respectively. 
The spin operators are replaced by 
$\J_x \to \frac{N}{2} \sqrt{1-z^2} \cos \varphi$, $\J_y \to \frac{N}{2} \sqrt{1-z^2} \sin \varphi$, and $\J_z \to \frac{N}{2} z$, 
where $\varphi = \varphi_a - \varphi_b$ is the relative phase between the two condensate modes, and
$z = (N_a - N_b)/N$ the fractional population difference $(-1 \leq z \leq 1)$. 
The Hamiltonian~\eqref{Eq.BJJ} becomes
\be \label{Eq.TwoModeHCL}
H(z,\varphi) = \frac{\Lambda z^2}{2} - \sqrt{1- z^2} \cos \varphi + \Delta E \, z, 
\ee
where energies are in units of $N\hbar \Omega/2$, $\Lambda=N\chi/\Omega$, and $ \Delta E = \delta/\Omega$.
In the following we mainly focus on the case $ \Delta E=0$, unless explicitly stated. 

A common method to extend the analysis beyond the mean-field approximation consists of quantizing the conjugate number and phase semiclassical variables \cite{LeggettFP1991}, 
\ie, replacing $z$ and $\varphi$ with operators $\op{z}$ and $\op{\varphi}$ obeying the commutation relation $[\op{z}, \op{\varphi}] = 2\ii/N$, 
where $2/N$ plays the role of a Planck constant. We can thus write $\op{\varphi}=-\frac{2\ii}{N}\op{\partial}_z$ and $\op{z}=\frac{2\ii}{N}\op{\partial}_{\varphi}$ by canonical quantization. 
Note that a rigorous phase operator is lacking \cite{CarruthersRMP1968, LynchPHYSREP1995}, and the above phase quantization may 
break down for large phase fluctuations.
Series-expanding Eq.~\eqref{Eq.TwoModeHCL} to second order in $z$ and $\varphi$ we have
$H(z, \varphi) = \tfrac{\Lambda+1}{2} z^2 + \tfrac{\varphi^2}{2}$.
By quantizing the conjugate variables we obtain 
\be \label{Eq.Hz}
\op{H}_z = - \frac{2}{N^2} \op{\partial}_z^2 + \frac{\Lambda+1}{2} \op{z}^2
\ee
in number representation, and
\be \label{Eq.Hphi}
\op{H}_\varphi = - \frac{2(\Lambda+1)}{N^2} \op{\partial}_\varphi^2 + \frac{1}{2} \op{\varphi}^2
\ee
in phase representation.
Equations~\eqref{Eq.Hz} and~\eqref{Eq.Hphi} predict Gaussian number and phase ground state wave functions with 
variances
\be \label{Eq.sigmasRabi}
\sigma_z^2 = \frac{1}{N \sqrt{1 + \Lambda}}
\text{ and }
\sigma_{\varphi}^2 = \frac{\sqrt{1 + \Lambda}}{N},
\ee
respectively \cite{ParaoanuJPB2001, SmerziPRA2000}. 
Improvements over Eqs.~\eqref{Eq.Hz} and~\eqref{Eq.Hphi} have been discussed in the literature. 
Considering a second-order expansion in phase around $\varphi=0$ of Eq.~\eqref{Eq.TwoModeHCL} one obtains the Hermitian Hamiltonian 
\be \label{Eq.WKBhamiltonian}
\op{H}_z  = - \frac{2}{N^2} \op{\partial}_z \sqrt{1-\op{z}^2}  \op{\partial}_z  + W_0(\op{z}), 
\ee
describing a fictitious quantum particle with $z$-dependent effective mass moving in an anharmonic 
one-dimensional potential
\be \label{Eq.Wz}
W_0(z)  = \frac{\Lambda z^2}{2} - \sqrt{1- z^2} + \Delta E \, z.
\ee
\textcite{ShchesnovichPRA2008} have derived the above Hamiltonian  
using a continuous approximation of the relative population difference, and avoiding phase-number commutation relations, see also \textcite{JavanainenPRA1999, SpekkensPRA1999}.
For $\Delta E=0$, the effective potential~\eqref{Eq.Wz} changes from harmonic for $\Lambda > -1$, to quartic at $\Lambda = -1$, to a double-well shape
with wells centered at $z_{\pm} = \pm \sqrt{1-1/\Lambda^2}$ for $\Lambda<-1$.
The fictitious quantum particle description is expected to give accurate results for large numbers of particles, 
provided that the wavefunction is sufficiently smooth and vanishes at the 
boundaries of $z$ \cite{Julia-DiazPRA2012}, and phase fluctuations remain small.
\textcite{AnglinPRA2001} proposed a complementary approach consisting of a
projection of the bosonic Josephson junction Hamiltonian~\eqref{Eq.BJJ} over the overcomplete Bargmann basis
$\ket{\varphi} = \sum_{m=-N/2}^{N/2} \tfrac{e^{\ii m \varphi}}{\sqrt{(N/2+m)!(N/2-m)!}} \ket{m_z}$.
This leads to an exact quantum phase model 
$\op{H}\si{BJJ} \ket{\psi} = \int_{-\pi}^{\pi} \tfrac{\dd \varphi}{2\pi} \ket{\varphi} e^{(N/2\Lambda)\cos \varphi} \op{H}\si{EQPM} \,\psi(\varphi)$, where 
$\ket{\psi}$ is an arbitrary two-mode state, $\psi(\varphi)$ is an effective phase wavefunction, and  
\be \label{Eq.EQPM}
\op{H}\si{EQPM} = - \frac{2 \Lambda}{N^2} \partial_{\varphi}^2 - \left( 1 + \frac{1}{N} \right) \cos(\varphi) - \frac{1}{4 \Lambda} \cos(2 \varphi).
\ee
The first two terms in Eq.~\eqref{Eq.EQPM} give the quantum phase model Hamiltonian that is widely used to study 
superconducting Josephson junctions \cite{BaroneBOOK1982}.
The quantum phase model is relevant in the Josephson and Fock regimes, see below.
The additional term proportional to $\cos(2\varphi)$ is important for weak interactions, in the Rabi regime.
Equation~\eqref{Eq.EQPM} is exact in the sense that the lowest $N+1$ frequencies in its eigenspectrum 
are exactly the spectrum of Eq.~\eqref{Eq.BJJ} \cite{AnglinPRA2001}.

Finally, it is worth pointing out that the spectrum of Eq.~\eqref{Eq.BJJ}---and of the more general Lipkin-Meshkov-Glick model---is 
analytically known in the thermodynamic limit \cite{RibeiroPRL2007, RibeiroPRE2008}.

\begin{figure*}
\begin{center}
\includegraphics[width=\textwidth]{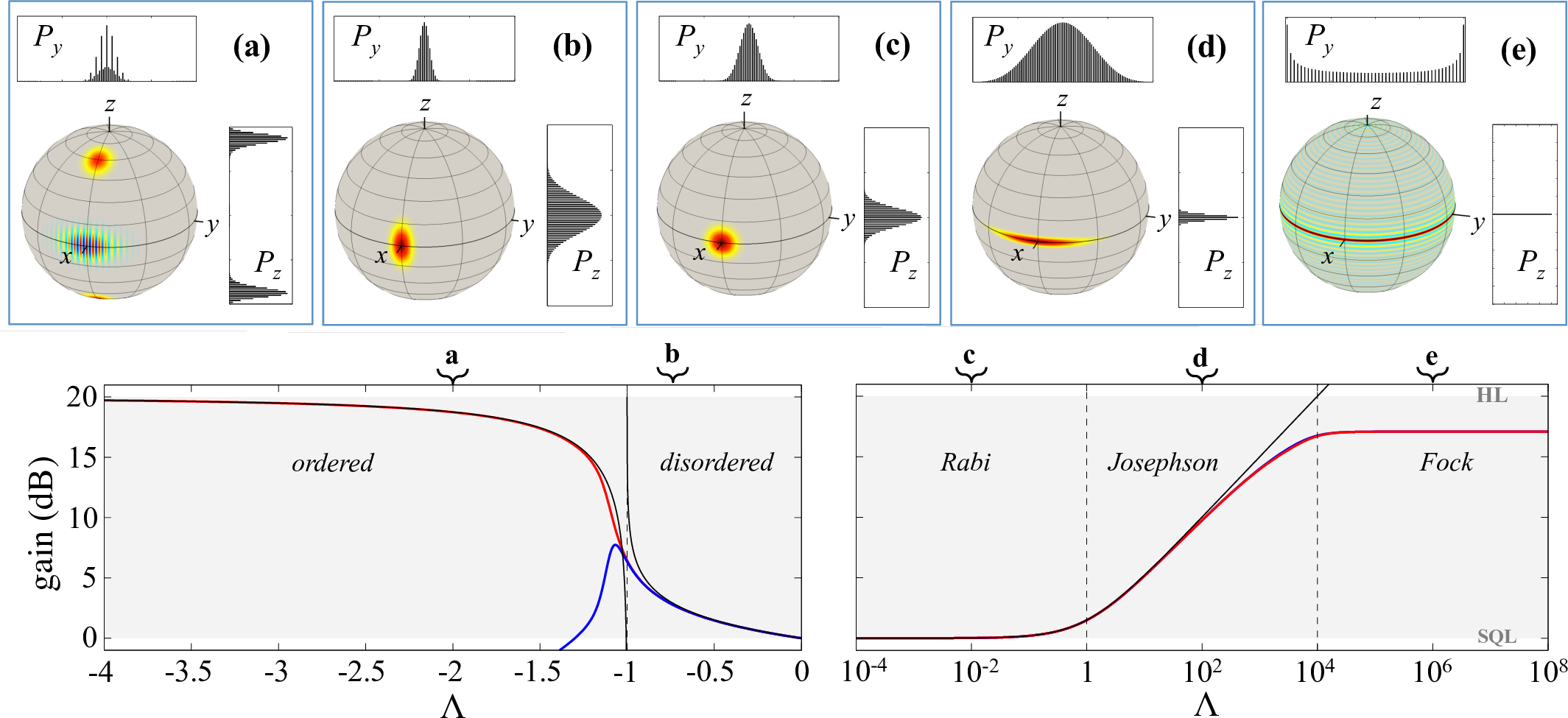}
\end{center}
\caption{{\bf Entanglement in the ground states of the BJJ.} 
Upper panels: normalized Wigner distributions of the ground state $\ket{\psi_\Lambda}$ of the Hamiltonian~\eqref{Eq.BJJ} in different regimes and for $\delta=0$.
The histograms are probability distributions $P_y(\di) = \abs{\scp{\di_y}{\psi_\Lambda}}^2$ and $P_z(\di) = \abs{\scp{\di_z}{\psi_\Lambda}}^2$.
Crossing from the Rabi to the Fock regime (for positive nonlinearities), $P_z$ narrows while $P_y$ broadens.
In the Fock regime, $P_y$ vanishes for odd values of $\di$.  
In the transition from the disordered to the ordered phase (for negative nonlinearities), $P_y$ narrows while $P_z$ broadens and splits 
when crossing the critical value $\Lambda=-1$.
Lower panels: normalized quantum Fisher information ($\Fish\si{Q}/N$, red line) and inverse spin-squeezing parameter ($1/\xiR^2$, blue line) as a function of $\Lambda$. 
For $\Lambda>0$, the solid black line is Eq.~\eqref{QFISS_Rabi}. 
For $\Lambda<0$, the solid black lines are Eq.~\eqref{QFISS_attractive} in the disordered regime, and 
Eq.~\eqref{QFI_ordered} in the ordered regime. 
Here, $N=100$ and the color scale of the Wigner distributions is as in Fig.~\ref{Fig:rotations}.}
\label{Fig:GS}
\end{figure*}

\subsubsection{Ground state for positive nonlinearities}
\label{SubSec.BJJRep}

A positive effective nonlinearity ($\Lambda>0$) makes number fluctuations between the two modes energetically unfavorable. 
The ground state  $\ket{\psi_\Lambda}$ of Eq.~\eqref{Eq.BJJ} with $\delta=0$
is thus characterized by a squeezed atom-number distribution at the expense of increased 
phase fluctuations \cite{LeggettPRL1998, JavanainenPRA1999, SpekkensPRA1999, SteelPRA1998}.
Following \textcite{LeggettRMP2001}, we distinguish three regimes, see Fig.~\ref{Fig:GS}.

\paragraph{Rabi regime, $0<\Lambda\ll1$.} 
At $\Lambda=0$ the ground state is given by the coherent spin state pointing in the positive-$x$ direction (for $\Omega>0$),
$\ket{\psi_{\Lambda=0}} \propto (\op{a}^\dag + \op{b}^\dag)^N \ket{\text{vac}}$,
with binomial occupation of each mode, where $\ket{\text{vac}}$ is the vacuum.
In the Rabi regime, the Josephson junction is dominated by tunneling, 
which keeps a well-defined relative phase between the two modes.
The coherence is high, $\mean{\op{J}_x} \approx N/2$, and number (phase)
fluctuations slightly decrease (increase) with respect to the noninteracting case. 
Using $(\Delta \op{J}_y)^2 = N^2 \sigma_{\varphi}^2/4$,  $(\Delta \op{J}_z)^2 = N^2 \sigma_{z}^2/4$ and Eq.~\eqref{Eq.sigmasRabi}, we find
\be \label{QFISS_Rabi}
\xiR^2  = \frac{1}{\sqrt{1+\Lambda}},
\quad \text{ and } \quad
\frac{\Fish\si{Q} \big[\ket{\psi_{\Lambda}}, \op{J}_y\big]}{N}= \sqrt{\Lambda +1}.
\ee  
In particular, $F\si{Q}/N = 1/\xiR^2$ in this regime.\footnote{In Eqs.~\eqref{QFISS_Rabi}-\eqref{Eq.TFodd}, \ie, for positive nonlinearities, 
the spin squeezing is calculated as $\xiR^2 = N (\Delta \op{J}_z)^2 / \mean{\J_x}^2$.
For negative nonlinearities, Eq.~\eqref{QFISS_attractive}, as $\xiR^2 = N (\Delta \op{J}_y)^2 / \mean{\J_x}^2$.}
The ground state of Eq.~\eqref{Eq.BJJ} is spin-squeezed along the $z$ axis, 
and useful for sensing rotations around the $y$ axis.

\paragraph{Josephson regime, $1 \ll \Lambda \ll N^2$.}
In the Josephson regime number (phase) fluctuations are further reduced (increased) while the coherence remains high. 
The approximations leading to Eq.~\eqref{Eq.sigmasRabi} are still very good.
We thus expect Gaussian number and phase distributions with width 
$\sigma_z^2 = 1/(N\sqrt{\Lambda})$ and 
$\sigma_{\varphi}^2 = \sqrt{\Lambda}/N$,
respectively, giving \cite{PezzePRA2005}
\be
\xiR^2 = \frac{1}{\sqrt{\Lambda}},
\quad \text{ and } \quad
\frac{\Fish\si{Q} \big[\ket{\psi_{\Lambda}}, \op{J}_y\big]}{N} = \sqrt{\Lambda}. 
\ee
For a fixed ratio $\chi/\Omega=\Lambda/N$, we have $F\si{Q}/N = 1/\xiR^2 =\sqrt{\chi/\Omega}\sqrt{N}$,
predicting a scaling of phase sensitivity $(\Delta \ps\si{QCR})^2 \sim N^{-3/2}$,
intermediate between the standard quantum limit and the Heisenberg limit. 

\paragraph{Fock regime, $\Lambda \gg N^2$.}
In the Fock regime interaction dominates over tunneling, giving rise to a fragmented Bose-Einstein condensate \cite{SpekkensPRA1999, JaaskelainenPRA2004}.
The phase becomes of the order of $2\pi$ and the approach leading to Eq.~\eqref{Eq.sigmasRabi} breaks down.
The ground state is approximatively obtained by putting an equal number of particles in both modes, with vanishing number fluctuations. 
For $\Lambda \to \infty$ and even values of $N$, we have
$\ket{\psi_{\Lambda}} = (\op{a}^\dag)^N (\op{b}^\dag)^N \ket{\text{vac}} = 
\ket{N/2}_a \ket{N/2}_b$. 
In the limit of large but finite $\Lambda$, we find 
\be \label{Eq.TFeven}
\xiR^2 = \frac{2}{N+2}, 
\quad \text{ and } \quad \frac{\Fish\si{Q}\big[\ket{\psi_{\Lambda}} , \op{J}_\perp \big]}{N} = \frac{N+2}{2},
\ee 
where $\perp$ is any axis on the equatorial plane of the Bloch sphere.
In this limit, both $(\Delta \J_z)^2$ and $\mean{\J_x}$ vanish, such the ratio $\xiR^2 = N (\Delta \J_z)^2/\mean{\J_x}$ is finite but very sensitive to experimental noise. 
For odd values of $N$, we have
$\ket{\psi_{\Lambda}} = \frac{1}{\sqrt{2}} \big( \ket{\frac{N+1}{2}}_a  \ket{\frac{N-1}{2}}_b + \ket{\frac{N-1}{2}}_a  \ket{\frac{N+1}{2}}_b \big)$
which is spin-squeezed with
\be \label{Eq.TFodd}
\xiR^2 = \frac{4N}{N(N+2)+1}, \text{ and } \frac{\Fish\si{Q}\big[\ket{\psi_{\Lambda}} , \op{J}_\perp \big]}{N} = \frac{(N^2-1)}{2N}+1.
\ee

\subsubsection{Ground state for negative nonlinearities}
\label{SebSec.GSattractive}

A negative effective nonlinearity ($\Lambda<0$) favors localization of particles in one mode and, in a symmetric Josephson junction ($\delta=0$), 
enhances number fluctuations. 
The interplay of linear coupling and nonlinear interaction in the Hamiltonian~\eqref{Eq.BJJ} gives rise to a second-order quantum phase 
transition with discrete symmetry breaking and mean-field critical exponents. 
This quantum phase transition is due to the competition between interaction and coupling \cite{UlyanovPR1992, BotetPRL1982, GilmoreNP1978}, and 
occurs in the thermodynamic limit, $N \to \infty$ and $\chi \to 0$ such that $\Lambda=-1$.
The order parameter is given by the absolute value of the population imbalance, $\abs{z}$, with $W_0(z)$ in Eq.~\eqref{Eq.Wz}
playing the role of an effective Ginzburg-Landau potential.
Entanglement \cite{BuonsantePRA2012, MaPRA2009, VidalPRA2004, OrusPRL2008, MazzarellaPRA2011}
and fluctuations of the order parameter \cite{ZinEPL2008} approaching the transition point have been extensively studied. 
This quantum phase transition has been experimentally investigated by \textcite{TrenkwalderNATPHYS2016} in a double-well potential.
The dynamics following the sudden quench into the vicinity of the quantum critical point (in a two-component Bose gas)
has been studied by \textcite{NicklasPRL2015}.

\paragraph{Disordered phase, $-1<\Lambda<0$.}
The approximations leading to Eq.~\eqref{Eq.sigmasRabi} remain valid also for weak attractive interactions. 
We thus find phase squeezing and number anti-squeezing \cite{SteelPRA1998}:
\be \label{QFISS_attractive}
\xiR^2  = \sqrt{1+\Lambda}, \text{ and }
\frac{\Fish\si{Q}\big[\ket{\psi_\Lambda}, \J_z\big]}{N} = \frac{1}{\sqrt{1+\Lambda}}.
\ee
The ground state is entangled and useful for sensing rotations around the $z$-axis, see Fig.~\ref{Fig:GS}.
At $\Lambda \to -1$ the harmonic-oscillator approximation breaks down and Eq.~\eqref{Eq.sigmasRabi} predicts a divergence of number fluctuations.
At the transition point one finds $F\si{Q}/N \approx 1/\xiR^2 = \mathcal{O}(N^{1/3})$.

\paragraph{Ordered phase,  $\Lambda < -1$.}
This phase is characterized by the presence of macroscopic-superposition state \cite{CiracPRA1998, HoJLTP2004, HuangPRA2006, LeePRL2006}. 
The mean spin length $\norm{\mean{\vectop{J}}}$ vanishes and spin squeezing is lost, see Fig.~\ref{Fig:GS}.
For strong nonlinearities, $\Lambda \ll -1$, the ground state is given by the symmetric superposition (\ie, a Schr\"odinger cat-like state) of states localized
on the right and the left wells of the effective double-well Ginzburg-Landau potential $W_0(z)$ for $\Delta E=0$ in Eq.~\eqref{Eq.Wz} \cite{ShchesnovichPRA2008}, giving 
\be \label{QFI_ordered}
\frac{\Fish\si{Q}\big[\ket{\psi_{\Lambda}} , \op{J}_z\big]}{N} = N \Big( 1 - \frac{1}{\Lambda^2}\Big).
\ee
For $\Lambda\ll-\sqrt{N}$, the ground state is approximatively given by a NOON state and 
the quantum Fisher information tends toward the ultimate bound $N^2$.
The macroscopic superposition is lost for and infinitesimally-small energy imbalance $\Delta E \neq 0$, 
breaking the symmetry of $W_0(z)$.

\begin{figure}[t!]
\begin{center}
\includegraphics[width=\columnwidth]{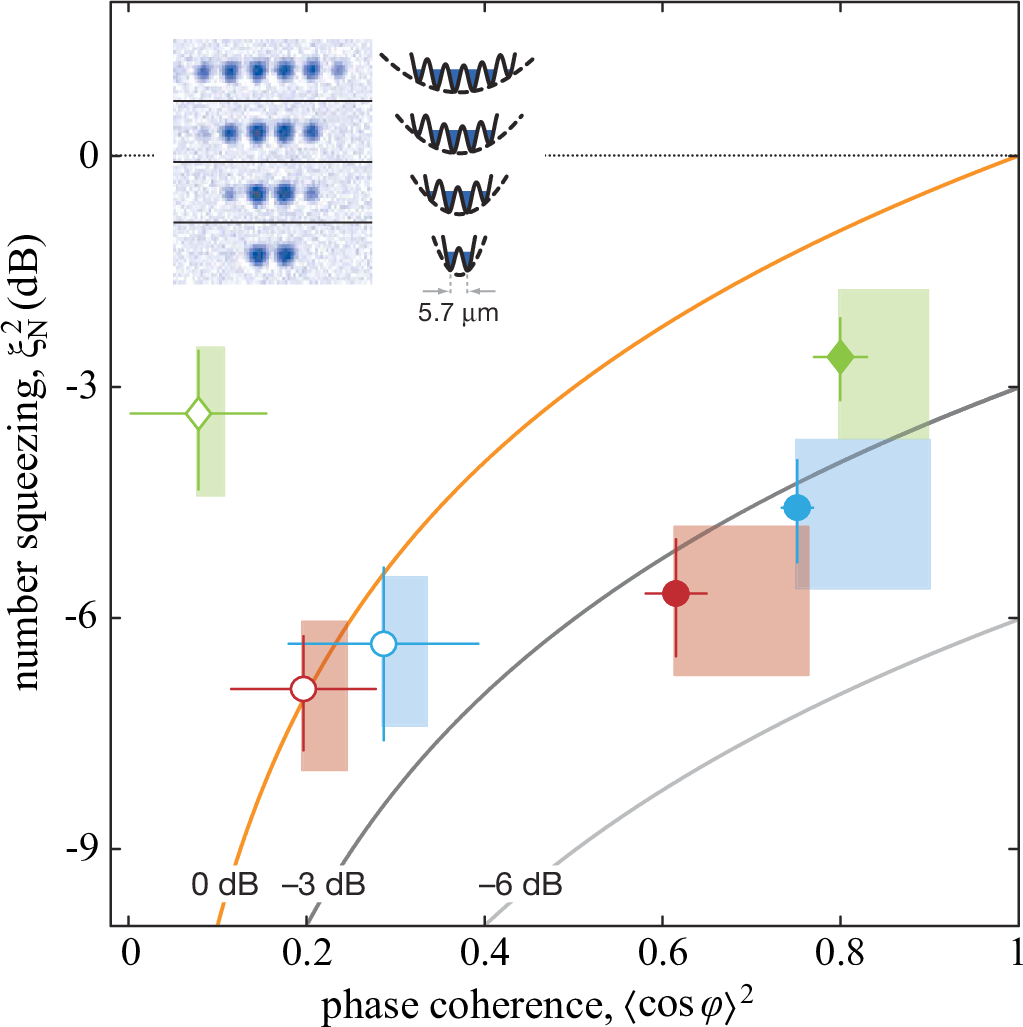}
\end{center}
\caption{{\bf Spin squeezing in the ground states of the bosonic Josephson junction.}
Symbols report number squeezing (with photon shot-noise subtracted) 
and phase coherence obtained by splitting a Bose-Einstein condensate.
Open and filled symbols correspond to different barrier heights (larger for the open symbols).
The shaded areas show systematic error bounds.
Solid lines are reference values for $\xiR^2$, the orange (0\,dB) line being the standard quantum limit. 
Measurements are shown for the two main well pairs of a six-well lattice (red/dark and blue/light circles) and for a double-well potential (green diamonds). 
The total atom number in each pair is approximately $N=2200$ in the six-well case and $N=1600$ in the double-well case. 
The inset shows single-site-resolving absorption images of atoms trapped in an optical lattice superposed on an atomic dipole trap.
Adapted from \textcite{EsteveNATURE2008}.}
\label{Fig:EsteveNATURE2008}
\end{figure}  

\begin{figure}[t!]
\begin{center}
\includegraphics[width=\columnwidth]{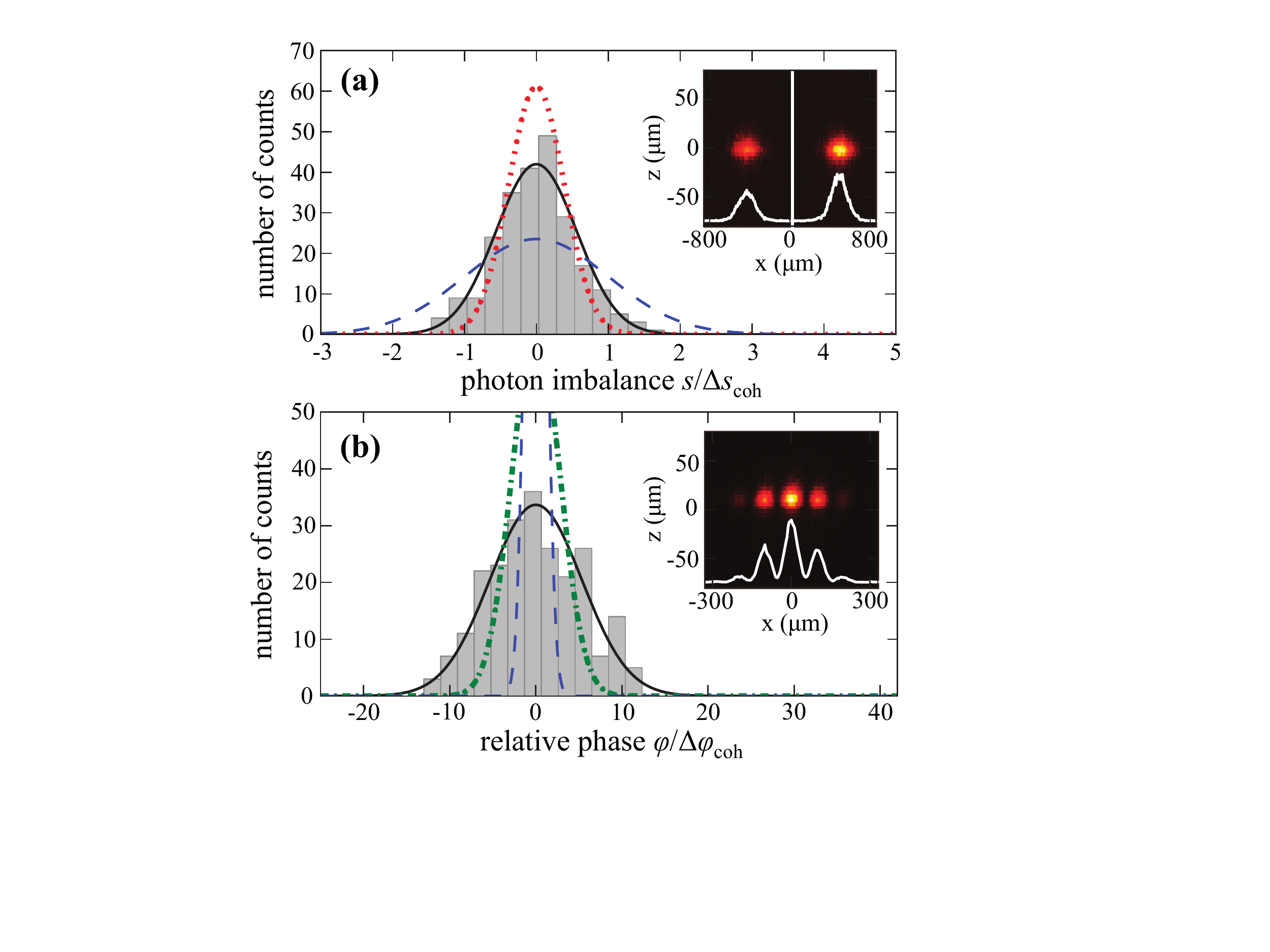}
\end{center}
\caption{ {\bf Number and phase distributions of a split Bose-Einstein condensate}. 
Number (a) and phase (b) distributions after the adiabatic splitting of a Bose-Einstein condensate in a double-well potential.
(a) Distribution of photon imbalance $s=s_L-s_R$ between fluorescence images of the left and right cloud (inset), 
proportional to the imbalance of atom numbers in the two wells.
The solid black line is the normal distribution corresponding to the measured number-squeezing factor, 
the dotted red line is the expected distribution when detection noise is subtracted, and the dashed blue line 
is the distribution expected for a coherent spin state in the absence of detection noise.
(b) The curves indicate a normal distribution with the measured $\Delta \varphi$ (solid black) and the distributions 
expected for a coherent state in the absence (dashed blue) 
and in the presence (dashed-dot green) of detection noise. 
The inset shows a typical matter-wave interference pattern from which the phase is extracted.
Adapted from \textcite{BerradaNATCOMM2013}.}
\label{Fig:BerradaNATCOMM2013}
\end{figure}  

\subsubsection{Adiabatic splitting}

Experimentally, the Rabi and Josephson regimes for repulsive interactions and the disordered phase for attractive interactions
can be reached by adiabatically splitting a Bose-Einstein condensate initially prepared in the ground state of the strong-tunneling (Rabi) 
regime \cite{JavanainenPRA1999, MenottiPRA2001, StreltsovPRL2007, PezzePRA2005, IsellaPRA2005, BodetPRA2010}. 
By varying the plasma frequency $\omega\si{p}=\Omega\sqrt{1+\Lambda}$ adiabatically in time, 
the system follows its ground state if \cite{JavanainenPRA1999, SchaffBOOK2014}
\be \label{Eq.adiab}
\left\lvert \frac{\dd \omega\si{p}}{\dd t} \right\rvert \lesssim \omega\si{p}^2.
\ee
Increasing the height of the potential barrier in order to reach the Fock regime,
the required timescale 
for adiabaticity eventually diverges, setting a limit to the attainable entanglement. 
Shortcut to adiabaticity \cite{LapertPRA2012, YustePRA2013, Julia-DiazPRA2012b} and 
optimal control techniques \cite{GrondPRA2008, HuangPRL2008, PichlerPRA2016}
for the fast production of highly-entangled states have been also studied. 
It should be noted that Eq.~\eqref{Eq.adiab} assumes that the gas is prepared in the ground states configuration. 
Experimentally, the finite temperature of the gas is the main factor limiting the squeezing via the 
adiabatic-splitting technique \cite{EsteveNATURE2008}.

\begin{figure*}
\begin{center}
\includegraphics[width=\textwidth]{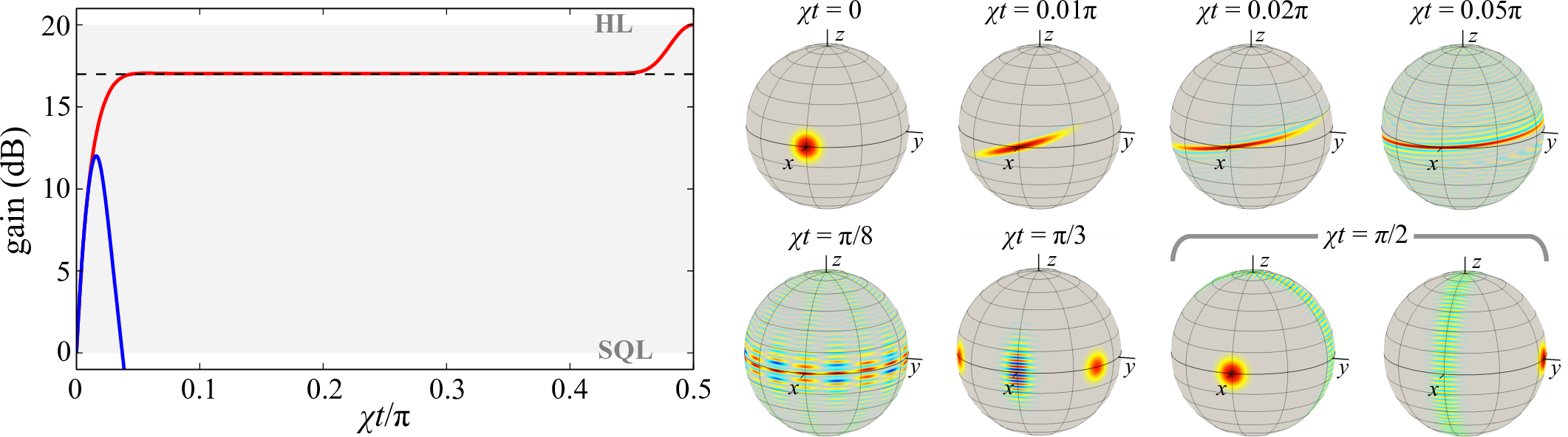}
\end{center}
\caption{{\bf One-axis twisting dynamics.} 
The left panel reports the inverse spin-squeezing parameter ($1/\xiR^2$, blue/lower line) and normalized
quantum Fisher information ($F\si{Q}/N$, red/upper line) as function of $\chi t/\pi$. 
The right panels show snapshots of the Wigner distribution at different times. 
For $\chi t =\pi/2$ we plot both the Wigner distributions for $N=100$ (left) and $N=101$ (right).
In all plots $N=100$ (unless specified) and the color scale is as in Fig.~\ref{Fig:rotations}.}
\label{Fig:OAT}
\end{figure*} 

\subsubsection{Experimental spin squeezing in the ground state of the BJJ}

\textcite{EsteveNATURE2008} reported the first direct experimental demonstration of spin squeezing and entanglement in the ground state of the bosonic Josephson junction.
A $^{87}$Rb  Bose-Einstein condensate confined in a shallow harmonic trap is split adiabatically by ramping up a one-dimensional optical lattice. 
The particles are distributed over two and six lattice sites, see inset of Fig.~\ref{Fig:EsteveNATURE2008}.  
Direct atom detection better than atomic shot noise is implemented by absorption imaging of the atomic density, 
with a spatial resolution well below the lattice spacing. 
This gives direct access to the number-squeezing parameter $\xiN^2 = 4(\Delta \J_z)^2/N$. 
Additionally, the phase coherence $\mean{\cos \varphi}^2$ is directly obtained from 
the interference pattern observed after releasing the atomic cloud from the trap.
Noticing that $\mean{\op{J}_x}^2 = \mean{\op{N}_a}\mean{\op{N}_b} \mean{\cos \varphi}^2$ (valid for $\mean{\op{N}_a}$,$\mean{\op{N}_b} \approx N/2\gg 1$ and $\sigma_\varphi^2 \ll 2 \pi$),
metrological spin squeezing is given by $\xiR^2 = \xiN^2 / \mean{\cos \varphi}^2$.
The simultaneous presence of number squeezing and high phase coherence allowed to reach an inferred spin squeezing of 
$\xiR^2 = -3.8$\,dB for the two main well pairs of a six-well lattice, and $\xiR^2=-2.3$\,dB for the double-well configuration. 
The results are summarized in Fig.~\ref{Fig:EsteveNATURE2008}. 

More recent experiments have studied number squeezing \cite{MaussangPRL2010} 
and spin squeezing \cite{BerradaNATCOMM2013} when splitting a Bose-Einstein condensate in a double-well trap on an atom chip. 
\textcite{MaussangPRL2010} realized purely magnetic double-well potential. 
This experiment reports a detailed investigation of 
the optimal splitting time, given by the interplay of the barrier raising rate (which must be slow in order  to 
avoid excitations to higher modes), heating effects and atom loss. 
In \textcite{BerradaNATCOMM2013} the splitting is based on radio-frequency dressing \cite{HofferberthNATPHYS2006} that dynamically deforms 
the static magnetic trap into a double-well potential \cite{SchummNATPHYS2005}.
A split Bose-Einstein condensate of $N=1200$ $^{87}$Rb atoms is characterized via the measurement 
of the number of particles in each well and the relative phase, see Fig.~\ref{Fig:BerradaNATCOMM2013}, 
showing number squeezing and phase anti-squeezing.
From these results, using $\xiR^2 = \xiN^2 / \mean{\cos \varphi}^2$, it is possible to obtain $\xiR^2 = -5.6$\,dB ($\xiR^2 = -7.8$\,dB inferred).
This corresponds to a useful (for metrology) entanglement depth of 150 atoms, excluding useful entanglement of less than 67 particles with 90\% probability.
\textcite{BerradaNATCOMM2013} have also investigated the reduced phase diffusion 
associated to number squeezing \cite{LewensteinPRL1996, CastinPRA1997, JavanainenPRL1997}
and the fast recombination \cite{JoPRL2007rec, NegrettiJPB2004, NegrettiPRA2008, ScottPRL2008, BerradaPRA2016}
of the condensates in the two wells.

\subsection{One-axis twisting dynamics}
\label{Sec.Atom-Atom.twisting}

One-axis twisting is a benchmark model for studying the generation of spin-squeezed states \cite{KitagawaPRA1993}, 
Schr\"odinger cats \cite{MolmerPRL1999} and useful entanglement for quantum metrology \cite{PezzePRL2009}.
The one-axis twisting Hamiltonian (here, along the $z$-axis) is 
\be \label{Eq:Jz2}
\op{H}\si{OAT} = \hbar \chi \op{J}_z^2.
\ee
This model can be realized via atom-atom elastic collisions in a Bose-Einstein condensate \cite{SorensenNATURE2001}, see Eq.~\eqref{Eq.BJJ},
trapped ions, see Sec.~\ref{Sec.Ions.metrology}, and cold atoms in an optical cavity, see Sec.~\ref{Sec.Atom-Light.effint}.

The quantum dynamics $e^{-\ii \op{H}\si{OAT} t/\hbar}$ of a localized spin wavepacket can be roughly viewed as a rotation 
$e^{-\ii \chi t \mean{\J_z} \J_z}$ around the $z$ axis. 
The $\mean{\J_z}$-dependent angular velocity (whose sign differs on the two hemispheres of the Bloch sphere)
leads to a twisting of the state on the Bloch sphere. 
To analyze this effect quantitatively, let us consider the dynamical evolution of an initial coherent spin state 
pointing along the positive $x$-axis, $\ket{\psi\si{OAT}(t)} = e^{-\ii \chi t \J_z^2} \ket{\pi/2,0,N}$, see Fig.~\ref{Fig:OAT}.
The coherent spin state initially stretches in the $y-z$ plane tangental to the Bloch sphere.
\textcite{KitagawaPRA1993} identified the squeezing angle $\delta=\frac12\arctan(\frac{B}{A})$ in terms of $A = 1 - \cos^{N-2}(2 \chi t)$ and 
$B = 4 \sin(\chi t) \cos^{N-2}(\chi t)$, as well as the squeezed and anti-squeezed spin components $\J\si{s} = \J_z \cos(\delta) - \J_y \sin(\delta)$ and $\J\si{as} = \J_y \cos(\delta) + \J_z \sin(\delta)$, respectively.
Squeezing is accompanied by loss of contrast,
$\mean{\J_x} = \frac{N}{2} \cos^{N-1}(\chi t)$, as the state spreads on the Bloch sphere.
The spin squeezing $\xiR^2 = N(\Delta \J\si{s})^2/\mean{\J_x}^2$ and 
the quantum Fisher information $\Fish\si{Q}\big[\ket{\psi\si{OAT}(t)}, \J\si{as} \big] = 4 (\Delta \J\si{as})^2$ are readily calculated, giving \cite{SorensenNATURE2001, KitagawaPRA1993}
\be \label{Eq:TwistxiR}
\xiR^2 = \frac{4+ (N-1) (A - \sqrt{A^2 + B^2})}{4 \cos^{2N - 2}(\chi t)}, 
\ee
and \cite{PezzePRL2009} 
\be \label{Eq:TwistQFI}
\frac{\Fish\si{Q}\big[\ket{\psi\si{OAT}(t)}, \J\si{as} \big]}{N} = 1 + \frac{(N-1)}{4} \big(A + \sqrt{A^2 + B^2}\big),
\ee
respectively.\footnote{The quantum Fisher information optimized over rotation directions in the Bloch sphere is given by the maximum between
$\Fish\si{Q}[\ket{\psi\si{OAT}(t)}, \J\si{as}]/N$, Eq.~\eqref{Eq:TwistQFI}, and $\Fish\si{Q}[\ket{\psi\si{OAT}(t)}, \J_x]/N = N [1 - \cos^{2N-2}(\chi t)] - (N-1)A/2$.}
For $\chi t \lesssim 1/\sqrt{N}$ the state is spin squeezed, $\xiR^2< 1$, reaching $\xiR^2 = \mathcal{O}(N^{-2/3})$ at an optimal time $\chi t=\mathcal{O}(N^{-2/3})$. 
For $\chi t \gtrsim 1/\sqrt{N}$ the states wraps around the Bloch sphere and spin squeezing is lost, $\xiR^2>1$. 
Yet, the state is still entangled. 
The quantum Fisher information reaches a plateau $\Fish\si{Q}[\ket{\psi\si{OAT}(t)}, \J\si{as}] = N(N+1)/2$ for $2/\sqrt{N} \lesssim \chi t \lesssim \pi/2 - 2/\sqrt{N}$
signaling entanglement, according to Eq.~\eqref{Fishent}, in the over-squeezed state.
The state evolves, at time $\chi t = \pi/n$, into a coherent superposition of $2 \leq n \lesssim \pi\sqrt{N}/2$ coherent spin states distributed evenly 
on the equator of the Bloch sphere \cite{AgarwalPRA1997}, see Fig.~\ref{Fig:OAT}. 
At $\chi t =\pi/2$ we observe the dynamical creation of a NOON state along the $x$ axis, 
$\ket{\psi\si{OAT}(\pi/2\chi)} = e^{-\ii \tfrac{\pi}{2} \J_y} \ket{\text{NOON}}$, if $N$ is even, and 
along the $y$ axis, $\ket{\psi\si{OAT}(\pi/2\chi)} = e^{-\ii \tfrac{\pi}{2} \J_x} \ket{\text{NOON}}$, if $N$ is odd \cite{AgarwalPRA1997, MolmerPRL1999}.
For the NOON state, the quantum Fisher information, optimized over the spin direction, reaches its maximum value $\Fish\si{Q}=N^2$, see Sec.~\ref{sec:NOONstate}.
For even (odd) values of $N$, the dynamics is reversed for $\chi t \geq \pi/2$ ($\chi t \geq \pi$) and we have a complete revival of the initial condition at $\chi t = \pi$
($\chi t = 2\pi$). 
The one-axis twisting dynamics is modified by particle loss \cite{LiPRL2008, SinatraPRL2011, SinatraEPL2013, SinatraEPJST2012, SpehnerEPJB2014} 
and other imperfections such as phase noise \cite{FerriniPRA2008, FerriniPRA2011} and finite temperature of the 
Bose gas \cite{SinatraPRL2011}, limiting the achievable squeezing. 
We will further comment on these works in Sec.~\ref{Sec.Working-Entanglement}.

\begin{figure}[t!]
\begin{center}
\includegraphics[width=\columnwidth]{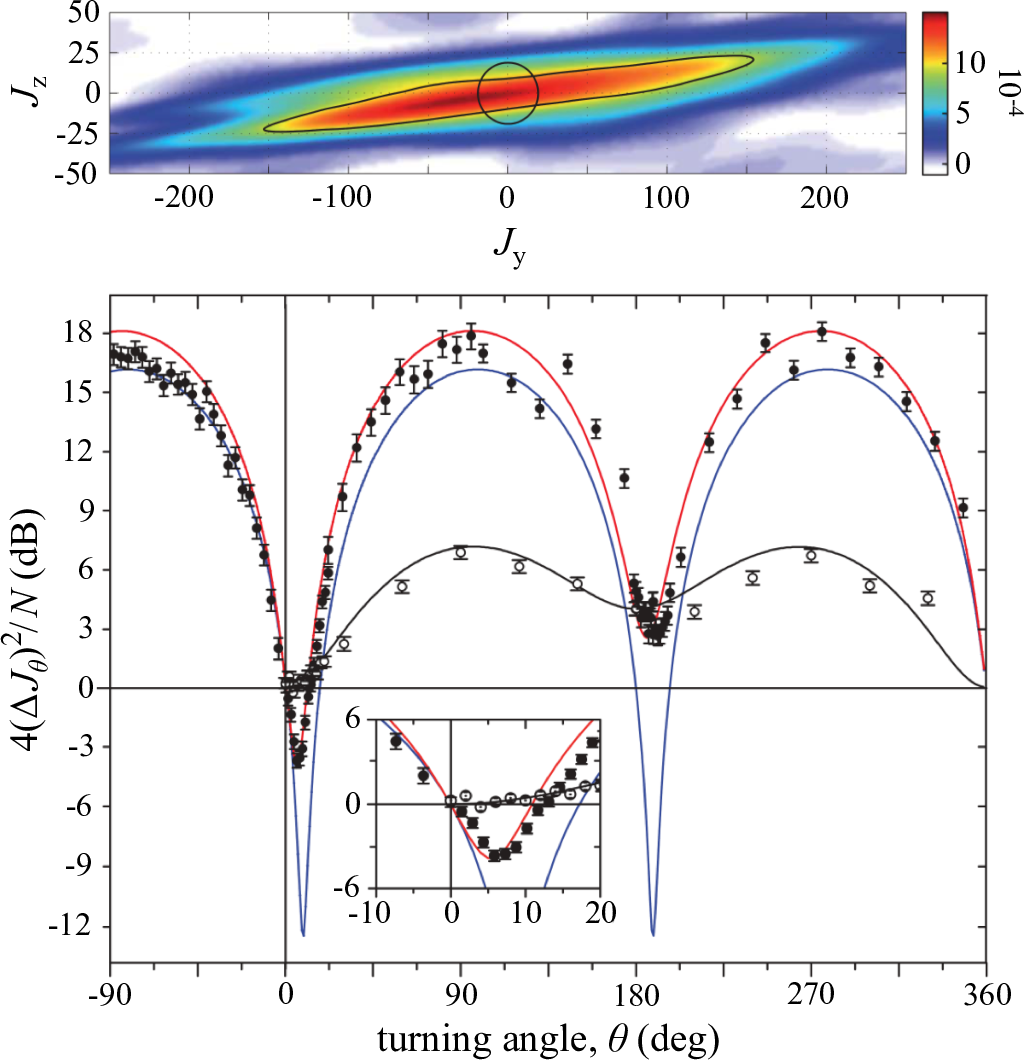}
\end{center}
\caption{{\bf Spin-noise tomography and reconstructed Wigner 
function of a spin-squeezed Bose-Einstein condensate.}
Top: Reconstructed Wigner distribution of a spin-squeezed state of $N \approx 1250$ atoms.
The black contour line indicates where the Wigner distribution has fallen to $1/e$ of its maximum. 
For comparison, the circular $1/e$ contour of an ideal coherent spin state is shown.
Bottom: Observed spin fluctuations of a spin-squeezed state (solid circles) 
and of a coherent spin state (open circles), 
as a function of the turning angle of the Bloch sphere.
Solid lines are results of dynamical simulations including losses and technical noise: 
blue (lowest), spin-squeezed state with losses but without technical noise; 
red (highest), spin-squeezed state with losses and technical noise; 
black, coherent state with losses and technical noise.
Adapted from \textcite{RiedelNATURE2010}.}
\label{RiedelNATURE2010_Fig2} 
\end{figure} 

\begin{figure}[t!]
\begin{center}
\includegraphics[width=\columnwidth]{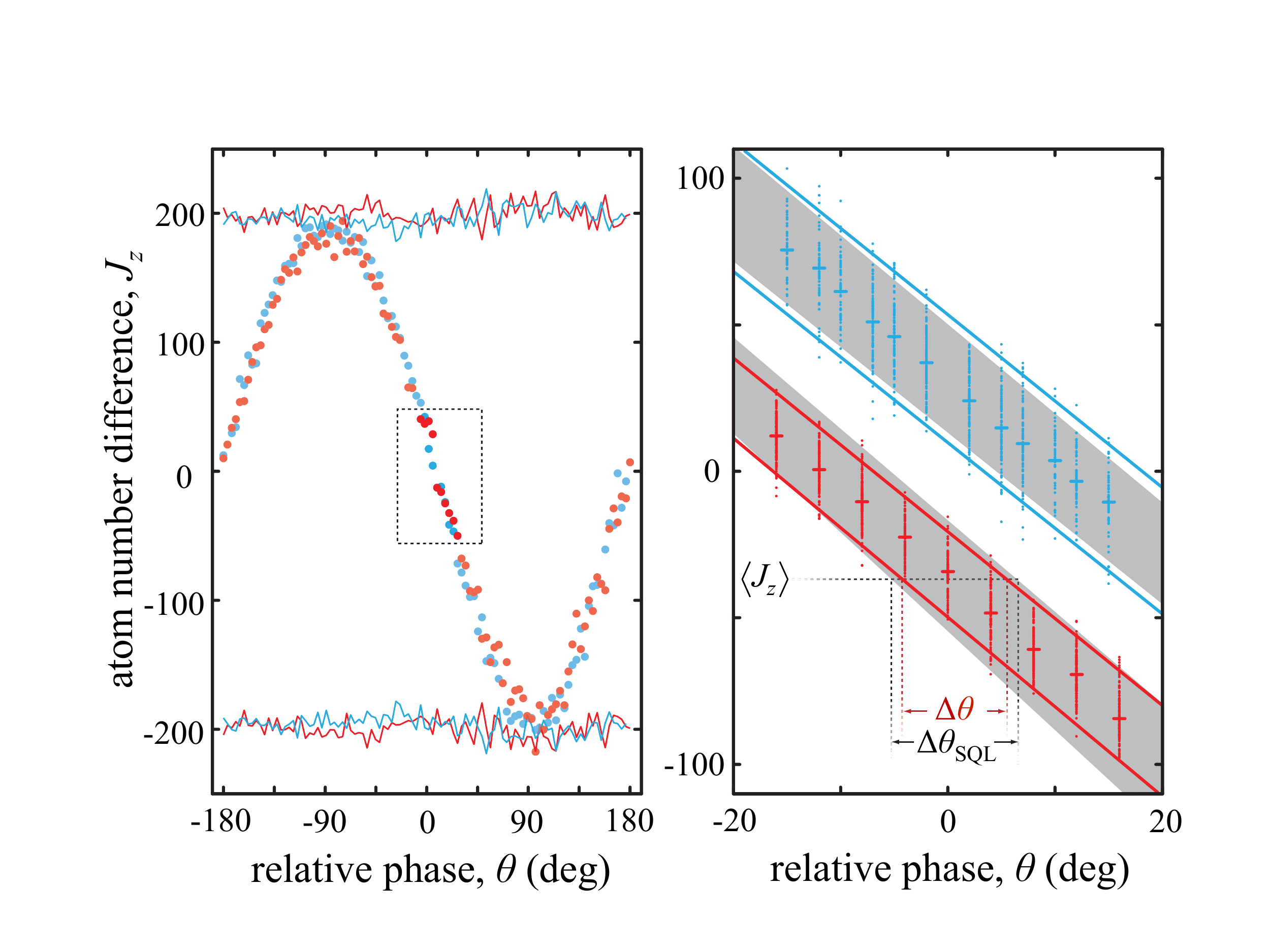}
\end{center}
\caption{{\bf Direct experimental demonstration of a phase sensitivity beyond the standard quantum limit.}
Left: Ramsey fringe scanned over a full $2\pi$ period. 
The blue (lighter) data are obtained with a coherent spin state, the red (darker) data with a spin-squeezed state showing a visibility of 92\%.
Solid lines are $\pm N/2$, measured for each phase setting, as a reference. 
Right: Ramsey fringe around the maximum-slope optimal point and
averaged over several experimental realizations.
The solid lines are fits through the lower and upper ends of the two standard deviation error bars. 
The grey shaded areas are the uncertainty regions for an ideal coherent spin state, they correspond to the standard quantum limit.
Red (darker) data show a phase error 15\% below the standard quantum limit. 
Adapted from \textcite{GrossNATURE2010}.}
\label{GrossNATURE2010_Fig2} 
\end{figure} 

\begin{figure}
\begin{center}
\includegraphics[width=\columnwidth]{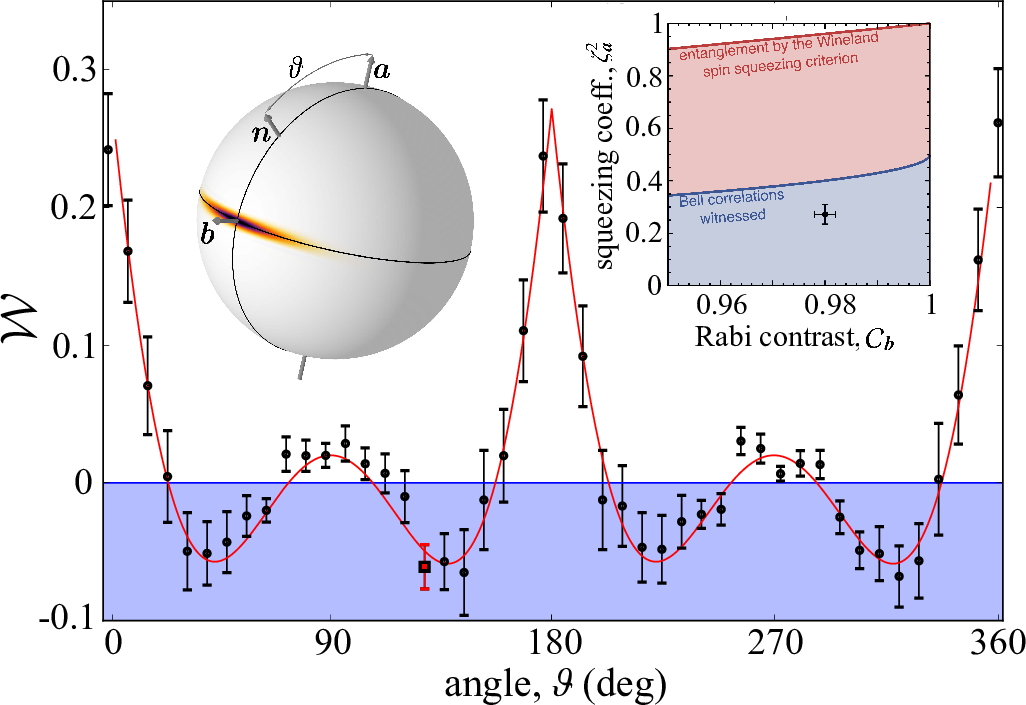}
\end{center}
\caption{{\bf Bell correlations in a Bose-Einstein condensate.}
Measurement of the Bell correlation witness $\mathcal{W}$ of Eq.~\eqref{eq:BellWitnessFull} on a 
spin-squeezed state as a function of the angle $\vartheta$ between the squeezing axis $\vect{a}$ 
and the axis $\vect{n}$ lying in the squeezing plane, with $1\sigma$ error bars. 
The red continuous line is the value of $\mathcal{W}$ computed from the measurement of 
$\zeta_{\vect{a}}^2=4\mean{\J_{\vect{a}}^2}/N$ and the fitted Rabi oscillation 
$\mathcal{C}_{\vect{n}}=\mathcal{C}_{\vect{b}}\sin(\vartheta)$ with $\mathcal{C}_{\vect{b}}=2\mean{\J_{\vect{b}}}/N$. 
Bell correlations are present in the blue shaded region.
Right inset: representation of the same data ($\zeta_{\vect{a}}^2$ and $\mathcal{C}_{\vect{b}}$) 
as a single black point, with $1\sigma$ error bars.
Red shaded region: entanglement detected by spin squeezing $\xiR^2<1$, Eq.~\eqref{Eq:xiWineland}.
Blue shaded region: Bell correlations detected by violation of the witness inequality~\eqref{eq:BellWitness}. 
Adapted from \textcite{SchmiedSCIENCE2016}.}
\label{fig:SchmiedSCIENCE2016} 
\end{figure} 

\subsubsection{Spin squeezing and particle entanglement}

Spin squeezing obtained via one-axis twisting dynamics in a Bose-Einstein condensate 
has been first experimentally demonstrated by \textcite{RiedelNATURE2010, GrossNATURE2010}.
These experiments used a $^{87}$Rb condensate prepared in a 
coherent spin state with equal mean populations in two hyperfine states. 
With $^{87}$Rb atoms, the interaction parameter $\hbar \chi = U_{aa} + U_{bb} -2 U_{ab}$, see Eq.~\ref{Eq.chiBJJ}, is small due to an almost perfect 
compensation of inter- and intra-species collisional interactions, \ie, $2 U_{ab} \approx U_{aa} + U_{bb}$.
Different approaches have been used to increase $\chi$ artificially: \textcite{RiedelNATURE2010} 
used a state-dependent trapping potential to change the wave function overlap 
between the two internal states and thus reduce $U_{ab}$ \cite{BohiNATPHYS2009, LiEPJB2008}, 
while \textcite{GrossNATURE2010} exploited a
Feshbach resonance \cite{ChinRMP2010} to change the scattering length $a\si{s}^{(a,b)}$.

\textcite{RiedelNATURE2010} used a spatially-inhomogeneous microwave field to manipulate the trapping potentials of the atom cloud
in the two hyperfine states $\ket{a} = \ket{F=1, \mf=-1}$ and $\ket{b} = \ket{2, 1}$ by microwave level shifts \cite{TreutleinPRA2006}.   
The trapping potential for atoms in one state is suddenly shifted by such a state-selective force, and the 
states coherently demix and begin to oscillate in space.
In this way the overlap of the wavefunctions of the two states changes dynamically and modulates collisional effects \cite{LiEPJB2008}:
according to Eq.~\eqref{Eq.UBJJ}, when the two components are spatially separated, the inter-species interaction $U_{ab}$ vanishes. 
The parameter $\chi$ is then determined solely by intra-species interactions and 
reaches sufficiently-high values to induce fast spin-squeezing dynamics.
After each full oscillation, the two states coherently remix and the collisional squeezing dynamics stops.
Spin-noise tomography and the reconstructed Wigner distribution of the squeezed state are shown in Fig.~\ref{RiedelNATURE2010_Fig2}.
The results demonstrate a squeezing parameter $\xiR^2 = -1.2$\,dB ($\xiR^2 = -2.5$\,dB inferred). 
More recent experiments using this technique achieved up to $\xiR^2 = -7.0$\,dB with detection 
noise subtracted \cite{OckeloenPRL2013, SchmiedSCIENCE2016}.

\textcite{GrossNATURE2010} used a Bose-Einstein condensate of $N\approx 400$ atoms in the
$\ket{a} = \ket{1, 1}$ and $\ket{b} = \ket{2, -1}$ hyperfine levels of the electronic ground state of Rubidium.
A bias magnetic field is used to bring the system near a Feshbach resonance, to reduce the inter-species s-wave scattering length
and thus enhance the effective nonlinearity $\chi$ leading to squeezing and entanglement. 
Spin-noise tomography leads to an inferred $\xiR^2 = -8.2$\,dB, 
predicting an entanglement depth excluding less than 80 particles within three standard deviations (and a mean of 170 entangled atoms).

\subsubsection{Quantum interferometry}
\textcite{GrossNATURE2010} have also demonstrated a full Ramsey interferometer sequence with spin-squeezed states, see Fig.~\ref{GrossNATURE2010_Fig2}.
Squeezed-state creation is followed by a rotation around the $x$ axis to prepare a phase-squeezed state (\ie, squeezed along the $y$ axis)---in effect constructing a nonlinear beam-splitter. 
After a short interrogation time $\chi t \ll 1$, during which the interaction is still active, a second $\pi/2$ beam splitter closes the Ramsey sequence.
The directly measured phase sensitivity gain is $(\Delta \ps/\Delta \ps\si{SQL})^2 = -1.4$\,dB, corresponding to  
a reduction of phase variance 15\% below the standard quantum limit.
In a more recent experiment, \textcite{MuesselPRL2014} have been able to scale the generation 
of spin-squeezed states up to $N =12500$ particles 
using a chain of trapped independent condensates and exploiting a differential estimation of the phase.
They demonstrated a spin squeezing $\xiR^2 = -3.4$\,dB and a sensitivity 24\% below the standard quantum limit.
Ramsey interferometry using spin-squeezed states generated via one-axis twisting in Bose-Einstein condensate has recently found application to sense 
magnetic fields \cite{OckeloenPRL2013, MuesselPRL2014}, as discussed in more detail in Sec.~\ref{Sec.Working-Entanglement}.

\subsubsection{Bell correlations}
\label{SubSec.BellCorr}

\textcite{SchmiedSCIENCE2016} have observed Bell correlations in spin-squeezed Bose-Einstein condensates of about $N=480$ $^{87}$Rb atoms trapped on an atom chip. Following the generation of the spin-squeezed state by one-axis twisting, the Bell correlation witness $\mathcal{W}$ of Eq.~\eqref{eq:BellWitnessFull} is determined 
by measuring the second spin moment along the squeezed direction and combining it with spin contrast measurements along other projection axes, see Fig.~\ref{fig:SchmiedSCIENCE2016}.
The measured values of $\mathcal{W}$ violate the inequality~\eqref{eq:BellWitnessFull} and thus also the inequalities~\eqref{eq:BellWitness} and~\eqref{eq:BellWitness2}.
This proves the presence of Bell correlations and implies that the spin-squeezed states can violate the many-particle Bell inequality of \textcite{TuraSCIENCE2014}.
The detection of Bell correlations indicates that spin-squeezed states are useful beyond quantum metrology: 
they also contain the resource for quantum information tasks such as certified randomness generation or quantum key distribution~ \cite{BrunnerRMP2014}.

While \textcite{SchmiedSCIENCE2016} observed Bell correlations in a spin-squeezed BEC, \textcite{EngelsenPRL2017} 
have confirmed Bell correlations in a spin-squeezed thermal ensemble. The interpretation of these experiments
relies on the common assumption that atoms do not communicate through unknown channels. This assumption can be relaxed if the atoms are spatially separated in an optical lattice potential, as proposed in \textcite{PelissonPRA2016}. Many-body Bell tests with Bose-Einstein condensates have also been proposed in 
\textcite{MullinPRA2008, LaloeEPJB2009}. In section \ref{Sec.Atom-Atom.separated} we discuss proposals to create entanglement 
between spatially separated spinor condensates, which could also be employed for Bell tests.

\subsection{Twist-and-turn dynamics}
\label{Sec.Atom-Atom.twistNturn}

The creation of entanglement in the one-axis twisting model is enriched---and accelerated, to some extent \cite{MuesselPRA2015, SorelliARXIV2015}---by 
the dynamical evolution using the turning term $\J_x$ simultaneously with the twisting term $J_z^2$ in the Hamiltonian~\eqref{Eq.BJJ}.
The resulting twist-and-turn dynamics is experimentally studied from the starting point of a coherent spin state pointing in the $+x$ or $-x$ direction. 
The interaction is suddenly switched to a finite value of $\Lambda$ in presence of linear coupling.

\begin{figure}[t!]
\begin{center}
\includegraphics[width=\columnwidth]{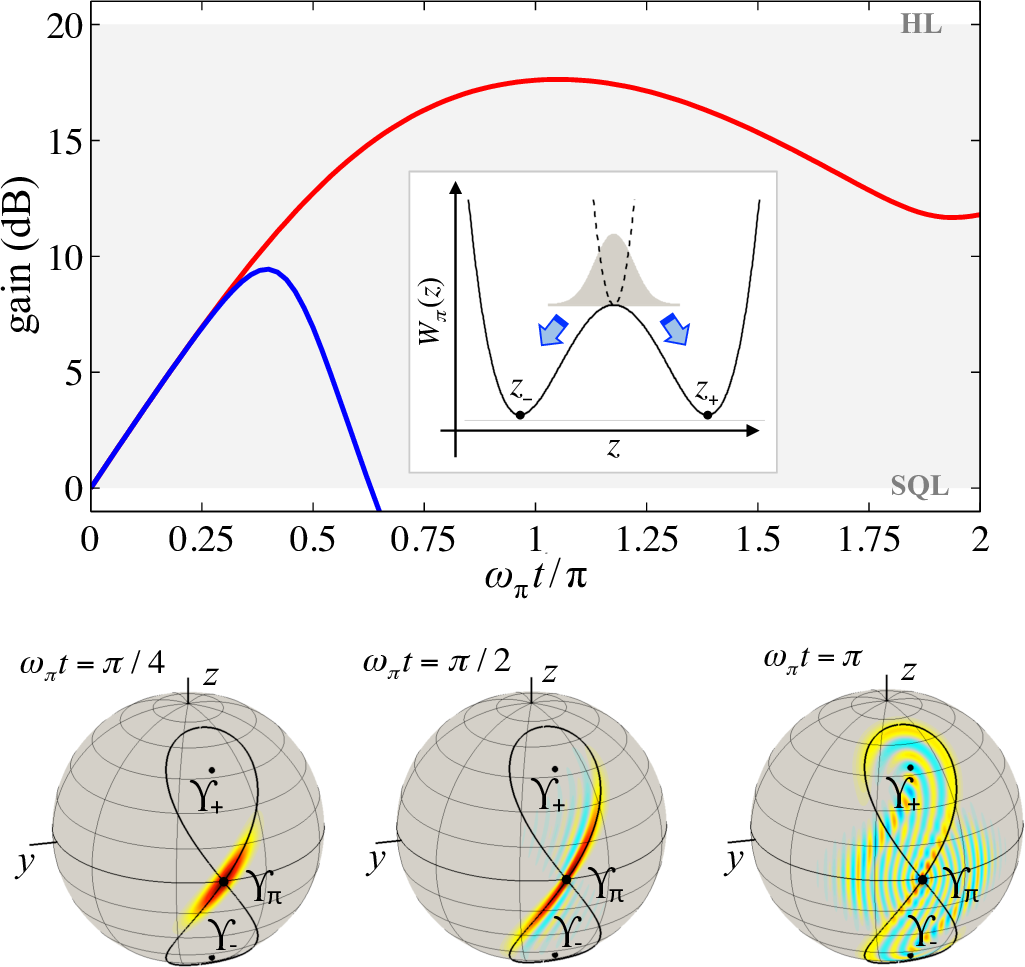}
\end{center}
\caption{{\bf Twist-and-turn dynamics.} 
Top: Normalized quantum Fisher information ($F\si{Q}/N$, red upper line) and inverse spin-squeezing 
parameter ($1/\xiR^2$, blue lower line) as a function of $\omega_\pi t/\pi$ for the twist-and-turn model,
where $\omega_\pi = \Omega\sqrt{\Lambda-1}$.
The inset shows the effective potential $W_{\pi}(z)$ initially (dotted line, $\Lambda=0$) and after a quench to a finite $\Lambda$ (solid line, $\Lambda=1.5$).
The initial coherent spin state corresponds to a Gaussian wavepacket located at the top of the barrier.
Bottom: Snapshots of the Wigner distribution at different times. 
The solid black line is the mean-field separatrix and the dots are fixed points of the semiclassical model.
Here $N=100$ and the color scale is as in Fig.~\ref{Fig:rotations}.}  
\label{Fig:TwistTurn}
\end{figure}

The twist-and-turn can be described as the dynamics of an effective relative-number wavepacket \cite{Julia-DiazPRA2012}. 
For an initial coherent spin state polarized along the positive $x$-axis, the effective Hamiltonian is given by Eq.~\eqref{Eq.WKBhamiltonian}.
$W_0(z)$ can be well approximated as a harmonic potential of frequency $\omega_0 = \Omega \sqrt{1 + \Lambda}$.
Suddenly switching the interaction from $\Lambda=0$ to a positive value 
(corresponding a tighter the harmonic oscillator potential, \ie, $\omega_0/\Omega > 1$)
gives rise to a breathing mode of the effective wavepacket with periodic squeezing of the relative population.
For a coherent spin state polarized along the negative $x$-axis, the effective Hamiltonian is 
$\op{H}_z  = - \frac{2}{N^2} \op{\partial}_z \sqrt{1-\op{z}^2}  \op{\partial}_z  + W_\pi(\op{z})$ with 
$W_\pi(z) = -\frac{\Lambda z^2}{2} - \sqrt{1- z^2}$.
For $0<\Lambda < 1$, $W_\pi(z)$ is harmonic with frequency $\omega_\pi = \Omega \sqrt{1 - \Lambda}$. 
Since $\omega_\pi/\Omega<1$, a sudden switch of the interaction leads to a breathing motion predicting number anti-squeezing (and phase squeezing).
Damping of the oscillations is observed for increasing values of $\Lambda$~ \cite{ChoiPRA2005, GordonPRA1999} 
corresponding to the wavepacket feeling the anharmonicity of the effective potential.

These predictions are recovered, for $0<\Lambda \ll 1$, by a frozen-spin approximation \cite{LawPRA2001}
that consists of neglecting fluctuations of the mean spin, $\op{J}_x$ being replaced by $\pm N/2$. The variances
\begin{subequations}
\begin{align} 
\label{Eq.TNTz} (\Delta \J_z)^2 =& \frac{N}{4} \Big(  \cos^2 (\omega t) + \frac{\Omega^2}{\omega^2} \sin^2 (\omega t) \Big), \\
\label{Eq.TNTy} (\Delta \J_y)^2 =& \frac{N}{4} \Big(  \cos^2 (\omega t) + \frac{\omega^2}{\Omega^2} \sin^2 (\omega t) \Big),
\end{align}
\end{subequations}
show periodic oscillations and the spin-squeezing parameter reaches $\xiR^2  = 1/(1 + \Lambda)$ 
($\xiR^2 = 1-\Lambda$) for the coherent spin state pointing along the positive (negative) $x$ axis.
 
The situation changes completely if $\Lambda>1$ \cite{MicheliPRA2003, GordonPRA1999}. 
The potential $W_{\pi}(z)$ turns from a single-well to a double-well shape, with minima at $z_{\pm} = \pm \sqrt{1-1/\Lambda^2}$.
The initial coherent spin state corresponds to a Gaussian wavepacket that sits at the top of the double-well barrier, see Fig.~\ref{Fig:TwistTurn}. 
The repulsive potential tends to split the state giving rise to a macroscopic superposition 
corresponding to the wavepacket localized in the left and right well of $W_\pi(z)$. 
The semiclassical equations of motion $\dot{z} = -\tfrac{\partial H(z,\varphi)}{\partial \varphi}$ and $\dot{\varphi} = \tfrac{\partial H(z,\varphi)}{\partial z}$, where 
$H(z,\varphi)$ is given by Eq.~\eqref{Eq.TwoModeHCL}, offer an alternative view \cite{SmerziPRL1997, RaghavanPRA1999}. 
The fixed point $\Upsilon_\pi \equiv (z=0, \varphi=\pi)$ becomes unstable for $\Lambda>1$ and  
two new stable fixed points $\Upsilon_{\pm} \equiv (z_{\pm}, \pi)$ appear. 
The exact quantum dynamics shows that, at short times, the state stretches along the semiclassical separatrix passing through $\Upsilon_\pi$, see Fig.~\ref{Fig:TwistTurn}.
At longer times the quantum state wraps around $\Upsilon_\pm$.
In this regime, spin squeezing is lost, while the quantum Fisher information---sensitive to the creation of macroscopic superposition of states---continues increasing. 
For $\Lambda = 2$ the semiclassical separatrix reaches the poles of the Bloch sphere: 
the dynamics creates NOON-like states on a time scale $\chi t \approx \ln(8N)/N$ \cite{MicheliPRA2003}. 
For $\Lambda > 2 $ the separatrix winds around the Bloch sphere, the maximum distance between the separatrix and the equator decreases,
and the dynamics resembles that of the one-axis twisting. 
 
\begin{figure}[t!]
\begin{center}
\includegraphics[width=\columnwidth]{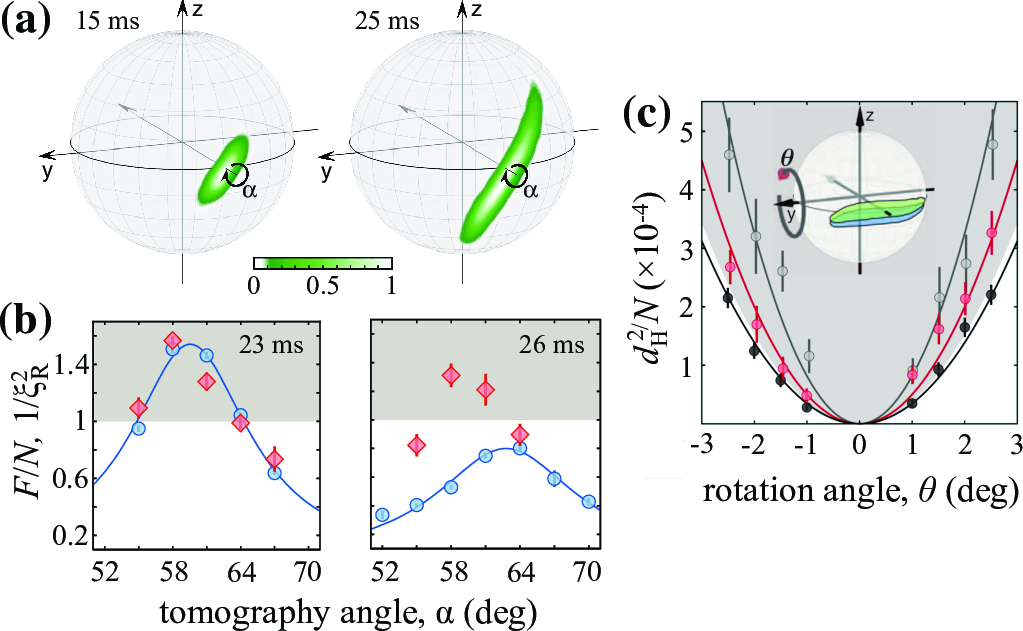}
\end{center}
\caption{{\bf Entanglement of non-spin-squeezed states.} 
(a) Experimental Husimi distributions at different evolution times of the twist-and-turn dynamics. 
(b) Normalized Fisher information ($F/N$, red diamonds) and inverse spin-squeezing parameter ($1/\xiR^2$, blue circles) 
as a function of the tomography angle [identified by the rotation angle $\alpha$ in panel (a)]. 
The grey shaded area is only accessible by separable states. 
At 26 ms, spin squeezing cannot witness the entanglement that is detected by the Fisher information.
(c) Squared Hellinger distance as a function of the rotation angle $\ps$ (see inset) at three evolution times: 
$t=0$ corresponding to a separable state (black points), $t=15$ ms corresponding to an optimal spin-squeezed state (light grey points), 
and $t=26$ ms corresponding to a state that is entangled but not spin squeezed (red [grey] points).
The curvature of the quadratic fits (solid lines) is proportional to $F/N$.
Adapted from \textcite{StrobelSCIENCE2014}.}  
\label{Fig:ExpTwistTurn}
\end{figure}
 
\subsubsection{Entanglement beyond spin squeezing}

Twist-and-turn dynamics has been investigated by \textcite{StrobelSCIENCE2014, MuesselPRA2015} using 
two internal hyperfine levels $\ket{b} = \ket{F=1, \mf=1}$ and $\ket{a} = \ket{2,-1}$ of $^{87}$Rb atoms. 
A narrow Feshbach resonance is used to reduce the inter-species interaction.
Coupling between the internal levels is provided by radiofrequency and microwave drive.
The experiments start with $N\approx 400$ atoms in the $\ket{b}$ mode, followed by a $\pi/2$ Rabi pulse that prepares a coherent spin state aligned with the $x$-axis.
Subsequently, the Rabi coupling is decreased to reach $\Lambda \approx 1.5$, and its phase adjusted to orient it along the negative $x$-axis.   
\textcite{MuesselPRA2015, StrobelSCIENCE2014} have investigated the short-time spin-squeezing dynamics.
For an evolution of 15\,ms, \textcite{StrobelSCIENCE2014} demonstrated $\xiR^2=-4.5$\,dB [and inferred $\xiR^2=-7.1$\,dB \cite{MuesselPRA2015}].
\textcite{MuesselPRA2015} also demonstrated a spin squeezing $\xiR^2=-4$\,dB using $\sim$30 independent condensates in parallel 
(each condensate experiencing independently a twist-and-turn dynamics), with a total of $N=10^4$ particles.
For longer times, spin squeezing is quickly lost and the experimental Husimi distribution of the reconstructed 
state shows the characteristic S-shape, see Fig.~\ref{Fig:ExpTwistTurn}.
\textcite{StrobelSCIENCE2014} extracted the Fisher information using a ``Hellinger method'', reaching values $F>N$ and thus demonstrating that the state is entangled.
The experimental extraction of the Fisher information requires to rotate the state around the $y$-axis, and to collect the probability distributions
of the relative particle number in $\ket{a}$ and $\ket{b}$ at different rotation angles. 
The Fisher information is obtained from a quadratic fit of the Hellinger distance~\eqref{Eq.Hell}, according to Eq.~\eqref{Eq.Hell.expansion},
between a reference distribution (at $\theta=0$) and the distribution obtained at a finite $\theta$.
Experimental results are shown in Fig.~\ref{Fig:ExpTwistTurn}.

\subsection{Entanglement of two spatially-separated spinor Bose-Einstein condensates}
\label{Sec.Atom-Atom.separated}

Creating entanglement between two spatially separated---individually addressable---spinor Bose-Einstein condensates offers interesting possibilities: 
in such a system, local manipulations and measurements can be performed on the spin state of each condensate separately and nonlocal 
quantum correlations between the measurement results can be directly observed. 
This is particularly relevant for experiments on EPR entanglement and Bell tests, where the spatial separation can be used to 
rule out unknown causal influences  between the clouds.

\textcite{HePRL2011,BarGillPRL2011,KurkjianPRA2013} have proposed schemes to generate EPR entanglement between two spatially separated 
Bose-Einstein condensates using elastic collisions. 
The condensates represent collective spins $\J^{(1)}$ and $\J^{(2)}$, respectively, which can be individually addressed and prepared in a coherent spin state. \textcite{HePRL2011,BarGillPRL2011} have considered schemes where EPR entanglement is generated by 
first spin-squeezing each condensate and then interfering the squeezed states on a beam splitter. 
In the scheme of \textcite{KurkjianPRA2013}, on the other hand, a state-dependent potential is turned on such that the wave function of state $\ket{a}$ of the first condensate overlaps with the wave function of state $\ket{b}$ of the second condensate, picking up a collisional phase shift. Dropping constant and linear terms in $\J^{(i)}$, this realizes the Hamiltonian
$H\si{2BEC} =  \chi_1 (\J_z^{(1)})^2 + \chi_2 (\J_z^{(2)})^2 - \chi_{12} \J_z^{(1)} \J_z^{(2)}$, which generates entanglement between the two condensates in addition to spin-squeezing in each condensate. After an interaction time, the state-dependent potential is turned off and the condensates are spatially separated again for detection, revealing EPR entanglement between them \cite{KurkjianPRA2013}. 
For long interaction times, macroscopic entangled states can be created, as also analyzed by \textcite{ByrnesPRA2013}, 
who investigated applications of entangled Bose-Einstein condensates in quantum information processing \cite{ByrnesPRA2012}. 
While such experiments have not yet been reported, we note that continuous-variable entanglement has been created 
between spatially separated atomic vapor cells using a measurement-based scheme \cite{JulsgaardNATURE2001}.

\section{Entanglement via atomic collisions: spin-mixing dynamics}
\label{Sec.Spin-Mix}

The  generation of correlated photon pairs via spontaneous parametric down-conversion 
in nonlinear crystals \cite{KwiatPRL1995}
is one of the most widely used sources of entangled quantum states in optics \cite{OBrienNATPHOT2009}.
When post-selecting $n$ photons in the pair distribution 
(let us indicate with $\ket{\pm1}$ the signal and idler modes), 
the corresponding state is a twin-Fock, $\ket{n}_{+1} \ket{n}_{-1}$.
The twin-Fock state has been used to overcome the standard quantum limit in an optical interferometer 
using up to $n = 2$ photons per pair \cite{KrischekPRL2011, NagataSCIENCE2007, XiangNATPHOT2011}.
Without post-selection, parametric down-conversion creates quadrature-squeezed 
light \cite{WallsNATURE1983, WuPRL1986, SlusherPRL1985, BreitenbachNATURE1997, OuPRL1992}. 
Following the proposal of \textcite{CavesPRD1981}, squeezed light has been successfully used for optical interferometry 
beyond the standard quantum limit \cite{VahlbruchPRL2005, VahlbruchPRL2016}, with direct application to gravitational wave detectors \cite{Aasi2013, SchnabelNATCOMM2010}. 
In the same spirit, the generation of correlated pairs of atoms has thus attracted large interest: many experiments have been proposed and performed.
In the following, we review the most successful of these techniques,
namely, spin-mixing dynamics in a spinor Bose-Einstein condensate \cite{Stamper-KurnRMP2013, KawaguchiPHYSREP2012}, see Sec.~\ref{Sec.Spin-Mix.spinmix}.
We discuss the creation of twin-Fock states in Sec.~\ref{Sec.Spin-Mix.pairs}, and quadrature squeezing in Sec.~\ref{Sec.Spin-Mix.squeezing}. 
Finally, in Sec.~\ref{Sec.Spin-Mix.alternative} we review alternative protocols for the creation of atom pairs. 

\subsection{Spinor Bose-Einstein condensates}
\label{Sec.Spin-Mix.spinmix}

\subsubsection{Spin-changing collisions}

\begin{figure} 
\begin{center}
\includegraphics[width=\columnwidth]{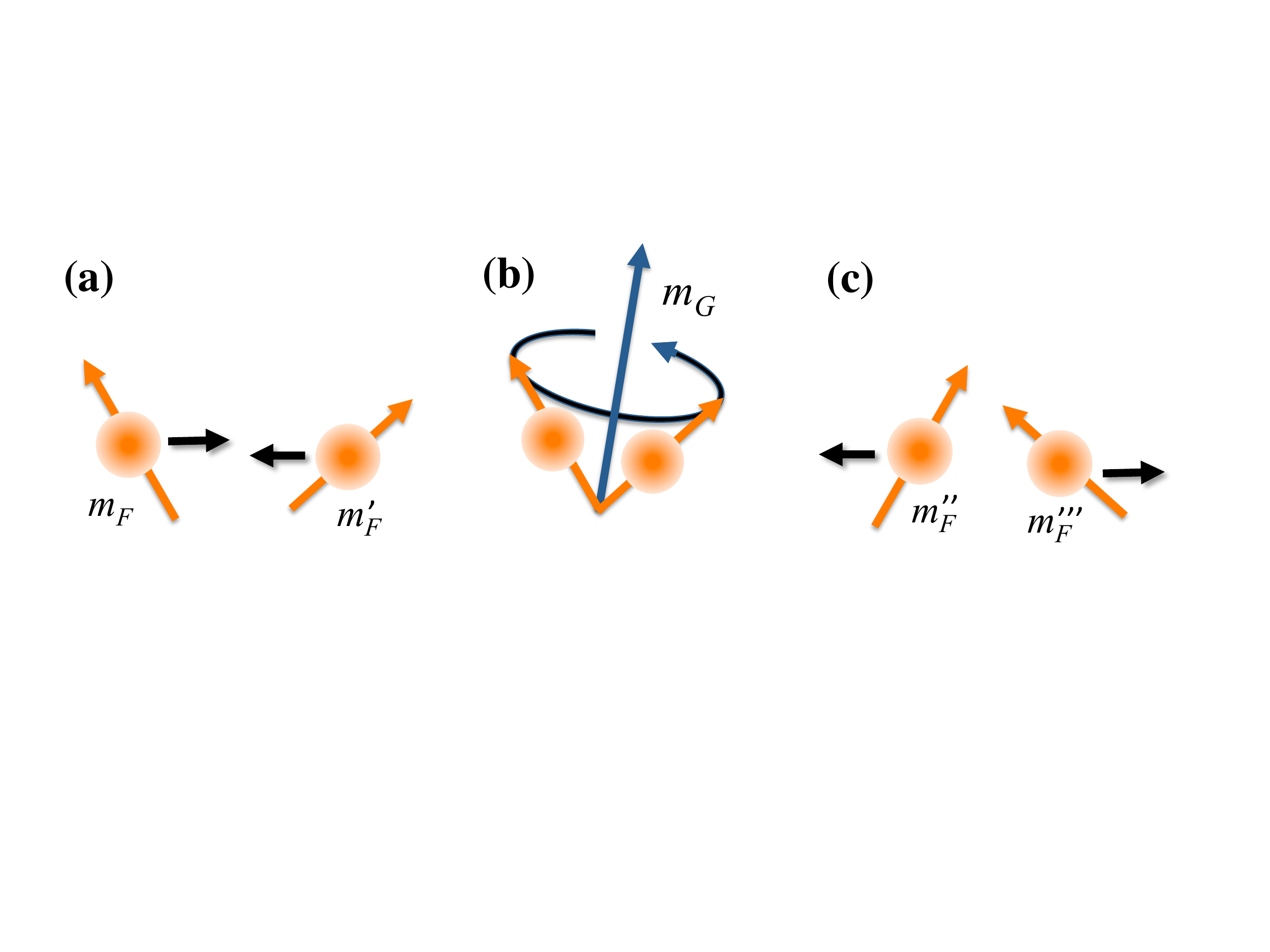}
\end{center}
\caption{{\bf Binary s-wave collision of two spin-$\f$ bosons.}
(a) When two spin-$\f$ bosons in internal states $\ket{\f,\mf}$ and $\ket{\f,\mf'}$ approach each 
other (here, $\f$ is the hyperfine spin and $\mf,\mf' = -\f, {-\f+1}, \dotsc, \f$ the magnetic quantum number), 
they couple to form a total spin $\vect{G} = \vect{\f} +\vect{\f}'$.
(b) The combined internal state is given by $\ket{G,m_G}$.
For bosons under elastic s-wave scattering, $G$ is restricted to even values satisfying $0\le G\le 2\f$, 
\eg, $G = 0,2$ for two $\f=1$ particles. 
(c)  After the collision the atoms fly apart and their internal states are again well described by the hyperfine spins $\f$. 
However, the magnetic quantum numbers may have changed: $\ket{\f,\mf}\ket{\f,\mf'}\mapsto\ket{\f,\mf''}\ket{\f,\mf'''}$.
The conservation of total angular momentum imposes $\mf+\mf'=\mf''+\mf'''$.}
\label{Fig:Spin01} \label{Fig:SpinColl}
\end{figure} 

When a Bose-Einstein condensate is confined in a far-off resonant optical dipole trap, 
the spin degree of freedom of the atoms evolves freely: 
spin-changing s-wave collisions, see Fig.~\ref{Fig:SpinColl}, give rise to a coherent redistribution of atomic 
populations among Zeeman sub-levels while preserving the total magnetization \cite{HoPRL1998, OhmiJPSJ1998}.
The quantum description of an ultracold gas of spin-$\f$ bosons requires introducing 
a vector order parameter with $2\f+1$ components, 
$\{ \op{\Psi}_{-\f}, \op{\Psi}_{-\f+1}, ...,  \op{\Psi}_{\f} \}$ \cite{Stamper-KurnRMP2013, KawaguchiPHYSREP2012}.
Here $\op{\Psi}_{\mf}$ is the atomic field annihilation operator associated 
with the hyperfine spin state $\ket{\f, \mf}$. 
The single-mode approximation \cite{LawPRL1998} assumes that the spatial atomic density distribution of all 
spin components is approximately equal, and not affected by the spin dynamics. 
The common scalar wavefunction $\phi(\vect{r})$ defining the spatial mode of the condensate is determined as the solution of the Gross-Pitaevskii equation,
neglecting contributions from the spin-dependent interactions.
In this approximation, the field operators are $\op{\Psi}_{\mf}(\vect{r}) = \phi(\vect{r}) \op{a}_{\mf}$, 
where $\op{a}_{\mf}$ are annihilation operators that obey the usual bosonic commutation relations.
We define as $\op{N}_{\mf} = \op{a}^\dag_{\mf} \op{a}_{\mf}$ the number of particles in the 
mode $\ket{\f,\mf}$, and $\op{N}=\sum_{\mf=-F}^F\op{N}_{\mf}$ the total number of particles.
The many-body Hamiltonian for a $\f=1$ Bose-Einstein condensate is \cite{LawPRL1998, PuPRA1999}
\begin{multline} \label{Hsmix}
\op{H}\si{SM} =   [q+ \lambda(2\op{N}_0 -1)] (\op{N}_{+1} + \op{N}_{-1}) 
+ 2 \lambda (\op{a}^\dag_{-1} \op{a}^\dag_{+1} \op{a}_{0} \op{a}_{0} + \op{a}_{0}^{\dag} \op{a}_{0}^{\dag} \op{a}_{-1} \op{a}_{+1} ),
\end{multline}
where we have neglected terms proportional to the conserved 
magnetization (we assume $\op{N}_{+1} - \op{N}_{-1}=0$ in the following) and total number of atoms.
Here, $q = (\Delta E_{1} + \Delta E_{-1})/2$ is an effective quadratic Zeeman shift, where $\Delta E_{\mf}  = E_{\mf}-E_0$ is the relative energy shift of the $\mf=\pm 1$ mode, 
which can be tuned by a magnetic 
field ($q$ being proportional to the square of the magnetic field) and/or near-resonant microwave dressing. The interaction parameter is 
$\lambda = \frac{g_2}{2} \int \dd^3 \vect{r} \abs{\phi(\vect{r})}^4$, 
with $g_2 = \tfrac{4 \pi \hbar^2 (a_2 - a_0)}{3M}$, and $a_G$ the scattering lengths for s-wave collisions in the $G=0,2$ allowed channels. 
The single-mode approximation is valid if the system size is much smaller than the spin healing length, 
$\sqrt{h^2/2M \abs{g_2} \rho_0}$, where $\rho_0$ is the density of the $\mf=0$ component, 
giving the minimum size of a spin-domain \cite{Stamper-KurnRMP2013}.
The last term in Eq.~\eqref{Hsmix} describes the coherent and reversible creation of a pair of atoms in the magnetic sublevels $\mf=\pm 1$ 
from the scattering of two atoms in $\mf = 0$.

It is useful to rewrite the Hamiltonian~\eqref{Hsmix} in terms of the spin operators 
\begin{align} \label{SAspin}
\op{\mathcal{S}}_x &= \frac{\op{a}_0^\dag \sa  + \op{a}_0 \sa^\dag}{2}, &
\op{\mathcal{A}}_x &= \frac{\op{a}_0^\dag \as  + \op{a}_0 \as^\dag}{2},\nonumber\\
\op{\mathcal{S}}_y &= \frac{\op{a}_0^\dag \sa  - \op{a}_0 \sa^\dag}{2\ii}, &
\op{\mathcal{A}}_y &= \frac{\op{a}_0^\dag \as  - \op{a}_0 \as^\dag}{2\ii},\nonumber\\
\op{\mathcal{S}}_z &= \frac{\op{a}_0^\dag \op{a}_0  - \sa^\dag \sa}{2}, &
\op{\mathcal{A}}_z &= \frac{\op{a}_0^\dag \op{a}_0  - \as^\dag \as}{2},
\end{align}
where $\sa = (\op{a}_{+1} + \op{a}_{-1})/\sqrt{2}$ and $\as = (\op{a}_{+1} - \op{a}_{-1})/\sqrt{2}$
are symmetric and antisymmetric combinations of $\op{a}_{+1}$ and $\op{a}_{-1}$.
The spin operators $\vectop{\mathcal{S}} = \{ \op{\mathcal{S}}_x, \op{\mathcal{S}}_y, \op{\mathcal{S}}_z\}$ and 
$\vectop{\mathcal{A}} = \{ \op{\mathcal{A}}_x, \op{\mathcal{A}}_y, \op{\mathcal{A}}_z\}$ 
do not commute with each other, and define two SU(2) subspaces.
With these operators, the Hamiltonian~\eqref{Hsmix} takes the compact form \cite{DuanPRA2002}
\be \label{HamSA}
\op{H}\si{SM} = \big( 4 \hbar \lambda \op{\mathcal{S}}_x^2 - \frac{2\hbar q}{3} \op{\mathcal{S}}_z \big) + \big( 4 \hbar \lambda \op{\mathcal{A}}_y^2 - \frac{2 \hbar q}{3} \op{\mathcal{A}}_z \big),
\ee 
which highlights the presence of nonlinear spin terms.
It should be noticed that a symmetric radio-frequency coupling between the $\mf=0,\pm 1$ modes with Rabi frequency $\Omega\si{rf}$ corresponds to a 
rotation of the $\op{\mathcal{\vect{S}}}$ vector around the $x$ axis, 
$\op{H}\si{rf} = \tfrac{\hbar \Omega\si{rf}}{2\sqrt{2}}(\op{a}_0^\dag \op{a}_{+1} + \op{a}_0^\dag \op{a}_{-1} + h.c.) = \hbar \Omega\si{rf} \op{\mathcal{S}}_x$.
A relative phase shift between the $\mf=0$ and $\mf=\pm 1$ modes corresponds to a rotation of both $\op{\mathcal{S}}$ and $\op{\mathcal{A}}$ around the $z$ axis, 
$e^{-\ii \ps (\op{N}_0 - \op{N}_{+1})/2}  e^{-\ii \ps (\op{N}_0 - \op{N}_{-1})/2} = e^{-\ii \ps \op{\mathcal{S}}_z } e^{-\ii \ps \op{\mathcal{A}}_z }$.

Another popular expression of Eq.~(\ref{Hsmix}) is \cite{LawPRL1998, ZhangPRL2013}
\be
\op{H}\si{SM} = \lambda \op{\vect{L}} - q \op{N}_0,
\ee
where $\op{\vect{L}} = \{\op{L}_x, \op{L}_y, \op{L}_z \}$, $\op{L}_x = 2 \op{\mathcal{S}}_x$, $\op{L}_y = 2 \op{\mathcal{A}}_y$
and $\op{L}_z = \op{N}_{-1} - \op{N}_{+1}$ obey angular momentum commutation relations.\footnote{Some authors have also 
studied spin-1 condensates using the 
a quadrupole tensor operator $\vectop{Q}$ \cite{HamleyNATPHYS2012, SauNJP2010, MustecapliogluPRA2002} with components
related to Eq.~(\ref{SAspin}) as
$\op{Q}_{yz} = -2 \op{\mathcal{S}}_y$, $\op{Q}_{xz} = 2 \op{\mathcal{A}}_x$, $\op{Q}_{zz} - \op{Q}_{yy} = -4 \op{\mathcal{S}}_z$, $\op{Q}_{xx} - \op{Q}_{zz} = 4 \op{\mathcal{A}}_z$ 
and $\op{Q}_{xy} = \ii (\op{a}_{-1}^\dag \op{a}_{+1}-\op{a}_{+1}^\dag \op{a}_{-1})$.}

\begin{figure}[t!]
\begin{center}
\includegraphics[width=\columnwidth]{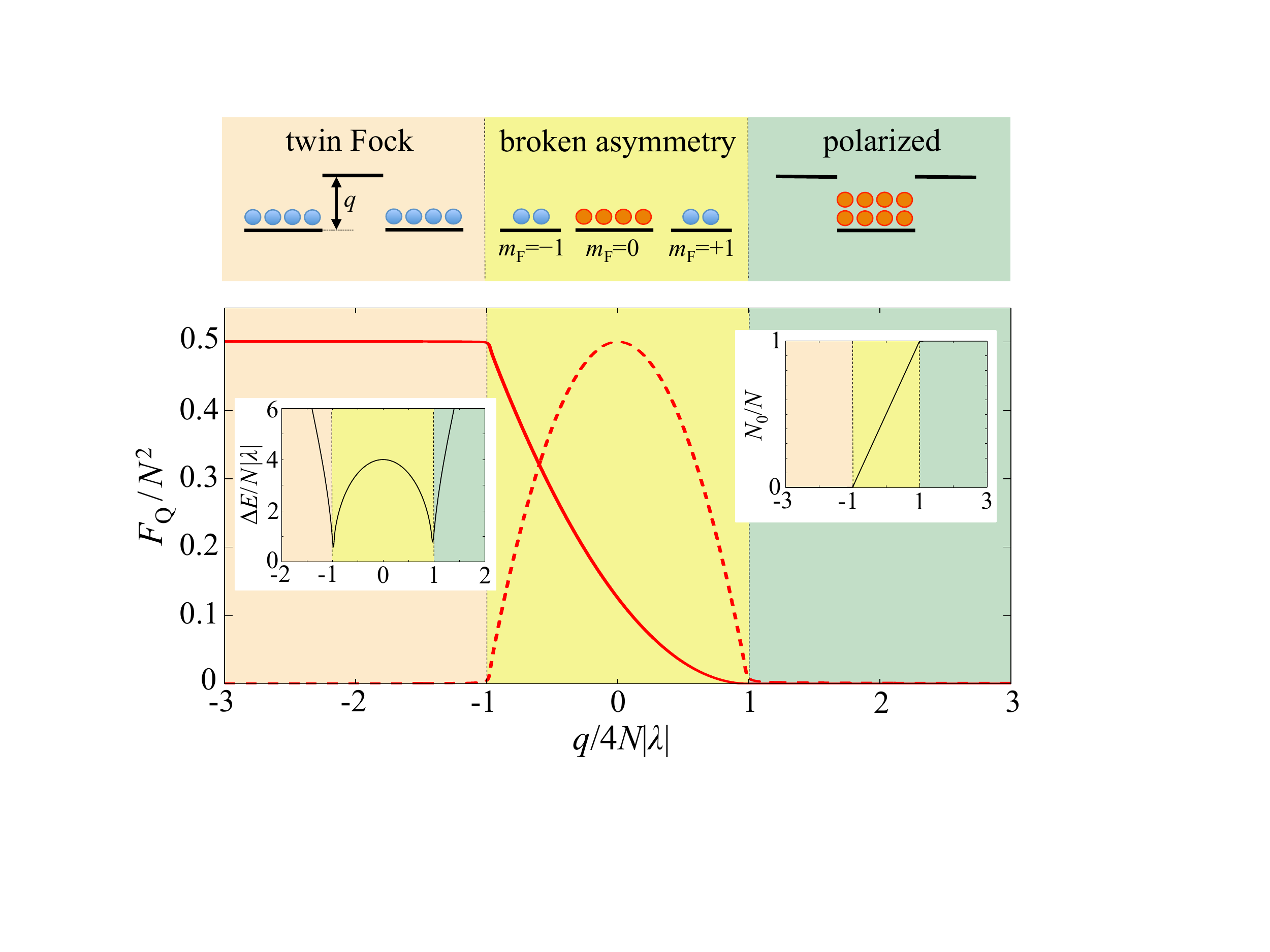}
\end{center}
\caption{{\bf Entanglement in the ground state of the spin-mixing Hamiltonian.}
The upper panel reports a schematic representation of the different phases obtained as a function of $q$ (linear Zeeman shifts are not shown).
The lower panel shows $F_Q[\ket{\psi_{\rm gs}}, \op{J}_{x}]/N^2$ (solid line) and $F_Q[\ket{\psi_{\rm gs}}, \op{\mathcal{S}}_x]/N^2$ (dashed line)
as a function of $q/4N|\lambda|$.
The left inset shows the energy gap $\Delta E$ between the ground state and the first excited state, closing at $q_c = \pm 4N|\lambda|$.
The right inset shows the normalized population of the $\mf=0$ mode, $N_0/N$.
Here $\lambda<0$ and $N=1000$.} 
\label{Fig:GSSM} 
\end{figure}

\subsubsection{Ground state of the spin-mixing Hamiltonian}
\label{Subsec4A2}

The ground state of the Hamiltonian~\eqref{Hsmix} is characterized by the competition between the 
energy shift, proportional to $q$, and the spin-dependent collisional interaction, proportional to $\lambda$ 
\cite{LamacraftPRL2007, MurataPRA2007, SadlerNATURE2006}.
The mean-field limit of Eq.~\eqref{Hsmix}, obtained by replacing $\op{a}_{\mf} = \sqrt{N_{\mf}} e^{-\ii \phi_{\mf}}$ and assuming $N_{+1}=N_{-1}$, 
gives $E\si{SM} = - q N_0 + 4 \hbar N_0 (N - N_0)  \cos^2(\phi/2)$, 
where $\phi = 2 \phi_0 - \phi_{+1} - \phi_{-1}$ \cite{ZhangPRA2005, Stamper-KurnRMP2013}.
In a full quantum treatment, the system is conveniently studied in the Fock basis $\{ \ket{k}_{-1} \ket{N-2k}_{0} \ket{k}_{+1} \}_{k=0,1,...,N/2}$, 
where $\ket{n}_{\mf}$ indicates a state of $n$ particles in the mode $\mf$, 
or in terms of the eigenstates $\ket{l,m}$ (here $m=0$) of $\op{\vect{L}}$ and $\op{L}_z$ \cite{LawPRL1998}.

For a ferromagnetic condensate, $\lambda<0$ (\eg, $^{87}$Rb) 
there are two quantum phase transitions with order parameter given by the mean population of the $\mf=0$ mode.
For $q > 4 N \abs{\lambda}$, the ground state is $\ket{0}_{-1} \ket{N}_{0} \ket{0}_{+1}$, corresponding to a polar phase with all particles in $\mf=0$ ($N_0 = N$).
For $\abs{q} < 4 N \abs{\lambda}$, the minimization leads to $\phi=0$ and $N_0 = (4 \abs{\lambda} N + q)/8 \abs{\lambda}$, corresponding to the co-called broken asymmetry phase.
The population of $N_0$ decreases linearly with $q$ and $\mf=\pm 1$ are (equally) populated.
For $q < -4 N \abs{\lambda}$, we have a twin-Fock phase with $N_0 = 0$ and the ground state given by the twin-Fock state $\ket{N/2}_{-1} \ket{0}_{0} \ket{N/2}_{+1}$.
Entanglement in the ground state for the spin-1 system has been studied by \textcite{ZhangPRL2013}.
\textcite{FeldmannPRA2018} have studied the quantum Fisher information of the ground state $\ket{\psi_{\rm gs}}$ 
of the Hamiltonian~\eqref{Hsmix} for different operators.
In particular, $F_Q[\ket{\psi_{\rm gs}}, \op{\mathcal{S}}_x] = F_Q[\ket{\psi_{\rm gs}}, \op{\mathcal{A}}_y] = N(N+1)/2$ at $q=0$ \cite{ZouArXiv2018},
associated to the presence of macroscopic superposition states \cite{PezzeArXiv2017}, 
and $F_Q[\ket{\psi_{\rm gs}}, \op{J}_{x,y}] =  N(N+2)/2$ in the twin-Fock phase, see Fig.~\ref{Fig:GSSM}, 
where $\op{J}_{x} = (\op{a}^\dag_{+1} \op{a}_{-1}+ \op{a}_{+1} \op{a}_{-1}^\dag)/2$ and 
$\op{J}_{y} = (\op{a}^\dag_{+1} \op{a}_{-1}- \op{a}_{+1} \op{a}_{-1}^\dag)/2\ii$.

For an anti-ferromagnetic condensate, $\lambda>0$ (\eg, $^{23}$Na).,
the mean field energy is minimized by taking $\phi=\pi$ and $N_0=N$ for $q>0$, and $N_0=N$ for $q<0$.
There is a quantum phase transition at $q=0$.
In the limit $N\to \infty$, for $q>0$ the ground state is given by the fully polarized state $\ket{0}_{-1} \ket{N}_{0} \ket{0}_{+1}$, 
while for $q<0$ by the twin-Fock state $\ket{N/2}_{-1} \ket{0}_{0} \ket{N/2}_{+1}$.
See \textcite{WuPRA2016} for a calculation of the quantum Fisher information of the ground state 
for $q=0$ and nonzero magnetization.

\subsubsection{Quantum spin-mixing dynamics in the low-depletion limit}
\label{SubSec.SMD}

The spin dynamics of an initial condensate in $\mf=0$ is  
formally analogous to optical spontaneous four-wave mixing \cite{GoldsteinPRA1999}:
atom-atom interaction plays the role of the nonlinear Kerr medium, the $\mf = 0$ condensate is equivalent to a coherent pump field with 
the external trapping potential corresponding to a high-finesse cavity, 
and the condensates in $\mf = \pm1$ can be identified as signal and idler.
Remarkably, spin changing collisions are unaffected by the linear Zeeman shift from a homogeneous magnetic field. 
Only higher-order effects, such as a quadratic Zeeman shift or a linear Zeeman shift from an inhomogeneous magnetic field, affect it.

The low-depletion limit, \ie, $N_0 \approx N \gg 1$,
is analyzed by replacing the mode operator $\op{a}_0$ in Eq.~\eqref{Hsmix} with $\sqrt{N_0}$. 
In this approximation, the condensate serves as an unlimited particle resource for the parametric amplification of the $\mf=\pm 1$ modes.
We obtain the quadratic Hamiltonian 
$\op{H}\si{SM} = \alpha (\op{a}^\dag_{+1}\op{a}_{+1} + \op{a}^\dag_{-1}\op{a}_{-1}) 
+ \beta (\op{a}^\dag_{-1} \op{a}^\dag_{+1} +  \op{a}_{-1} \op{a}_{+1} )$, 
where $\alpha = q + \lambda (2N_0-1)$ and $\beta = 2 \lambda N_0$,
that can be diagonalized by a Bogoliubov transformation \cite{DuanPRL2000, PuPRL2000}, see also \textcite{TruaxPRD1985}.
The unitary evolution $\ket{\psi\si{SM}(t) } = e^{-\ii \op{H}\si{SM} t/\hbar} \ket{\text{vac}}$ can be calculated exactly, giving
\be \label{Eq.TMSVgeneral}
\ket{\psi\si{SM}(t) } = \sum_{n=0}^{+\infty} 
\frac{[-\ii \beta e^{-2 \ii \phi_0} \tau \sin (\tfrac{t}{\tau})]^n}{[\cos (\tfrac{t}{\tau})  + \ii \alpha \tau \sin(\tfrac{t}{\tau})]^{n+1}}
 \ket{n}_{+1} \ket{n}_{-1},
\ee
where $\tau = \hbar/\sqrt{\alpha^2-\beta^2}$, and $\ket{\text{vac}} = \ket{0}_{+1}\ket{0}_{-1}$ indicates empty $\mf=\pm1$ modes.
This state has a vanishing population difference $\Delta^2 (\op{N}_{+1}-\op{N}_{-1})=0$, while the $\mf=\pm 1$ modes are nonempty, 
$\mean{\op{N}_{\pm 1}} =  \tfrac{\beta^2}{\alpha^2-\beta^2} \sin^2(t\sqrt{\alpha^2-\beta^2}/\hbar)$, and characterized by large (super-Poissonian) population fluctuations, 
$\Delta^2 N_{\pm 1} = \mean{\op{N}_{\pm 1}}(\mean{\op{N}_{\pm 1}}+1)$.
It is also interesting to consider the quadratures
\begin{subequations}
\begin{align}
	2(\Delta \op{Q}_{\mathcal{s}})^2 &= V_Q \cos^2 ( \phi  - \phi_Q ) + \frac{1}{V_Q} \sin^2 ( \phi  - \phi_Q ),\label{EqQs}\\
	2(\Delta \op{Q}_{\mathcal{a}})^2 &= V_Q \sin^2 ( \phi  - \phi_Q ) +  \frac{1}{V_Q} \cos^2 ( \phi  - \phi_Q ),\label{EqQa}
\end{align}
\end{subequations}
where $\op{Q}_{\mathcal{s}} = \cos \phi \op{X}_{\mathcal{s}} + \sin \phi \op{P}_{\mathcal{s}}$, 
$\op{X}_{\mathcal{s}} = (\sa+\sa^\dag)/\sqrt{2}$, $\op{P}_{\mathcal{s}} = (\sa-\sa^\dag)/\ii\sqrt{2}$, 
and an analogous definitions of $\op{Q}_{\mathcal{a}}$ in terms of the operators $\as$ and $\as^\dag$ introduced above.
The time-dependent coefficients are $V_Q = 1 + 2 \mean{\op{N}_{\pm 1}} - 2 \Delta \op{N}_{\pm 1}$ and 
$\cos(2 \phi_Q) = \tfrac{\alpha}{\beta} \tfrac{ \mean{\op{N}_{\pm 1}} }{ \Delta \op{N}_{\pm 1} }$. 

For $\alpha^2<\beta^2$, $\tau$ is imaginary and the condensate is dynamically unstable. 
The instability is characterized by the exponential increase of population in the $\mf = \pm 1$ modes, 
$\mean{\op{N}_{\pm1}} = \tfrac{\beta^2}{\beta^2-\alpha^2} \sinh^2(t\sqrt{\beta^2-\alpha^2}/\hbar)$.
On resonance ($\alpha = 0$) the spin-mixing Hamiltonian becomes 
$\op{H}\si{SM} = 2 \lambda N_0 ( \op{a}^\dag_{-1} \op{a}^\dag_{+1} + \op{a}_{-1} \op{a}_{+1})$, 
which generates the familiar two-mode squeezed-vacuum state \cite{WallsBOOK},
\be \label{Eq.TMSV}
\ket{\psi\si{SM}(t) } = \sum_{n=0}^{+\infty} \frac{(-\ii  \tanh r)^{n}}{\cosh r} \ket{n}_{+1} \ket{n}_{-1},
\ee
where  $r = \abs{\beta} t/\hbar = 2 \abs{\lambda} N_0 t / \hbar$ and $\mean{\op{N}_{\pm 1}}= \sinh^2 r$.
The Hamiltonian $\op{H}\si{SM}$ (for $\alpha=0$) can be rewritten
using the operators $\sa$ and $\as$ as
$\op{H}\si{SM} = \tfrac{\beta}{2} (\sa^\dag \sa^\dag + \sa \sa ) - \tfrac{\beta}{2} (\as^\dag \as^\dag + \as \as )$,
which generates a single-mode squeezed-vacuum \cite{WallsBOOK} in each mode.
It can also be rewritten as
$\op{H}\si{SM} = 2 \lambda [ (\op{\mathcal{S}}_x^2 - \op{\mathcal{S}}_y^2) - (\op{\mathcal{A}}_x^2 - \op{\mathcal{A}}_y^2) ]$,
corresponding to two-axis counter-twisting \cite{KitagawaPRA1993, DuanPRA2002, AndersPRA2018} for the $\mathcal{S}$ and $\mathcal{A}$ spins.
Note that $\mathcal{S}$ and $\mathcal{A}$ commute in the low-depletion limit.
For $\alpha=0$, Eq.~\eqref{EqQs} reduces to $2(\Delta \op{Q}_{\mathcal{s}})^2 = e^{2r} \sin^2 ( \phi - \frac{\pi}{4} ) + e^{-2r} \cos^2 ( \phi - \frac{\pi}{4} )$, and 
Eq.~\eqref{EqQa} to $2(\Delta \op{Q}_{\mathcal{a}})^2 = e^{2r} \cos^2 ( \phi - \frac{\pi}{4} ) + e^{-2r} \sin^2 ( \phi - \frac{\pi}{4} )$.
The quadrature $(\Delta \op{Q}_{\mathcal{s}})^2$ [$(\Delta \op{Q}_{\mathcal{a}})^2$] is squeezed 
by a factor $e^{-2r}$ below the vacuum level at an optimal angle $\phi=\pi/4$ [$\phi=3\pi/4$], and 
anti-squeezed by a factor $e^{2r}$ at $\phi=3\pi/4$ [$\phi=\pi/4$].
For a sufficiently long evolution time, $t \gtrsim \hbar/(2\abs{\lambda} N_0)$ spin dynamics is modified by the depletion of the condensate. 
The initially exponential growth of population in $\mf=\pm 1$ slows down and finally stops \cite{LawPRL1998, MiasPRA2008}
 
For $\alpha^2 \geq \beta^2$ the spin-mixing dynamics is stable and the population of $\mf=\pm 1$ oscillates in time, 
$\mean{\op{N}_{\pm1}} = \tfrac{\beta^2}{\alpha^2-\beta^2} \sin^2(t\sqrt{\alpha^2-\beta^2}/\hbar)$.
For $\alpha^2 \gg \beta^2$ the amplitude of this oscillation becomes negligible and the system remains stable in $\mf=0$.

Typical experiments exploring the quantum spin-mixing dynamics \cite{LesliePRA2009, LiuPRL2009, 
KlemptPRL2010, ChangNATPHYS2005, SchmaljohannPRL2004, PeiseNATCOMM2015b} start with a condensate in the stable configuration. 
A sudden quench brings the condensate in the dynamical unstable configuration.
After spin-mixing dynamics, the trap is turned off and a Stern-Gerlach field is used to separate the $\mf$ components during a time-of-flight expansion.

Finally, it is worth recalling that the discussion so far has been focused on one spatial eigenmode.
A rich resonance structure is expected when taking into account many spatial modes, as experimentally shown by \textcite{KlemptPRL2010, SchererPRL2010}.

\subsection{Population correlations and twin-Fock state} 
\label{Sec.Spin-Mix.pairs}
 
\begin{figure}[t!]
\begin{center}
\includegraphics[width=\columnwidth]{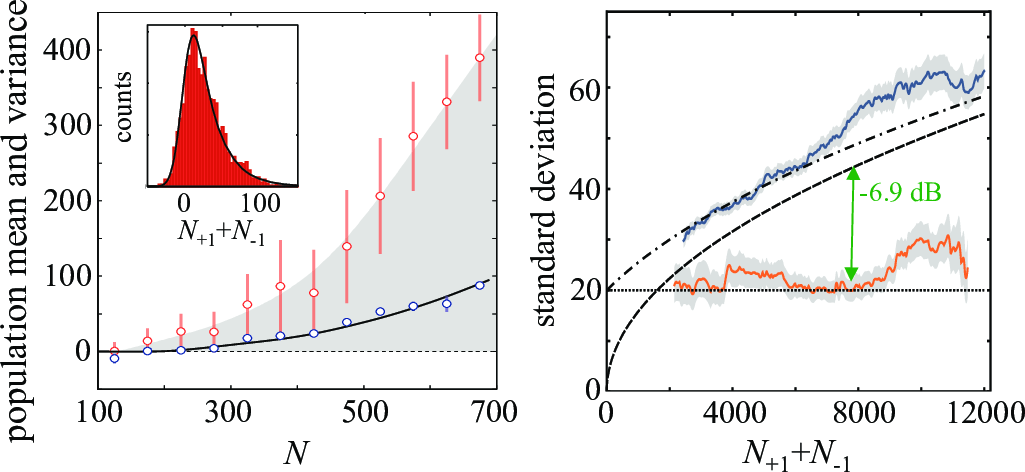}
\end{center}
\caption{{\bf Population correlations after spin-mixing dynamics.}
Left panel: $\mean{\op{N}_{+1} + \op{N}_{-1}}$ (red/light circles) and $\Delta^2 (\op{N}_{+1} - \op{N}_{-1})$ (blue/dark circles)
as a function of the total atom number $N$. 
Error bars are fluctuations (one standard deviation) and the grey area corresponds to the sub-Poisson regime.
The black line is a theoretical model including particle loss due to spin relaxation.
The inset shows the distribution of $N_{+1} + N_{-1}$ for $250<N < 300$. 
The black line is a fitted squeezed-vacuum distribution corresponding to $r \approx 2$.
Adapted from \textcite{GrossNATURE2011}.
Right panel: standard deviation of the population difference $(\op{N}_{+1} - \op{N}_{-1})/2$ (red/lower line),
compared to the projection noise $\sqrt{N_{+1} + N_{-1}}/2$ (dashed line), and $\sqrt{(N_{+1} + N_{-1})/4+\sigma\si{dn}^2}$ (dot-dashed line)
taking into account a number-independent detection noise $\sigma\si{dn} = 20$ (dotted line).
The blue/upper line is the experimental result for unentangled atoms.
The shaded area indicates the standard deviation.
Adapted from \textcite{LuckeSCIENCE2011}} 
\label{Fig:MV} 
\end{figure}

\begin{figure}[t!]
\begin{center}
\includegraphics[width=\columnwidth]{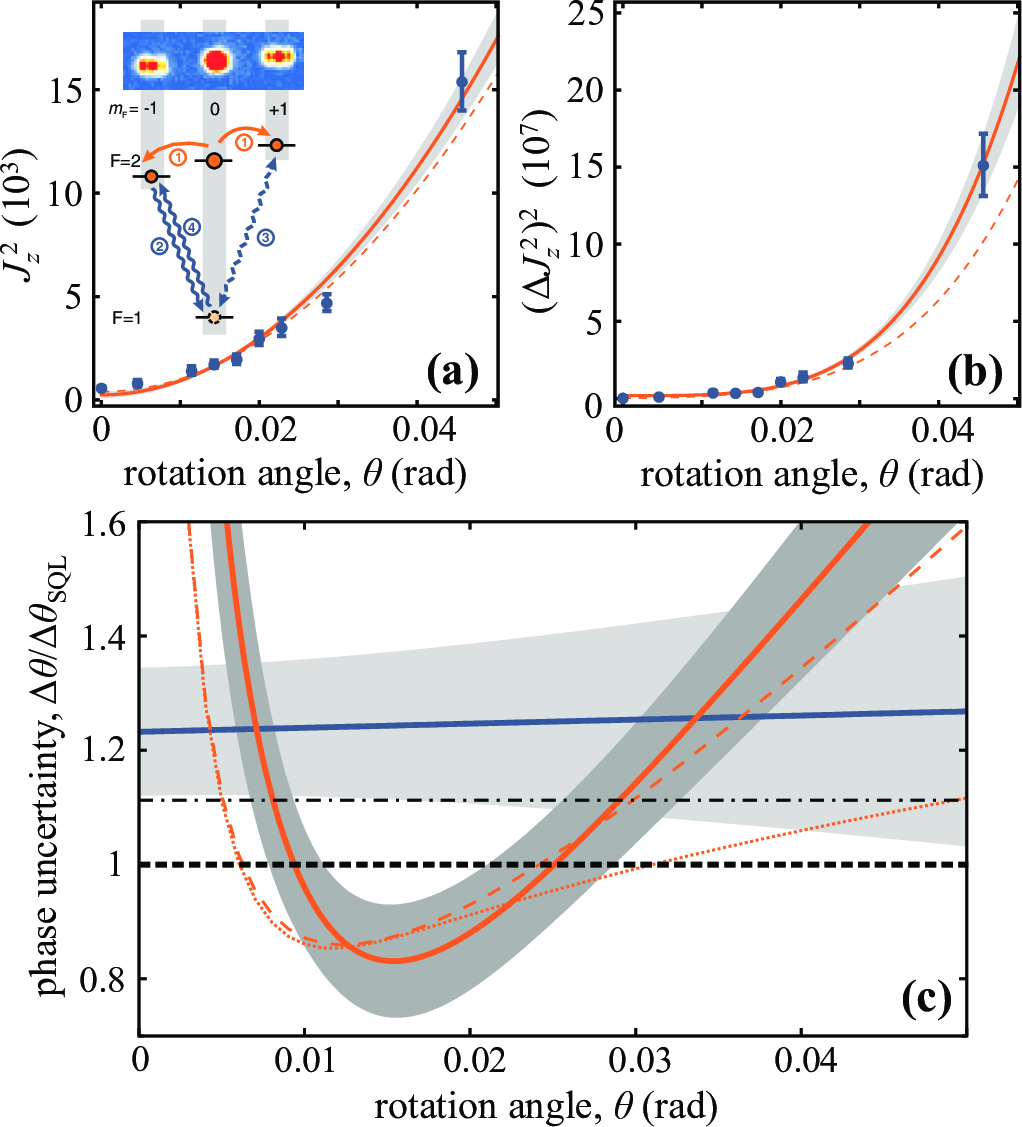}
\end{center}
\caption{{\bf Twin-Fock interferometry.} 
The inset of panel (a) shows the experimental operations: (1) spin dynamics, (2) a resonant microwave $\pi$ pulse 
between $\ket{2,-1}$ and $\ket{1,0}$, (3) a pulse of variable duration, defining $\ps$, and (4) a second $\pi$ pulse.
a) Second moment $\mean{\J_z^2}$ of the population imbalance $\J_z = (\op{N}_{+1}-\op{N}_{-1})/2$ 
as a function of $\ps$ (dots with error bars) for post-selected numbers of atoms between $6400$ and $7600$. 
The solid line is a polynomial fit (with grey uncertainty region); the dotted line is the theoretical prediction 
including detection noise. 
b) Same for $(\Delta \J_z^2)^2=\mean{\J_z^4}-\mean{\J_z^2}^2$.
c) Phase estimation uncertainty obtained via error propagation, $\Delta \ps = (\Delta \J_z^2)/\abs{\dd \mean{\J_z^2} / \dd \ps}$ 
(solid orange line with grey uncertainty region) compared to the theoretical prediction (dashed orange line)
and Cram\'er-Rao bound (dotted orange line) including detection noise only. 
Around $\ps=0.015$ rad the phase variance lies 1.61\,dB below the standard quantum limit (black dashed line).
Adapted from \textcite{LuckeSCIENCE2011}.} 
\label{Fig:LuckeSCIENCE2012}
\end{figure} 
 
Several experiments have explored the creation of the
twin-Fock state $\ket{n}_{+1}\ket{n}_{-1}$ in the $\mf =\pm 1$ Zeeman modes.
This can be accessed either by quantum spin-mixing dynamics
(which can turn a large fraction of an atomic Bose-Einstein condensate into a mixture of perfectly correlated pairs of atom, with large fluctuations of $n$) 
or by adiabatic preparation.
The twin-Fock state is particle entangled and useful in a Ramsey interferometer to reach sensitivities beyond the standard quantum limit, see Sec.~\ref{sec:DickeStates}. 
  
\subsubsection{Number squeezing} 
 
The presence of correlations in the populations of $\mf =\pm 1$ states after spin dynamics has 
been first demonstrated by \textcite{BookjansPRL2011, LuckeSCIENCE2011, GrossNATURE2011}. 
Number squeezing $\xiN^2 = \Delta^2 (\op{N}_{+1} - \op{N}_{-1})/(N_{+1} + N_{-1})$ up to
$\xiN^2\approx -7$\,dB [$\xiN^2=-12.4$\,dB has been reported more recently by \textcite{LueckePRL2014}] 
below the projection noise level has been reported, see Figs.~\ref{Fig:MV}.
The main limitation is given by detection noise for short-time evolutions, and particles losses for long time.  
 
\subsubsection{Quantum interferometry with twin-Fock states} 
\label{SubSec.TF}

\textcite{LuckeSCIENCE2011} have investigated the phase sensitivity of the paired atoms.
The experimental sequence starts with the spin-mixing dynamics of a $^{87}$Rb condensate prepared in $\ket{F= 2,\mf = 0}$.
The generated atom pairs in $\ket{2, \pm 1}$ 
(due to the short evolution time, the populations in $\ket{2, \pm 2}$ can be neglected due to the quadratic Zeeman shift detuning) are 
coupled via a series of microwave pulses of variable duration, see Fig.~\ref{Fig:LuckeSCIENCE2012}. 
This forms an internal-state beam splitter $e^{-\ii \ps \J_x}$, where
the rotation angle $\ps$ is estimated via the measurement of $\J_z^2=(\op{N}_{+1}-\op{N}_{-1})^2/4$.
The output state is post-selected to a total number $N_{+1}+N_{-1} \approx 10^4$.  
Ideally, this post-selection would correspond to a twin-Fock state as input of the beam splitter operation.
Applying error propagation $\Delta \ps = (\Delta \J_z^2)/\abs{\dd \mean{\J_z^2} / \dd \ps}$, it was possible to demonstrate
a sensitivity 1.61\,dB below the standard quantum limit at an optimal rotation angle, see Fig.~\ref{Fig:LuckeSCIENCE2012}.

\subsubsection{Particle entanglement} 
 
\textcite{LueckePRL2014} have investigated particle entanglement in the atomic twin-Fock state.
After spin-mixing dynamics of a $^{87}$Rb condensate, the number of particles in $\mf= \pm 1$ 
is post-selected for a total number of $N\approx 8000$.
Entanglement is witnessed by the inequality $\xiD^2 <1$, Eq.~\eqref{Eq.xiE}, 
further extended to detect an arbitrary $k$-particle entanglement \cite{LueckePRL2014}.  
Experimental data exclude an entanglement depth of less than 28 atoms with two standard deviations of confidence, 
and show an average entanglement depth of 68 atoms.
It should be noted that $\xiD^2 <1$ signals entanglement independently from its metrological usefulness. 
Useful $k$-particle entanglement is a more demanding resource than algebraic $k$-particle entanglement:
in \textcite{ApellanizNJP2015}, similar experimental data as those reported by \textcite{LueckePRL2014} show a useful entanglement depth of about 3 atoms.

\subsubsection{Adiabatic state preparation}

Twin-Fock states can be accessed, for an initial condensate prepared in $\mf=0$ at high $q$ values, 
via an adiabatic passage through the quantum phase transition(s) \cite{ZhangPRL2013, LuoSCIENCE2017} discussed in Sec.~\ref{Subsec4A2}. 
The ferromagnetic condensate is favorable since the energy gap closes as $\Delta E\si{SM}/N \hbar \abs{\lambda} = \mathcal{O}(N^{-1/3})$, while 
the energy gap for the anti-ferromagnetic condensate closes as $\Delta E\si{SM}/N \hbar \lambda = \mathcal{O}(N^{-1})$.
The condensate is initially prepared via optical pumping with all $N$ atoms in $\mf=0$ and, ideally, an adiabatic ramp of $q$
would transform the polarized state to a twin-Fock state with $N/2$ atoms in $\mf=\pm 1$ \cite{ZhangPRL2013}.
\textcite{HoangPNAS2016} have characterized the amplitude excitations and measured the energy gap 
of the $q>0$ quantum phase transition in a $^{87}$Rb condensate.
The entangled state preparation has been experimentally investigated by \textcite{LuoSCIENCE2017} for a 
$^{87}$Rb condensate of $N\sim 10^4$ atoms.
This experiment demonstrates---despite the diabatic ramp of the parameter and the finite loss rate---a deterministic and almost perfect conversion to the $\mf=\pm 1$ modes [$(N_{+1}+N_{-1})/N = 96 \pm 2 \%$ at the end of the ramp],
high squeezing [$\xi_N^2 = -10.7$ dB] and high coherence.  
The measurements demonstrate an average entanglement depth of $910$ atoms, with more than $450$ atoms at the confidence level of 
one standard deviation.
In a successive experiment, \textcite{ZouArXiv2018} have investigated the adiabatic preparation of the ground state at $q=0$, 
corresponding to a spin-1 Dicke state $\ket{l=N, m=0}$ of the $\hat{\vect{L}}$ manifold.
The prepared state has been used to estimate the angle of a Rabi rotation with a sensitivity 2.42 dB beyond the standard quantum limit.

\begin{figure} 
\begin{center}
\includegraphics[width=\columnwidth]{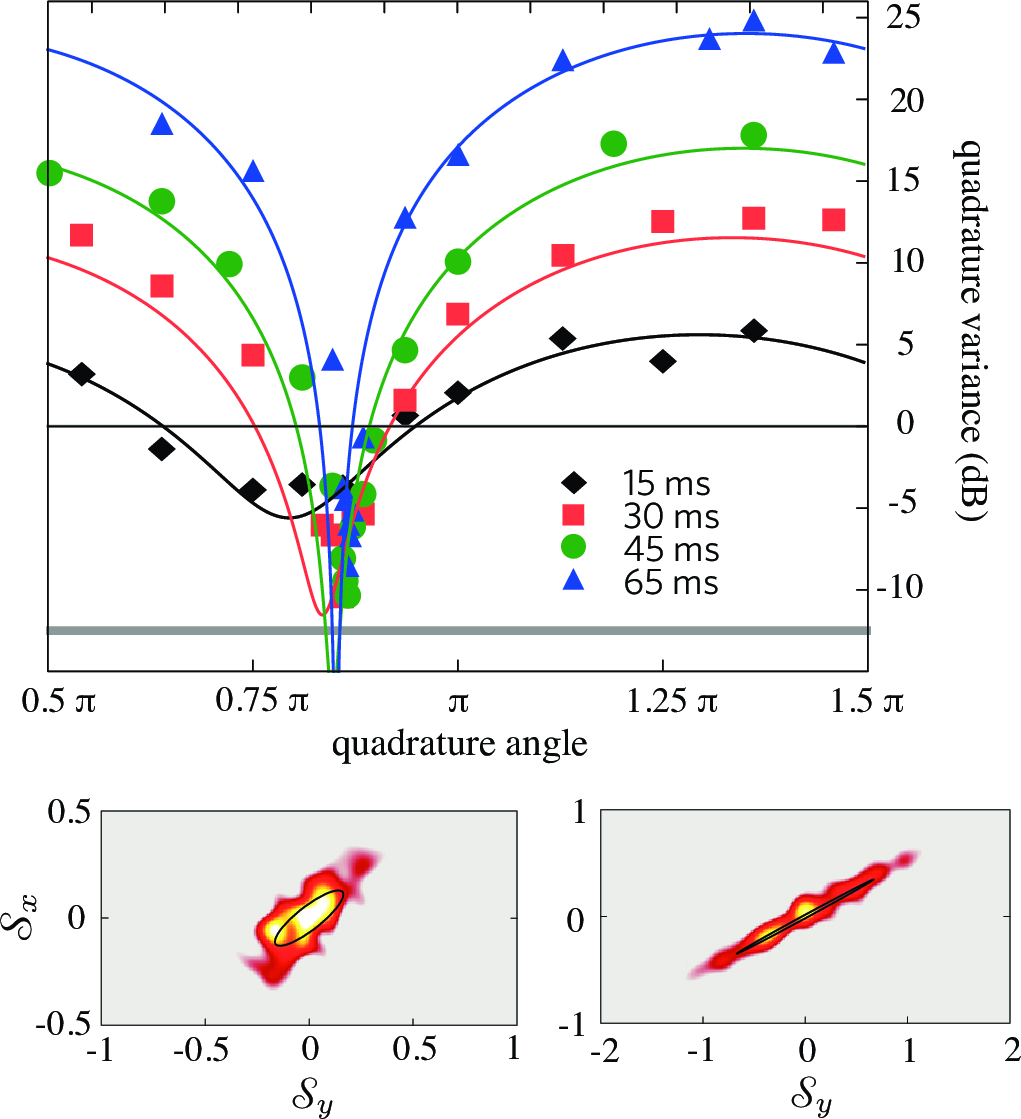}
\end{center}
\caption{{\bf Quadrature squeezing.}
Top: quadrature variance $4 \Delta^2 (\op{\mathcal{S}}_x \cos \phi + \op{\mathcal{S}}_y \sin \phi)/N$ 
after variable duration of the spin-mixing dynamics (colored lines and symbols) as functions of the quadrature angle $\phi$.
Symbols are experimental results obtained via state tomography, whereas the solid lines are theoretical predictions. 
Bottom: reconstructed phase space distributions at $t=15$\,ms (left) and $t=45$\,ms (right); the black ellipses are the $1/\sqrt{e}$ uncertainty ellipse
predicted theoretically.
Adapted from \textcite{HamleyNATPHYS2012}.}
\label{Fig:NemSpin}
\end{figure} 

\subsection{Quadrature squeezing and squeezed-vacuum state}
\label{Sec.Spin-Mix.squeezing}

\subsubsection{Quadrature squeezing} 
 
The experiments of \textcite{HamleyNATPHYS2012, GrossNATURE2011, PeiseNATCOMM2015}
have investigated quadrature squeezing in the state generated via spin-mixing dynamics in the low-depletion limit.
In this regime, the quadrature variance Eq.~\eqref{EqQs} can be rewritten as
$2 (\Delta \op{Q}_{\mathcal{s}})^2  = N \Delta^2 (\op{\mathcal{S}}_x \cos \phi + \op{\mathcal{S}}_y \sin \phi)/\mean{\op{\mathcal{S}}_z}^2$
and $\mean{\op{\mathcal{S}}_z} \approx N/2$ [and analogous relation between Eq.~\eqref{EqQa} and the $\vectop{\mathcal{A}}$ spin]:
quadrature squeezing thus corresponds to spin squeezing.  
Experimentally, the quadrature modes are accessed via atomic homodyne detection first realized by \textcite{GrossNATURE2011}, see also \textcite{PeiseNATCOMM2015}.
In analogy to standard techniques in quantum optics \cite{ScullyBOOK, OuPRL1992}, 
it consists of a symmetric radiofrequency coupling between the $\mf = 0, \pm 1$ modes. 
In the limit of low transfer and $N_0 \gg N_{\pm 1}$, this corresponds to a displacement operation, 
$e^{-\ii \Omega\si{rf} t \op{\mathcal{S}}_x }= e^{- \ii \Omega\si{rf} t \sqrt{N} (\sa + \sa^\dag)/2}$
where the condensate in $\mf=0$is used as local oscillator. 
\textcite{HamleyNATPHYS2012} have used $N=4.5 \times 10^4$ $^{87}$Rb atoms and characterized the 
spin-squeezed states via noise tomography, see Fig.~\ref{Fig:NemSpin}. 
Values up to $\xiR^2=-8.3$\,dB ($-10.3$\,dB inferred) below the SQL  are reported.

\subsubsection{Continuous variable and EPR entanglement} 
 
\begin{figure} 
\begin{center}
\includegraphics[width=\columnwidth]{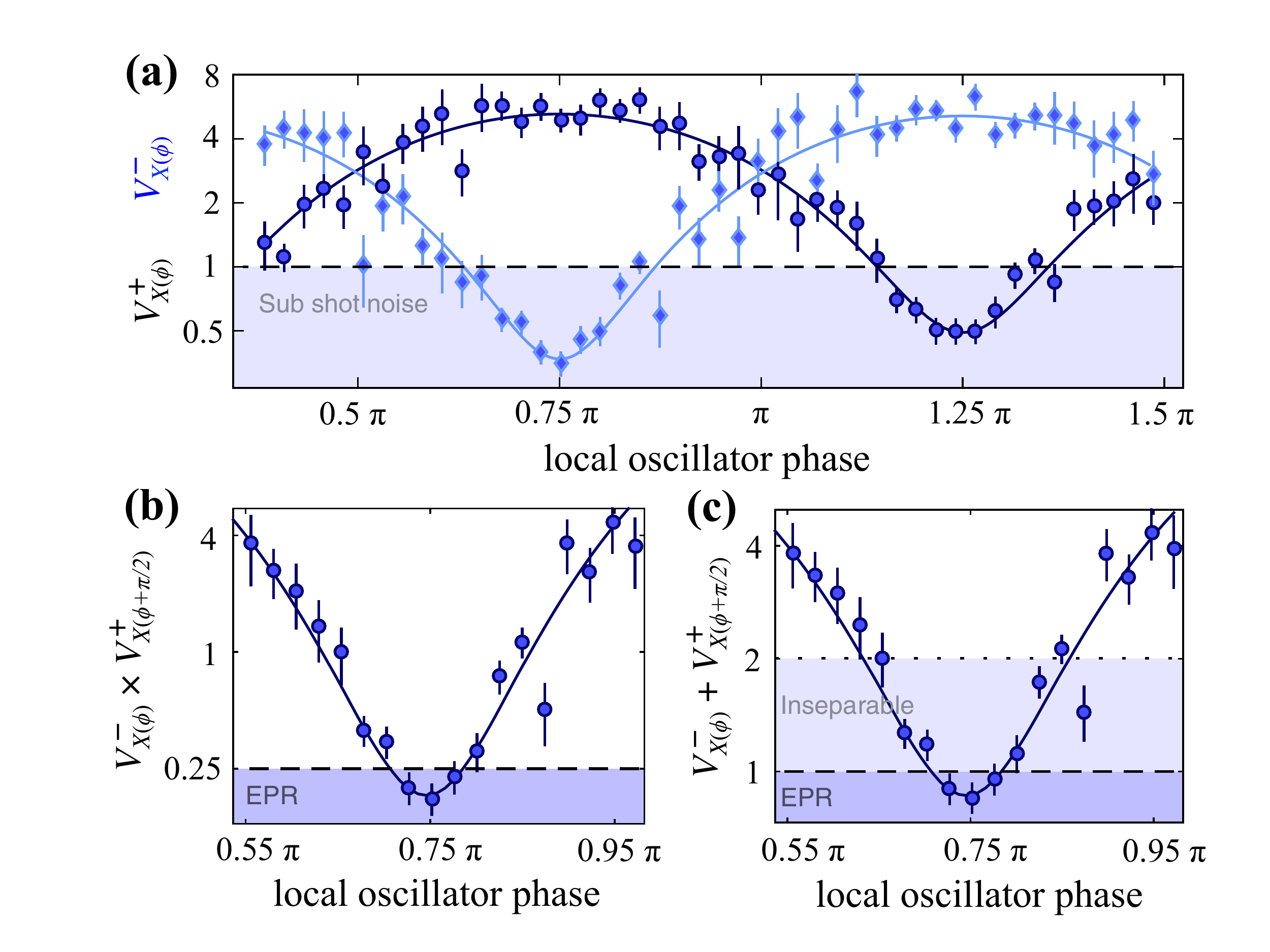}
\end{center}
\caption{{\bf Einstein-Podolsky-Rosen entanglement with spinor condensates.} 
a) Experimental variances $V_{X(\phi)}^+$ and $V_{X(\phi)}^-$ as functions of the local oscillator phase $\phi$. 
The dashed line is the quadrature variance for the vacuum state.
The lowest measured quadrature variance is $V_{X}^- = 0.42$, corresponding to a squeezing of 3.77\,dB below the vacuum limit. 
b) Product $V_{X(\phi)}^- \times V_{X(\phi+\frac{\pi}{2})}^+$: 
data points below the dashed line violate the inequality~\eqref{Eq.EPRcriterion} and thus signal Einstein-Podolsky-Rosen entanglement. 
Data reach $V_{X(\phi)}^- \times V_{X(\phi+\frac{\pi}{2})}^+ = 0.18(3)$,  which is 2.4 standard deviations below the limit of $1/4$.
c) Sum $V_{X(\phi)}^+ + V_{X(\phi+\frac{\pi}{2})}^-$: 
data points below the dotted line violate the inequality~\eqref{Eq.SEPcriterion} and thus signal entanglement between the $\mf=\pm 1$ modes, 
data below the dashed line signal Einstein-Podolsky-Rosen entanglement.
Data reach $V_{X(\phi)}^+ + V_{X(\phi+\frac{\pi}{2})}^- = 0.85(8)$.
Adapted from \textcite{PeiseNATCOMM2015}.} 
\label{Fig:Carsten}
\end{figure} 

The two-mode squeezed-vacuum state~\eqref{Eq.TMSV} realizes---in field modes---the position-momentum correlations 
that are at the heart of the Einstein-Podolsky-Rosen (EPR) criterion \cite{ReidPRA1989}, see Sec.~\ref{sec:EPR}.  
We recall that\footnote{Notice the relations $\op{Q}_{\mathcal{s}}(\phi) = [\op{X}_{+1}(\phi)+\op{X}_{-1}(\phi)]/\sqrt{2} $ and 
$\op{Q}_{\mathcal{a}}(\phi) = [\op{X}_{+1}(\phi) - \op{X}_{-1}(\phi)]/\sqrt{2}$, 
where $\op{X}_{\pm1}(\phi) = (\op{a}_{\pm 1} e^{-\ii \phi} + \op{a}_{\pm 1}^\dag e^{\ii \phi})/\sqrt{2}$.
In particular $V_{X(\phi)}^{+} = \Var[\op{X}_{+1}(\phi)+\op{X}_{-1}(\phi)] = 2 (\Delta \op{Q}_{\mathcal{s}})^2$ and 
$V_{X(\phi)}^{-} = \Var[\op{X}_{+1}(\phi)-\op{X}_{-1}(\phi)] = 2 (\Delta \op{Q}_{\mathcal{a}})^2$.} 
$V_{X(\phi)}^{\pm} = \cosh (2 r) \mp \sin (2 \phi) \sinh (2r)$. 
As the angle $\phi$ is varied, these quadrature variances oscillate between a maximum value $e^{2r}$ and a minimum value $e^{-2r}$, 
which is below the value of 1 of the unsqueezed vacuum state.
In the limit $r \to \infty$, at $\phi=3\pi/4$ ($\phi=\pi/4$), we have perfect correlations 
between $\op{X}_{+ 1}$ and $\op{X}_{- 1}$ ($\op{X}_{+ 1}$ and $-\op{X}_{- 1}$), 
as well as between $\op{P}_{+ 1}$ and $-\op{P}_{- 1}$ ($\op{P}_{+ 1}$ and $\op{P}_{- 1}$), where $\op{P}_{\pm 1}(\phi) = \op{X}_{\pm 1}(\phi + \pi/2)$.
According to the criteria~\eqref{Eq.SEPcriterion} and~\eqref{Eq.EPRcriterion} discussed in Sec.~\ref{Sec.EPRBELL}, 
the two-mode squeezed-vacuum state 
produced via spin-mixing dynamics in spinor condensates \cite{DuanPRL2000, PuPRL2000}
is mode entangled for $r >  0$ and fulfills the EPR criterion for $r> \ln \sqrt{2}$ \cite{ReidRMP2009}.
\footnote{In the case of a finite number of particles in the $\mf = 0$ mode \cite{FerrisPRA2008, RaymerPRA2003}, 
the inequalities~\eqref{Eq.SEPcriterion} and~\eqref{Eq.EPRcriterion} generalize to 
$V_{X(\phi)}^{\pm} + V_{P(\phi)}^{\mp} \geq 2 - \tfrac{N_{+1} + N_{-1}}{N_0}$, and 
$V_{X(\phi)}^{-} \times V_{P(\phi)}^{+} \geq \tfrac{1}{4} ( 1 - \tfrac{N_{-1}}{N_0} )^2$, respectively.}
The quadratures $X_{\pm}(\phi)$ are proportional to the number of particles $\op{N}_{+1} \pm \op{N}_{-1}$ measured after a radiofrequency pulse coupling 
$\mf=0$ with $\mf=\pm 1$ \cite{GrossNATURE2011}. 
With spinor condensates \cite{DuanPRA2002}, continuos variable entanglement, \eg, the violation of Eq.~\eqref{Eq.SEPcriterion}, 
was first demonstrated by \textcite{GrossNATURE2011}, 
while EPR entanglement, \eg, the violation of Eq.~\eqref{Eq.EPRcriterion} was observed by \textcite{PeiseNATCOMM2015}, see Fig.~\ref{Fig:Carsten}.
The experimental violation of Eq.~\eqref{Eq.EPRcriterion} demonstrates a form of entanglement intrinsically connected with local realism. 
The demonstration of the continuous-variable EPR paradox
with massive particles would additionally require space-like separation of measurements to rule out causal influences, which has not yet been experimentally achieved. 

\subsubsection{Interferometry with squeezed vacuum}

Quadrature squeezing is a resource for quantum interferometry, see Sec.~\ref{Sec.QuadSq}.
Within spin-mixing dynamics, squeezing occurs in the quadratures corresponding to the 
symmetric $\ket{s} =( \ket{1,+1} + \ket{1,-1})/\sqrt{2}$ and the anti-symmetric $\ket{a} =( \ket{1,+1} - \ket{1,-1})/\sqrt{2}$
combinations of the $\ket{\f=1,\mf=\pm1}$ modes.
\textcite{Kruse} have realized an atomic clock that exploits the squeezed-vacuum states created in the $\ket{s}$ mode. 
An initial condensate of $N \approx 10^4$ atoms undergoes spin-mixing dynamics in the $F=1$ manifold of $^{87}$Rb, 
in the low-depletion regime ($\mean{\op{N}_0} \gg \mean{\op{N}_{\pm 1}}$ after spin dynamics).
The interferometer consists of a rf coupling (corresponding to a balanced beam splitter between the condensate in $\ket{1,0}$ and the squeezed vacuum in $\ket{s}$, 
$\op{H}\si{rf} = \hbar \Omega\si{rf}\op{\mathcal{S}}_x$), 
a relative phase shift $\ps$ implemented by a detuned $\pi$ pulse between $\ket{1,0}$ and $\ket{2,0}$, and a second balanced beam splitter between $\ket{1,0}$ and $\ket{s}$. 
The anti-symmetric mode $\ket{a}$ is left unchanged by the interferometer transformation (up to an overall phase), while the $\ket{s}$
is rotated in the Bloch sphere identified by the $\op{\mathcal{\vect{S}}}$ manifold.
The experiment of \textcite{Kruse} reports Ramsey fringes as a function of $\ps$
reaching a clock sensitivity 2.05\,dB below $\Delta \ps\si{SQL} = 1/\sqrt{N}$.
In the absence of noise, the expected sensitivity is 
$\Delta \ps = e^{-r}/\sqrt{N}$ \cite{CavesPRD1981} in the low squeezing regime, and
can reach the Heisenberg limit $\Delta \ps\si{HL} = 1/N$ when the input states have the same population on average \cite{PezzePRL2008}.

\subsubsection{Nonlinear SU(1,1) interferometry}
\label{SU11}

Spin-mixing dynamics can be also used to realize a SU(1,1) interferometer, as first proposed by \textcite{YurkePRA1986} in optics 
[recently realized with a bright laser source by \textcite{HudelistNATCOMM2014} and with a hybrid atom-light system by \textcite{ChenPRL2015b}] 
and further analyzed by \textcite{MarinoPRA2012, GabbrielliPRL2015} for spinor condensates.
The basic idea of this interferometric scheme is to 
replace the linear beam splitters of a standard Mach-Zehnder scheme with nonlinear beam splitters 
that create/annihilate pairs of particles, as implemented via spin-mixing dynamics.
Let us indicate with $\pazocal{N}$ the total average number of atoms
transferred in pairs from a condensate prepared in $\mf=0$ to the initially empty $\mf=\pm1$ modes.
After spin mixing, the system acquires a relative phase $\ps = 2 \ps_0 - (\ps_{+1} + \ps_{-1})$ between $\mf=0$ and $\mf=\pm1$ modes.
A second spin-mixing dynamics closes the interferometer. 
The final populations in $\mf=0,\pm 1$ depend on $\ps$.
When treating the condensate in $\mf=0$ as an undepletable source of atomic pairs, the predicted phase sensitivity is \cite{YurkePRA1986}
\be \label{SU11Yurke}
\Delta \ps = \sqrt{\frac{\pazocal{N}(\pazocal{N}+2) \cos^2 (\ps/2) + 1}{\pazocal{N}(\pazocal{N}+2) \sin^2 (\ps/2)}}.
\ee
Equation~\eqref{SU11Yurke} reaches $\Delta \ps = 1/\sqrt{\pazocal{N}(\pazocal{N}+2)}$ 
at the optimal working point $\ps=\pi$ (a dark fringe, where no particle is found in the output $\mf=\pm1$ states) and shows that
only the particles outcoupled from the $\mf=0$ mode contribute to the phase sensitivity.

An analysis beyond the low-depletion limit shows that the SU(1,1) interferometer scheme can overcome the standard quantum limit with respect to the total number of particles $N$ in the initial condensate \cite{GabbrielliPRL2015} and reach $\Delta\ps = \mathcal{O}(1/N)$. The sensitivity can be further enhanced by an additional linear coupling of the three modes before and after phase imprinting, a scheme called ``pumped-up'' SU(1,1) interferometer \cite{SzigetiPRL2017}.

\textcite{LinnemannPRL2016} have experimentally realized a nonlinear SU(1,1) interferometer within the $F = 2$ manifold of $^{87}$Rb. 
The experiment is performed in the low-depletion regime,  $\pazocal{N} =  2.8$, using $N \approx 400$ atoms. 
The interferometer is probed by tuning the relative phase between the $\mf=0$ condensate and the $\mf=\pm 1$ 
modes to the dark fringe.
The phase $\ps$ is imprinted via a second-order Zeeman shift by applying a magnetic field for varying times, and
is read out from the mean total number of atoms in the $\mf=\pm 1$ modes. 
Results demonstrate a sensitivity $\Delta \ps$ close to the theoretical prediction of Eq.~\eqref{SU11Yurke}.

\subsection{Other protocols to create correlated atomic pairs}
\label{Sec.Spin-Mix.alternative}

Spin-mixing dynamics is not the only possibility to create correlated atom pairs. 
Alternative methods have been studied and implemented experimentally. 
Relative number squeezing between two outgoing Bose-Einstein condensates with opposite momenta 
was obtained in the collisional de-excitation of a one-dimensional quasi-condensate, 
where the reduced dimensionality restricted the number of available modes \cite{BucknerNATPHYS2011}. 
The atoms are initially prepared in a highly non-equilibrium state such that the only allowed de-excitation channel is a two-particle collision process, emitting atom pairs. 
\textcite{BucknerNATPHYS2011} observed a number squeezing of up to $-4.3$\,dB.
\textcite{JaskulaPRL2010} observed sub-Poissonian atom number fluctuations, with 0.5\,dB of number squeezing, between 
outgoing modes of opposing momenta in the halo produced by the s-wave scattering of two Bose-Einstein condensates of metastable $^4$He.
The amount of detected squeezing was limited by the large number of outgoing collisional modes 
(isotropically distributed over the scattering sphere) and the correspondingly small number of atoms per mode.
With the same scheme \cite{Lewis-SwanNATCOMM2014} it was possible to demonstrate Hong-Ou-Mandel interference \cite{HOM1987}.
Each atom of the pair is in the motional degree of freedom is sent to the input channel of a beam splitter
realized using Bragg scattering on an optical lattice \cite{BonneauPRA2013}. 
When two inputs are indistinguishable, they emerge together in one of the output channels,
demonstrating two-particle interference \cite{LopesNATURE2015}. 

\section{Entanglement creation via atom-light interaction}
\label{Sec.Atom-Light}

The light-matter quantum interface finds important applications in several 
areas of quantum information processing \cite{HammererRMP2010, KimbleNATURE2008}.
Here we review the many successful experiments and proposals that exploit the coupling between atoms and light
for the creation of useful entangled states for quantum metrology. 
This quest started in the late 1990s with pioneering experiments using room-temperature vapor 
cells \cite{KuzmichPRA1999, KuzmichPRL2000} and laser-cooled atomic gases \cite{HaldPRL1999}, 
which demonstrated the reduction of atomic spin (or pseudospin) fluctuations. 
In these experiments, the atom-light interface was implemented in free space using optically-thick atomic ensembles \cite{KuzmichBOOK}. 
Optical cavities can be used to effectively increase the optical depth of the atomic sample and represent a 
versatile system with rich possibilities for creating entanglement 
between atoms \cite{MillerJPB2005, TanjiSuzukiAAMOP2011, RitschRMP2013}. 
These techniques have allowed the demonstration of metrological spin-squeezing, 
squeezed-state atomic clocks 
and the creation of highly non-classical states.

Key theoretical proposals and experiments on atom-light interfaces for quantum metrology can be classified in three broad categories:
(a) nondestructive measurements, including quantum nondemolition measurements and Zeno dynamics, Sec.~\ref{Sec.Atom-Light.nd}; 
(b) light-mediated coherent interaction between distant atoms, Sec.~\ref{Sec.Atom-Light.effint}; and 
(c) transfer of squeezing to atoms via absorption of non-classical light, Sec.~\ref{Sec.Atom-Light.state-transfer}. 
Atom-light interaction currently represents the most successful method for producing large amounts of 
squeezing and entanglement in atomic ensembles. 

\subsection{Quantum state preparation using nondestructive measurements}
\label{Sec.Atom-Light.nd}

\begin{figure}
\begin{center}
\includegraphics[width=\columnwidth]{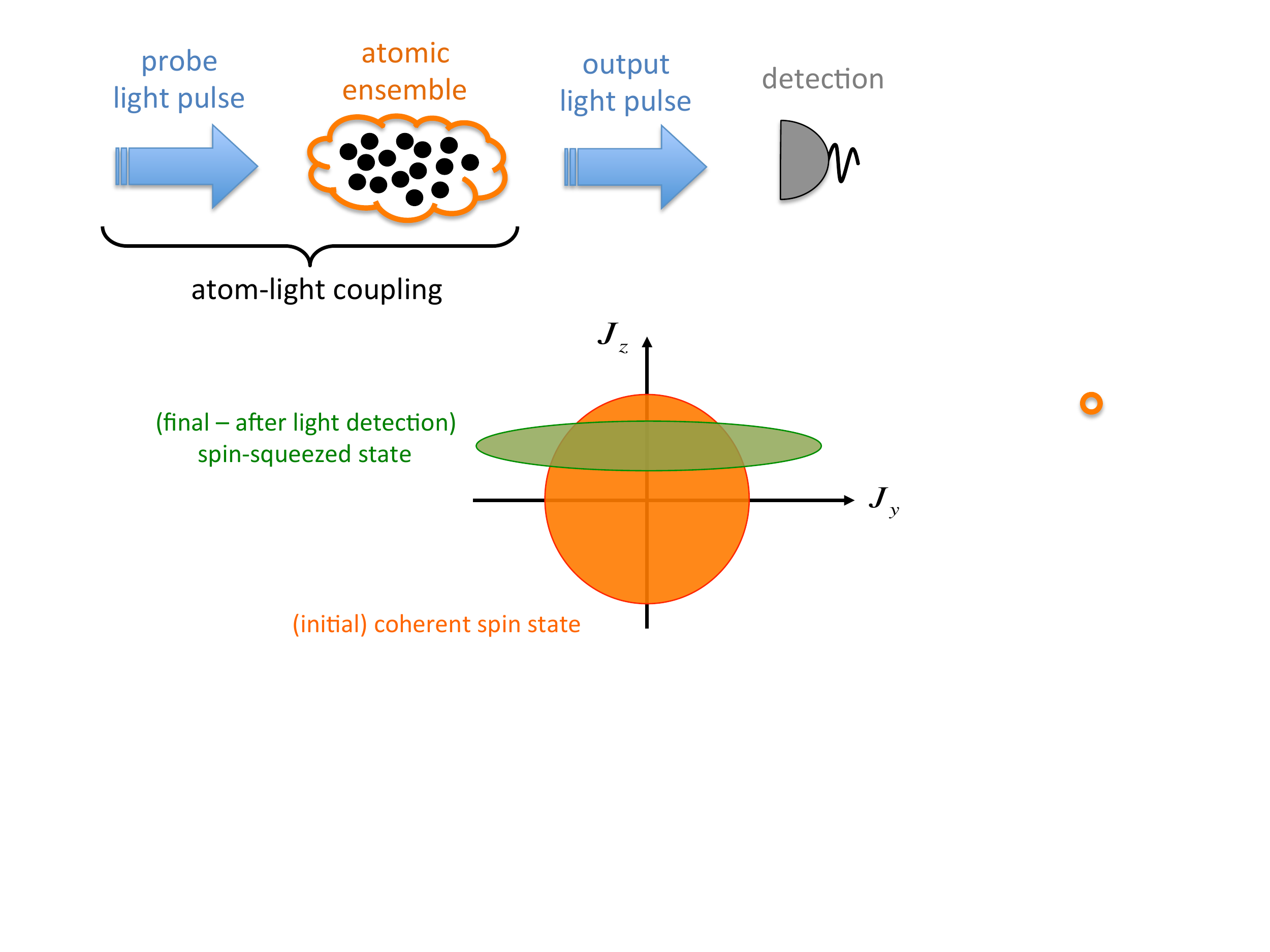}
\end{center}
\caption{{\bf Conditional spin squeezing via QND measurements.}
A two-mode light beam (blue arrow) propagates through the atomic ensemble. 
The light beam becomes entangled with the atomic ensemble. 
By measuring the light field, one gains information about the atomic state.
The atomic distribution is plotted here in the $y-z$ plane.
The initial coherent spin state (orange circle) changes after the detection of the light beam (green ellipse).  
It becomes squeezed in $\J_z$ and shifted (conditioned by the measurement result), 
according to Eqs.~\eqref{Eq.JzQND} and~\eqref{Eq.DJzQND}.} 
\label{Fig:QND}
\end{figure} 

\subsubsection{Quantum nondemolition measurements in free space}
\label{Sec.QND}

In a quantum nondemolition (QND) measurement, a \emph{system} (\eg, an atomic cloud) and 
\emph{meter}  (a measurement degree of freedom, \eg, a light beam) are coupled, 
such that a direct measurement of the meter provides indirect information 
about an observable of the system \cite{BraginskySCIENCE1980, BraginskyBOOK, GrangierNATURE1998}.
The measurement is called QND if the system and the measured value of the observable are conserved after the measurement.
QND measurements are a resource for quantum metrology as they allow the preparation of  
entangled and spin-squeezed states of many atoms.   
This is well illustrated by a model Hamiltonian describing a far-off resonant dispersive interaction 
between the collective spin of an ensemble of atoms and a two-mode light beam in free space \cite{KuzmichEPL1998,TakahashiPRA1999}:
\be \label{Eq.QND}
\op{H}\si{QND} =  (\hbar k /t\si{p} )  \Sp_z \J_z.
\ee
Here, $t\si{p}$ is the light pulse duration, $\vectop{J}=\{ \J_x, \J_y, \J_z\}$ is the atomic collective spin,
and $\vectop{S}=\{ \Sp_x, \Sp_y, \Sp_z\}$ is the Stokes vector operator of the light, with components 
$\Sp_x = (\op{a}^\dag_+ \op{a}_- + \op{a}^\dag_- \op{a}_+)/2$,
$\Sp_y = (\op{a}^\dag_+ \op{a}_- - \op{a}^\dag_- \op{a}_+)/(2\ii)$, and 
$\Sp_z = (\op{a}^\dag_+ \op{a}_+ - \op{a}^\dag_- \op{a}_-)/2$.
$\op{a}_{\pm}$ can be two polarization modes \cite{TakahashiPRA1999, HammererPRA2004}, 
in which case Eq.~\eqref{Eq.QND} describes the paramagnetic Faraday rotation of light; they can also be
two spatial modes of an optical Mach-Zehnder interferometer where atoms are placed in one arm and 
phase-shift the light \cite{OblakPRA2005,ChaudhuryPRL2006, WindpassingerPRL2008}.  
The dimensionless interaction strength in Eq.~\eqref{Eq.QND} is $k \propto \tfrac{\sigma}{A} \tfrac{\Gamma}{\Delta}$, where 
$\sigma$ is the resonant photon scattering cross section of the probe transition, 
$\Gamma$ is the spontaneous emission rate, 
$A$ is the spatial cross section of the atomic ensemble illuminated by the pulse, and 
$\Delta$ the detuning of the light from resonance \cite{HammererRMP2010}. 
The Hamiltonian~\eqref{Eq.QND} satisfies the back-action evasion condition $[\J_z, \op{H}\si{QND}]=0$ 
such that $\J_z$ is a constant of motion.
During the interaction, $\vectop{S}$ precesses around the $z$-axis by an angle $k \J_z$.
To lowest order in this angle, we have
\begin{subequations}
\begin{align}
& \Sp_x\se{out} \approx \Sp_x\se{in} - k \Sp_y\se{in} \J_z\se{in}, \,\,  \J_x\se{out} \approx \J_x\se{in} - k \J_y\se{in} \Sp_z\se{in},  \\ 
& \Sp_y\se{out} \approx \Sp_y\se{in} + k \Sp_x\se{in} \J_z\se{in}, \,\, \J_y\se{out} \approx \J_y\se{in} + k \J_x\se{in} \Sp_z\se{in}, \\
& \Sp_z\se{out} = \Sp_z\se{in}, \qquad \qquad \quad  \J_z\se{out} = \J_z\se{in}.
\end{align}
\end{subequations}
A measurement of $\Sp_x\se{out}$ or $\Sp_y\se{out}$ thus realizes a QND measurement of $\J_z$ while preserving the system's quantum coherence.
As an illustration, we take atomic and optical systems both initially prepared in coherent spin states polarized along the $x$-axis.
Assuming $\mean{\Sp_y\se{in}}=0$ and $\mean{\J_z}=0$, the average phase precession is zero, $\mean{\Sp_y\se{out}}=0$, 
while the variance $(\Delta \Sp_y\se{out})^2 = n(1+ \kappa^2)/4$ increases with $\kappa^2 = nN k^2/4$, where
$N$ is the number of atoms, and $n$ the number of photons in the pulse.
The mean value and variance of $\J_z$ after the measurement of the light spin $\Sp_y$ (with result $m_y$) are
\begin{subequations}
\begin{align} 
\mean{\J_z\se{out}}\rvert_{m_y} &= \frac{\kappa}{1+\kappa^2} \sqrt{\frac{N}{n}} m_y, \label{Eq.JzQND} \\
(\Delta \J_z\se{out})^2\rvert_{m_y} &= \frac{1}{1+\kappa^2} \frac{N}{4}, \label{Eq.DJzQND}
\end{align}
\end{subequations}
respectively, see Fig.~\ref{Fig:QND}.
Although the final atomic state {depends on the result of the measurement of the light beam (and it is thus termed conditional), 
the reduction in spin noise according to Eq.~\eqref{Eq.DJzQND} is completely deterministic, with no post-selection necessary.
Unconditional spin squeezing can be achieved via quantum feedback that compensates, via a spin rotation, 
the stochastic shift of $\mean{\J_z}$ in Eq.~\eqref{Eq.JzQND} due to the random measurement outcome \cite{ThomsenPRA2002, BerryPRA2002}.
\textcite{InouePRL2013,CoxPRL2016} have experimentally demonstrated such unconditional spin squeezing via feedback control, 
see also \textcite{VanderbruggenPRL2013}.
In the QND scheme, the squeezing is due to the projective measurement 
of the atomic state upon detection of the light, and its efficacy is determined by the performance of the detector.
Yet, the above discussion is highly idealized: 
far-off resonant dispersive probing is unavoidably accompanied by decoherence due to Raman spin flips.\footnote{In the literature, 
non-ideal measurements are still termed QND provided that the loss of coherence in the atomic ensemble remains small.}
Notice that $\kappa^2 = \eta  \alpha_0$, where $\alpha_0 \propto N \sigma/A$ is the resonant optical depth and 
$\eta \propto n \tfrac{\sigma}{A} (\tfrac{\Gamma}{\Delta})^2$ is the average photon scattering rate per atom 
during the probe pulse.
Scattered photons carry information about the spin state of the atoms: 
these photons reduce the spin length as $\abs{\mean{\J_x\se{out}}}\approx(1-\eta)\abs{\mean{\J_x}}$ for $\eta\ll 1$ \cite{HammererPRA2004}]
and add noise that counteracts the reduction of spin variance \cite{HammererPRA2004, MadsenPRA2004, EchanizJOB2005}---unless a suitable choice of atomic levels is used to avoid this effect \cite{ChenPRA2014, SaffmanPRA2009}.
The successful implementation of QND measurements is thus based on a proper choice of light power (increasing $n$ reduces the photon shot noise of light detection but increases Raman scattering) 
and detuning [exploiting the favorable scaling of $k \propto \Gamma/\Delta$ over $\eta \propto  (\Gamma/\Delta)^2$].
The most important figure of merit of an atom-light interface is the resonant optical depth $\alpha_0$,
which should be much larger than unity.

\begin{figure}
\begin{center}
\includegraphics[width=\columnwidth]{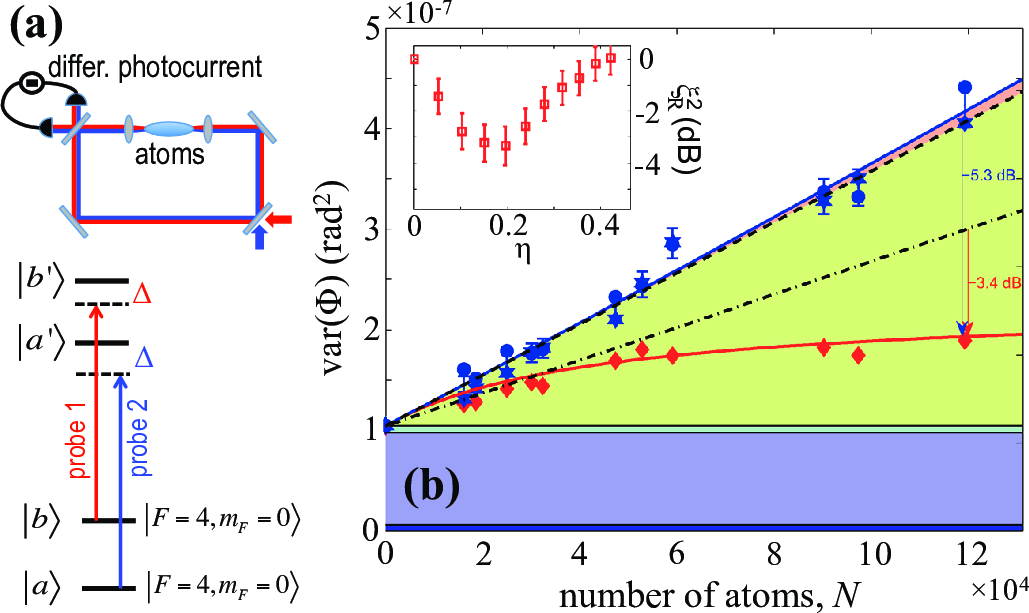}
\end{center}
\caption{{\bf Spin squeezing via quantum nondemolition measurements in free space.}
a) A cloud of cold Cs atoms is confined in an elongated dipole trap placed in one arm of an optical Mach-Zehnder interferometer and
aligned with the largest resonant optical depth.
The relative populations of the clock levels $\ket{a}$ and $\ket{b}$ are measured via the phase shift $\phi$ 
accumulated by two detuned probe beams propagating through the atomic cloud.
b) Red diamonds are $\Var(\Phi)$ (with $\Phi= \phi_1 - \zeta \phi_2$) obtained after two consecutive QND measurements of covariance $\zeta$. 
Blue symbols are $\Var(\Phi)$ obtained for a coherent spin state ($\Phi=\phi_1$, dots, and $\Phi=\phi_2$, stars). 
The solid line is a quadratic fit and the dashed line is the expected linear scaling with the atom number $N$ due to projection noise.
The dot-dashed line is the projection noise scaled down by the factor $(1-\eta)^2=0.64$, corresponding to the reduction 
by the measured observed loss of atomic coherence.
Different color regions are the optical shot noise (light blue), detector noise (dark blue) and projection noise (green/light).
The inset illustrates the trade-off between spin squeezing $\xi_R^2$ and loss of coherence $\eta$ due to spontaneous emission.
Adapted from \textcite{AppelPNAS2009}.} 
\label{Fig:Polzik}
\end{figure} 

\textcite{AppelPNAS2009} reported the first experimental demonstration of metrological spin squeezing 
(reaching $\xiR<1$) via QND measurements in free space.
This experiment used $N = 1.2 \times 10^5$ Cs atoms (with measured $\kappa^2 \approx 3.2$ and optical depth $\alpha_0 \approx 16$).
Two equally intense and linearly polarized laser beams of different frequencies enter the arms of an optical Mach-Zehnder interferometer, 
see Fig.~\ref{Fig:Polzik}(a). 
The beams off-resonantly probe different atomic transitions and  
experience phase shifts proportional to the number of atoms in the probed levels \cite{SaffmanPRA2009}. 
The detection of the relative phase shift $\phi$ accumulated in the optical path
performs a QND measurement of the relative population in the two atomic levels \cite{KuzmichEPL1998,SaffmanPRA2009}.  
Spin squeezing is quantified by correlations between two consecutive QND measurements. 
One finds $\Var(\phi_2 - \zeta \phi_1) = \frac{1}{n} + \frac{\kappa^2}{1+\kappa^2} \tfrac{N}{4}$, where 
$\phi_1$ and $\phi_2$ refer to the first and second phase shift detection, respectively, and the covariance
$\zeta = \Cov(\phi_1, \phi_2) / \Var( \phi_1) = \tfrac{\kappa^2}{1+\kappa^2}$
expresses the correlations between the two measurements.
The results reveal a spin squeezing $\xiR^2 = -3.4$\,dB, see Fig.~\ref{Fig:Polzik}(b).
Furthermore, \textcite{LouchetChauvetNJP2010} have used a similar apparatus to perform a Ramsey sequence (between the two QND measurements) 
and demonstrated an atomic clock with a measured phase sensitivity 1.1\,dB below the standard quantum limit.

Spin squeezing has also been experimentally performed by measuring the polarization rotation of probe light passing through a cloud of atoms. 
This technique was  pioneered by \textcite{KuzmichPRA1999, KuzmichPRL2000}, and 
more recently investigated in ensembles of 
spin-1/2 $^{171}$Yb \cite{TakanoPRL2009, TakanoPRL2010, InouePRL2013} 
and spin-1 $^{87}$Rb atoms \cite{SewellPRL2012, SewellNATPHOT2013, SewellPRX2014}.
In \textcite{SewellPRL2012, SewellPRX2014} spin squeezing was achieved using a 
two-polarization probing technique \cite{KoschorreckPRL2010a,KoschorreckPRL2010}.
These experiments demonstrated $\xiR^2=-1.5$\,dB \cite{SewellPRX2014} [$\xiR^2=-2$\,dB inferred \cite{SewellPRL2012}]
using $N \approx 5 \times 10^5$ atoms, mainly limited by the photon shot noise of the readout light.
Polarization-based QND measurements find direct application in entanglement-assisted optical magnetometry, 
see Sec.~\ref{Sec.Working-Entanglement.measurements}.  
 
\textcite{PuentesNJP2013} have proposed to generate planar spin squeezing (see~\ref{Sec.PQSintro}) in spin-1 atomic ensembles via sequential QND measurements of two orthogonal spin components. Measurements by \textcite{ColangeloNATURE2017,ColangeloPRL2017} in cold $F=1$ $^{87}$Rb atoms show phase and amplitude squeezing simultaneously, as well as entanglement between the atoms' spins \cite{VitaglianoPRA2018}.

\subsubsection{Cavity-based quantum nondemolition measurements in the dispersive regime}
\label{Sec5A_cavity}

The strength of the interaction between the light and the individual atoms is usually weak, but 
can be enhanced by placing the atoms inside an optical cavity.
This method is very promising as the squeezing factor increases with the cavity finesse, which can be pushed to large values. 
Moreover, a small single-pass optical depth is advantageous for applications in atomic clocks, since it allows to 
reduce atomic-density-dependent atom losses, dephasing, and systematic errors.

\begin{figure}[t!]
\begin{center}
\includegraphics[width=\columnwidth]{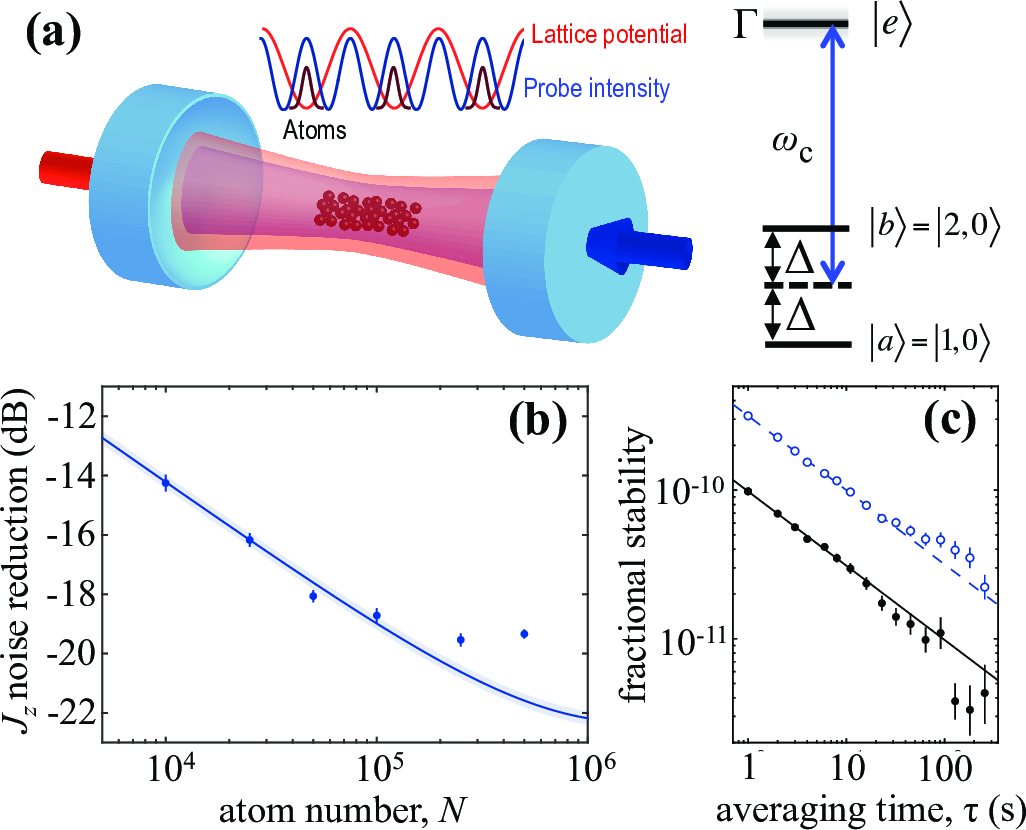}
\end{center}
\caption{{\bf Spin squeezing via cavity-based quantum nondemolition measurements in the dispersive regime.} 
(a) Laser-cooled atoms are optically trapped in a standing wave (red/lighter line) inside an optical resonator.
The cavity resonance is shifted in proportion to the relative population of two clock levels. 
The shift is measured from the transmission of a probe beam (blue/darker line).
(b) Measured spin-noise reduction $4(\Delta \J_z)^2/N$ (dots) normalized to the coherent spin state projection noise.
The solid line is a model fit. 
(c) Allan deviation of a clock that uses the generated squeezed states 
(black dots; the solid line indicates $9.7\times10^{-11}\,\text{s}^{1/2}/\sqrt{\tau}$) or coherent spin states 
(blue circles; the dashed line being the standard quantum limit).
Adapted from \textcite{HostenNATURE2016}.} 
\label{Fig:Vuletic_cavity}
\end{figure} 

The essential features of dispersive atom-light interaction in a cavity \cite{VernacPRA2000, NielsenPRA2008, BouchoulePRA2002b, MadsenPRA2004}
are captured by a simplified model comprising $N$ three-levels atoms, see Fig.~\ref{Fig:Vuletic_cavity}(a).
Each atom has two hyperfine levels $\ket{a}$ and $\ket{b}$ of energy difference $\hbar \omega$, and an excited state $\ket{e}$ with linewidth $\Gamma$
(spontaneous decay rate into free space). 
The atoms are placed in an optical cavity with resonance frequency $\omega\si{c}$ and linewidth $\kappa_c$, driven resonantly 
with a single-atom-single-photon effective intra-cavity Rabi frequency $2g$.
The detuning of the cavity from the $\ket{a} \leftrightarrow \ket{e}$ and $\ket{b} \leftrightarrow \ket{e}$ transitions is chosen 
of equal magnitude $\Delta= \pm \omega/2$.
Assuming homogeneous interaction (see Sec.~\ref{sec:effectivespin}), low intra-cavity photon number ($n_c = \mean{\op{c}^\dag \op{c}} \ll \Delta^2/g^2$), and 
large detuning ($\Delta \gg \kappa_c, \Gamma, \sqrt{N}g$), the coupling Hamiltonian is \cite{Schleier-SmithPRA2010}
\be \label{cavityH}
\op{H} =  \hbar \omega_c \op{c}^\dag \op{c} + \hbar \frac{2 g^2}{\Delta} \op{c}^\dag \op{c} \J_z + \hbar \omega \J_z,
\ee
where $\op{c}$ and $\op{c}^\dag$ are cavity mode operators.
The effect of the light on the atoms is an ac Stark shift of the transition frequency $\delta \omega = \tfrac{2 g^2 }{\Delta} n_c$ between $\ket{a}$ and $\ket{b}$. 
Atoms in $\ket{a}$ ($\ket{b}$) increase (decrease) the index of refraction seen by the probe light, so that the net effect 
is a shift of the cavity resonance by $\delta \omega\si{c} = \tfrac{2 g^2 }{ \Delta} \tfrac{N_a - N_b}{2} = \tfrac{2 g^2 }{\Delta} J_z$, 
where $N_a$ and $N_b$ are the numbers of atoms in $\ket{a}$ and $\ket{b}$, respectively. 
This shift can be probed by injecting a laser into the cavity, providing a QND measurement of $\J_z$.
A detailed analysis including decoherence associated with free-space scattering of the probe light 
shows the possibility to achieve \cite{ChenPRA2014, HostenNATURE2016}
\be
\xiR^2 \approx  \frac{1+NC(\Gamma/\omega)^2}{\sqrt{NC}},
\ee
where $C=\tfrac{(2 g)^2}{\kappa_c \Gamma}$ is the single-atom cavity cooperativity.   
$C$ is the ratio between the number of photons scattered into the cavity mode and those scattered into free space, and 
quantifies the optical depth of an atom with respect to the cavity mode \cite{TanjiSuzukiAAMOP2011}. 
Note that $C$ depends on the cavity geometry and is proportional to the cavity finesse \cite{TanjiSuzukiAAMOP2011}.

Spin squeezing via cavity-based QND measurement has been first demonstrated by \textcite{Schleier-SmithPRL2010}
using magnetically-insensitive clock states of $^{87}$Rb atoms.
This experiment reached a spin squeezing $\xiR^2 = -1.45$\,dB with respect to 
$(\Delta \ps\si{SQL})^2 = 1/N\si{eff}$, referring to $N\si{eff} \approx 0.66 N$ uncorrelated atoms 
($N\si{eff}$ accounts for spatial variation in the atom-light coupling due to the
trapping lattice being incommensurate with the cavity mode used for probing, and $N= 5 \times 10^4$).
This result is mainly limited by inhomogeneous dephasing due to the cavity locking light.
\textcite{HostenNATURE2016} have used a cavity of higher cooperativity
and exploited probing and trapping beams of commensurate frequencies, achieving a uniform atom-light coupling, see Fig.~\ref{Fig:Vuletic_cavity}(a). 
This avoids the need for spin-echo techniques required for non-uniformly coupled systems.
This experiment demonstrated $10.5$\,dB of improved phase sensitivity with respect to $(\Delta \ps\si{SQL})^2=1/N$, 
with $N=1 \times 10^5$ $^{87}$Rb atoms, and a spin squeezing $\xiR^2 = -18.5$\,dB, see Fig.~\ref{Fig:Vuletic_cavity}(b).
This represents the highest value in expected metrological spin squeezing and measured phase sensitivity gain to date.

\begin{figure}[t!]
\begin{center}
\includegraphics[width=\columnwidth]{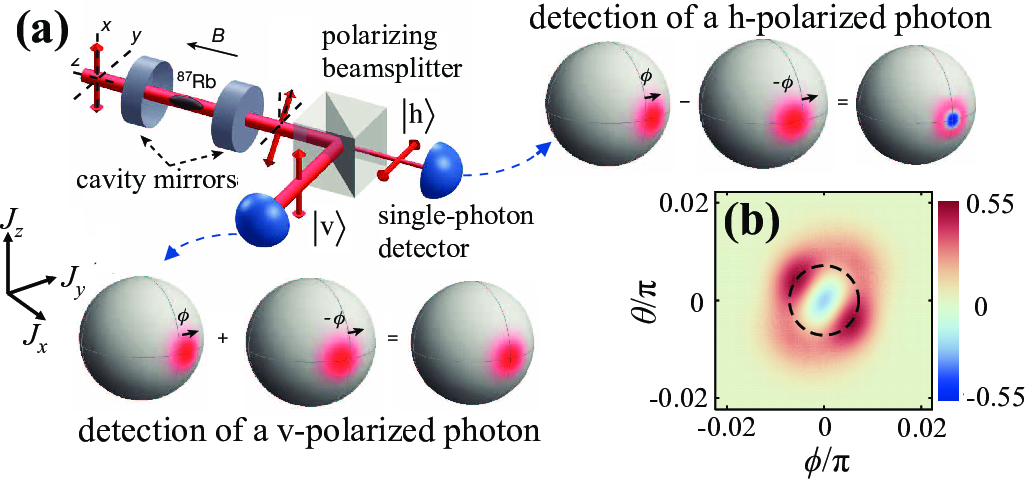}
\end{center}
\caption{{\bf Entangled states generated by the detection of a single photon.}
(a) An atomic ensemble confined in a cavity interacts with vertically polarized light. 
The detection of an outgoing horizontally (vertically) polarized single photon projects the atoms into a state with a negative (positive) Wigner distribution.
(b) Experimental Wigner distribution of the heralded atomic state 
obtained from the least-squares-fitted atomic density matrix via Eq.~\eqref{eq:sphericalquasiprobability}.
It shows negative parts.
The dashed line is the contour at which the Wigner distribution of an $N$-atom coherent spin state is equal to 
$1/\sqrt{e}$ of its maximum value. 
Adapted from \textcite{McConnellNATURE2015}.}
\label{Fig:singphot}
\end{figure} 

\subsubsection{Heralded atomic entanglement created by single photon detection}

In the preceding sections we have discussed the interaction of an atomic ensemble with an intense 
laser pulse ($10^5{-}10^7$ photons in typical experiments), which is subsequently detected by a standard linear photodetector that does not resolve individual photons. The light in these experiments is effectively a continuous-variable quantum system. 
A different approach uses weak laser pulses in combination with single-photon counting, in close analogy to the techniques developed for atomic ensemble-based quantum repeaters \cite{DuanNATURE2001, SangouardRMP2011}. These techniques allow to create a range of highly nonclassical states of a large number of atoms in a heralded way, \ie, conditioned on the detection of a single photon. 
The interaction of a linearly polarized photon with the atoms according to Eq.~\eqref{Eq.QND} produces a weak Faraday rotation of the photon polarization.  
Denoting $\ket{\text{v}}$ and $\ket{\text{h}}$ the vertical and horizontal polarizations, respectively, an atom-light system initialized in 
$\ket{\text{v}} \otimes \ket{\vartheta, \varphi, N}$ [where $\ket{\vartheta, \varphi, N}$ is a coherent spin state, see Sec.~\ref{Sec.Fundamentals.entanglement}] evolves by the QND interaction~\eqref{Eq.QND} to the state \cite{McConnellPRA2013}
\begin{multline}
\ket{\psi} = \tfrac{1}{2} \ket{\text{v}} \otimes \big( \ket{\vartheta, \varphi_+,N} +  \ket{\vartheta, \varphi_-,N} \big) +  \\
\qquad \qquad \qquad + \tfrac{1}{2} \ket{\text{h}} \otimes \big( \ket{\vartheta, \varphi_+,N} -  \ket{\vartheta, \varphi_-,N} \big),
\end{multline}
where $\varphi_\pm  = \varphi \pm \phi$ and $\phi$ is the phase accumulated during the interaction.
The detection of a vertically polarized photon projects the atomic ensemble to
$\ket{\psi\si{v}} \propto \ket{\vartheta, \varphi_+,N} +  \ket{\vartheta, \varphi_-,N}$, whereas
the detection of a horizontally polarized photon prepares 
$\ket{\psi\si{h}} \propto \ket{\vartheta, \varphi_+,N} -  \ket{\vartheta, \varphi_-,N}$.
The entanglement properties strongly depend on the phase $\phi$, and the quantum Fisher information is
(without loss of generality we assume here that $\vartheta=\pi/2$ and $\varphi=0$)
\be \label{QFI_hv}
\frac{\Fish\si{Q}\big[\ket{\psi\si{h,v}}, \J_y\big]}{N}  = \frac{1 +(N-1) \sin^2 \phi \pm \cos^N \phi}{ 1 \pm \cos^N \phi},
\ee 
where the plus (minus) sign holds for the detection of a vertically (horizontally) polarized photon.
To leading order in $N$, we have $\Fish\si{Q}[\ket{\psi\si{v}}, \J_y ]/N = 1 + N\phi^2/2 + \mathcal{O}(N^2\phi^4)$
and $\Fish\si{Q}[\ket{\psi\si{h}}, \J_y ]/N = 3 + N\phi^2/2 + \mathcal{O}(N^2\phi^4)$: for $\abs{\phi} \lesssim 2/\sqrt{N}$, 
the state $\ket{\psi\si{h}}$ is metrologically more useful than $\ket{\psi\si{v}}$. 
In particular, for $\abs{\phi} \ll 1/\sqrt{N}$, the (rare) detection of a horizontally polarized photon heralds the 
generation  of $\ket{\psi\si{h}} = \ket{(N/2-1)_x}$, which is the one-excitation Dicke state.
For $\abs{\phi} \gtrsim 2/\sqrt{N}$, we have $\cos^N \phi \approx 0$ in Eq.~\eqref{QFI_hv} and the two states 
give approximatively the same quantum Fisher information.
At $\phi = \pm\pi/2$, we get the NOON state $\ket{\psi\si{h,v}} = (\ket{(N/2)_y} \pm \ket{(-N/2)_y})/\sqrt{2}$.
In the realistic scenario $\abs{\phi} \ll 1/\sqrt{N}$, 
the detection of a sequence of photons conditionally prepares more and more entanglement. 
For instance, disregarding losses and decoherence, the detection of  $n\si{h}$ h-polarized photons prepares the NOON 
state $(\ket{(N/2)_z} + \ket{(-N/2)_z})/\sqrt{2}$ when $n\si{h} \times \abs{\phi} \gtrsim \sqrt{N}$.
Finally, \textcite{ChenPRL2015} have generalized the above method showing the possibility to generate a broad class of 
entangled states of many atoms using single photons having a tailored frequency spectrum and time-resolved detection.

Experimentally, the use of single-photon counting for preparing entangled states of an 
atomic ensemble has been pioneered by \textcite{ChristensenNJP2013, ChristensenPRA2014}.
These experiments have shown the expected increase of spin fluctuations due to the single-photon detection. 
\textcite{McConnellNATURE2015} have demonstrated the generation of entanglement in an atomic ensemble 
of $N=3100$ laser-cooled $^{87}$Rb atoms prepared in an optical cavity 
and probed with an off-resonant weak pulse (of about 200 photons) of vertically polarized light, see Fig.~\ref{Fig:singphot}.
The accumulated phase is $\phi = \tfrac{4 g^2}{\Delta \kappa_c}$, where $2/\kappa_c$ is the characteristic atom-photon interaction time. 
For the experiment of \textcite{McConnellNATURE2015} we have $\phi\sqrt{N} \approx 0.03$.
This experiment has investigated the entangled state produced by the detection of a horizontally polarized photon, while the detection 
of vertically polarized photons gives, for these experimental parameters, only a slight spin squeezing.
From the tomographic reconstructed density matrix it is possible to obtain a Wigner distribution with negative areas, see Fig.~\ref{Fig:singphot}, and an 
entanglement depth (not related to metrological usefulness) of 2900 atoms. 
This work demonstrates how the information carried by a single (or a few) photon(s) can create entanglement in a large atomic ensemble.
In \textcite{HuPRA2017} the same authors have re-analyzed these data to show that the generated atomic state violates classical 
physics even if no assumptions related to quantum mechanics are made. 
This conclusion \cite{KotPRL2012}, similarly to Bell inequalities \cite{BrunnerRMP2014}, observes that marginal probability distributions measured for non-commuting 
observables do not always come from a joint probability distribution identified with a classical state. As in \textcite{SchmiedSCIENCE2016} 
(see Fig.~\ref{fig:SchmiedSCIENCE2016}), \textcite{HuPRA2017} show that classical physics is insufficient for describing 
mesoscopic entangled states of atoms.

\subsubsection{Cavity-based quantum nondemolition measurements in the normal mode splitting regime}}

\begin{figure}
\begin{center}
\includegraphics[width=\columnwidth]{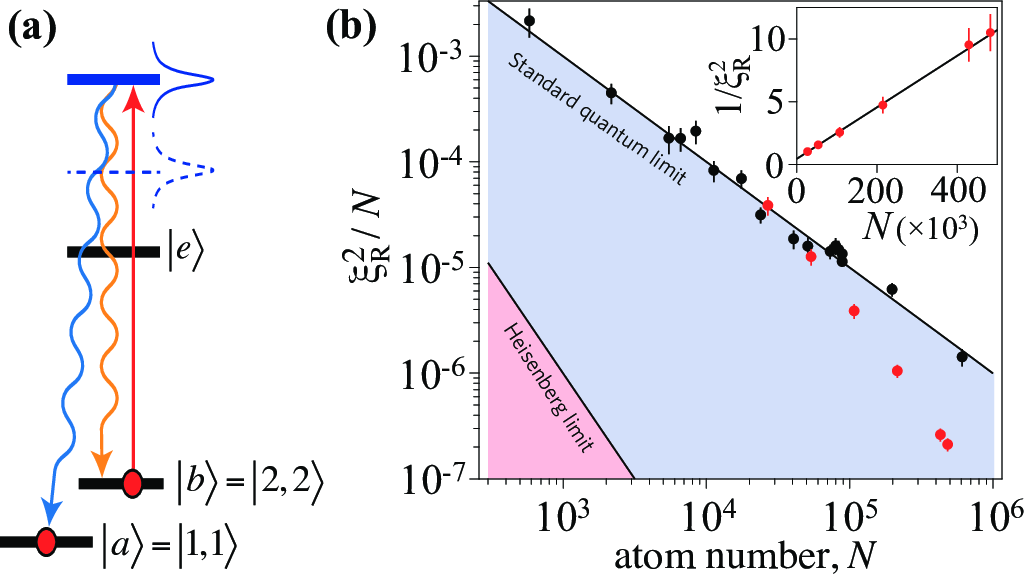}
\end{center}
\caption{{\bf Spin squeezing via quantum nondemolition measurements in the normal mode splitting regime.} 
(a) Energy-level diagram showing the cavity shift due to atoms in level $\ket{b}$. 
(b) $\xiR^2/N$ as a function of atom number $N$ for uncorrelated particles (black dots) and spin-squeezed states (red [grey] dots).
The inset shows $1/\xiR^2$ of the spin-squeezed state (red dots) as a function of $N$.
The black line is a fit to the data according to a $1/N^2$ scaling.
Adapted from \textcite{BohnetNATPHOT2014}.} 
\label{Fig:Thompson}
\end{figure}

The experiments discussed in the previous sections operate in the dispersive regime, where the cavity is far detuned from atomic resonance. 
In this section, we discuss quantum state preparation schemes where the cavity is tuned near an atomic resonance 
with respect to the atomic transition $\ket{b} \leftrightarrow\ket{e}$, see Fig.~\ref{Fig:Thompson}(a).
Let us assume that 
atoms in $\ket{a}$ do not to interact with the cavity mode, the number of atoms in $\ket{b}$ is
$N_b \gg 1$, and the system is driven weakly so that the mean number of atoms in $\ket{e}$ 
is small with respect to $N_b$. 
We have that atoms in $\ket{b}$ cause a splitting $\omega_c \rightarrow \omega_c^{\pm}$ of the cavity resonance frequency \cite{ChenPRA2014}
that depends on $N_b$: 
\be
\label{eq:ChenSplitting}
\omega_c^{\pm} = \omega_c - \frac{\delta_c \pm \sqrt{\delta_c^2 + 4 g^2 N_b} }{2},
\ee
where $\delta_c = \omega_c - \omega_{be}$ is the cavity detuning from the atomic transition $\omega_{be}$, 
$\omega_c$ is the bare cavity frequency, and $g$ is the single-atom-single-photon coupling rate (uniform coupling is assumed).
At $\delta_c=0$, the appearance of two well-resolved resonances separated by $2 g \sqrt{N_b}$ is referred to as collective vacuum Rabi splitting. 
In the limit of collective strong coupling $g \sqrt{N_b} \gg \kappa_c, \Gamma$ the normal modes of the coupled system are well resolved:
a measurement of $\omega_c^{\pm}$ thus allows to determine $N_b$.

In the experiments of \textcite{BohnetNATPHOT2014, ChenPRL2011, CoxPRL2016},
the atom number $N_b$ is measured non-destructively by 
probing the cavity resonant frequency with an additional probe laser injected into the cavity. 
The number of particles in level $\ket{a}$ can be probed by first applying a resonant $\pi$ 
pulse between $\ket{a}$ and $\ket{b}$ and then repeating the QND measurement.
A detailed analysis of the impact of decoherence in this system (mainly due to scattering 
of cavity photons into free space) can be found in \textcite{ChenPRA2014}.
In \textcite{ChenPRL2011} the optical cavity is tuned to resonance, $\delta_c=0$, with $NC \approx 1400$.  
\textcite{ChenPRL2011} reported a spin squeezing $\xiR^2 = -1$\,dB ($\xiR^2 = -3.3$\,dB inferred) relative to the standard quantum limit
using $N\si{eff} = 7 \times 10^5$ $^{87}$Rb atoms ($N\si{eff}$ is an effective atom number taking into account inhomogeneous coupling in the cavity, see Sec.~\ref{sec:effectivespin}).
In \textcite{BohnetNATPHOT2014, CoxPRL2016} noise effects were considerably reduced
taking advantage of a cycling transition \cite{ChenPRA2014, SaffmanPRA2009},
and higher collective cooperativity $NC \approx 6000$ was reached. 
Experimental results demonstrate $\xiR^2 = -10.1$\,dB \cite{BohnetNATPHOT2014}, see Fig.~\ref{Fig:Thompson}, 
and $\xiR^2 = -17.7$\,dB \cite{CoxPRL2016}, using  $N\si{eff} = 5 \times 10^5$ $^{87}$Rb atoms. 
The achieved spin squeezing witnesses the presence of useful $170 \pm 30$ particle entanglement \cite{CoxPRL2016}.
This squeezing technique has been further combined with quantum feedback, reaching $\xiR^2 = -7.4$\,dB of deterministic spin squeezing \cite{CoxPRL2016}.

Resonant atom-light coupling in an optical cavity also offers the possibility to create entangled states beyond spin squeezing. 
\textcite{HaasSCIENCE2014} have used an atom chip 
integrated with a high-finesse fiber-optical cavity in the strong-coupling regime \cite{VolzNATURE2011} 
to prepare a generalized $W$ state \cite{DurPRA2000} of $N \approx 40$ cold $^{87}$Rb atoms.  
In this experiment the cavity and probe laser tuned on resonance with the $\ket{ b} \leftrightarrow \ket{e}$ transition. 
The single-atom strong coupling $g \gg \kappa_c, \Gamma$ guarantees that the cavity only transmits light if all $N$ atoms are in level $\ket{a}$: 
a single atom in $\ket{b}$ is sufficient to block the cavity transmission. 
The experiment of \textcite{HaasSCIENCE2014} starts with all atoms in level $\ket{a}$, and a weak microwave pulse
coupling $\ket{a}$ and $\ket{b}$ prepares the state $\sqrt{1-p} \ket{N}_a \ket{0}_b + \sqrt{p} \ket{N-1}_a \ket{1}_b$, 
where $p \ll 1$ to avoid multi-particle excitations. 
The cavity transmission is then measured with a probe-light beam.
A low transmission heralds the QND preparation of the W state $\ket{N-1}_a \ket{1}_b$.
The resulting state is characterized by state tomography, from which it is possible to extract a lower bound on the 
entanglement depth of $13$ atoms, at least.

\begin{figure}[t!]
\begin{center}
\includegraphics[width=\columnwidth]{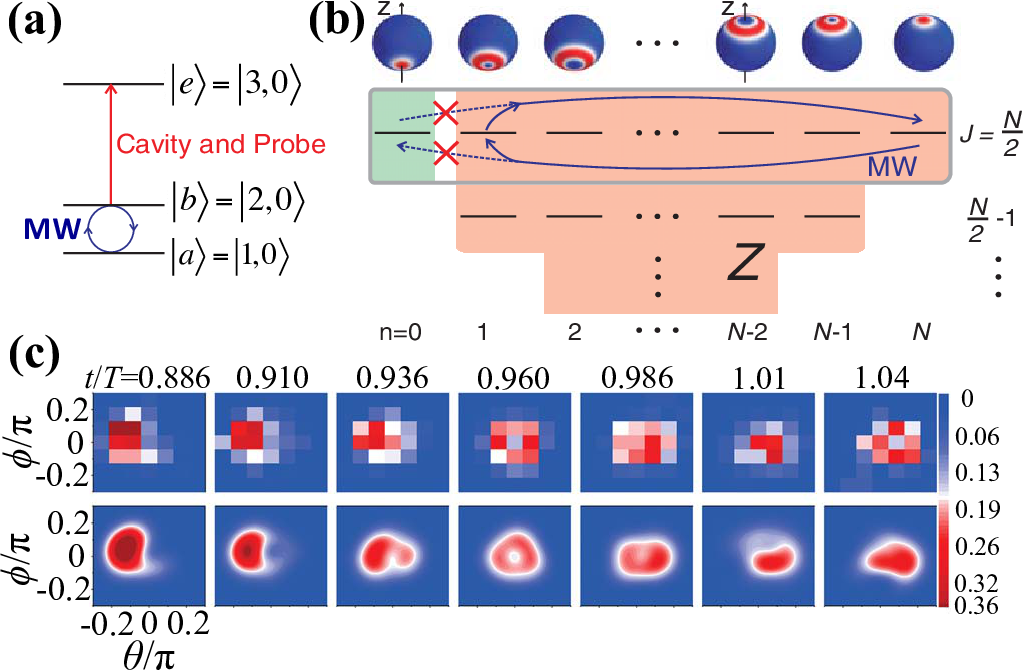}
\end{center}
\caption{{\bf State engineering via quantum Zeno dynamics.}
(a) $\ket{a}$ and $\ket{b}$ are coupled by a microwave field at Rabi frequency $\Omega$, while the optical cavity transition
$\ket{b} \leftrightarrow \ket{e}$ is probed with a coupling rate much faster than $\Omega$.
(b) Starting with all atoms in $\ket{b}$, microwave Rabi coupling and simultaneous measurement of the cavity transmission leads to a coherent evolution 
restricted to the subspace highlighted by the orange/right shaded area. 
The preparation of the state with no atoms in $\ket{b}$ (green/left shaded area) is forbidden by the quantum Zeno effect.
The upper row shows the Husimi distributions of symmetric Dicke states.
(c) Experimental Husimi distributions for different evolution times
$t/T$, where $T=\pi/\Omega$ is the time for a Rabi $\pi$-pulse in the absence of cavity probing.
The upper row shows direct measurements, while the lower row the reconstructed density matrix. 
Adapted from \textcite{BarontiniSCIENCE2015}.} 
\label{Fig:QZD}
\end{figure} 

\subsubsection{Quantum state preparation via quantum Zeno dynamics}

Quantum Zeno dynamics combines the coherent evolution of a system with the measurement of an 
initially unoccupied quantum state \cite{FacchiJPA2008}:
if the measurement is performed with high enough frequency, the measured state remains unoccupied. 
This profoundly modifies the system dynamics and allows engineering specific entangled states.  
In \textcite{BarontiniSCIENCE2015} a cavity-based measurement in the regime $C\gg 1$ \cite{VolzNATURE2011}
is used to prepare a generalized $W$ state of 36 $^{87}$Rb atoms via quantum Zeno dynamics. 
The $N$ atoms are all initialized in level $\ket{b}$ and coupled to the $\ket{a}$ level by a microwave drive at Rabi frequency $\Omega$.
The cavity is in resonance with the $\ket{b} \leftrightarrow \ket{e}$ transition and 
transmits only if there are no atoms in level $\ket{b}$. 
The cavity transmission is probed at a rate much larger than $\Omega$. 
When no measurement is performed, the microwave Rabi coupling prepares the state $\ket{N}_a \ket{0}_b$ after a time $T=\pi/\Omega$.
In presence of the cavity measurement, the quantum Zeno effect forbids the preparation of $ \ket{N}_a \ket{0}_b$ and deterministically prepares 
the W state $\ket{N-1}_a \ket{1}_b$, see Fig.~\ref{Fig:QZD}.  
From the density matrix extracted via state tomography, it is possible to obtain the quantum Fisher information.
Values up to $\Fish\si{Q}/N=1.51$ are reported by \textcite{BarontiniSCIENCE2015}. 

\subsection{Light-mediated coherent interaction between distant atoms}
\label{Sec.Atom-Light.effint}

In contrast to the measurement-based scheme of Sec.~\ref{Sec.Atom-Light.nd}, 
off-resonant atom-light coupling can also be used to 
realize a light-mediated coherent interaction between distant atoms. 
When the Stokes operator $\Sp_z$ in Eq.~\eqref{Eq.QND} is proportional to $\J_z$, the Hamiltonian
reduces to $\op{H}\si{QND} = \hbar k \J_z^2$ \cite{AgarwalPRA1997, ZhangPRA2003b}, 
corresponding to an effective one-axis twisting nonlinearity
that generates unconditional spin squeezing \cite{KitagawaPRA1993}.
The condition $\Sp_z \propto \J_z$ can be realized in an optical cavity, see below.
As a main difference with respect to contact interaction discussed in Sec.~\ref{Sec.Atom-Atom.twisting}, 
this light-mediated interaction can be applied to a dilute atomic sample, which is 
preferred for minimizing systematic errors in precision measurements.
Moreover, in contrast to spin squeezing obtained from QND measurements, 
it deterministically produces known entangled states, independently from the detector performance.
Furthermore, atom-cavity coupling can be easily switched on and off.

\begin{figure}[t!]
\begin{center}
\includegraphics[width=\columnwidth]{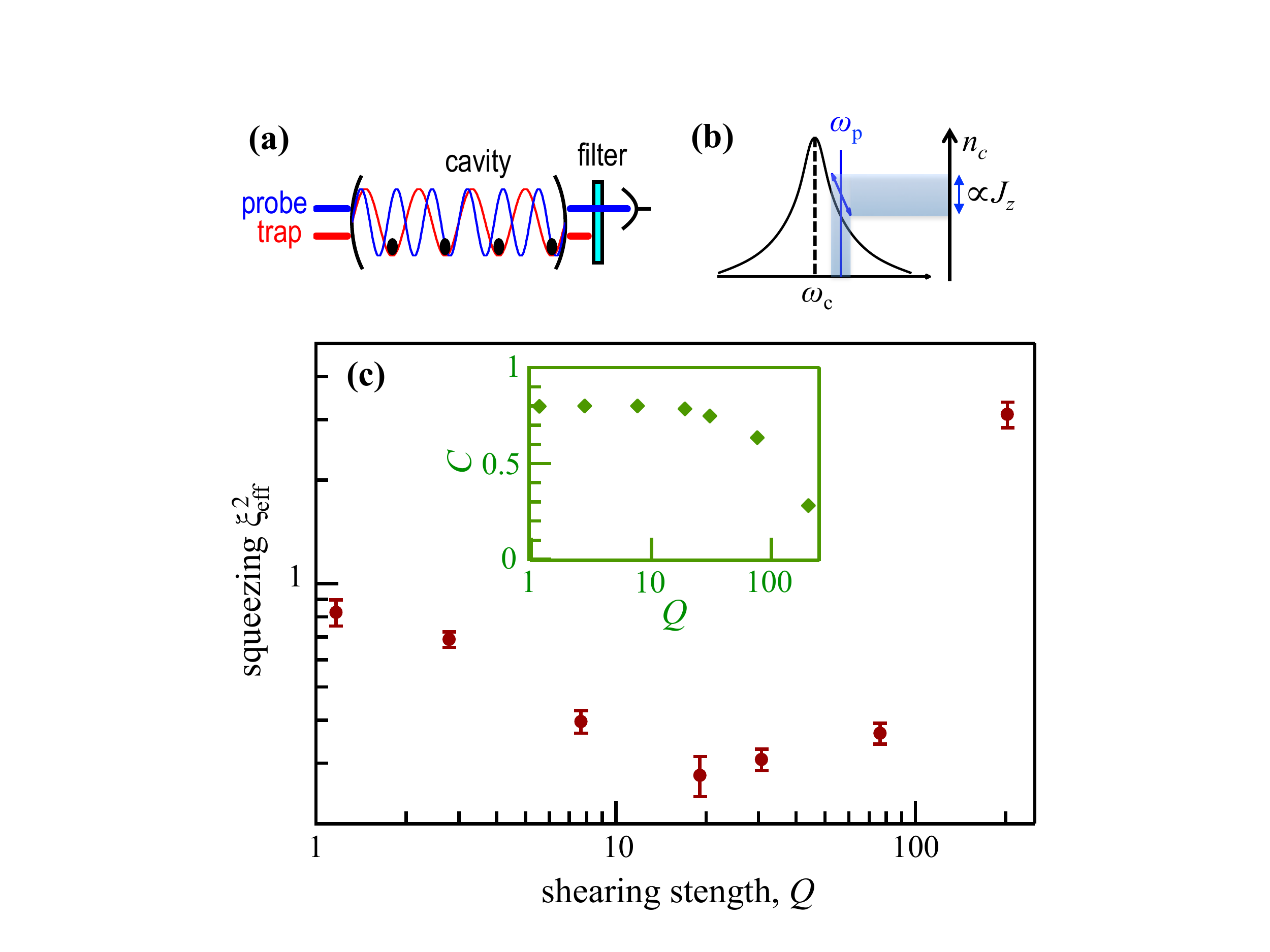}
\end{center}
\caption{{\bf Squeezing via light-mediated effective interaction between atoms.} 
(a) The atoms are trapped in a standing-wave dipole trap inside an optical resonator. 
(b) The probe laser is detuned from cavity resonance by half a linewidth, 
so that atom-induced shifts of the cavity frequency change the transmitted power by an amount proportional to $J_z$.
(c) Effective spin squeezing $\xi\si{eff}^2 = \sigma^2(Q) \mathcal{C}(0)/\mathcal{C}(Q)^2$ giving the expected gain 
with respect to the sensitivity experimentally achieved in absence of entanglement,
as a function of the shearing strength $Q$ (proportional to the photon number in the cavity).
Here $\sigma^2(Q) = (\Delta \J_z)^2/(N\si{eff}/4)$, and $\mathcal{C}(Q)$ is the contrast of Rabi oscillations of the sheared state, shown in inset.
The effective spin squeezing is related to the metrological spin squeezing Eq.~\eqref{Eq:xiWineland} as $\xi\si{eff}^2 = \mathcal{C}(0) \xiR^2$.
An effective squeezing $\xi\si{eff}^2=-5.6$\,dB is reached, corresponding to $\xiR^2 = -4.6$\,dB.
Adapted from \textcite{LerouxPRL2010a}.} 
\label{Fig:Vuletic_twisting}
\end{figure} 

There are different proposals for the realization of an effective interaction between distant atoms.
\textcite{TakeuchiPRL2005} have studied a double-pass Faraday interaction
in free space where a polarized light beam passes twice
through an atomic ensemble (after the first passage the
polarization is properly rotated such that $\Sp_z \propto \J_z$ at the second passage).
An improved version of this proposal has been discussed by \textcite{TrailPRL2010} using a
quantum eraser to remove residual spin-probe entanglement.
\textcite{Schleier-SmithPRA2010,ZhangPRA2015} have proposed the
realization of light-induced atom-atom interaction in an optical cavity (also called cavity squeezing or cavity feedback). 
This relies on the dispersive interaction between one mode of the cavity and an ensemble of three-level atoms.
As shown by Eq.~\eqref{cavityH} the atoms modify the cavity resonance frequency by an amount  proportional to $\J_z$. 
When the cavity is driven by a probe laser of frequency $\omega_p$ tuned to the slope of the cavity resonance
($\omega\si{p} = \kappa_c$), the intracavity photon number
$n_c=\mean{\op{c}^\dag \op{c}}$ in Eq.~\eqref{cavityH}, being detuning-dependent, changes linearly
with the index of refraction of the atomic cloud (which, in turn, is proportional to $\J_z$), see Fig.~\ref{Fig:Vuletic_twisting}(b).
Notice that tracing out the light field (which carries information about the atomic spin) 
yields a dissipative dynamics that limits the attainable squeezing \cite{LerouxPRA2012}.
\textcite{PawlowskiEPL2016} analyzed the spin squeezing attainable with the cavity feedback method in presence 
of cavity losses and spontaneous emission, finding that $\xiR^2 \propto (\Gamma/\Delta)^2$ in the limit $N \to \infty$.
Finally, an alternative proposal to realize an effective one-axis twisting atomic Hamiltonian uses a four-wave Raman scheme 
to directly drive the atomic ensemble, where atoms communicate with one another by exchanging photons via a weakly-occupied cavity mode \cite{SorensenPRA2002}.  

One axis twisting by light-mediated coherent interaction has been experimentally implemented in an optical cavity \cite{LerouxPRL2010a, LerouxPRL2010b}, 
using an effective number $N\si{eff}=3 \times 10^4$ of $^{87}$Rb atoms. 
The experiment implements the scheme of \textcite{Schleier-SmithPRA2010}.
The squeezed state is characterized via spin-noise tomography and the contrast of Rabi oscillations.
The results of \textcite{LerouxPRL2010a} are summarized in Fig.~\ref{Fig:Vuletic_twisting}(c) and demonstrate a gain 
$\xiR^2 = -4.6$\,dB over $(\Delta \ps\si{SQL})^2 = 1/N\si{eff}$
($\xi\si{eff}^2 = -5.6$\,dB over the sensitivity experimentally reached in absence of entanglement).
\textcite{LerouxPRL2010b} have used the squeezed states generated with this method to realize an atomic clock with sensitivity 
$4.5$\,dB below the standard quantum limit, see Sec.~\ref{Sec.Working-Entanglement}. 

\subsection{Quantum state transfer from non-classical light to atoms}
\label{Sec.Atom-Light.state-transfer}

Spin-squeezed states can be created via atom-light interaction by transferring quadrature squeezing of light to
atomic spin squeezing \cite{AgarwalPRA1990, KuzmichPRL1997, HammererRMP2010}. 
This effect can be understood from the Jaynes-Cummings model~Hamiltonian
\be
\op{H}\si{JC}= \hbar \Omega (\op{c} \J_+ + \op{c}^\dag \J_-),
\ee 
describing the interaction of nondecaying two-level atoms with a light mode.
Here $\op{c}^\dag$ and $\op{c}$ are the light creation and annihilation operators, respectively, 
$\J_{\pm}$ are spin rising and lowering operators, see Sec.~\ref{Sec.Fundamentals.spin}, 
and $\Omega$ is a coupling constant.
The Heisenberg equations of motion for $\op{c}$ and $\op{J}_{\pm}$ can be found analytically 
within a frozen-spin approximation, assuming $\J_z$ constant and equal to $N/2$. 
We obtain \cite{WinelandPRA1992, WinelandPRA1994}
\begin{subequations}
\begin{align}
\xiR^2(t) &= \xiR^2(0) \cos^2 (\Omega_N t) + \xi_{p}^2(0) \sin^2 (\Omega_N t) \\
\xi_{p}^2(t) &= \xi_{p}^2(0) \cos^2 (\Omega_N t) + \xiR^2(0) \sin^2 (\Omega_N t),
\end{align}
\end{subequations}
where $\xi_{p}^2 = 2 (\Delta \op{P}_c)^2$, $\op{P}_c = \frac{\op{c} - \op{c}^\dag}{\ii\sqrt{2}}$, 
$\xiR^2 = \frac{N (\Delta \J_x)^2}{\mean{\J_z}^2}$, and $\Omega_N = k\sqrt{N}$.
Similar expressions can be obtained relating $\xiR^2 = \frac{N (\Delta \J_y)^2}{\mean{\J_z}^2}$ with $\xi_x^2 = 2 (\Delta \op{X}_c)^2$, 
where $\op{X}_c = \frac{\op{c} + \op{c}^\dag}{\sqrt{2}}$.
These equations predict that atomic spin squeezing $\xiR<1$ can be achieved by first squeezing the light field, $\xi_{p}^2<1$, and then 
transferring this quadrature squeezing onto the spins. Optimal spin squeezing is reached at $t=\pi / (2 \Omega_N)$.
For strong squeezing the contrast $\mean{\J_z}$ decreases, and the frozen-spin approximation breaks down. 
Since the squeezed light is the source of spin squeezing in this method, the degree of squeezing is determined by the quality of the 
quadrature squeezed light.

Quantum state transfer from light to atoms has been first experimentally demonstrated by \textcite{HaldPRL1999} using a cloud of $10^7$ cold Cs atoms,
following the theoretical proposal of \textcite{KuzmichPRL1997}.
This experiment used a V-level scheme consisting of three atomic hyperfine levels \cite{HaldPRL1999, HaldJMO2000}.
The reduction of atomic spin noise below the projection noise of uncorrelated atoms is generated by the absorption of 
polarized coherent and squeezed vacuum light with opposite circular polarizations.
The atomic fluctuations are read out by the differential photocurrent of a probe beam at the output ports of a polarizing beam splitter.  

An alternative approach, mapping a quantum state of light onto an atomic state via electromagnetically induced transparency 
has been proposed by \textcite{FleischhauerPRL2000}, see also \textcite{PhillipsPRL2001, LiuNATURE2001}.
Further studies include the mapping quantum states of light into a Bose-Einstein condensate with application 
to atom interferometers \cite{SzigetiPRA2014} and atom lasers \cite{FleischhauerPRL2002, HainePRA2005}.

\section{Entangled states of trapped ions}
\label{Sec.Ions}

Ensembles of trapped ions confined in electromagnetic traps and manipulated with laser beams
are one of the most successful systems to generate and exploit entanglement \cite{BlattNATURE2008,WinelandRMP2013}.
Direct applications include quantum information and computation \cite{HaffnerPHYSREP2008,RoosBOOK2014}, quantum networks \cite{DuanRMP2012},
quantum simulations \cite{BlattNATPHYS2012,SchneiderRPP2012,MartinezNATURE2016} and, as reviewed here, quantum metrology.
Trapped ions are well isolated from the environment, 
can be coherently manipulated, individually addressed and detected with almost unit efficiency \cite{LeibfriedRMP2003}. 

\begin{figure}[t!]
\begin{center}
\includegraphics[width=\columnwidth]{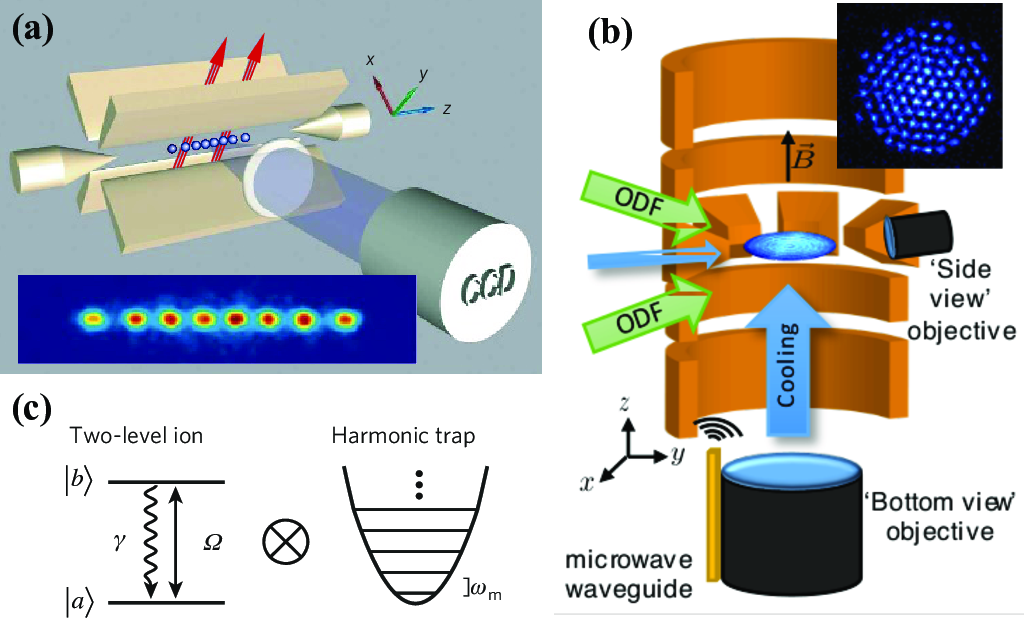}
\end{center}
\caption{{\bf Common ion-trapping setups.}
a) Linear Paul trap. Ions are trapped using radio-frequency oscillating electric fields combined with static electric control fields.
In tight radial confinement, laser-cooled ions form a linear string (see inset image for 8 ions) with a spacing determined 
by the balance between external confining fields and ion-ion Coulomb repulsion.
Taken from \textcite{BlattNATURE2008}.
b) Penning trap. Ions are confined in the $xy$ plane with a strong homogeneous magnetic 
field and axially by a quadrupole electric field. 
The green ODF arrows indicate bi-chromatic light fields used to generate entanglement. 
The inset shows an image of a 2D crystal of $91$ ions.
Taken from \textcite{BohnetSCIENCE2016}.
c) Each trapped ion can be modeled as an effective internal two-level system---$\ket{a}$ and $\ket{b}$ 
interacting with a radiation characterized by Rabi frequency $\Omega$ and decay rate $\gamma$---and a shared external harmonic oscillator potential with equally spaced energy levels of mode frequency $\omega_m$.}
\label{Fig:IonTrap}
\end{figure}

Typical experiments \cite{LeibfriedRMP2003} confine ions in a linear Paul trap 
[Fig.~\ref{Fig:IonTrap}(a)], surface-electrode Paul trap \cite{SterlingNATCOMM2014,MielenzNATCOMM2016}, 
or Penning trap [Fig.~\ref{Fig:IonTrap}(b)], followed by ground-state cooling.
So far, up to 20 ions have been entangled in a Paul trap \cite{MonzPRL2011} and up to about 200 ions in a Penning trap 
\cite{BrittonNATURE2012, BohnetSCIENCE2016}.  
One- and two-dimensional ion arrangements are well suited for individually controlling and reading out internal states of the ions. 
We note that individual particle manipulation and detection are
crucial for many quantum-computation and quantum-information tasks, 
but they are not necessary for quantum-enhanced metrology. 
Interesting metrological experiments can thus be also performed in Penning traps
(where local manipulation is currently not available), which 
are well suited to store, manipulate, and entangle large numbers of ions.  

In trapped-ion experiments, the qubit is formed by two internal (usually hyperfine) states $\ket{a}$ and $\ket{b}$ of each ion. 
This internal qubit can be coupled by an electromagnetic field to external (motional, vibrational) degrees of freedom, see Fig.~\ref{Fig:IonTrap}(c).
A key aspect is the local detection of the internal ion's state with extremely high efficiency. 
This is usually achieved by resonant fluorescence. 
Detection errors arise if the qubit changes its state during the detection interval or if photons scattered off trap electrodes are mistaken 
for fluorescence photons. In \textcite{MyersonPRL2008,HartyPRL2014} qubit detection errors smaller than $10^{-4}$ are reported.

\subsection{Generation of metrologically useful entangled states of many ions}
\label{Sec.Ions.states}

The generation of entanglement between different qubits is one major challenge in quantum information processing with trapped ions \cite{WinelandREP1998}. 
The Coulomb force pushes the ions apart to a distance much larger than the Bohr radius and thus prevents state-dependent ion-ion 
interactions in absence of external electromagnetic fields.
However, as a consequence of the Coulomb force, the normal modes of motion in the trap are shared among all ions and 
can provide a means for entangling the qubits.
Most of the schemes for entangling trapped ion qubits are based on internal-state-dependent forces acting on the ions, 
generated by external electromagnetic fields.
These forces can be arranged in time and frequency in such a way that an effective entangling operation is carried out between 
two or more qubits and no entanglement persists between the qubits and the vibrational degrees of freedom.

The entangling gates can be broadly classified in three categories, according to the way the electromagnetic fields interact with the ions:
\emph{i)} Quantum gates induced by a laser beam that interacts with a single ion at a time as originally proposed by \textcite{CiracPRL1995}. 
The ion is entangled with a vibrational mode of the ion string and the entanglement is subsequently transferred from the vibrational mode 
to the internal state of a second ion. 
\emph{ii)} Quantum gates induced by an electromagnetic field that simultaneously addresses two or more ions, either 
through optical ac Stark shifts \cite{LeibfriedNATURE2003,GarciaRipollPRA2005,KimNATURE2010,SorensenPRA2000,MilburnFP2000,LeeJOB2005,KimPRL2009,MolmerPRL1999,SolanoPRA1999,SorensenPRL1999}, through static magnetic-field gradients in combination with homogeneous radio-frequency 
fields \cite{MintertPRL2001,WelzelEPJD2011,JohanningJPB2009,JohanningPRL2009}, or 
with radio-frequency field gradients \cite{OspelkausPRL2008,OspelkausNATURE2011}.
As an important advantage,  these gates do not require to cool all ion normal modes to their ground states, 
but only the ion motion to be well within the Lamb-Dicke regime $\mean{\op{z}^2} \ll (\lambda/2\pi)^2$, 
where $z$ is the displacement of an ion along the laser propagation direction and $\lambda$ the laser wavelength.
The basic idea is to drive a phonon (normal) mode of frequency $\nu$ by a periodic state-dependent force with frequency $\omega$. 
In the rotating-wave approximation, the Hamiltonian describing this conditional interaction is \cite{RoosBOOK2014}
\be
\op{H} = \ii \hbar k (\op{c}^\dag e^{\ii \delta t} - \op{c} e^{-\ii \delta t}) \op{O}, 
\ee
where $\op{O}$ is an operator acting on the (internal) qubit states, $\op{c}^\dag$ and $\op{c}$ are the harmonic 
oscillator creation and annihilation operators of the addressed normal mode, respectively, $k$ is a coupling strength, and $\delta =\omega - \nu$.
The propagator $\op{U}$ corresponding to this (time-dependent) Hamiltonian is 
\be \label{Eq.propagator}
\op{U}(t) = \exp[\alpha(t) \op{O} \op{c}^\dag - \alpha^*(t) \op{O}^\dag \op{c}] \exp[\ii \phi(t) \op{O}^2],
\ee
where $\alpha(t) = \ii (k/\delta) (1 - e^{\ii \delta t})$ and $\phi(t) = (k/\delta)^2 (\delta t - \sin \delta t)$.
If the interaction time $t=\tau$ is chosen such that $\alpha(\tau)=0$, then the propagator reduces to $U(\tau) = \exp[\ii \phi(\tau) \op{O}^2]$.
A vibrational mode becomes transiently entangled with the qubits before getting disentangled at the end of the 
 gate operation, resulting in an effective nonlocal operation capable of entangling the ions.  
The conditional phase gate \cite{LeibfriedNATURE2003} is obtained by setting 
$\op{O} = \tfrac{1}{2}\sum_{i=1}^N \op{\sigma}_z^{(i)} = \J_z$ in Eq.~\eqref{Eq.propagator}, and
is realized by placing the ions in a qubit-state-dependent potential created by two counter-propagating laser beams. 
The M\o lmer-S\o rensen gate \cite{MolmerPRL1999, SorensenPRA2000} is obtained setting $\op{O} = \tfrac{1}{2}\sum_{i=1}^N \op{\sigma}_y^{(i)} = \J_y$
(or $\op{O} =\J_x$).
This gate is realized by a bichromatic laser field tuned close to the upper and lower motional sideband of the qubit transition \cite{BenhelmNATPHYS2008, SackettNATURE2000}.
In both gates, the propagator is of the kind $\op{U} = \exp[\ii \phi(\tau) \J_{\vect n}^2]$, formally equivalent to Kitagawa-Ueda one-axis twisting, see Sec.~\ref{Sec.Atom-Atom.twisting}.
Spin-dependent optical dipole forces generated by a pair of off-resonance laser beams with different frequencies, addressing many normal modes simultaneously,
have been recently used to engineer tunable Ising spin-spin coupling \cite{KimPRL2009, PorrasPRL2006, LeeJOB2005}. 
The resulting propagator is \cite{PorrasPRL2006}
\be \label{Eq.UIsing}
\op{U}(t) = \exp \bigg[ \ii \sum_{i,j=1}^N V_{i,j} \op{\sigma}_z^{(i)} \op{\sigma}_z^{(j)}  \bigg],
\ee
where $V_{ij} \propto d^{-\alpha}_{ij}$, $ d_{ij}$ is the distance between the $i$th and $j$th ions and $\alpha$ can be tuned by adjusting 
the laser frequencies in the range $0\leq \alpha \leq3$ \cite{BrittonNATURE2012, SchmiedNJP2011b}. 
Ising spin-spin couplings have been implemented in a linear Paul trap with up to $\sim$20 ions \cite{IslamSCIENCE2013, JurcevicNATURE2014, RichermeNATURE2014}
and in a Penning trap forming a 2D Coulomb crystal
of $\sim$200 ions \cite{BrittonNATURE2012, BohnetSCIENCE2016}, see Fig.~\ref{Fig:IonTrap}(b).
\emph{iii)} A third way to generate ion entanglement is based on performing joint measurements on photons that are first entangled 
with ion qubits \cite{MoehringNATURE2007, StuteNATURE2012, CasabonePRL2013}.
This scheme enables the generation of entanglement between ions separated by large distances and does not require the ions to be in the Lamb-Dicke regime. 

\subsection{Quantum metrology with trapped ions}
\label{Sec.Ions.metrology}

\subsubsection{Quantum metrology with two ions}

The first deterministic generation of entanglement of two trapped ions was reported by \textcite{TurchettePRL1998}.
This experiment was readily followed by the demonstration of a phase sensitivity below the standard quantum limit \cite{SackettNATURE2000, MeyerPRL2001}.
\textcite{MeyerPRL2001} reported the creation of the state
\be \label{Eq.Mayer}
\ket{\psi} = \cos \beta \ket{a}^{\otimes 2} + \ii \sin \beta \ket{b}^{\otimes 2}
\ee
of two ${}^9$Be$^+$ ions via a M\o lmer-S\o rensen gate in a linear Paul trap.
Here $\beta$ is a tunable parameter proportional to the laser pulse duration.
The state~\eqref{Eq.Mayer} is spin-squeezed, $\xiR^2 = \tfrac{N (\Delta \J_\perp)^2}{\mean{\J_z}^2} 
= \tfrac{1-\sin 2\beta}{\cos^2 2\beta}<1$ for $0 < \beta < \pi/2$. It reaches the Heisenberg limit $\xiR^2=1/2$  
at $\beta=\pi/4$, corresponding to the creation of a maximally entangled state.
However, at $\beta=\pi/4$, $\mean{\J_z} = 0$ and any technical noise prevents detecting the maximally entangled state as spin squeezed.
The experiment of \textcite{MeyerPRL2001} reached $\xiR^2=0.85$ at the experimentally optimal working point $\beta=\pi/10$.
A phase estimation with sensitivity below $\Delta \ps\si{SQL}$ with the maximally entangled state ($\beta=\pi/4$) was demonstrated 
by applying a spin rotation $\exp(-\ii \ps \J_z)$ followed by $\exp(-\ii \frac{\pi}{2} \J_x)$.
Measuring the parity $\op{\Pi}=\op{\sigma}_z^{(1)}\op{\sigma}_z^{(2)}$ 
led to a sensitivity $(\Delta \ps)^2/(\Delta \ps\si{SQL})^2 =0.70$, where $\Delta \ps = \Delta \op{\Pi}/\abs{\dd \mean{\op{\Pi}}/\dd\ps}$.
Many experiments with two ions have have reported parity oscillations from which it is possible to extract the phase sensitivity \cite{LeibfriedNATURE2003}:
in particular, in magnetic-field-insensitive clock states of $^{111}$Cd$^+$ \cite{HaljanPRA2005}, 
and using an optical transition in ${}^{40}$Ca$^+$ \cite{HomeNJP2006,NoguchiPRL2012,MonzPRL2011} and ${}^{88}$Sr$^+$ \cite{NavonPRA2014} ions.
A very high visibility of the parity signal has been reported with 
${}^{40}$Ca$^+$ \cite{BenhelmNATPHYS2008} and
$^{43}$Ca$^+$ \cite{BallancePRL2016} ions, see also \textcite{GaeblerPRL2016} for the high-fidelity generation of the two-qubit GHZ states with ${}^9$Be$^+$ ions.
Finally, \textcite{MeyerPRL2001} also reported a full Ramsey sequence,
where the maximally entangled state is generated between two Ramsey pulses and the phase $\ps$ is
proportional to the interrogation time. 
In this case, the initial probe state is a symmetric Dicke (or twin-Fock) state.  
By measuring the parity, it was possible to reach a spectroscopic sensitivity $(\Delta \ps)^2/(\Delta \ps\si{SQL})^2 =0.77$. 
\textcite{NoguchiPRL2012} reported the creation of symmetric Dicke states of two and four $^{40}$Ca$^+$ ions.

\begin{figure}[t!]
\begin{center}
\includegraphics[width=\columnwidth]{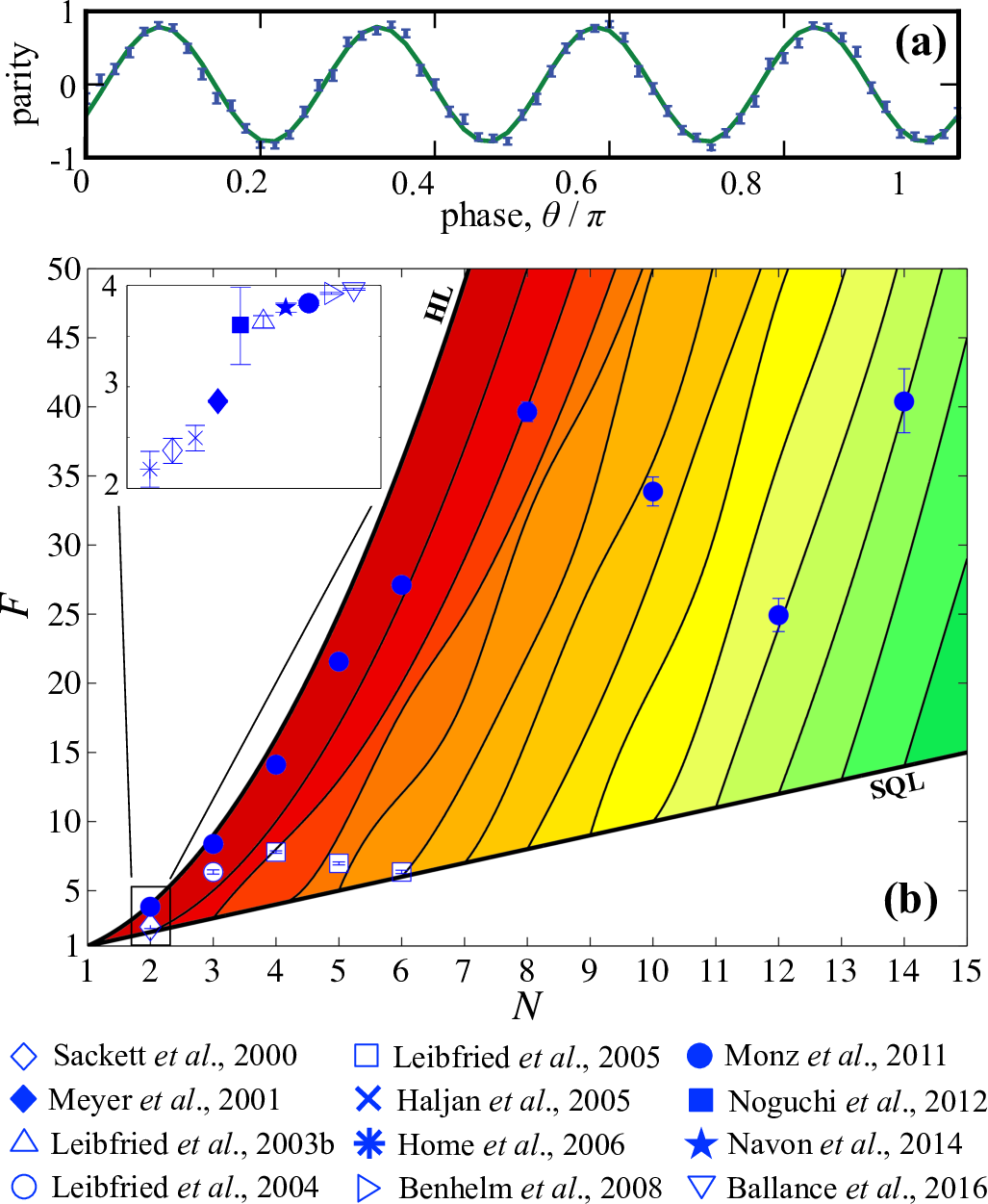}
\end{center}
\caption{{\bf Phase sensitivity of ion Schr\"odinger cat states.}
a) Typical parity oscillations obtained with cat states. These have a characteristic period $2\pi/N$ (here for $N=8$).
Taken from \textcite{MonzPRL2011}.
b) Summary of the experimental achievements (symbols). 
Here we show the Fisher information as a function of the number of qubits $N$, $F=V^2N^2$, obtained from the extracted experimental visibilities $V$. 
The upper thick line is the Heisenberg limit $F=N^2$, the lower thick line is the standard quantum limit, $F=N$. 
The thin lines are bounds for useful $k$-particle entanglement, Eq.~\eqref{Eq.FQIk}: they delimit from below a shaded region corresponding to 
$(k+1)$-particle entanglement.
In particular, the darker red region stands for useful genuine $N$-particle entanglement.
The next lighter red region stands for useful $(N-1)$-particle entanglement, and so on.
For instance, the point at $N=10$ reveals useful 4-particle entanglement.
The inset is a zoom for $N=2$ ions. Adapted from \textcite{PezzePNAS2016}.}
\label{FigGHZ}
\end{figure} 

\begin{figure}[t!]
\begin{center}
\includegraphics[width=\columnwidth]{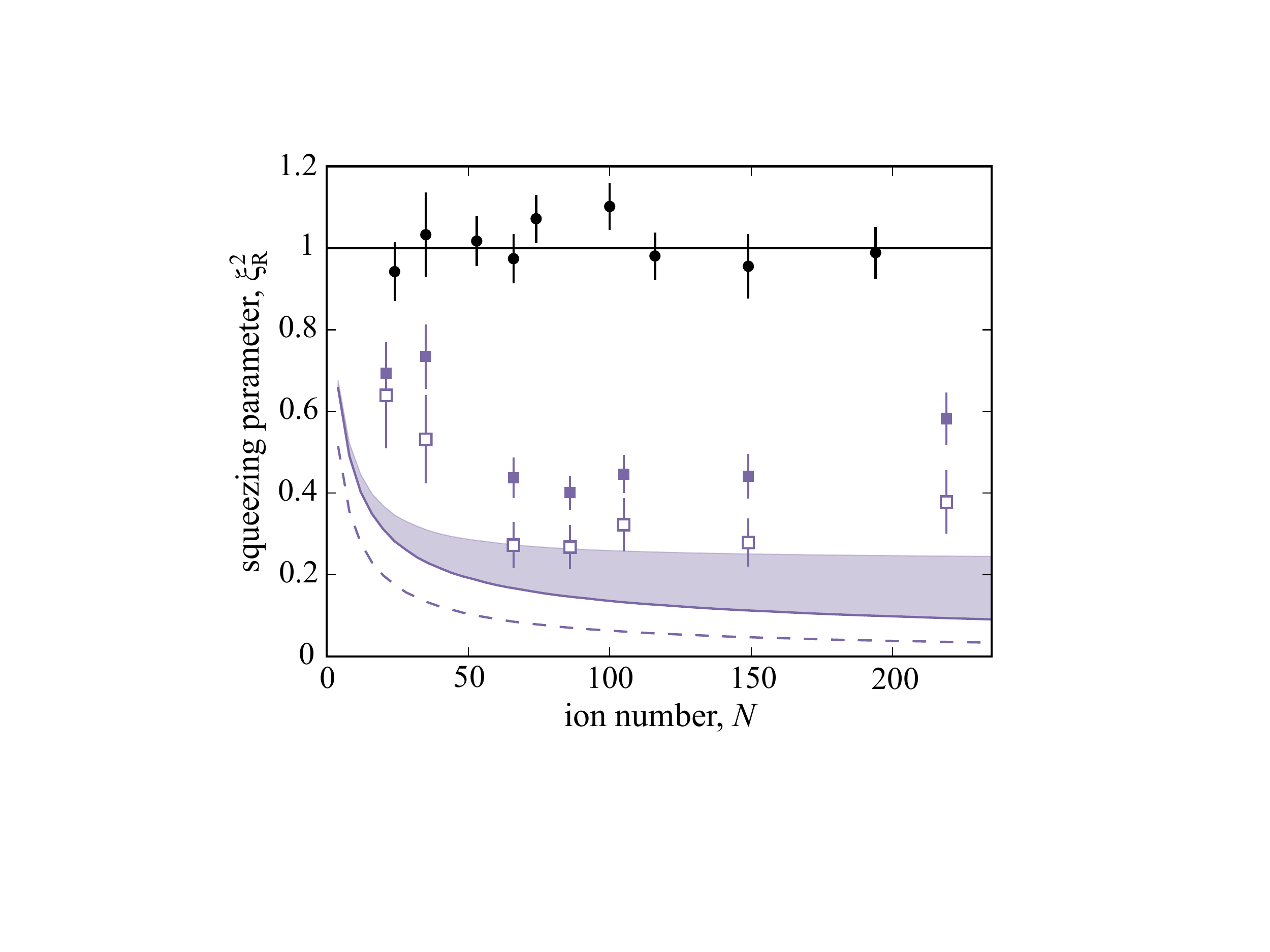}
\end{center}
\caption{{\bf Spin squeezing of hundreds of trapped ions.}
Metrological spin-squeezing parameter as a function of the number of ions.
The black points are obtained for a coherent spin state. 
The purple symbols are obtained with spin-squeezed states (optimized over probe preparation time):
with (open squares) and without (solid squares)
subtraction of photon shot noise. 
The dashed (solid) line is a theoretical prediction (including spontaneous emission).
Taken from \textcite{BohnetSCIENCE2016}.}
\label{Fig.Bohnet}
\end{figure} 

\subsubsection{Quantum metrology with GHZ states of several ions}
\label{GHZions}

The maximally entangled (or ``Schr\"odinger cat'') state 
\be \label{Eq.Ncat}
\ket{\psi_{N, \text{cat}}} = \frac{\ket{a}^{\otimes N} + \ket{b}^{\otimes N}}{\sqrt{2}}
\ee 
has been created with up to $N=6$ ${}^9$Be$^+$ ions \cite{LeibfriedNATURE2006} and 
$N=14$ $^{40}$Ca$^+$ ions \cite{MonzPRL2011} in a linear Paul trap.
This state is formally obtained from $\op{U}\si{cat}\ket{b}^{\otimes N}$, where 
$\op{U}\si{cat} =\exp(-\ii \frac{N\pi}{2}\J_x) \exp(\ii \frac{\pi}{2} \J_x^2)$, see Sec.~\ref{Sec.Atom-Atom.twisting}. 
When applying a spin rotation $\exp(-\ii \ps \J_z)$, the state~\eqref{Eq.Ncat} transforms as
$\ket{\psi_{N,\text{cat}}(\ps)} = [\exp(\ii \ps N/2) \ket{a}^{\otimes N} + \exp(-\ii \ps N/2) \ket{b}^{\otimes N}]/\sqrt{2}$, 
such that the relative phase between $\ket{a}^{\otimes N}$ and $\ket{b}^{\otimes N}$ is $N\ps$. 
The state can be detected after a $\pi/2$ Rabi rotation around the $x$-axis.
The probability to obtain $N_a$ atoms in level $a$ and  $N_b=N-N_a$ atoms in level $b$ is 
$P(N_a | N, \ps) = 2^{-N} \binom{N}{N_a} \left\{1+(-1)^{N_a}\cos\left[N(\ps+\pi/2)\right]\right\}$.
These probabilities oscillate with a frequency $N$ times faster than that of a single qubit.
Therefore, most observable properties of these states,
such as the occurrence of $N$ atoms in level $a$ or $b$, or collective properties such as the parity of the relative number of particles 
in the output state \cite{BollingerPRA1996}, feature this rapid oscillation.
Notably, applying the nonlinear transformation $\op{U}\si{cat}$ again after phase encoding, 
one obtains only two detection events ($N$ atoms in level $a$ or $b$) that can be experimentally distinguished 
with high efficiency \cite{LeibfriedSCIENCE2004, LeibfriedNATURE2006}. 
Experiments exploring quantum metrology with states~\eqref{Eq.Ncat} 
generally report dichotomic events---the parity of the relative atom number, in \textcite{LeibfriedSCIENCE2004, MonzPRL2011}, 
or the detection of all atoms in the same level in \textcite{LeibfriedSCIENCE2004, LeibfriedNATURE2006}.
Figure~\ref{FigGHZ}(a) shows an example of parity oscillations, 
$P(\pm | \ps) = \frac{1 \pm V \cos N\ps}{2}$, 
where the visibility $0\leq V\leq 1$ accounts for experimental noise and imperfect state preparation.
The corresponding the Fisher information,
$F(\ps) =  \tfrac{1}{P(+ | \ps)} \big( \tfrac{d P(+ | \ps)}{d \ps} \big)^2 + \tfrac{1}{P(- | \ps)} \big( \tfrac{d P(- | \ps)}{d \ps} \big)^2$, is
\be \label{Eq.FV}
F(\ps) = \frac{V^2 N^2 \sin^2 N\ps}{1 - V^2 \cos^2 N\ps}.
\ee
It reaches the maximum value $F =V^2N^2$ when $\sin N \ps=\pm1$.
It is thus possible to detected useful entanglement and obtain sensitivities below the standard quantum limit when $V^2 > 1/N$ 
(corresponding to $F>N$). 
Metrologically-useful maximally entangled states are detected when $V^2 > ( 1 - \tfrac{1}{N} )^2 +\tfrac{1}{N^2}$ 
[corresponding to $F>(N-1)^2 +1$, obtained from Eq.~\eqref{Eq.FQIk} with $k=N-1$].
Figure \ref{FigGHZ}(b) gives an overview of the experimental results obtained with cat states of trapped ions.  
In \textcite{MonzPRL2011} the fringe visibility of parity oscillations was sufficiently high that the Fisher information witnesses 
metrologically useful genuine $N$-particle entanglement up to $N=6$ ions \cite{PezzePNAS2016}.  

Note that the fast oscillations of the output probabilities obtained with the state~\eqref{Eq.Ncat}
seem to prevent the unambiguous estimate of a phase larger than $2\pi/N$ \cite{BerryPRA2009, PezzeEPL2007, BotoPRL2000, MitchellSPIE2005}.
A phase estimation at the Heisenberg limit with no constraint on $\ps$ in $[-\pi, \pi]$ can be obtained 
using states~\eqref{Eq.Ncat} with different numbers of particles.
The optimal sequence $N=1,2,4,\dotsc,2^p$ provides $\Delta \ps = 2.55/N\si{T}$ (at $\ps=0$),
where $N\si{T}=\sum_{n=0}^p 2^n = 2^{p+1}-1$ is the total number of particles used \cite{PezzeEPL2007}.

It is finally worth pointing out that W states (that, ideally, are less useful than cat states for metrological applications)
have been created experimentally with up to eight ions \cite{RoosSCIENCE2004, HaffnerNATURE2005}.

\subsubsection{Spin squeezing and useful entanglement of many ions}

Recently, entanglement and metrological spin-squeezing of hundreds of trapped $^9$Be$^+$ ions have been demonstrated
in a Penning trap \cite{BohnetSCIENCE2016}.
The experiment realized a homogeneous Ising spin-spin interaction \cite{KitagawaPRA1993} (\ie, $\alpha \approx 0$ in Eq.~\ref{Eq.UIsing})
corresponding to the one-axis twisting model,
see Sec.~\ref{Sec.Atom-Atom.twisting}.
Figure~\ref{Fig.Bohnet} shows $\xiR^2$, optimized over time, as a function of the number of ions
ranging from $N=20$ to $220$. 
For long interaction times, the state wraps around the Bloch sphere and spin squeezing is lost (see Sec.~\ref{Sec.Atom-Atom.twisting}).
In this case, the experimental data of \textcite{BohnetSCIENCE2016} still show $\Fish/N>2.1$ signaling useful entanglement, 
where the Fisher information has been extracted following the Hellinger method of \textcite{StrobelSCIENCE2014}.
For an analysis of decoherence in this system see \textcite{Foss-FeigPRA2013}.

\section{Working entanglement-enhanced interferometers and prospects for applications}
\label{Sec.Working-Entanglement}

Atomic ensembles are routinely used as precision sensors of inertial forces or external fields, or as atomic clocks, 
which usually operate as Ramsey interferometers \cite{WynandsMETROLOGIA2005, WynandsLectNotesPhys2009, CroninRMP2009, KitchingIEEESensors2011, BordeMETROLOGIA2002, BarrettPS2016}.
The accumulated phase is $\ps = \omega T\si{R}$, where $T\si{R}$ is the Ramsey interrogation time (not the overall measurement cycle time) and 
$\omega$ is the parameter to be estimated. 
In an atomic sensor, $ \omega = \Delta E/\hbar$, where  $\Delta E$ is the energy shift between the two interferometer modes,
induced by an external field, an acceleration, a rotation, or a force.
In an atomic clock, $\omega = \omega\si{LO}-\omega_0$, where $\omega\si{LO}$ is the instantaneous frequency of a local oscillator to be locked to the transition frequency $\omega_0$ between the two atomic levels that define the clock. 
The standard quantum limit for the estimation of the frequency $\omega$ is, from Eq.~\eqref{SNbound},
\be \label{Eq.Dw}
\Delta \omega\si{SQL} = \frac{1}{T\si{R} \sqrt{N \nu}}
=\frac{1}{T\si{R} \sqrt{N}} \times \sqrt{\frac{T\si{cycle}}{\tau}},
\ee
where we have expressed the number of measurements $\nu = \tau/T\si{cycle}$ in terms of the total averaging time $\tau$ and the cycle time $T\si{cycle}$ of the experiment, with $T\si{cycle} \ge T\si{R}$.
Using separable states, if quantum noise dominates over technical noise and $T\si{cycle} \approx T\si{R}$ is optimized, the sensitivity can thus only be increased by increasing 
the interrogation time $T\si{R}$, the number $N$ of particles, and/or the total averaging time $\tau$.  
In realistic setups this is not always possible: every sensor has an optimal working range for $T\si{R}$ and $N$, and 
drifts always limit the duration $\tau$ over which identical state preparation can be assumed \cite{BakPRL1987}.  
In general, increasing the atom number leads to systematic errors due to collisions that spoil the sensitivity.
Ramsey interrogation times are limited by mechanical restrictions (\eg, the limited ballistic flight time in fountain clocks),
by the desired temporal resolution (\eg, when measuring high-frequency magnetic fields in broadband magnetometers), or decoherence 
(\eg, the phase coherence time of the local oscillator in atomic clocks).
Can entangled states improve the performance of atomic sensors and, if so, under what conditions?

Entangled states are typically more susceptible to decoherence than separable states during phase encoding.
Thus their gain (\ie, their more favorable scaling of phase sensitivity with $N$) 
is often counterbalanced by the requirement of shorter interrogation times \cite{HuelgaPRL1997}. 
Whether or not an improvement is possible can only be decided after a detailed analysis of a specific sensor 
taking its actual limitations into account, and ultimately hinges on an experimental demonstration.
As an example we note that squeezing of light has recently been used to improve the performance of 
a laser-interferometer gravitational wave observatory \cite{Schnabel2011,Aasi2013}.
The aim of this section is to review the state of the art of entanglement-enhanced measurements with 
atomic ensembles beyond proof-of-concept demonstrations, and to discuss prospects for applications.

\subsection{The influence of noise and decoherence}
\label{Sec.Working-Entanglement.decoherence}

A common and central aspect of quantum-enhanced metrology is the fragility of useful 
entangled states when coupled to the environment. 
Noise affects the preparation of the probe state, phase encoding, and detection. 
The literature has focused mainly on specific noise models and only few general results are available. 
An important challenge is to devise protocols (\eg, differential schemes, or error correction), 
special phase sensing situations (\eg, non-Markovian, or transverse noise) and detection schemes 
that make metrology below the standard quantum limit robust against noise.

\subsubsection{General results}

Phase encoding in the presence of noise can be formally described by a 
$\ps$-dependent completely-positive trace-preserving map $\Lambda_\ps$ \cite{HuelgaBOOK}, 
such that $\rhops = \Lambda_\ps[\op{\rho}_0]$, 
where $\op{\rho}_0$ and $\rhops$ are the probe and output state, respectively. 
The evolution from $\op{\rho}_0$ to $\rhops$ can be conveniently studied by introducing additional degrees of freedom
that are not under control and play the role of an ``environment''.
The action of any quantum channel $ \Lambda_\ps$ can be described via its Kraus representation
$\rhops = \sum_\ell \op{K}_\ell \op{\rho}_0 \op{K}_\ell^\dag$, where 
$\op{K}_\ell \equiv \bra{\ell\si{E}} \op{U}\si{SE}(\ps) \ket{0\si{E}}$ are Kraus operators, 
$\ket{\ell\si{E}}$ is an orthonormal and complete set of states ($\sum_\ell \ket{\ell\si{E}} \bra{\ell\si{E}} = \1\si{E}$ guarantees 
$\sum_\ell \op{K}_\ell  \op{K}_\ell^\dag=1$), and
$\ket{0\si{E}}$ is the initial state  of the environment (taken pure, without loss of generality).
We recall that a Krauss representation is not unique:
we can generate a new set by a unitary 
(eventually $\ps$-dependent) transformation \cite{NielsenBOOK}.

In the enlarged Hilbert space $\mathcal{H}\si{S} \otimes \mathcal{H}\si{E}$, given by the product of the Hilbert space of the system and of the environment, 
phase encoding is modeled by a unitary operation $\op{U}\si{SE}(\ps)$.
Decoherence is taken into account by tracing over the environment:
$\rhops = \Tr\si{E}[ \ket{ \psi\se{SE}_\theta } \bra{ \psi\se{SE}_\theta } ]$, 
where $\ket{ \psi\se{SE}_\theta } = \op{U}\si{SE}(\ps) \ket{\psi\se{SE}_0}$ is a purification of $\rhops$ and
$\ket{\psi\se{SE}_0}$ a purification of the probe state $\op{\rho}_0 =  \Tr\si{E}[ \ket{ \psi\se{SE}_0 } \bra{ \psi\se{SE}_0 } ]$.
We recall that two purifications $\ket{ \psi\se{SE}_\ps }$ and $\ket{ \tilde{\psi}\se{SE}_\ps }$
of the same state are related via unitary operations acting only on the environment space \cite{NielsenBOOK}, 
$\ket{ \psi\se{SE}_\theta } = \op{U}\si{E}(\ps) \ket{ \tilde{\psi}\se{SE}_\ps }$.
Discarding part of the information regarding the unitary phase encoding in $\mathcal{H}\si{S} \otimes \mathcal{H}\si{E}$
motivates the inequality 
\be \label{FKineq}
F\si{Q}[\rhops] \leq F\si{Q}\big[\ket{ \psi\se{SE}_\theta }\big],
\ee
valid for all purifications $\ket{\psi\se{SE}_\ps}$ of $\rhops$, where
\be \label{FKvar}
F\si{Q}\big[\ket{\psi\se{SE}_\ps}\big] = 4
\Big( \scp{\partial_\theta \psi\se{SE}_\ps}{\partial_\ps \psi\se{SE}_\ps} - \abs{\scp{\partial_\ps \psi\se{SE}_\ps}{\psi\se{SE}_\ps}}^2 \Big).
\ee
Equation~\eqref{FKineq} suggests to search for the minimum of $F\si{Q}[\ket{ \psi\se{SE}_\theta } ]$ over all purifications of $\rhops$
\cite{EscherNATPHYS2011, EscherPRL2012}.
Following Uhlmann's theorem \cite{UhlmannRMP1976}, \textcite{EscherNATPHYS2011} have indeed shown that 
\be \label{FKeq}
F\si{Q}[\rhops] = \min_{\{ \ket{ \psi\se{SE}_\ps } \} } F\si{Q}\big[ \ket{ \psi\se{SE}_\theta } \big],
\ee
see also \textcite{FujiwaraJPA2008}.
In other words, there is always an environment that---when monitored together with the system---does not lead to more information 
about $\ps$ than monitoring the system alone.
In particular, for an initial pure state $\op{\rho}_0 = \ket{\psi_0} \bra{\psi_0}$, 
Eq.~\eqref{FKeq} can be rewritten as \cite{EscherNATPHYS2011}
\be \label{FisherKraus}
\Fish\si{Q}[\rhops] = 4 \min_{\{ \op{K}_\ell \}}  \sum_\ell \mean{ (\partial_\theta \op{K}^\dag_\ell ) (\partial_\theta \op{K}_\ell ) } - 
\Big(\ii \sum_\ell \mean{ \op{K}_\ell^\dag \partial_\theta \op{K}_\ell} \Big)^2,
\ee
where the mean values are calculated over $\ket{\psi_0}$ and the 
minimization runs over all Kraus representations of the quantum channel.
It is worth pointing out that, even if the optimal purification in Eq.~\eqref{FKeq}, or the optimal set of Kraus operators in Eq.~\eqref{FisherKraus}, 
are difficult to find, specific choices may yields non-trivial upper bounds to the quantum Fisher information \cite{EscherNATPHYS2011}.

A special case of the above formalism is that of a $\ps$-independent quantum channel $\Lambda$ where noise acts after phase encoding.
In this case the quantum Fisher information
never increases \cite{PetzJPA2002, FujiwaraPRA2001}: 
\be \label{Eq.QFI.CPTP}
\Fish\si{Q}\big[\Lambda ( \op{\rho}_\ps)  \big] \leq \Fish\si{Q}\big[\op{\rho}_\ps \big].
\ee
For example, the partial trace operation can only decrease the quantum Fisher information, 
consistent with the intuition that ignoring part of a system can only 
decrease the information about the estimated parameter.
The equality in Eq.~\eqref{Eq.QFI.CPTP} is always obtained if $\Lambda$ is unitary, because
unitary $\ps$-independent transformations can be absorbed into a redefinition of the optimal measurement
saturating the quantum Fisher information.  

\subsubsection{Uncorrelated decoherence}
A typical example of the fragility of frequency estimation with respect to noise has been
discussed by \textcite{HuelgaPRL1997}. 
Let us consider a single atom prepared in the superposition $(\ket{a} + \ket{b})/\sqrt{2}$ that evolves freely for a time $T\si{R}$, according to the 
Hamiltonian $\omega \op{\sigma}_z$, and acquires a phase $\omega T\si{R}$ in the presence of local Markovian dephasing. 
The probability to find the probe in its initial state is 
$P=[1+e^{-\gamma T\si{R}} \cos (\omega T\si{R})]/2$, where $\gamma\geq 0$ is the dephasing rate. 
If the phase estimation is repeated $\nu$ times in parallel (using a total of $N=\nu$ particles), 
a calculation of the Cram\'er-Rao bound gives $\Delta \omega = \sqrt{2 \gamma e/(N T\si{R})} 2\gamma\sqrt{e/N}$ at the optimal Ramsey time $T\si{R} = 1/(2\gamma)$, with a total measurement time of $\tau=T\si{R}=1/(2 \gamma)$.
If we take a GHZ state $(\ket{a}^{\otimes N} + \ket{b}^{\otimes N})/\sqrt{2}$ in the presence of the same source of (single-particle) dephasing, 
the probability to find the probe in the initial state is 
$P = [1+e^{-\gamma N T\si{R}} \cos (\omega N T\si{R})]/2$: it oscillates $N$ times faster than in the single-particle case
(this is typical for GHZ states, see Sec.~\ref{sec:NOONstate}), but the visibility of these oscillations decays $N$ times quicker. 
A calculation of the Cram\'er-Rao bound gives $\Delta \omega = \sqrt{2 \gamma e/(N T\si{R})}=2\gamma\sqrt{e}$, 
reached already at the optimal time $T\si{R} = 1/(2 N \gamma)$.
Repeating this experiment $N$ times in the same total measurement time $\tau=1/(2 \gamma)$ (neglecting experimental dead-time, $T\si{cycle}\approx T\si{R}$) allows for a reduction of the measurement uncertainty to $\Delta \omega = 2\gamma\sqrt{e/N}$.
In this example, the ideal sensitivity enhancement offered by GHZ states for $\gamma=0$
disappears in the presence of arbitrarily small uncorrelated dephasing ($\gamma>0$).  
Although the overall absolute sensitivity is not improved, the GHZ state reaches the sensitivity limit $N$ with times faster $T\si{R}$ with respect to the uncorrelated atoms:
this may be of practical interest when experimental constraints require $T\si{R} \ll 1/\gamma$ \cite{HuelgaPRL1997, ShajiPRA2007}.
An optimization over probe states shows that at most a constant factor of $1/\sqrt{e}$
in the absolute frequency error can be gained in the presence of single-particle dephasing \cite{HuelgaPRL1997, EscherNATPHYS2011, Ulam-OrgikhPRA2001}.

Bounds to the quantum Fisher information in presence of more general uncorrelated decoherence have been discussed in the literature.
Let us consider $N$ particles prepared in a state $\op{\rho}_0$ and a noisy channel $\Lambda_\ps$
that acts independently on each particle, such that $\op{\rho}_\ps = \Lambda_\ps^{\otimes N} ( \op{\rho}_0)$. 
The maximum of the quantum Fisher information over all possible probe states is bounded as \cite{FujiwaraJPA2008}
\be \label{Fujibound}
\max_{\op{\rho}_0} \Fish\si{Q}\big[ \Lambda_\ps^{\otimes N} ( \op{\rho}_0) \big] 
\leq 4 \min_{\{ \op{K}_\ell \}} \big\{ N \norm{\alpha_{K}} + N(N-1) \norm{\beta_{K}}^2 \big\},
\ee
where $\alpha_{K} = \sum_\ell (\partial_\ps \op{K}_\ell^\dag) (\partial_\ps \op{K}_\ell)$,
$\beta_K = \ii \sum_\ell (\partial_\ps \op{K}_\ell^\dag) \op{K}_\ell$, 
$\norm{\cdot}$ denotes the operator norm, and the minimization runs over all 
equivalent Kraus representations of the channel $\Lambda_\ps$.
In particular, if there exists a Kraus representation such that $\beta_K =0$, 
then the second term in Eq.~\eqref{Fujibound} vanishes and 
$\Fish\si{Q} [ \Lambda_\ps^{\otimes N} ( \op{\rho}_0) ]$ has---asymptotically in $N$---a bound that scales linearly with $N$.
Relevant single-particle quantum channels fulfill the condition\footnote{A notable class of channels that fulfill $\beta_K=0$ is that of \emph{quantum simulable} channels
\cite{KolodinskyNJP2013}, \ie, those that can be written as 
$\Lambda_\ps(\op{\rho}) = \Phi(\op{\rho} \otimes \op{\sigma}_\ps)$, where $\Phi$ is a $\ps$-independent channel
\cite{MatsumotoArXiv2010}.
$\Phi$ acts on an enlarged space including 
the auxiliary state $\op{\sigma}_\ps$ that contain all information about the parameter.
We have
$\Fish\si{Q}\big[\Lambda_\ps^{\otimes N} ( \op{\rho})  \big] \leq \Fish\si{Q}\big[\op{\sigma}_\ps^{\otimes N} \big] = N \Fish\si{Q}\big[\op{\sigma}_\ps \big]$,
that follows from Eq.~\eqref{Eq.QFI.CPTP} and the additivity of the quantum Fisher information.
A quantum simulable channel may admit several decompositions, the optimal one being that giving the smallest 
value $\Fish\si{Q}[\op{\sigma}_\ps]$.
When $\op{\sigma}_\ps$ has a diagonal form, $\op{\sigma}_\ps = \sum_i p_{\ps, i} \ket{e_i} \bra{e_i}$, 
where $\ket{e_i}$ is some basis of the enlarged Hilbert space, the channels is said to be \emph{classical simulable} \cite{MatsumotoArXiv2010}. 
In this case we have $\Fish\si{Q}[\op{\sigma}_\ps] = 1 / \epsilon_+ \epsilon_-$ \cite{DemkowiczNATCOMM2012}, 
where $\epsilon_\pm$ can be found from the geometric properties of the convex space of quantum channels \cite{BengtssonBOOK2006}.
In particular, $\epsilon_\pm>0$ for channels that lie into the set of completely-positive trace-preserving maps away from its boundary, 
including full-rank channels.} $\beta_K =0$ \cite{DemkowiczNATCOMM2012,KolodinskyNJP2013,EscherNATPHYS2011}.
This means that, for those channels, the phase sensitivity achievable with an arbitrary 
probe state is $\Delta \ps \geq 1/\sqrt{\m \alpha N}$, with $\alpha = 4 \min_{\{ \op{K}_\ell \}; \beta_K=0} \norm{\alpha_{K}}$.
Therefore, asymptotically in the number of particles, the optimal achievable phase sensitivity has a scaling $N^{-1/2}$ and 
the possible gain over the standard quantum limit is only limited to a prefactor (when $\alpha>1$). 
We notice however that this is an asymptotic result for $N\to\infty$: entangled states may still provide a scaling of phase 
sensitivity better than the standard quantum limit, up to the Heisenberg limit,
even for relatively large $N$.

According to Eq.~\eqref{Fujibound} a necessary condition to overcome the asymptotic scaling $N^{-1/2}$ of precision is to have $\beta_K  \neq 0$.
It has been shown that this condition can be achieved for relatively short interrogation times in the case of 
non-Markovian noise \cite{ChinPRL2012, MatsuzakiPRA2011} and 
also when considering dephasing noise perpendicular to phase encoding evolution \cite{ChavesPRL2013}.
In these cases, quantum-enhanced frequency estimation can be found in the limit $N \to \infty$.
In particular, \textcite{ChinPRL2012, MatsuzakiPRA2011} have shown that a frequency variance $(\Delta \omega)^2 = \mathcal{O} (N^{-3/2})$
can be reached for an interrogation time $T\si{R} = \mathcal{O}(N^{-1/2})$ under general models of non-Markovian
phase noise; see also \textcite{MacieszczakPRA2015, BerradaPRA2013, SzankowskiPRA2014, SmirnePRL2016}.
Moreover, \textcite{ChavesPRL2013} have shown that if Markovian dephasing is acting along a spin direction perpendicular to the phase encoding, 
in the limit $N \to \infty$
a sensitivity $(\Delta \omega)^2 = \mathcal{O}(N^{-5/3})$ can be obtained for interrogation times $T\si{R} = \mathcal{O}(N^{-1/3})$.
The possibility to reach the Heisenberg limit $(\Delta \omega)^2 = \mathcal{O}(N^{-1})$ in this case has been discussed by \textcite{KesslerPRL2014b, DurPRL2014} 
when making use of quantum error correction techniques (see below).
The orientation-dependent lifetime of spin-squeezed states has been investigated experimentally by \textcite{LerouxPRL2010b}.

\subsubsection{Correlated phase noise, differential interferometry, and decoherence-free subspaces}

In many systems, a significant source of noise is correlated dephasing, where all atoms are subject to the same stochastic fluctuation of the phase shift. 
For instance, correlated dephasing is relevant in experiments with ions stored in linear Paul traps \cite{MonzPRL2011, RoosNATURE2006}:
phase fluctuations are caused by noisy stray fields inducing random energy shifts of the atomic levels.
Correlated phase noise is modeled as
\be \label{pnmodel}
\op{\rho}_\ps = \int \dd \varphi P(\varphi | \ps) \op{U}_\varphi \op{\rho}_0 \op{U}_\varphi^\dag, 
\ee
where $P(\varphi | \ps)$ describes the phase fluctuations around $\ps$ and $\op{U}_\varphi$ is the phase-encoding unitary tramsformation.
The probability of a generic detection event $\mu$ is 
\be
P\si{pn}(\mu | \theta) =  \int \dd \varphi P(\mu | \varphi) P(\varphi | \ps),
\ee
where $P(\varphi | \ps) = \bra{\mu} \op{U}_\varphi \op{\rho}_0 \op{U}_\varphi^\dag \ket{\mu}$. 
Inserting $P\si{pn}(\mu | \theta)$ into Eq.~\eqref{Eq.FI} allows to calculate the 
Cram\'er-Rao bound in presence of arbitrary phase noise. 
Taking $P(\varphi | \ps) = e^{-(\varphi - \ps)^2 / (2\sigma\si{pn}^2)} / \sqrt{2 \pi \sigma\si{pn}^2}$ for $\sigma\si{pn} \ll 2 \pi$, 
$\op{U}_\varphi = e^{-\ii \varphi \J_z}$, and considering
states symmetric under particle exchange, we obtain 
\be \label{Eq.rhopn}
 \bra{m_z} \op{\rho}_\ps \ket{n_z} =  \bra{m_z} \op{\rho}_0 \ket{n_z} e^{- \frac{\sigma^2\si{pn}}{2} (m-n)^2} e^{- \ii \ps (m - n)},
\ee
where $\ket{m_z}$ is an eigenstate of $\J_z$ with eigenvalue $m$. 
As a consequence of correlated phase noise, off-diagonal elements $\bra{m} \op{\rho}_\ps \ket{n}$ are exponentially suppressed at a rate 
proportional to $(m-n)^2$.
Equation~\eqref{Eq.rhopn} predicts that the coherence of a $N$-qubit GHZ state decays faster than that of a single qubit by a factor $N^2$.
This effect, also known as super-decoherence, has been demonstrated experimentally
with maximally entangled states of $^{40}$Ca$^+$ ions by adding a variable delay time between creation and coherence investigation \cite{MonzPRL2011}. 
Correlated phase noise is more dramatic than uncorrelated dephasing discussed above.
Indeed, taking the sum of (the smallest possible) quantum and phase noise [see also \textcite{EscherPRL2012, GenoniPRL2011}], we have 
\be \label{pnlimit}
\Delta \ps \geq \frac{1}{\sqrt{\m}}\sqrt{\sigma^2\si{pn} + \frac{1}{N^2}},
\ee
which does not scale with $N$ for $N \gg 1/\sigma\si{pn}$, in contrast with uncorrelated dephasing where $\Delta \ps \geq \alpha/\sqrt{\m N}$.

\begin{figure}[t!]
\begin{center}
\includegraphics[width=\columnwidth]{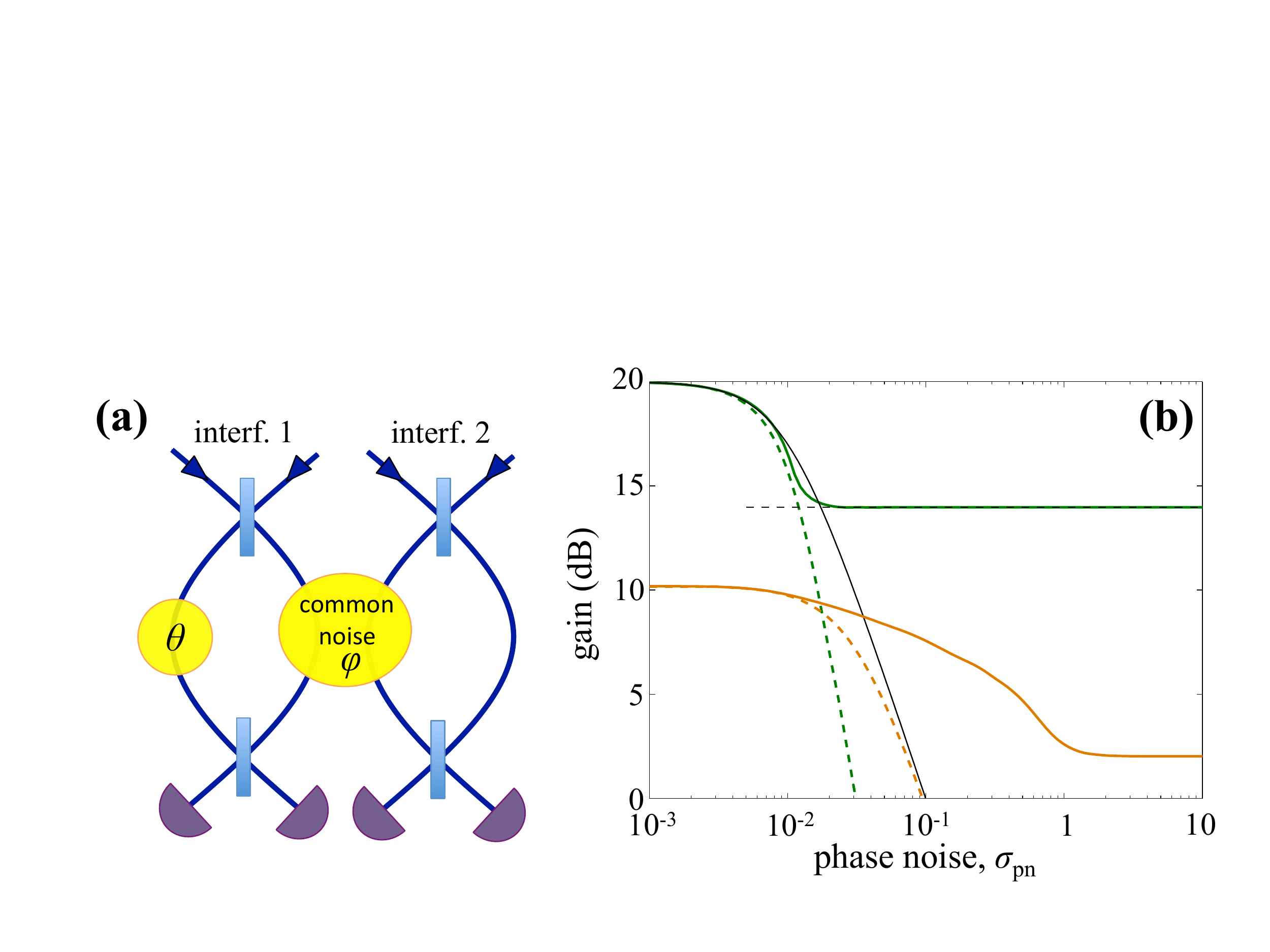}
\end{center}
\caption{{\bf Differential interferometry.} 
(a) Differential interferometry scheme: two Mach-Zehnder interferometers working in parallel are coupled to
the shot-to-shot fluctuating phase noise $\varphi$.
The signal to be estimated is $\ps$.
(b) Normalized Fisher information ($F/N$, gain over the shot noise) as a function of the phase noise, 
modeled according to Eq.~\eqref{pnmodel} with $P(\varphi | \ps) \sim e^{-(\varphi-\ps)^2/(2\sigma\si{pn}^2)}$.
The green (darker) lines refer to NOON states in each interferometer, while the orange (lighter) lines to spin-squeezed states obtained via Kitagawa-Ueda evolution 
$e^{-\ii \chi t \J_z^2} \ket{0,\pi/2}$ of a coherent spin state $\ket{0,\pi/2}$, with $\chi t=0.01\pi$. 
The thick solid lines are obtained with a differential interferometer, 
while the dashed lined with a single interferometer.
The thin black solid line is Eq.~\eqref{pnlimit}.
The horizontal dashed line is $N^2/2$, giving the quantum Fisher information of a differential NOON-state interferometer
in presence of arbitrary large phase noise.
Here $N=100$.
}
\label{Fig:L} 
\end{figure} 

Correlated phase noise can be counteracted by 
a differential interferometer scheme \cite{LandiniNJP2014}, see Fig.~\ref{Fig:L}(a),
that consists of two interferometers running in parallel. 
The phase shift in the first interferometer is $\ps+\varphi_1$, while that in the second interferometer is $\varphi_2$, 
where $\ps$ is the signal phase to be estimated and $\varphi_{1,2}$ are shot-to-shot random phases.
Taking a Gaussian $P(\varphi_{1,2} | \ps)$ as above, 
$\op{U}_\varphi = e^{-\ii \varphi \J_z^{(1)}} \otimes e^{-\ii \varphi \J_z^{(2)}}$
and perfectly correlated interferometers with $\varphi_1=\varphi_2$, we have
\be \label{Eq.rhopndiff}
\bra{\vect{m}_z} \op{\rho}_\ps \ket{\vect{n}_z} =  \bra{\vect{m}_z} \op{\rho}_0 \ket{\vect{n}_z} e^{- \tfrac{\sigma\si{pn}^2}{2} 
 (m_1 + m_2 -n_1 - n_2)^2} e^{- \ii \ps (m_1 - n_1)}, 
\ee
where $\ket{\vect{m}_z} = \ket{m_z^{(1)}} \otimes \ket{m_z^{(2)}}$ and $\ket{m_z^{(1,2)}}$ are eigenstates of 
$\J_z^{(1,2)}$ with eigenvalues $m_{1,2}$.
Components with $m_1+m_2 = n_1+n_2$
are insensitive to phase noise and define a decoherence-free subspace \cite{ZanardiPRL1997, LidarPRL1998}.  
This condition is met by nontrivial states achieving the Heisenberg limit of phase sensitivity \cite{DornerNJP2012, LandiniNJP2014, JeskeNJP2014}.
Figure~\ref{Fig:L}(b) illustrates examples showing the superior performance of a differential scheme (thick solid lines) with respect to a single interferometer (dashed lines)
affected by the phase noise with the same amplitude $\sigma\si{pn}$.
For a differential scheme where each interferometer is fed with NOON states and relative particle number measurement, 
the Fisher information saturates to $F\si{Q} = N^2/4$ for $N \sigma\si{pn} \lesssim 1$ \cite{LandiniNJP2014}.
Finally, noise correlations between the two interferometers are crucial: achieving the Heisenberg limit in a differential interferometer
requires $\Delta(\varphi_1-\varphi_2) \gg 1/N$ \cite{LandiniNJP2014}.
\textcite{RoosNATURE2006} demonstrated experimentally the preparation of a designed entangled state of two ion qubits in a 
decoherence-free subspace and proved its robustness by measuring the electric field quadrupole shift while removing sensitivity to a noisy magnetic field environment.
In \textcite{MonzPRL2011} decoherence-free states with up to 8 ions were constructed, achieving coherence times of $\sim$100, see also
\textcite{PruttivarasinNATURE2015}.
In specific cases, entanglement can be preserved in the presence of correlated dephasing even if the state is not in a decoherence-free subspace \cite{CarnioPRL2015}. 
 
\subsubsection{Error correction}

Quantum error correction techniques are crucial in quantum computing \cite{NielsenBOOK, ShorPRA1995, SteanePRL1996} and can be used to counteract the 
effect of noise in quantum metrology \cite{KesslerPRL2014b, DurPRL2014, ArradPRL2014, OzeriArXiv2013, PlenioPRA2016, HerreraPRL2015}.
These schemes 
correct the imprinting of phase information in a quantum state against decoherence by 
employing ancilla qubits that neither interact with the parameter nor are subject to noise.
Noise operators map the state to ancilla Hilbert subspaces orthogonal to the phase encoding subspace. 
Error correction is accomplished by projective measurements into the orthogonal subspaces, and then applying a correction sequence. 
These techniques can be used to extend the coherence time and/or to achieve the Heisenberg limit for 
certain noise models \cite{KesslerPRL2014b, DurPRL2014, ArradPRL2014, OzeriArXiv2013}.
\textcite{LuNATCOMM2015} have provided conditions under which the quantum Fisher information,
rather than the full quantum state, can be protected under a class of noisy channels. 

Finally, it has been shown that entanglement between the probe and an ancillary system in the preparation and measurement stage can be useful in the presence of uncorrelated noise \cite{DemkowiczPRL2014, HainePRA2015a, HainePRA2015b, ZixinPRA2016}.
  
\subsubsection{Particle losses}

The incoherent loss of particles can strongly impact the usefulness of a state for quantum metrology.
A paradigmatic example is the NOON state: the loss of a single particle transforms Eq.~\eqref{MaxEnt} into the incoherent mixture 
$\tfrac{1}{2} \ket{N-1}_a\bra{N-1} \otimes \ket{0}_b \bra{0}+ \tfrac{1}{2} \ket{0}_a \bra{0} \otimes \ket{N-1}_b\bra{N-1}$, 
which is useless for phase sensing. 
Losses can be modeled by a beam splitter of transmission coefficient 
$0 \leq \eta \leq 1$ that equally couples each interferometer arms to environment modes. 
This leads to the bound of sensitivity
\be \label{lossbound}
\Delta \theta \geq \frac{1}{\sqrt{\m} N} \sqrt{1+ \frac{1-\eta}{\eta} N},
\ee
that is valid for any probe state \cite{EscherNATPHYS2011}: when $N \ll \eta/(1-\eta)$, $\Delta \theta$ is bounded by the Heisenberg limit, 
while for $N \gg \eta/(1-\eta)$ one obtain $\Delta \theta \geq \tfrac{1}{\sqrt{\m N}} \sqrt{\tfrac{1-\eta}{\eta}}$.
In this case, one recovers the $1/\sqrt{N}$ scaling with at best, a prefactor that is $\sqrt{\tfrac{1-\eta}{\eta}} < 1$ for $\eta>1/2$. 

For ultracold atoms, collisional losses are especially relevant. 
One-body losses are due to collisions of trapped atoms with residual hot atoms due to imperfect vacuum, 
whereas two- and three-body losses are caused by inelastic collisions within the trapped cloud and are relevant in dense samples.
In particular, \textcite{LiPRL2008} have studied the impact of atom losses
during the preparation of spin-squeezed states via one-axis twisting (see Sec.~\ref{Sec.Atom-Atom.twisting}) and
have shown that $\xiR^2 = \mathcal{O}(N^{-4/15})$ for one-body losses, assuming $N \to \infty$
and an optimal evolution time. For two-body losses, the optimal $\xiR^2$ does not depend on $N$, 
while for three-body losses $\xiR^2 = \mathcal{O}(N^{4/15})$ and there is a finite optimal number of particles for squeezing.

\subsubsection{Finite detection efficiency}  
\label{sec:finitedetectionefficiency}

Low efficiency in the detection of large atom numbers is one of the main limitations in many current quantum-enhanced metrology experiments.
Finite detection efficiency blurs the interferometer signal and thus degrades its phase sensitivity.
A noisy detector can be modeled as a beam splitter of transmission coefficient $\eta$ followed a perfect detector.
In this case (assuming equal efficiency for both detectors), the bound~\eqref{lossbound} applies.
More generally, we can model an imperfect detection by replacing ideal probabilities $P(\mu | \ps) = \bra{\mu}\op{\rho}_\ps \ket{\mu}$---here $\mu$ is the result of a measurement (we restrict for simplicity to projective measurements 
but the discussion can be straightforwardly extended to generalized measurements, see footnote \ref{POVMfootnote})---with
\be \label{Pdn}
P\si{dn}(\mu | \ps) = \sum_{\tilde{\mu}} P(\mu | \tilde{\mu}) P(\tilde{\mu} | \ps),
\ee
where $P(\mu | \tilde{\mu})$ is a convolution function giving the probability to obtain the result $\mu$ when the ``true'' value is $\tilde{\mu}$.
In practice, one can use a normalized Gaussian
$P(\mu | \tilde{\mu}) = e^{-(\mu - \tilde{\eta} \tilde{\mu})^2/(2\sigma\si{dn}^2)}/\sum_\mu e^{-(\mu - \tilde{\eta} \tilde{\mu})^2/(2\sigma\si{dn}^2)}$, where 
$\sigma\si{dn}$ accounts for the detection noise (independent on the detection signal, for simplicity) and $0 \leq \tilde{\eta} \leq 1$ for the attenuation of the signal.
Finite detection efficiency limits the distinguishability of the probability distribution when changing the parameter and thus decreases the Fisher information, 
see Sec.~\ref{SubSecStatDist}.

\begin{figure}[t!]
\begin{center}
\includegraphics[width=\columnwidth]{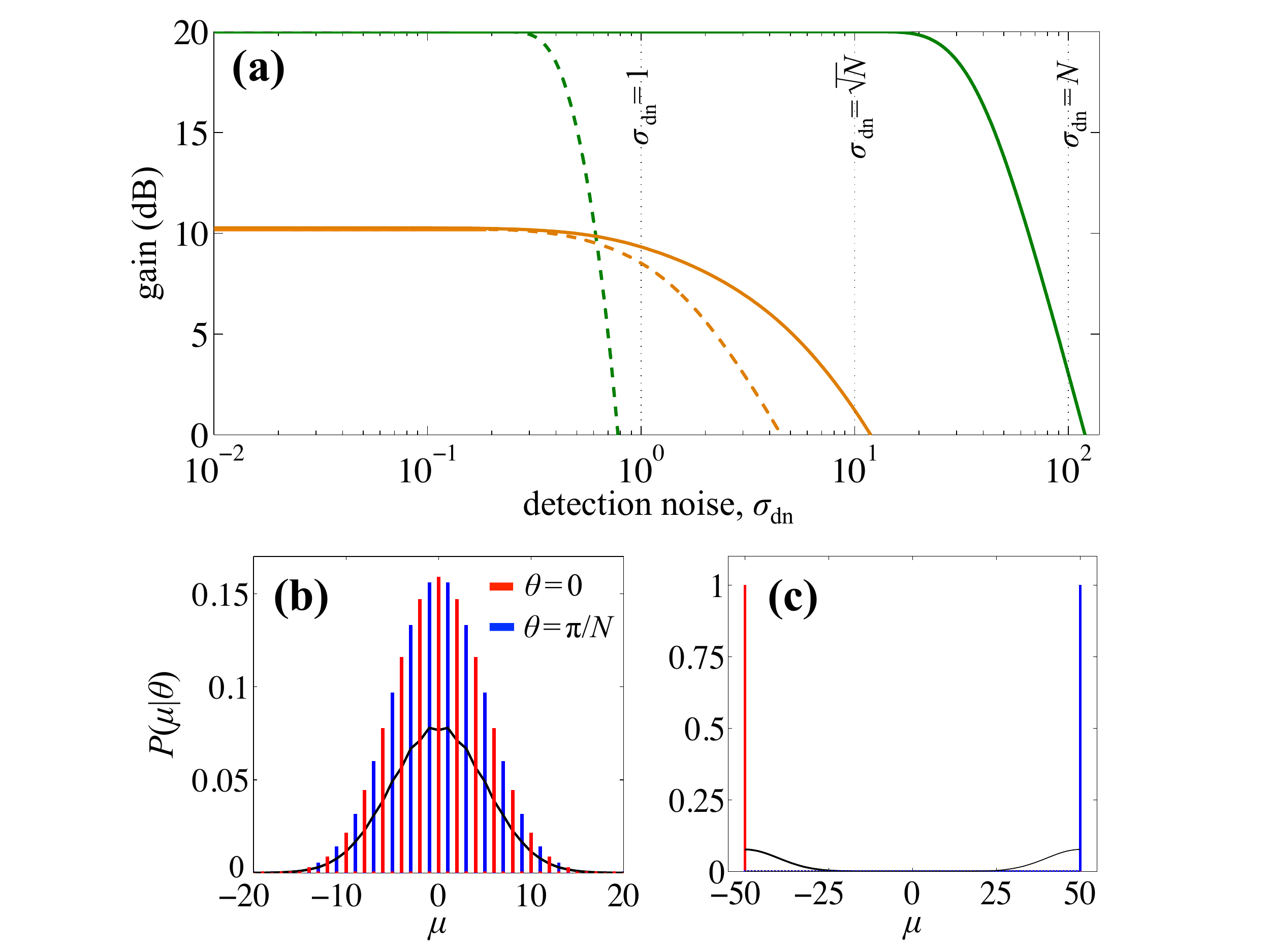}
\end{center}
\caption{{\bf Interaction based readout.} 
(a) Normalized Fisher information ($F/N$, gain over the shot noise) as a function of the detection noise, 
modeled according to Eq.~\eqref{Pdn} with a Gaussian $P(\mu | \tilde{\mu})$ of width $\sigma\si{dn}$ and $\tilde{\eta}=1$.
The green (darker) lines refer to a NOON state, and the orange (lighter) lines to a spin-squeezed state obtained via Kitagawa-Ueda evolution 
$e^{-\ii \tau \J_z^2} \ket{0,\pi/2}$ of a coherent spin state $\ket{0,\pi/2}$, 
with $\tau=0.01\pi$. The solid lines are obtained with an interaction based readout, 
and the dashed lines with a standard particle number detection.
(b) Probability distributions of a NOON state, $P(m | \theta) = \abs{\me{m_y}{e^{-\ii \ps \J_z}}{\text{NOON}}}^2$,
for $\theta=0$ (red/lighter histogram) and $\theta=\pi/N$ (blue/darker histogram). 
Parity oscillations of the probability distributions are washed out by a detection noise of just one atom ($\sigma=1$, black line).
(c) Probability distribution for an interaction based readout, $P(m | \theta) = \abs{\me{m_y}{e^{-\ii \pi/2 \J_x^2} e^{-\ii \ps \J_z}}{\text{NOON}}}^2$.
The distribution for $\theta=0$ [$P(m | \theta) = \delta_{m, -N/2}$, red/lighter histogram] and that for 
$\theta=\pi/N$ [$P(m | \theta) = \delta_{m, N/2}$, blue/darker histogram] are maximally distinguishable and thus robust against detection noise:
the thick and thin black lines are $P\si{dn}(m | \theta)$ for $\sigma\si{dn}=\sqrt{N}$ for $\theta=0$ and $\pi/N$, respectively, 
showing that the two distributions do not overlap even in presence of large detection noise. 
Here $N=100$.}
\label{Fig:nonlinnreadout} 
\end{figure} 

An interaction-based readout can make phase estimation robust against detection noise, 
removing the requirement of single-atom resolution to reach a precision approaching the Heisenberg limit 
\cite{DavisPRL2016, FrowisPRL2016, NolanPRL2017, AndersPRA2018}.
In an interaction-based readout, some nonlinear unitary evolution $\op{U}\si{nl}$ 
(\eg, generated by the interaction between the particles) is applied before the readout measurement, 
such that the probability of a result $\mu$ is given by $P(\mu | \ps) = \bra{\mu} \op{U}\si{nl} \op{\rho}_\ps  \op{U}\si{nl}^\dag \ket{\mu}$.
For instance, $\op{U}\si{nl}$ can be the inverse of the nonlinear transformation that generates the entangled probe state, 
thus realizing an echo sequence \cite{MacriPRA2016, GarttnerNATPHYS2017}.
Notice that SU(1,1) interferometers, see Sec.~\ref{SU11}, take advantage of an interaction-based readout
and their robustness to detection noise has been emphasized by \textcite{MarinoPRA2012, OuPRA2012}.
An example of the advantage offered by the interaction-based readout is illustrated in Fig.~\ref{Fig:nonlinnreadout}(a).
This effect can be understood as a phase magnification \cite{HostenSCIENCE2016}: output probability distributions of phase-shifted states become 
more distinguishable (and thus less prone to detection noise) when applying a nonlinear evolution prior to readout.
As a simple example, following \textcite{LeibfriedNATURE2006}, let us consider phase estimation using the NOON state~\eqref{MaxEnt}.
Applying a phase encoding $\exp(-\ii \ps \J_z)$ followed by a $\pi/2$ Rabi rotation around the $x$-axis, 
one can measure the relative number of particles in $\ket{a}$ and $\ket{b}$.
The phase information is included in fine structure of this probability distribution.
These structures are washed out by the detection noise of just one atom, see Fig.~\ref{Fig:nonlinnreadout}(b).
However, applying  the transformation $\exp(-\ii \frac{N\pi}{2}\J_x) \exp(\ii \frac{\pi}{2} \J_x^2)$ after phase encoding 
(\ie, to $\exp(-\ii \ps \J_z) \ket{\text{NOON}}$) one obtains
$\cos\left(\frac{N\ps}{2}-\frac{\pi}{4}\right)\ket{N,0}+\ii \cos\left(\frac{N\ps}{2}+\frac{\pi}{4}\right)\ket{0,N}$. 
Only two detection events are possible ($N$ atoms in $\ket{a}$ or $\ket{b}$) that can be experimentally distinguished 
even for a large detection noise (\eg, $\sigma\si{dn} \approx \sqrt{N}$), see Sec.~\ref{Fig:nonlinnreadout}(c).

A nonlinear readout has been exploited by \textcite{LeibfriedSCIENCE2004, LeibfriedNATURE2006} for 
quantum-enhanced metrology with GHZ states of $N \lesssim 6$ trapped ions (see Sec.~\ref{Sec.Ions.metrology}), and 
more recently, by \textcite{LinnemannPRL2016} for the realization of a SU(1,1) interferometer with a spinor Bose-Einstein condensate (See Sec.~\ref{SU11}).
In the experiment of \textcite{HostenSCIENCE2016} the collective spin of $N=5 \times 10^5$ $^{87}$Rb atoms is first squeezed via a light-mediated 
interaction in an optical cavity (see Sec.~\ref{Sec.Atom-Light.effint}), and then rotated by an angle $\ps$. 
After a second period of collective spin interactions in the cavity, the state is detected via fluorescence imaging,
demonstrating a phase sensitivity 8\,dB below the standard quantum limit, 
using a detection with a technical noise floor 10\,dB above the projection noise of uncorrelated atoms.

\subsection{Atomic clocks}
\label{Sec.Working-Entanglement.clocks}

Passive atomic clocks operate by locking---via a feedback loop---the frequency of a local oscillator $\omega\si{LO}$ to the 
transition frequency $\omega_0$ between two levels $\ket{a}$ and $\ket{b}$ of an atom
\cite{WynandsMETROLOGIA2005, WynandsLectNotesPhys2009, LudlowRMP2015,KohlhaasPRX2015}.
The standard clock configuration is based on Ramsey  spectroscopy \cite{RamseyBOOK1990}. 
Atoms are initially prepared in the clock state $\ket{a}$.
A near-resonant $\pi/2$ pulse from the local oscillator prepares a superposition of $\ket{a}$ and $\ket{b}$. 
Finally, after a time $T\si{R}$ of free evolution (Ramsey interrogation time), a second $\pi/2$ pulse is applied to the atoms. 
The difference $\omega\si{LO}-\omega_0$ is estimated from a measurement of the atom number in the two levels 
and used for a feedback loop that steers the $\omega\si{LO}$ toward $\omega_0$.
Atomic clocks are characterized by their 
accuracy and stability \cite{VanierBOOK}. 
Accuracy refers to the frequency offset from the ideal value, whereas stability describes the fluctuations of 
the instantaneous frequency $\omega\si{LO}(t)$ from $\omega_0$.
Improving the stability of time-keeping is one of the primary targets of quantum-enhanced metrology. 
It also allows for faster evaluation of systematic errors, which in turn can improve the accuracy \cite{NicholsonNATCOMM2015}.

The standard figure of merit for quantifying the stability of a clock is the Allan standard deviation of the relative frequency fluctuations 
$y(t)=(\omega\si{LO}(t) - \omega_0)/\omega_0$ \cite{VanierBOOK},
\be
\sigma(\tau) = \sqrt{ \frac{\sum_{k=1}^{\nu-1} (y_{k+1} - y_{k})^2}{2(\nu-1)} }, \text{ with } y_k = \int_{t_k}^{t_k+\tau} \frac{\dd t}{\tau} \, y(t),
\ee 
where $y_{k+1} - y_{k}$ is the relative change of the clock frequency between two successive averaging bins of duration $\tau$. 
The Allan deviation quantifies the discrepancy between two consecutive observations of the clock frequency averaged for a time $\tau$,
including dead times in preparation of the sample and data acquisition. 
For an atomic clock operating at the standard quantum limit with uncorrelated (shot-to-shot) measurements,
we have
\be  \label{AllanSQL}
\sigma\si{SQL} (\tau)= \frac{1}{\omega_0 T\si{R} \sqrt{N}} \sqrt{\frac{T\si{cycle}}{\tau}},
\ee
where $T\si{cycle} \geq T\si{R}$ is the clock cycle duration 
($\tau/T\si{cycle}$ being the number of measurements performed in a time $\tau$).  
Currently, the stability of atomic clocks is nearly \cite{NicholsonPRL2012, HinkleySCIENCE2013, BloomNATURE2014} 
or already---as in the case of ion spectroscopy \cite{ItanoPRA1993} and fountain clocks \cite{SantarelliPRL1999}---limited by Eq.~\eqref{AllanSQL}. 
In principle, the clock stability is optimized by extending the interrogation time until atomic decoherence dominates in the interferometer output signal. 
In practice, a main limitations of current clocks---in particular of optical clocks using trapped atoms---are random fluctuations of the local oscillator frequency.  
These limit the stability of frequency comparisons to interrogation times of the order of a second, 
well short of the limits imposed by atomic decoherence, such as excited-state decay \cite{WinelandREP1998}.
 
Theoretical studies of entanglement-enhanced atomic clocks have focused on the impact of 
collective dephasing caused by fluctuations of the local oscillator frequency, 
taking into account temporal noise correlations as well as the feedback 
mechanism that controls the local oscillator frequency.
State and measurement optimization \cite{BuzekPRL1999, AndrePRL2004, RosebandICOLS2011} 
as well as adaptive schemes \cite{KesslerPRL2014, BorregaardPRL2013} have been considered.
\textcite{AndrePRL2004} have shown the possibility of achieving an entanglement-enhanced 
stability $\sigma(\tau)  \propto 1/(N^{2/3}\sqrt{\tau})$ by optimizing over a family of moderately squeezed states. 
More recently, \textcite{BorregaardPRL2013} have shown that using squeezed states and weak output measurements 
it is possible to achieve $\sigma(\tau)  \propto 1/(N \sqrt{\tau})$, thus reaching the Heisenberg scaling.
The adaptive optimization of probe state and output measurement has been studied 
for low atom number by \textcite{RosebandICOLS2011} and \textcite{MullanPRA2014}.
The adaptive use of GHZ states in an atomic clock---in the presence of local oscillator noise only---has been considered in \textcite{KesslerPRL2014}, see also \textcite{LerouxMETROLOGIA2017}.
Overall, these protocols show that entangled states can be can useful to track and stabilize the fluctuations of the local oscillator over long time scales.
Finally, quantum bounds of frequency stability have been discussed by \textcite{ChabudaNJP2016,FraasCMP2016}.

\subsubsection{Entanglement-assisted atomic clocks}

\begin{figure}[t!]
\begin{center}
\includegraphics[width=\columnwidth]{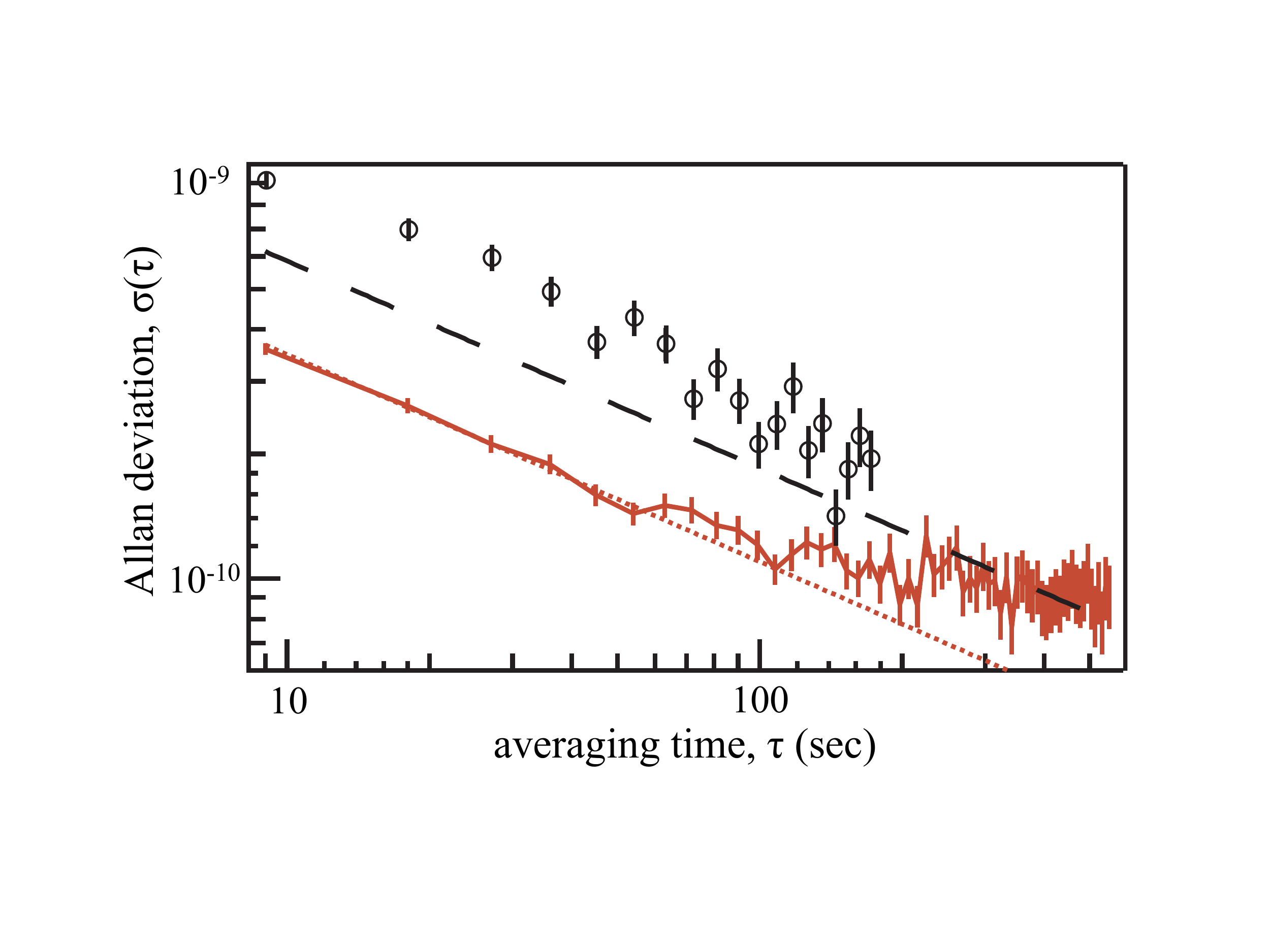}
\end{center}
\caption{{\bf Stability of an atomic clock beating the standard quantum limit.} 
Allan deviation of an atomic clock with a spin-squeezed input state 
(red data with error bars, the solid red line being a guide to the eye), with $T\si{R} = 200\,\mu$s, $N=3.5\times 10^4$ and $T\si{cycle}=9$\,s.
The red dotted line, $\sigma (\tau) = 1.1\times 10^{-9}\,\text{s}^{1/2}/\sqrt{\tau}$ is a factor 
2.8 (corresponding to 4.5 dB) below the standard quantum limit (black dashed line). 
Open black circles: reference measurements with uncorrelated atoms. 
Taken from \textcite{LerouxPRL2010b}.}
\label{Fig:Allan} 
\end{figure}

Atomic clocks with a performance beating the standard quantum limit have been experimentally demonstrated with two trapped ions \cite{MeyerPRL2001}, ensembles of 
cold thermal atoms \cite{LouchetChauvetNJP2010, LerouxPRL2010b, HostenNATURE2016} and 
Bose-Einstein condensates \cite{GrossNATURE2010,Kruse,OckeloenPRL2013}. 
The neutral atom experiments have realized microwave-frequency clocks on magnetic field-insensitive hyperfine transitions with spin-squeezed states as input.\footnote{Note that, to date, squeezing has not yet been realized on an optical transition. \textcite{MonzPRL2011} have created GHZ states on the optical clock transition of $^{40}$Ca$^+$ ions.}
\textcite{LerouxPRL2010b} measured an Allan deviation of $\sigma (\tau) = 1.1\times 10^{-9}\,\text{s}^{1/2}/\sqrt{\tau}$ for their squeezed clock, corresponding to an improvement of 4.5\,dB in variance beyond the standard quantum limit, see Fig.~\ref{Fig:Allan}.
More recently, \textcite{HostenNATURE2016} performed a similar measurement, reaching $\sigma (\tau) = 9.7\times 10^{-11}\,\text{s}^{1/2}/\sqrt{\tau}$, or 10.5\,dB in variance beyond the standard quantum limit.
While these experiments are impressive proof-of-principle demonstrations of entanglement-enhanced atomic clocks, they do not yet reach the frequency stability of state-of-the-art fountain clocks. This is mostly due to a much shorter Ramsey interrogation time of $T\si{R}\sim 200\,\mu$s, limited by the noise of the microwave local oscillator.
For comparison, current fountain clocks using uncorrelated atoms operate with $T\si{R} \sim 1$\,s and reach  
$\sigma (\tau) \sim 10^{-14}\,\text{s}^{1/2}/\sqrt{\tau}$ \cite{WynandsLectNotesPhys2009}.

\subsubsection{Proposals for entanglement-assisted lattice clocks}
 
Optical atomic clocks that interrogate narrow optical transitions in ensembles of atoms trapped in an optical lattice are currently the most precise and accurate measurement devices \cite{LudlowRMP2015, YeSCIENCE2008}, reaching stabilities of $\sigma (\tau) \sim 10^{-16}\,\text{s}^{1/2}/\sqrt{\tau}$, integrating down to the lower $10^{-18}$ level after a few thousand seconds of averaging \cite{NicholsonNATCOMM2015}.
Their stability is within a factor of two above the quantum projection noise limit \cite{LudlowSCIENCE2008,HinkleySCIENCE2013, BloomNATURE2014}. 
Since, in these systems, there is a limit to the exploitable number of atoms, using entanglement to increase the sensitivity is of considerable interest. 
So far, only theoretical studies are available. 
\textcite{MeiserNJP2008} have studied the creation of spin-squeezed states in a neutral-atom optical lattice clock through a cavity-based QND measurement. \textcite{WeinsteinPRA2010} considered the creation of GHZ states through the on-site interaction of an atom moving across the lattice.
 A further interesting possibility is to exploit long-range Rydberg-dressing interactions between atoms at different lattice 
sites induced by laser excitations \cite{BouchoulePRA2002a, HenkelPRL2010, PupilloPRL2010, OpatrnyPRA2012, MohammadsadeghPRA2016} 
[see \textcite{BrowaeysAMOP2016, SaffmanRMP2010} for reviews] 
to generate spin-squeezed \cite{GilPRL2014} and non-Gaussian entangled states \cite{MacriPRA2016}.
 
\subsection{Optical magnetomenters}
\label{Sec.Working-Entanglement.measurements}

Optical magnetometers \cite{BudkerNATPHYS2007, BudkerBOOK2013} exploit the interaction of light with a 
spin-polarized atomic ensemble (either thermal atoms in a vapor cell at room temperature or laser-cooled atom in an optical trap\footnote{Vapor cell magnetometers are currently the most sensitive measuring devices for low-frequency magnetic fields, 
reaching sensitivity levels below $1\,\text{fT}/\sqrt{\text{Hz}}$ with probe volumes of $10^{6}$-$10^{12}\,(\mu\text{m})^3$ and large atom numbers  $N=10^{11}$-$10^{15}$. 
Laser-cooled atoms in an optical trap contain a much smaller number of atoms $N=10^5$-$10^8$ but are more compact and better suited for 
field measurements with high spatial resolution.}) 
to measure the strength of a magnetic field \cite{BudkerRMP2002}. 
Generally, the device consists of $N$ spin-$F$ atoms initially optically pumped into a fully polarized state 
(pointing along the $x$-axis, for instance), 
such that the collective spin of the ensemble has length $\mean{\J_x}=FN$.
A weak magnetic field along the $y$-axis causes the precession of the collective spin in the $x-z$ plane at a rate $g \mu\si{B} B/\hbar$, 
where $g$ is the gyromagnetic ratio, $\mu\si{B}$ the Bohr magneton, and $B$ the magnetic field strength.
The atomic spin precession is measured by observing the polarization rotation of the probe light 
transmitted through the atomic sample. 
The interrogation time is limited by spin-relaxation due 
to collisions of the atoms with the cell walls that enclose the vapor or, for high density gases, due to spin-exchange collisions \cite{BudkerNATPHYS2007}.
The sensitivity of current optical magnetometers is fundamentally limited by quantum noise in the form of
atomic projection noise and photon (polarization) shot-noise.
The back-action of light onto the atoms is a further limiting factor.
Current optical magnetometers are approaching quantum noise limits \cite{KominisNATURE2003, WasilewskiPRL2010}.

Polarization squeezing of the probe light \cite{WolfgrammPRL2010, HorromPRA2012} or 
spin squeezing of the atomic ensemble \cite{AuzinshPRL2004, GeremiaPRL2003, PetersenPRA2005, SewellPRL2012} 
can be used to reduce the quantum noise. 
Spin-squeezing induced by continuous quantum nondemolition measurements \cite{AuzinshPRL2004} can enhance the 
sensitivity, and a scaling $N^{-3/4}$ is predicted when using shot-noise-limited light. 
This can be further pushed to the Heisenberg limit $1/N$ if squeezed light is used. 
However, single-atom spin-relaxation modifies this picture, and a reduction of quantum noise below the standard quantum limit is 
expected only for interrogation times shorter than the spin coherence time \cite{AuzinshPRL2004}.  
Spin-squeezed states can thus increase the measurement bandwidth 
(\ie, they are useful for short interrogation times) without loss of sensitivity,
with important applications in biology and medicine. 
For long interrogation times, entanglement-enhanced sensitivity using spin squeezed states is expected for high-density ensembles 
due to the suppression of spin-relaxation noise \cite{KominisPRL2008, VasilakisPRL2011}.  
 
\begin{figure}[t!]
\begin{center}
\includegraphics[width=\columnwidth]{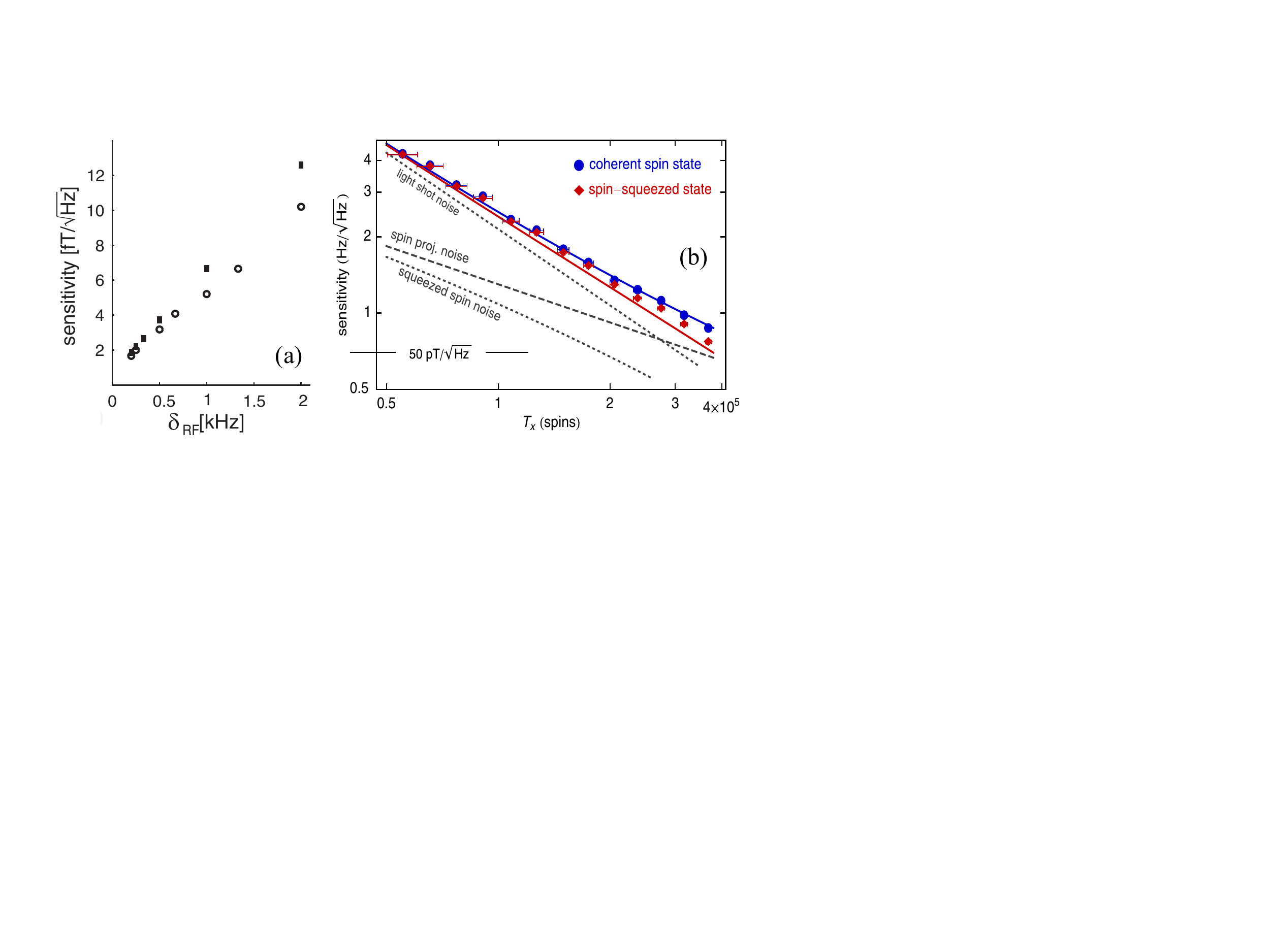}
\end{center}
\caption{{\bf Entanglement-assisted optical magnetometry.}
(a) Measurement sensitivity obtained in absence (squares) and in presence (open circles) of entanglement between atomic ensembles in a two-cell
radio-frequency magnetometer.
An improvement is demonstrated for short radio-frequency pulse durations $\tau$, where $\delta\si{RF} = 1/\tau$. 
Taken from \textcite{WasilewskiPRL2010}.
(b) Measurement sensitivity as a function of the number of atoms with a coherent spin state (blue circles) and spin-squeezed (red diamonds) probe.\
Solid lines are expected results while dashed lines are different contributions to the sensitivity.
Taken from \textcite{SewellPRL2012}.}
\label{Fig:Magn}
\end{figure}  
  
\subsubsection{Entanglement-assisted optical magnetometers}
  
An improvement in magnetic field sensing via the reduction of the optical polarization 
noise has been demonstrated by \textcite{WolfgrammPRL2010, HorromPRA2012} using squeezed probe light.
Atomic projection noise limited and entanglement-assisted magnetometry has been demonstrated in the experiment of 
\textcite{WasilewskiPRL2010}, who realized a pulsed RF magnetometer made of two atomic ensembles 
(two vapor cells with opposite atomic polarizations) with $N=3.6 \times 10^{12}$ Cs atoms. 
A first light pulse is used to entangle the two ensembles via a quantum nondemolition measurement, as first demonstrated by \textcite{JulsgaardNATURE2001}.
This is a useful resource to improve the sensitivity of the magnetometer for large bandwidths, as shown in Fig.~\ref{Fig:Magn}(a).
The improvement is observed for interrogation times shorter than the entanglement lifetime of 4\,ms.
The increase of measurement bandwidth in DC magnetometers has been shown by \textcite{ShahPRL2010} using continuous 
quantum nondemolition measurements.
\textcite{SewellPRL2012} generated spin-squeezed states in ensembles of cold spin-1 $^{87}$Rb atoms via 
quantum nondemolition measurements \cite{KoschorreckPRL2010, KoschorreckPRL2010a} 
and applied them to optical magnetometry. 
Results corresponding to interrogation pulses of 5\,$\mu$s (giving a measurement bandwidth of 200\,kHz) are shown in Fig.~\ref{Fig:Magn}(b).

\subsection{Scanning-probe magnetometers using Bose-Einstein condensates}
\label{Sec.Working-Entanglement.probe}
  
\begin{figure}[t!]
\begin{center}
\includegraphics[width=\columnwidth]{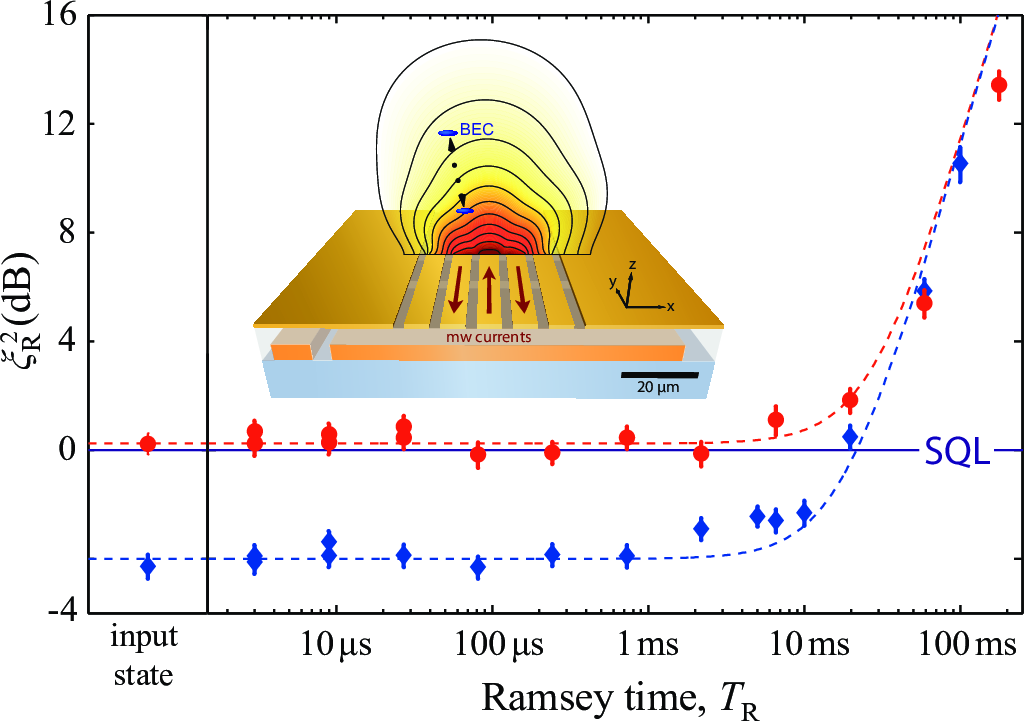}
\end{center}
\caption{{\bf Entanglement-enhanced Bose-Einstein condensate interferometer}. 
Performance of a spin-squeezed Bose-Einstein condensate interferometer expressed in terms of $\xiR^2$ as a function of varying interrogation times $T\si{R}$ for squeezed (blue diamonds) and coherent (red circles) input states. 
Dashed lines model constant performance of $\xiR^2=-4.0$\,dB (squeezed state) and $+0.2$\,dB (coherent state) plus technical noise due to shot-to-shot frequency fluctuations. 
Inset: scanning probe measurements of microwave fields. The Bose-Einstein condensate (blue) is translated near the surface of an atom chip to measure the spatial dependence of microwave magnetic near-fields. Adapted from \textcite{OckeloenPRL2013}.}
\label{Fig:OkeloenPRL2013} 
\end{figure} 
  
Trapped Bose-Einstein condensates are particularly well suited for sensing applications requiring high spatial resolution, 
taking advantage of their small size, high degree of coherence, and the availability of sophisticated techniques for precise positioning of the atoms. 
Current Bose-Einstein condensate magnetometers reach sensitivities of the order of $10\,\text{pT}/\sqrt{\text{Hz}}$ 
with probe volumes of $10^{2}$-$10^{4}\,(\mu\text{m})^3$ \cite{EtoPRA2013, VengalattorePRL2007, WildermuthAPL2006,AignerSCIENCE2008}. 
In these systems, the required small probe size gives rise to an upper bound on the atom number due to density-dependent collisional trap losses. 
It is thus crucial to use entangled states to increase the sensitivity without increasing the atom number. 
  
\textcite{OckeloenPRL2013} have demonstrated a scanning probe atom interferometer operating beyond the standard quantum limit. 
Bose-Einstein condensates of $^{87}$Rb atoms were prepared on an atom chip and spin-squeezed by means of collisions in 
a spin-dependent trapping potential (see Sec.~\ref{Sec.Atom-Atom.twisting}). 
Using the spin-squeezed state as input, a Ramsey interferometer sequence was implemented, achieving a performance of 4\,dB below the 
standard quantum limit with $N=1400$ atoms for interrogation times up to $T\si{R}=10$\,ms, see Fig.~\ref{Fig:OkeloenPRL2013}. 
This interferometer is sensitive to microwave magnetic fields, with a sensitivity of 77\,pT/$\sqrt{\text{Hz}}$ using a small probe volume of $20\,(\mu\text{m})^3$. 
The interferometer was operated as a scanning probe, by translating the Bose-Einstein condensate on the atom chip to 
measure the spatial dependence of an on-chip microwave near-field. 

The magnetometer of \textcite{MuesselPRL2014} consists of $\sim$30 independent Bose-Einstein condensates in a one-dimensional optical lattice each containing about $N=400$ $^{87}$Rb atoms. Spin squeezing is created via one-axis twisting (see Sec.~\ref{Sec.Atom-Atom.twisting}), with interactions tuned by a Feshbach resonance. Subsequently, a microwave $\pi$-pulse transfers the squeezing to a magnetic-field-sensitive hyperfine transition. 
This experiment demonstrated sub-SQL magnetometry (3.4\,dB below the standard quantum limit) 
with interrogation times up to $340\,\mu$s, reaching a sensitivity of 1.9\,nT/$\sqrt{\text{Hz}}$ for static magnetic fields. 
 The array configuration is particularly well suited for differential measurements and magnetic field gradiometry. 
 
Taking the small probe volume into account, the experiments of \textcite{OckeloenPRL2013,MuesselPRL2014} already achieve 
state-of-the-art sensitivity \cite{BudkerBOOK2013}.

\subsection{Nonlocal phase encoding}
\label{Sec.Working-Entanglement.nonlocal}

Usually, the coupling between an atomic ensemble and an external field is local in the particles. 
The phase-encoding transformation of standard (or linear) interferometers is thus modeled by
$\exp( -\ii \ps \sum_{i=1}^N \op{\sigma}_{\vect{n}}^{(i)}/2)$, where $\op{\sigma}^{(i)}_{\vect{n}}$ 
is a single-particle Pauli operator (see Sec.~\ref{Sec.Fundamentals}).
Yet in some cases (see below) the coupling between the atoms and the external field to be measured 
may be associated with a nonlocal Hamiltonian that involves interactions between the particles. 
Nonlinear interferometry has been first proposed by \textcite{LuisPLA2004} and extensively studied 
in the literature \cite{LuisPRA2007, RoyPRL2008, RivasPRL2010, BoixoPRL2007, BoixoPRA2008, ChoiPRA2008}.
In this case, the notions of a standard quantum limit (\ie, the highest phase sensitivity achievable by separable states)
and a Heisenberg limit (\ie, the ultimate allowed phase sensitivity) discussed in Sec.~\ref{Sec.Fundamentals.entanglement} still hold.
However, the scalings of these bounds with the number of particles in the probe state depend on the specific phase-encoding Hamiltonian. 
For instance, \textcite{BoixoPRL2007, ChoiPRA2008, BoixoPRA2008} have considered phase encoding of the kind 
$\exp[ -\ii \ps ( \sum_{i=1}^N \op{\sigma}_{\vect{n}}^{(i)}/2 )^k]$. 
For $N\gg1$, the standard quantum limit and Heisenberg limit become 
\be
\Delta \ps\si{SQL} = \frac{\alpha_k}{N^{k-1/2} \sqrt{\m}} \text{ and } \Delta \ps\si{HL} = \frac{\beta_k}{N^{k} \sqrt{\m}},
\ee
respectively, where $\alpha_k$ and $\beta_k$ are constant prefactors.
For $k > 1$ these bounds have faster scalings of phase sensitivity with the number of particles 
than $\Delta \ps\si{SQL}  = 1/\sqrt{N \m}$ and $\Delta \ps\si{HL}  =  1/(N\sqrt{\m})$, discussed in Sec.~\ref{Sec.Fundamentals} for linear interferometers.
Several systems have been proposed to observe these scalings of phase sensitivity, including 
Kerr nonlinearities \cite{BeltranPRA2005}, collisions in Bose-Einstein condensates \cite{ChoiPRA2008, BoixoPRA2008, ReyPRA2007, BoixoPRA2009}, 
nonlinearities in nano-mechanical resonators \cite{WoolleyNJP2008},
double-pass effective nonlinearities with a cold atomic ensemble \cite{ChasePRA2009}, 
topological excitations in nonlinear systems \cite{NegrettiPRA2008}, and 
atom-photon interactions in cold atom systems \cite{NapolitanoNJP2010}.
Nonlinear interferometers find applications in optical magnetometry \cite{ChasePRA2009, SewellPRX2014} and in the 
measurement of atomic scattering properties \cite{ReyPRA2007}. 

\textcite{NapolitanoNATURE2011} have engineered an effective atom-light Hamiltonian $\op{H}\si{eff} = \alpha \J_z \Sp_z + \beta \J_z \Sp_z N\si{ph}/2$
using an optical pulse passing through a cold atom ensemble of $^{87}$Rb atoms.
Here, $\vectop{J}$ is collective atomic spin, $\vectop{S}$ is the Stokes vector of the light, and $N\si{ph}$ is the photon number.
$\op{H}\si{eff}$ describes a paramagnetic (nonlinear) Faraday rotation of the light beam \cite{NapolitanoNJP2010} 
with rotation angle proportional to $\mean{\op{J}_z}$.
The coefficients $\alpha$ and $\beta$ strongly depend on the detuning of the light beam, and experimental 
conditions for which $\alpha=0$ can be achieved \cite{NapolitanoNJP2010}. 
The second term in $\op{H}\si{eff}$ accounts for nonlinear photon-photon interactions. 
The measurement uncertainty achieved with unentangled photons is $\Delta J_z = 1/(A  N\si{ph}^{1/2}+ B  N\si{ph}^{3/2})$ 
where $A\propto \alpha$ and $B\propto \beta$.
A scaling $\Delta J_z \propto  N\si{ph}^{-3/2}$, obtained by varying the photon number 
between $5 \times 10^5$ and $5 \times 10^7$ has been demonstrated by \textcite{NapolitanoNATURE2011}. 
\textcite{SewellPRX2014}, using a nonlinear Faraday rotation based on alignment-to-orientation conversion \cite{BudkerPRL2000},
have demonstrated $\Delta J_z \propto  N\si{ph}^{-3/2}$,
surpassing the sensitivity achievable by a linear measurement with the same photon number. 

\section{Outlook}
\label{Sec.Outlook}

The possibility to achieve phase sensitivities beyond the standard quantum
limit using atomic ensembles is fueling a vivid and exciting research
activity that focuses on the engineering, characterization and
manipulation of entangled many-body quantum states. On the theoretical
side, a comprehensive and convincing conceptual framework has been
developed for quantum metrology with entangled states, centered around key
concepts such as spin-squeezing and Fisher information. Connections between
the concepts of quantum metrology and other fields of quantum information 
(such as Bell correlations, quantum Zeno dynamics, and Einstein-Podolsky-Rosen entanglement)
can shed new light on quantum technologies. Current active research trends
are centered on quantum-enhanced metrology, taking into account relevant
experimental imperfections and searching for protocols where the fragile
entanglement is protected against noise sources. 

On the experimental side,
many of the key concepts that were proposed during the past decades, such
as spin-squeezing created via particle-particle or atom-light interactions, have
recently been implemented in proof-of-principle experiments. The progress---in terms of sensitivity gain with respect to the standard quantum limit---has been
extremely fast, in particular when compared to squeezing of light. In
atomic ensembles, less than ten years after the first spin-squeezing
experiments, impressive gains up to 20\,dB have been achieved. The field
is now at the verge of moving from proof-of-principle experiments
to technological applications in entanglement-enhanced precision measurements of time,
external fields, and forces.

\section*{Acknowledgments}
We are indebted to our colleagues and collaborators 
with whom we have shared many useful discussions over the past years.
In particular, we would like to thank B.~Allard, M.~Fadel, M.~Fattori, C.~Klempt, W.~Muessel, L.~Santos, H.~Strobel, G.~T\'oth, and T.~Zibold.
R.S. and P.T. acknowledge funding from the Swiss National Science Foundation.
All authors acknowledge financial support by EU-STREP Project QIBEC, No.\ FP7-ICT-2011-C.

\bibliography{QuantumMetrology}

\end{document}